\documentclass[a4paper,oneside]{book}
\usepackage{graphics,setspace,bm,amsmath,amsfonts}
\usepackage{natbib}

\newenvironment{quoting}{\begin{quote}\singlespacing}{\end{quote}}

\newcommand{\ket}[2]{|#1\rangle _{#2}}
\newcommand{\bra}[2]{\langle _{#1}#2|}
\newcommand{\up}{\!\uparrow}
\newcommand{\down}{\!\downarrow}

\newtheorem{proposition}{Proposition}

\newtheorem{theorem}{Theorem}
\newtheorem{definition}{Definition}

\begin{document}


\frontmatter

\begin{titlepage}
\begin{center}
{\Huge Quantum Information Theory and The Foundations of Quantum Mechanics}

\vspace{2\baselineskip}

{\LARGE Christopher Gordon Timpson\\
The Queen's College}

\vspace{8\baselineskip}

\begin{figure}[h]
	\begin{center}
		\scalebox{1.5}{	
			\includegraphics*{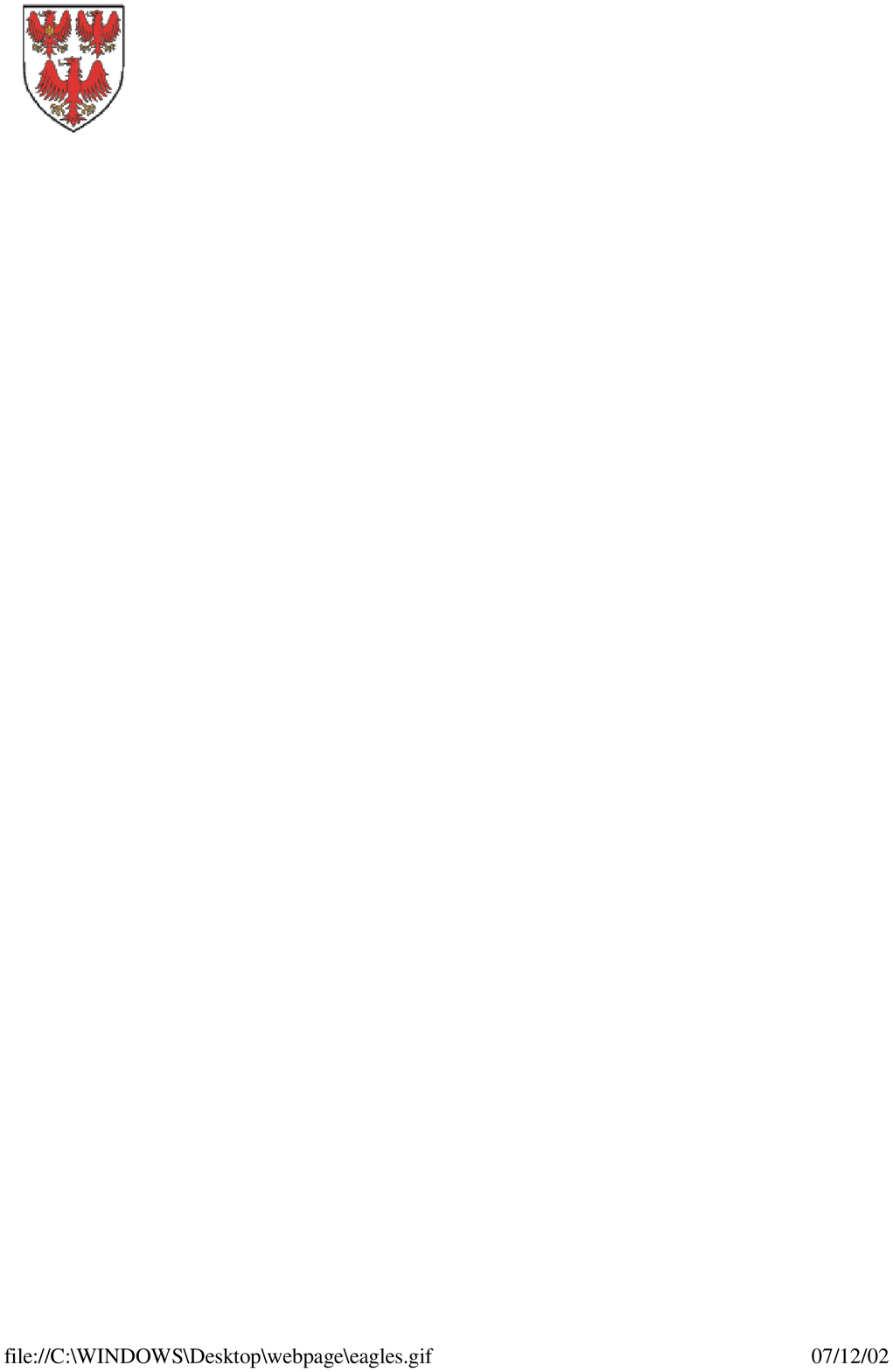}}
	\end{center}
\end{figure}

\vspace{6\baselineskip}

{\large A thesis submitted for the degree of Doctor of Philosophy\\ at the University of Oxford}

\vspace{1\baselineskip}

{\large Trinity Term 2004}

\end{center}
\end{titlepage}


\begin{titlepage}\enlargethispage*{1000pt}

\begin{center}\vspace*{-.7in}
{\Large Quantum Information Theory and the Foundations of Quantum Mechanics\\}
\vspace{1\baselineskip}
{\large Christopher Gordon Timpson, The Queen's College\\
Oxford University, Trinity Term 2004\\}
\vspace{1\baselineskip}
{\Large Abstract of Thesis Submitted for the Degree of Doctor of
Philosophy}
\vspace{1\baselineskip}
\end{center}

\begin{onehalfspace}
This thesis is a contribution to the debate on the implications of quantum information theory for the foundational problems of quantum mechanics.

In Part I an attempt is made to shed some light on the nature of information and quantum information theory.
It is emphasized that the everyday notion of information is to be firmly distinguished from the technical notions arising in information theory; however it is maintained that in both settings `information' functions as an abstract noun, hence does not refer to a particular or substance. The popular claim `Information is Physical' is assessed and it is argued that this proposition faces a destructive dilemma. Accordingly, the slogan may not be understood as an ontological claim, but at best, as a methodological one. A novel argument is provided against Dretske's (1981) attempt to base a semantic notion of information on ideas from information theory.

The function of various measures of information content for quantum systems is explored and the applicability of the Shannon information in the quantum context maintained against the challenge of Brukner and Zeilinger (2001). The phenomenon of quantum teleportation is then explored as a case study serving to emphasize the value of recognising the logical status of `information' as an abstract noun: it is argued that the conceptual puzzles often associated with this phenomenon result from the familiar error of hypostatizing an abstract noun.

The approach of Deutsch and Hayden (2000) to the questions of locality and information flow in entangled quantum systems is assessed.  
It is suggested that the approach suffers from an equivocation between a conservative and an ontological reading; and the differing implications of each is examined. Some results are presented on the characterization of entanglement in the Deutsch-Hayden formalism.

Part I closes with a discussion of some philosophical aspects of quantum computation. In particular, it is argued against Deutsch that the Church-Turing hypothesis is not underwritten by a physical principle, the Turing Principle. Some general morals are drawn concerning the nature of quantum information theory.

In Part II, attention turns to the question of the implications of quantum information theory for our understanding of the meaning of the quantum formalism. Following some preliminary remarks, two particular information-theoretic approaches to the foundations of quantum mechanics are assessed in detail. It is argued that Zeilinger's (1999) Foundational Principle is unsuccessful as a foundational principle for quantum mechanics. The information-theoretic characterization theorem of Clifton, Bub and Halvorson (2003) is assessed more favourably, but the generality of the approach is questioned and it is argued that the implications of the theorem for the traditional foundational problems in quantum mechanics remains obscure.

\end{onehalfspace}
\end{titlepage}

\begin{titlepage}
\begin{center}
{\Large Acknowledgements}
\end{center}
\pagestyle{empty}
\begin{onehalfspace}
It is my pleasant duty to thank a large number of people, and more than one institution, for the various forms of help, encouragement and support that they have provided during the time I have been working on this thesis.

The UK Arts and Humanities Research Board kindly supported my research with a postgraduate studentship for the two years of my BPhil degree and a subsequent two years of doctoral research. I should also like to thank the Provost and Fellows of The Queen's College, Oxford for the many years of support that the College has provided, both material and otherwise. \textit{Reginae erunt nutrices tuae}: no truer words might be said. A number of libraries have figured strongly during the time I have been at Oxford: I would like in particular to thank the staff at the Queen's and Philosophy Faculty libraries for their help over the years.

On a more personal note, I would like to extend my thanks and appreciation to my supervisor Harvey Brown, whose good example over the years has helped shape my approach to foundational questions in physics and who has taught me much of what I know. I look forward to having the opportunity in the future to continue working with, and learning from, him.

Another large debt of thanks is due to John Hyman, my earliest teacher in philosophy, who has continued to offer a great deal of assistance and encouragement over the years; and whose fearsome questioning helped show me what it is to do philosophy (and, incidentally, alerted me to the dangers of pernicious theorising).  

Jon Barrett and I started out on the quest to understand the foundations and philosophy of physics at the same time, just about a decade ago, now. Since then, we have shared much camaraderie and many conversations, several of which have found their way into this thesis at one point or another. And Jon is still good enough to check my reasoning and offer expert advice.

I would like to thank Jeremy Butterfield, Jeff Bub, Chris Fuchs and Antony Valentini, all of whom have been greatly encouraging and who have offered useful comments on and discussion of my work. In particular, I should single out Jos Uffink for his unstinting help in sharing his expertise in quantum mechanics, uncertainty and probability; and for providing me with a copy of his unpublished PhD dissertation on measures of uncertainty and the uncertainty principle. My understanding of measures of information has been heavily influenced by Jos's work.
 
The (rest of the) Oxford philosophy of physics mob are also due a great big thank-you: one couldn't hope for a more stimulating intellectual environment to work in. So thanks especially to Katharine Brading, Guido Bacciagaluppi, Peter Morgan, Justin Pniower, Oliver Pooley, Simon Saunders and David Wallace for much fun, support and discussion (occasionally of the late-night variety).   

\newpage
A little further afield, I would like to thank Marcus Appleby, Ari Duwell, Doreen Fraser, Hans Halvorson, Michael Hall, Leah Henderson, Clare Hewitt-Horsman (in particular on the topic of Chapter 5), Richard Jozsa, James Ladyman, Owen Maroney, Michael Seevink, Mauricio Suarez, Rob Spekkens and Alastair Rae, amongst others, for stimulating conversations on information theory, quantum mechanics and physics.       

Finally I should like to thank my parents, Mary and Chris Timpson, \textit{sine qua non}, bien s\^{u}r; and my wife Jane for all her loving support, and her inordinate patience during the somewhat extended temporal interval over which this thesis was finally run to ground. (Oh, and she made most of the pictures too!)
\end{onehalfspace}
\end{titlepage}


\tableofcontents
\vspace*{-.3in}

\chapter{Introduction}


\begin{doublespacing}



Much is currently made of the concept of information in physics, following the rapid growth of the fields of quantum information theory and quantum computation. These are new and exciting fields of physics whose interests for those concerned with the foundations and conceptual status of quantum mechanics are manifold. On the experimental side, the focus on the ability to manipulate and control individual quantum systems, both for computational and cryptographic purposes, has led not only to detailed realisation of many of the \textit{gedanken}-experiments familiar from foundational discussions (see e.g. \citet{zeilinger:review}), but also to wholly new demonstrations of the oddity of the quantum world \citep{rome,inns,caltech}. Developments on the theoretical side are no less important and interesting. Concentration on the possible ways of using the distinctively quantum mechanical properties of systems for the purposes of carrying and processing information has led to considerable deepening of our understanding of quantum theory. The study of the phenomenon of entanglement, for example, has come on in leaps and bounds under the aegis of quantum information (see e.g. \cite{bruss} for a review of recent developments).

The excitement surrounding these fields is not solely due to the advances in the physics, however. It is due also to the seductive power of some more overtly philosophical (indeed, controversial) theses. There is a feeling that the advent of quantum information theory heralds a new way of doing physics and supports the view that information should play a more central r\^ole in our world picture. In its extreme form, the thought is that information is perhaps the fundamental category from which all else flows (a view with obvious affinities to idealism)\footnote{Consider, for example, Wheeler's infamous `It from Bit' proposal, the idea that every physical thing (every `it') derives its existence from the answer to yes-no questions posed by measuring devices:
`No element in the description of physics shows itself as closer to primordial than the elementary quantum phenomenon...in brief, the elementary act of observer participancy...
It from bit symbolizes the idea that every item of the physical world has at bottom---at a very deep bottom, in most instances---an immaterial source and explanation; that which we call reality arises in the last analysis from the posing of yes-no questions that are the registering of equipment evoked responses; in short that all things physical are information-theoretic in origin and this is a \textit{participatory universe}.' \citep[p.3,5]{wheeler}}, and that the new task of physics is to discover and describe how this information evolves, manifests itself and can be manipulated. Less extravagantly, we have the ubiquitous, but baffling, claim that `Information is Physical' \citep{landauer:1996} and the widespread hope that quantum information theory will have something to tell us about the still vexed questions of the interpretation of quantum mechanics. 

These claims are ripe for philosophical analysis. To begin with, it seems that the seductiveness of such thoughts appears to stem, at least in part, from a confusion between two senses of the term `information' which must be distinguished: `information' as a technical term which can have a legitimate place in a purely physical language, and the everyday concept of information associated with knowledge, language and meaning, which is completely distinct and about which, I shall suggest, physics has nothing to say. The claim that information is physical is baffling, because the everyday concept of information is reliant on that of a person who might read or understand it, encode or decode it, and makes sense only within a framework of language and language users; yet it is by no means clear that such a setting may be reduced to purely physical terms; while the mere claim that some physically defined quantity (information in the technical sense) is physical would seem of little interest. 
The conviction that quantum information theory will have something to tell us about the interpretation of quantum mechanics seems natural when we consider that the measurement problem is in many ways the central interpretive problem in quantum mechanics and that measurement is a transfer of information, an attempt to gain knowledge. But this seeming naturalness only rests on a confusion between the two meanings of `information'.

My aim in this thesis is to clarify some of the issues raised here. In Part~I, I attempt to shed some light on the question of the nature of information and quantum information theory, emphasising in particular the distinction between the technical and non-technical notions of information; in Part~II, I turn to consider, in light of the preceding discussion, the question of what r\^ole the concept of information, and quantum information theory in particular, might have to play in the foundations of quantum mechanics. What foundational implications might quantum information theory have? 

In Chapter~\ref{concepts of info} I begin by describing some features of the everyday notion of information and indicate the lines of distinction from the technical notion of information deriving from the work of \citet{shannon}; I also highlight the important point that `information' is an abstract noun. Some of the distinctive ideas of quantum information theory are then introduced, before I turn to consider the dilemma that faces the slogan `Information is Physical'. The claim that the everyday and information-theoretic notions of information are to be kept distinct is defended against the view of \citet{dretske:1981}, who sought to base a semantic notion of information on Shannon's theory. I present a novel argument against Dretske's position.

One of the more prominent proposals that seeks to establish a link between information and the foundations of quantum mechanics is due to \citet{foundationalprinciple}, who puts forward an information-theoretic foundational principle for quantum mechanics. As a part of this project, \citet{conceptualinadequacy} have criticised Shannon's measure of information, the quantity fundamental to the discussion of information in both classical and quantum information theory. I address these arguments in Chapter~\ref{supposed conceptual inadequacy} and show their worries to be groundless. \textit{En passant} the function of various notions of information content and total information content for quantum systems, including measures of mixedness, is investigated.

Chapter~\ref{study} is a case study whose purpose is to illustrate the value of recognising clearly the logico-grammatical status of the term `information' as an abstract noun: in this chapter I investigate the phenomenon of quantum teleportation. While teleportation is a straightforward consequence of the formalism of non-relativistic quantum mechanics, it has nonetheless given rise to a good deal of conceptual puzzlement. I illustrate how these puzzles generally arise from neglecting the fact that `information' is an abstract noun. When one recognises that `the information' does not refer to a particular or to some sort of pseudo-substance, any puzzles are quickly dispelled. One should not be seeking, in an information-theoretic protocol---quantum or otherwise---for some particular `the information', whose path one is to follow, but rather concentrating on the physical processes by which the information is transmitted, that is, by which the end result of the protocol is brought about. When we bear this in mind for teleportation, we see that the only remaining source for dispute over the protocol is the quotidian one regarding what interpretation of quantum mechanics one wishes to adopt. 

Chapter~\ref{dh} continues some of the themes from the preceding chapter. In it I discuss the important paper of \citet{dh}, which would appear to have significant implications for the nature and location of quantum information: Deutsch and Hayden claim to have provided an account of quantum mechanics which is particularly local, and which finally clarifies the nature of information flow in entangled quantum systems. I provide a perspicuous description of their formalism and assess these claims. It proves essential to distinguish, as Deutsch and Hayden do not, between two ways of interpreting their formalism. On the first, conservative, interpretation, no benefits with respect to locality accrue that are not already available on either an Everettian or a statistical interpretation; and the conclusions regarding information flow are equivocal. The second, ontological interpretation, offers a framework with the novel feature that global properties of quantum systems are reduced to local ones; but no conclusions follow concerning information flow in more standard quantum mechanics.

In Chapter~\ref{entdh} I investigate the characterization of bi-partite entanglement in the Deutsch-Hayden formalism. The case of pure state entanglement is, as one would expect, straightforward; more interesting is mixed state entanglement. The Horodecki's positive partial transpose condition \citep{horodeckisPLA:1996} provides necessary and sufficient conditions in this case for $2\otimes 2$ and $2 \otimes 3$ dimensional systems, but it remains an interesting question how their condition may be understood in the geometrical setting of the Deutsch-Hayden formalism. I provide some sufficient conditions for mixed state entanglement which may be formulated in a simple geometrical way and provide some concrete illustrations of how the partial transpose operation can be seen to function from the point of view of the Deutsch-Hayden formalism.   

Chapter~\ref{comp} is a discussion of some of the philosophical questions raised by the theory of quantum computation. First I consider whether the possibility of exponential speed-up in quantum computation provides an argument for a more substantive notion of quantum information than I have previously allowed, concluding in the negative, before moving on to consider some questions regarding the status of the Church-Turing hypothesis in the light of quantum computation. In particular, I argue against Deutsch's claim that a physical principle, the Turing Principle, underlies the Church-Turing hypothesis; and consider briefly the question of whether the Church-Turing hypothesis might serve as a constraint on the laws of physics.

Chapter~\ref{morals} brings together some morals from Part~I. 

Part~II begins with Chapter~\ref{talk} wherein I outline some preliminary considerations that are pertinent when assessing approaches to the foundational questions in quantum mechanics that appeal to information. One point noted is that if all that appeal to information were to signify in a given approach is the advocacy of an instrumentalist view, then we are not left with a very interesting, or at least, not a very distinctive, position. 

The most prominent lines of research engaged in bringing out implications of quantum information theory for the foundations of quantum mechanics have been concerned with establishing whether information-theoretic ideas might finally provide a perspicuous conceptual basis for quantum mechanics, perhaps by suggesting an axiomatisation of the theory that lays our interminable worrying to rest. That one might hope to make progress in this direction is a thought that has been advocated persuasively by \citet{fuchs:paulian}, for example.   
In the final chapter, I investigate some proposals in this vein, in particular, Zeilinger's Foundational Principle and the information-theoretic characterization theorem of Clifton, Bub and Halvorson \citep{cbh}. I show that Zeilinger's Foundational Principle (`\textit{An elementary system represents the truth value of one proposition}') does not in fact provide a foundational principle for quatum mechanics and fails to underwrite explanations of the irreducible randomness of quantum measurement and the existence of entanglement, as Zeilinger had hoped. The assessment of the theorem of Clifton, Bub and Halvorson is more positive: here indeed an axiomatisation of quantum mechanics has been achieved. However, I raise some questions concerning the $C^{*}$-algebraic starting point of the theorem and argue that it remains obscure what implications for the standard interpretational questions of quantum mechanics this axiomatisation might have. \vspace*{-.3in}

\end{doublespacing}




\mainmatter
\pagestyle{headings}


\part{What is Information?}


\begin{quoting}
To suppose that, whenever we use a singular substantive, we are, or ought to be, using it to refer to something, is an ancient, but no longer a respectable, error.\\ \citet{strawson}
\end{quoting}


\chapter{Concepts of Information}\label{concepts of info}










\begin{doublespacing}

\section{How to talk about information: Some simple ways}

The epigraph to this Part is drawn from Strawson's contribution to his famous 1950 symposium with Austin on truth. Austin's point of departure in that symposium provides also a suitable point of departure for us, concerned as we are with information. 

Austin's aim was to de-mystify the concept of truth, and make it amenable to discussion, by pointing to the fact that `truth' is an abstract noun. So too is `information'. This fact will be of recurrent interest in the first part of this thesis.


`` `What is truth?' said jesting Pilate, and would not stay for an answer." Said Austin: ``Pilate was in advance of his time." 

As with truth, so with\footnote{Due apologies to Austin.} information:

\begin{quoting}
For `truth' \texttt{[`information']} itself is an abstract noun, a camel, that is of a logical construction,
which cannot get past the eye even of a grammarian.

We approach it cap and categories in hand: we ask ourselves whether Truth \texttt{[Information]} is a substance (the Truth \texttt{[the information]}, the Body of Knowledge),
or a quality (something like the colour red, inhering in truths \texttt{[in messages]}),
or a relation (`correspondence' \texttt{[`correlation']}).

But philosophers should take something more nearly their own size to strain at. 
What needs discussing rather is the use, or certain uses, of the word `true' \texttt{[`inform']}. \citep[p.149]{austin:1950}
\end{quoting}  

A characteristic feature of abstract nouns is that they do not serve to denote kinds of entities having a location in space and time. An abstract noun may be either a count noun (a noun which may combine with the indefinite article and form a plural) or a mass noun (one which may not). `Information' is an abstract mass noun, so may usefully be contrasted with a \textit{concrete} mass noun such as `water'; and with an abstract \textit{count} noun such as `number'\footnote{An illuminating discussion of mass, count and abstract nouns may be found in \citet[\S\S 27-29]{rundle:grammar}.}. Very often, abstract nouns arise as nominalizations of various adjectival or verbal forms, for reasons of grammatical convenience. Accordingly, their function may be explained in terms of the conceptually simpler adjectives or verbs from which they derive; thus Austin leads us from the substantive `truth' to the adjective `true'. Similarly, `information' is to be explained in terms of the verb `inform'. Information, we might say, is what is provided when somebody is informed of something. If this is to be a useful pronouncement, we should be able to explain what it is to inform somebody without appeal to phrases like `to convey information', but this is easily done. To inform someone is to bring them to know something (that they did not already know).  

Now, I shall not be seeking to present a comprehensive overview of the different uses of the terms `information' or `inform', nor to exhibit the feel for philosophically charged nuance of an Austin. It will suffice for our purposes merely to focus on some of the broadest features of the concept, or rather, concepts, of information.

The first and most important of these features to note is the distinction between the everyday concept of information and technical notions of information, such as that deriving from the work of \citet{shannon}. The everyday concept of information is closely associated with the concepts of knowledge, language and meaning; and it seems, furthermore, to be reliant in its central application on the the prior concept of a person (or, more broadly, language user) who might, for example, read and understand the information; who might use it; who might encode or decode it. 

By contrast, a technical notion of information is specified using a purely mathematical and physical vocabulary and, \textit{prima facie}, will have at most limited and deriviative links to semantic and epistemic concepts\footnote{For discussion of Dretske's opposing view, however, see below, Section~\ref{dretske}.}.

A technical notion of information might be concerned with describing correlations and the statistical features of signals, as in communication theory with the Shannon concept, or it might be concerned with statistical inference \citep[e.g.][]{fisher,kullback:leibler,savage, kullback}. Again, a technical notion of information might be introduced to capture certain abstract notions of structure, such as complexity (algorithmic information, \citet{chaitin,kolmogorov:1965,solomonoff}) or functional r\^ole (as in biological information perhaps, cf. \citet{jablonka} for example\footnote{N.B. To my mind, however, Jablonka overstates the analogies between the technical notion she introduces and the everyday concept.}).

In this thesis our concern is information theory, quantum and classical, so we shall concentrate on the best known technical concept of information, the Shannon information, along with some closely related concepts from classical and quantum information theory. The technical concepts of these other flavours I mention merely to set to one side\footnote{Although it will be no surprise that one will often find the same sorts of ideas and mathematical expressions cropping up in the context of communication theory as in statistical inference, for example. There are also links between algorithmic information and the Shannon information: the average algorithmic entropy of a thermodynamic ensemble has the same value as the Shannon entropy of the ensemble \citep{bennett:1982}.}.

With information in the everyday sense, a characteristic use of the term is in phrases of the form: `information \textit{about} $p$', where $p$ might be some object, event, or topic; or in phrases of the form: `information \textit{that} $q$'. Such phrases display what is often called \textit{intentionality}. They are directed towards, or are about something (which something may, or may not, be present). The feature of intentionality is notoriously resistant to subsumption into the bare physical order.

As I have said, information in the everyday sense is intimately linked to the concept of knowledge. Concerning information we can distinguish between possessing information, which is to have knowledge; acquiring information, which is to gain knowledge; and containing information, which is sometimes the same as containing knowledge\footnote{Containing information and containing knowledge are not always the same: we might, for example say that a train timetable contains information, but not knowledge.}. Acquiring information is coming to possess it; and as well as being acquired by asking, reading or overhearing, for example, we may acquire information via perception. If something is said to contain information then this is because it provides, or may be used to provide, knowledge. As we shall presently see, there are at least two importantly distinct ways in which this may be so. 
    
It is primarily a person of whom it can be said that they possess information, whilst it is objects like books, filing cabinets and computers that contain information \citep[cf.][]{hacker:1987}. In the sense in which my books contain information and knowledge, I do not. To contain information in this sense is to be used to store information, expressed in the form of propositions\footnote{Or perhaps expressed pictorially, also.}, or in the case of computers, encoded in such a way that the facts, figures and so on may be decoded and read as desired. 

On a plausible account of the nature of knowledge originating with Wittgenstein \citep[e.g.][\S 150]{wittgenstein:investigations} and \citet{ryle:concept}, and developed, for example by \citet{white:1982}, \citet{kenny:1989}  and \citet{hyman:1999}, to have knowledge is to possesses a certain capacity or ability, rather than to be in some state. On this view, the difference between possessing information and containing information can be further elaborated in terms of a category distinction: to possess information is to have a certain ability, while for something to contain information is for it to be in a certain state (to possess certain occurrent categorical properties). We shall not, however, pursue this interesting line of analysis further here (see \citet[p.108]{kenny:1989} and \citet[\S 2.1]{timpson:bphil} for discussion).   

In general, the grounds on which we would say that something contains information, and the senses in which it may be said that information is contained, are rather various. One important distinction that must be drawn is between containing information \textit{propositionally} and containing information \textit{inferentially}. If something contains information propositionally, then it does so in virtue of a close tie to the expression of propositions. For example, the propositions may be written down, as in books, or on the papers \textit{in} the filing cabinet. Or the propositions might be otherwise recorded; perhaps encoded, on computers, or on removable disks. The objects said to contain the information in these examples  are the books, the filing cabinet, the computers, the disks.

That these objects can be said to contain information \textit{about} things, derives from the fact that the sentences and symbols inscribed or encoded, possess meaning and hence themselves can be about, or directed towards something. Sentences and symbols, in turn, possess meaning in virtue of their r\^ole within a framework of language and language users.

If an object $A$ contains information about $B$\footnote{Which might be another object, or perhaps an event, or state of affairs.} in the second sense, however, that is, \textit{inferentially}, then $A$ contains information about $B$ because there exist correlations between them that would allow inferences about $B$ from knowledge of $A$. (A prime example would be the thickness of the rings in a tree trunk providing information about the severity of past winters.) Here it is the possibility of our \textit{use} of $A$, as part of an inference providing knowledge, that provides the notion of information \textit{about}\footnote{Such inferences may become habitual and in that sense, automatic and un-reflected upon.}. And note that the concept of knowledge is functioning prior to the concept of containing information: as I have said, the concept of information is to be explained in terms of the provision of knowledge.

It is with the notion of containing information, perhaps, that the closest links between the everyday notion of information and ideas from communication theory are to be found. The technical concepts introduced by Shannon may be very helpful in describing and quantifying any correlations that exist between $A$ and $B$. But note that describing and quantifying correlations does not provide us with a concept of why $A$ may contain information (inferentially) about $B$, in the everyday sense. 
Information theory can describe the facts about the existence and the type of correlations; but to explain \textit{why} $A$ contains information inferentially about $B$ (if it does), we need to refer to facts at a different level of description, one that involves the concept of knowledge. A further statement is required, to the effect that: `Because of these correlations, we can learn something about $B$'.
Faced with a bare statement: `Such and such correlations exist', we do not have an explanation of why there is any link to information. It is because correlations may sometimes be used as part of an inference providing knowledge, that we may begin to talk about containing information. 



While I have distinguished possessing information (having knowledge) from containing information, there does exist a very strong temptation to try to explain the former in terms of the latter. However, caution is required here. We have many metaphors that suggest us filing away facts and information in our heads, brains and minds; but these \textit{are} metaphors. If we think the possession of information is to be explained by our containing information, then this cannot be `containing' in the straightforward sense in which books and filing cabinets contain information (propositionally), for our brains and minds do not contain statements written down, nor even encoded. As we have noted, books, computers, and so on contain information about various topics because they are used by humans (language users) to store information. As Hacker remarks:
\begin{quoting}
...we do not \textit{use} brains as we use computers. Indeed it makes no more sense to talk of storing information in the brain than it does to talk of having dictionaries or filing cards in the brain as opposed to having them in a bookcase or filing cabinet. \citep[p.493]{hacker:1987}
\end{quoting} 
We do not stand to our brains as an external agent to an object of which we may make use to record or encode propositions, or on which to inscribe sentences. 

A particular danger that one faces if tempted to explain possessing information in terms of containing it, is of falling prey to the \textit{homunculus fallacy} \citep[cf.][]{kenny:1971}.

The homunculus fallacy is to take predicates whose normal application is to complete human beings (or animals) and apply them to parts of animals, typically to brains, or indeed to any insufficiently human-like object. 
The fallacy properly so-called is attempting to argue from the fact that a person-predicate applies to a person to the conclusion that it applies to his brain or \textit{vice versa}. This form of argument is non-truth-preserving as it ignores the fact that the term in question must have a different meaning if it is to be applied in these different contexts. 

`Homunculus' means `miniature man', from the Latin (the diminutive of \textit{homo}). This is an appropriate name for the fallacy, for in its most transparent form it is tantamount to saying that there is a little man in our heads who sees, hears, thinks and so on. Because if, for example, we were to try to explain the fact that a person sees by saying that images are produced in his mind, brain or soul (or whatever) then we would not have offered any explanation, but merely postulated a little man who perceives the images. For exactly the same questions arise about what it is for the mind/brain/soul to perceive these images as we were trying to answer for the whole human being. This is a direct consequence of the fact that we are applying a predicate---`sees'---that applies properly only to the whole human being to something which is merely a part of a human being, and what is lacking is an explanation of what the term means in this application. It becomes very clear that the purported explanation of seeing in terms of images in the head is no explanation at all, when we reflect that it gives rise to an infinite regress. If we see in virtue of a little man perceiving images in our heads, then we need to explain what it is for him to perceive, which can only be in terms of another little man, and so on.

The same would go, \textit{mutatis mutandis}, for an attempt to explain possession of information in terms of containing information propositionally. Somebody is required to read, store, decode and encode the various propositions, and peruse any pictures; and this leads to the regress of an army of little men. Again, the very same difficulty would arise for attempts to describe possessing information as containing information inferentially: now the miniature army is required to draw the inferences that allow knowledge to be gained from the presence of correlations.   

This last point indicates that a degree of circumspection is required when dealing with the common tendency to describe the mechanisms of sensory perception in terms of information reaching the brain. 
In illustration \citep[cf.][]{hacker:1987}, it has been known since the work of Hubel and Weisel (see for example \citet{hubel:weisel}) that there exist systematic correlations between the responses of groups of cells in the visual striate cortex and certain specific goings-on in a subject's visual field. It seems very natural to describe the passage of nerve impulses resulting from retinal stimuli to particular regions of the visual cortex as visual information reaching the brain. This is unobjectionable, so long as it is recognised that this is not a passage of information in the sense in which information has a direct conceptual link to the acquisition of knowledge. In particular, the visual information is not information for the subject about about the things they have seen. The sense in which the brain contains visual information is rather the sense in which a tree contains information about past winters.

Equipped with suitable apparatus, and because he knows about a correlation that exists, the neurophysiologist may make, from the response of certain cells in the visual cortex, an inference about what has happened in the subject's visual field. But the brain is in no position to make such an inference, nor, of course, an inference of any kind.
Containing visual information, then, is containing information inferentially, and trying to explain a person's possession of information about things seen as their brain containing visual information would lead to a homunculus regress: who is to make the inference that provides knowledge? 

This is not to deny the central importance and great interest of the scientific results describing the mechanisms of visual perception for our understanding of how a person can gain knowledge of the world surrounding them, but is to guard against an equivocation. The answers provided by brain science are to questions of the form: what are the causal mechanisms which underlie our ability to gain visual knowledge? This is misdescribed as a question of how information flows, if it is thought that the information in question is the information that the subject comes to possess.
One might have `information flow' in mind, though, merely as a picturesque way of describing the processes of electrochemical activity involved in perception, in analogy to the processes involved in the transmission of information by telephone and the like. This use is clearly unproblematic, so long as one is aware of the limits of the analogy. (We don't want the question to be suggested: so who answers the telephone? This would take us back to our homunculi.)

\section{The Shannon Information and related concepts}\label{shannon concepts}

The technical concept of information relevant to our discussion, the Shannon information, finds its home in the context of communication theory. We are concerned with a notion of \textit{quantity} of information; and the notion of quantity of information is cashed out in terms of the resources required to transmit messages (which is, note, a very limited sense of quantity).
I shall begin by highlighting two main ways in which the Shannon information may be understood, the first of which rests explicitly on Shannon's 1948 noiseless coding theorem.

\subsection{Interpretation of the Shannon Information}

It is instructive to begin by quoting Shannon:
\begin{quoting}\samepage
The fundamental problem of communication is that of reproducing at one point either exactly or approximately a message selected at another point. Frequently these messages have \textit{meaning}...These semantic aspects of communication are irrelevant to the engineering problem. \citep[p.31]{shannon}
\end{quoting}
The communication system consists of an information source, a transmitter or encoder, a (possibly noisy) channel, and a receiver (decoder). It must be able to deal with \textit{any} possible message produced (a string of symbols selected in the source, or some varying waveform), hence it is quite irrelevant whether what is actually transmitted has any meaning or not, or whether what is selected at the source might convey anything to anybody at the receiving end. It might be added that Shannon arguably understates his case: in the \textit{majority} of applications of communication theory, perhaps, the messages in question will not have meaning. For example, in the simple case of a telephone line, what is transmitted is not \textit{what is said} into the telephone, but an analogue signal which records the \textit{sound waves} made by the speaker, this analogue signal then being transmitted digitally following an encoding.

It is crucial to realise that `information' in Shannon's theory is not associated with individual messages, \textit{but rather characterises the source of the messages}. The point of characterising the source is to discover what capacity is required in a communications channel to transmit all the messages the source produces; and it is for this that the concept of the Shannon information is introduced. The idea is that the statistical nature of a source can be used to reduce the capacity of channel required to transmit the messages it produces (we shall restrict ourselves to the case of discrete messages for simplicity).

Consider an ensemble $X$ of letters $\{ x_{1},x_{2}, \ldots ,x_{n} \}$ occurring with probabilities $p(x_{i})$. This ensemble is our source\footnote{More properly, this ensemble \textit{models} the source.}, from which messages of $N$ letters are drawn. We are concerned with messages of very large $N$. For such messages, we know that typical sequences of letters will contain $Np(x_{i})$ of letter $x_{i}$, $Np(x_{j})$ of $x_{j}$ and so on. The number of distinct typical sequences of letters is then given by \[ \frac{N!}{Np(x_{1})!Np(x_{2})! \ldots Np(x_{n})!} \] and using Stirling's approximation, this becomes $2^{NH(X)}$, where
\begin{equation}
H(X)=-\sum_{i=1}^{n}\!p(x_{i})\log p(x_{i}), \label{shannon}
\end{equation}
is the Shannon information (logarithms are to base 2 to fix the units of information as binary bits).

Now as $N \rightarrow \infty$, the probability of an atypical sequence appearing becomes negligible and we are left with only $2^{NH(X)}$ equiprobable typical sequences which need ever be considered as possible messages. We can thus replace each typical sequence with a binary code number of $NH(X)$ bits and send that to the receiver rather than the original message of $N$ letters ($N\log n$ bits).

The message has been compressed from $N$ letters to $NH(X)$ bits ($\leq N\log n$ bits). Shannon's noiseless coding theorem, of which this is a rough sketch, states that this represents the optimal compression (Shannon 1948). The Shannon information is, then, appropriately called a measure of information because it represents the maximum amount that messages consisting of letters drawn from an ensemble $X$ can be compressed.

One may also make the derivative statement that the information \textit{per letter} in a message is $H(X)$ bits, which is equal to the information of the source. But `derivative' is an important qualification: we can only consider a letter $x_{i}$ drawn from an ensemble $X$ to have associated with it the information $H(X)$ if we consider it to be a member of a typical sequence of $N$ letters, where $N$ is large, drawn from the source. 

Note also that we must strenuously resist any temptation to conclude that because the Shannon information tells us the maximum amount a message drawn from an ensemble can be compressed, that it therefore tells us the irreducible meaning content of the message, specified in bits, which somehow possess their own intrinsic meaning. This idea rests on a failure to distinguish between a code, which has no concern with meaning, and a language, which does (cf. \citet{harris:1987}).

\subsubsection{Information and Uncertainty}

Another way of thinking about the Shannon information is as a measure of the amount of information that we \textit{expect} to gain on performing a probabilistic experiment.
The Shannon measure is a measure of the uncertainty of a probability distribution as well as serving as a measure of information. A measure of uncertainty is a quantitative measure of the lack of concentration of a probability distribution; this is called an uncertainty because it measures our uncertainty about what the outcome of an experiment completely described by the probability distribution in question will be. \citet{jos} provides an axiomatic characterisation of measures of uncertainty, deriving a general class of measures, $U_{r}(\vec{p})$, of which the Shannon information is one (see also Maassen and Uffink 1989). The key property possessed by these measures is Schur concavity (for details of the property of Schur concavity, see Uffink (1990), Nielsen (2001) and Section~\ref{differentnotions} below).

Imagine a random probabilistic experiment described by a probability distribution $\vec{p}=\{p(x_{1}),\ldots,p(x_{n})\}$. The intuitive link between uncertainty and information is that the greater the uncertainty of this distribution, the more we stand to gain from learning the outcome of the experiment.
In the case of the Shannon information, this notion of how much we gain can be made more precise.

Some care is required when we ask `how much do we know about the outcome?' for a probabilistic experiment. In a certain sense, the shape of the probability distribution might provide no information about what an individual outcome will actually be, as any of the outcomes assigned non-zero probability can occur. 
However, we can use the probability distribution to put a \textit{value} on any given outcome. If it is a likely one, then it will be no surprise if it occurs, so of little value; if an unlikely one, it is a surprise, hence of higher value. A nice measure for the value of the occurrence of outcome $x_{i}$ is $-\log p(x_{i})$, a decreasing function of the probability of the outcome. We may call this the `surprise' information associated with outcome $x_{i}$; it measures the value of 
having observed this outcome of the experiment (as opposed to: not bothering to observe it at all) given that we know the probability distribution for the outcomes\footnote{Of course, this is a highly restricted sense of `value'. It does not, for example, refer to how much might be implied by this particular outcome having occurred, nor to the value of what might be learnt from it, nor the value of what it conveys (if anything); these ideas all lie on the `everyday concept of information' side that is not being addressed here. The distinction between the surprise information and the everyday concept becomes very clear when one reflects that what one learns from a particular outcome may well be, in fact generally will be, quite independent of the probability assigned to it.}.

If the information (in this restricted sense) that we would gain if outcome $x_{i}$ were to occur is $-\log p(x_{i})$, then before the experiment, the amount of information we expect to gain is given by the expectation value of the `surprise' information, $\sum_{i}p(x_{i})(-\log p(x_{i}))$; and this, of course, is just the Shannon information $H$ of the probability distribution $\vec{p}$. Hence the Shannon information tells us our expected information gain.  

More generally, though, any of the measures of uncertainty $U_{r}(\vec{p})$ may be understood as measures of information gain; and   
a similar story can be told for measures of `how much we know' given a probability distribution. These will be the inverses of an uncertainty: we want a measure of the concentration of a probability distribution; the more concentrated, the more we know about what the outcome will be; which just means, the better we can predict the outcome. 
(To say in this way that we have certain amount of information (knowledge) about what the outcome of an experiment will be, therefore, is not to claim that we have partial knowledge of some predetermined fact about the outcome of an experiment.)  


\subsubsection{The minimum number of questions needed to specify a sequence}

The final common interpretation of the Shannon information is as the minimum average number of binary questions needed to specify a sequence drawn from an ensemble (Uffink 1990; Ash 1965), although this appears not to provide an interpretation of the Shannon information actually independent of the previous two.

Imagine that a long sequence $N$ of letters is drawn from the ensemble $X$, or that $N$ independent experiments whose possible outcomes have probabilities $p(x_{i})$ are performed, but the list of outcomes is kept from us. Our task is to determine what the sequence is by asking questions to which the guardian of the sequence can only answer `yes' or `no'; and we choose to do so in such a manner as to minimize the average number of questions needed. We need to be concerned with the \textit{average} number to rule out lucky guesses identifying the sequence.

If we are trying to minimize the average number of questions, it is evident that the best questioning strategy will be one that attempts to rule out half the possibilities with each question, for then whatever the answer turns out to be, we still get the maximum value from each question. Given the probability distribution, we may attempt to implement this strategy by dividing the possible outcomes of each individual experiment into classes of equal probability, and then asking whether or not the outcome lies in one of these classes. We then try and repeat this process, dividing the remaining set of possible outcomes into two sets of equal probabilities, and so on. It is in general not possible to proceed in this manner, dividing a finite set of possible outcomes into two sets of equal probabilities, and it can be shown that in consequence the average number of questions required if we ask about each individual experiment in isolation is greater than or equal to $H(X)$. However, if we consider the $N$ repeated experiments, where $N$ tends to infinity, and consider asking joint questions about what the outcomes of the independent experiments were, we can always divide the classes of possibilities of (joint) outcomes in the required way. Now we already know that for large $N$, there are $2^{NH(X)}$ typical sequences, so given that we can strike out half the possible sequences with each question, the minimum average number of questions needed to identify the sequence is $NH(X)$. (These last results are again essentially the noiseless coding theorem.)

It is not immediately obvious, however, why the minimum average number of questions needed to specify a sequence should be related to a notion of information. (Again, the tendency to think of bits and binary questions as irreducible meaning elements is to be resisted.) It seems, in fact that this is either just another way of talking about the maximum amount that messages drawn from a given ensemble can be compressed, in which case we are back to the interpretation of the Shannon information in terms of the noiseless coding theorem, or it is providing a particular way of characterising how much we stand to gain from learning a typical sequence, and we return to an interpretation in terms of our expected information gain. 

\subsection{More on communication channels}\label{communication channels}

So far we have concentrated on only one aspect of describing a communication system, namely, on characterising the information source. The other important task is to characterise the communication channel.

A channel is defined as a device with a set $\{x_{i}\}$ of input states, which are mapped to a set $\{y_{j}\}$ of output states. If a channel is noisy then this mapping will not be one-to-one. A given input could give rise to a variety of output states, as a result of noise. The basic type of channel---the \textit{discrete memoryless channel}---is characterised in terms of the conditional probabilities $p(y_{j}|x_{i})$: given that input $x_{i}$ is prepared, what is the probability that output $y_{j}$ will be produced?

\sloppypar If the distribution, $p(x_{i})$, for the probability with which the various inputs will be prepared is also specified, then we may calculate the joint distribution $p(x_{i}\wedge y_{j})$. We may consider which input state is prepared on a given use of the channel to be a random variable $X$, with $p(X=x_{i})=p(x_{i})$; which output produced to be a random variable $Y$, $p(Y=y_{j})=p(y_{j})$; and we may consider also the joint random variable $X\wedge Y$, where $p(X\wedge Y = x_{i}\wedge y_{j})=p(x_{i}\wedge y_{j})$. 

The joint distribution $p(x_{i}\wedge y_{j})$ allows us to define the joint uncertainty
\begin{equation}\label{joint}
H(X\wedge Y)=-\sum_{i,j} p(x_{i}\wedge y_{j})\log p(x_{i}\wedge y_{j}),
\end{equation}
and an important quantity known as the `conditional entropy':
\begin{equation}\label{conditional entropy}
H(X|Y)=\sum_{j}p(y_{j})\bigl(-\sum_{i}p(x_{i}|y_{j})\log p(x_{i}|y_{j})\bigr).
\end{equation}
The scare quotes are significant, as this quantity is not actually an entropy or uncertainty itself, but is rather the \textit{average} of the uncertainties of the conditional distributions for the input, given a particular $Y$ output. It measures the average of how uncertain someone will be about the $X$ value when they have observed an output $Y$ value.  

As \citet[\S1.6.6]{jos} notes, it pays to attend to the fact that $H(X|Y)$ is not a measure of uncertainty. It is easy to show \citep[e.g.][Thm.1.4.3--5]{ash} that  
\begin{equation}\label{uncertainty decrease}
H(X|Y) \leq H(X),\; \text{with equality \textit{iff} $X$ and $Y$ are independent};
\end{equation}
and it is often held that this is a particularly appealing feature of the Shannon measure of information because it captures the intuitive idea that by learning the value of $Y$, we gain some information about $X$, therefore our uncertainty in the value of $X$ should go down (unless the two are independent). Thus, Shannon describes the inequality (\ref{uncertainty decrease}) as follows:
\begin{quoting}
The uncertainty of $X$ is never increased by knowledge of $Y$. It will be decreased unless $Y$ and $X$ are independent events, in which case it is not changed. \citep[p.53]{shannon} 
\end{quoting}

But this description is highly misleading. As Uffink remarks, one's uncertainty certainly \textit{can} increase following an observation: increasing knowledge need not lead to a decrease in uncertainty. This is well illustrated by Uffink's `keys' example: my keys are in my pocket with a high probability, if not, they could be in a hundred places all with equal (low) probability. This distribution is highly concentrated so my uncertainty is low. If I look, however, and find that my keys are not in my pocket, then my uncertainty as to their whereabouts increases enormously. An increase in knowledge has led to an increase in uncertainty.

This does not conflict with the inequality (\ref{uncertainty decrease}), of course, as the latter involves an average over post-observation uncertainties. Uffink remarks, against \citet[p.186]{jaynes:1957} for example, that
\begin{quoting}
...there is no paradox in an increase of uncertainty about the outcome of an experiment as a result of information about its distribution. The confusion is caused by a liberal use of the multifaceted term information, and also by the deceptive name of conditional entropy for what is actually an average of the entropies of conditional distributions. \citep[p.83]{jos}
\end{quoting} 

To see why the conditional entropy is important, consider a very large number $N$ of repeated uses of our channel. There are $2^{NH(X)}$ typical $X$ (input) sequences that could arise, $2^{NH(Y)}$ typical output sequences that could be produced, and $2^{NH(X\wedge Y)}$ typical sequences of pairs of $X,Y$ values that could obtain. Suppose someone observes which $Y$ sequence has actually been produced. If the channel is noisy, then there is more than one input $X$ sequence that could have given rise to it. The conditional entropy measures the number of possible input sequences that could have given rise to the observed output (with non-vanishing probability).

If there are $2^{NH(X\wedge Y)}$ typical sequences of pairs of $X,Y$ values, then the number of typical $X$ sequences that could result in the production of a given $Y$ sequence will be given by   
\[\frac{2^{NH(X\wedge Y)}}{2^{NH(Y)}}= 2^{N(H(X\wedge Y)-H(Y))}.\]
Due to the logarithmic form of $H$, $H(X\wedge Y) = H(Y) + H(X|Y)$, and it follows that the number of input sequences consistent with a given output sequence will be $2^{NH(X|Y)}$. 

\citet[\S 12]{shannon} points out that this means that if one is trying to use a noisy channel to send a message, then the conditional entropy specifies the number of bits per letter that would need to be sent by an auxiliary \textit{noiseless} channel in order to correct all the errors that have crept into the transmitted sequence, as a result of the noise. If input and output states are perfectly correlated, i.e., there is no noise, then obviously $H(X|Y)=0$. 

Another most important quantity is the \textit{mutual information}, $H(X:Y)$, defined as 
\begin{equation}\label{mutual}
H(X:Y)= H(X)-H(X|Y).
\end{equation}
It follows from Shannon's \textit{noisy coding theorem} (1948) that the mutual information $H(X:Y)$ governs the rate at which information may be sent over a channel with input distribution $p(x_{i})$, with vanishingly small probability of error. 

The following sorts of heuristic interpretations of the mutual information may also be given: With a noiseless channel, an output $Y$ sequence would contain as much information as the input $X$ sequence, i.e., $NH(X)$ bits. If there is noise, it will contain less. We know, however, that $H(X|Y)$ measures the number of bits per letter needed to correct an observed $Y$ sequence, therefore the amount of information this sequence actually contains will be $NH(X)- NH(X|Y)=NH(X:Y)$ bits.

Or again, we can say that $NH(X:Y)$ provides a measure of the amount that we are able learn about the identity of an input $X$ sequence from observing the output $Y$ sequence: There are $2^{NH(X|Y)}$ input sequences that will be compatible with an observed output sequence, and
the size of this group, as a fraction of the total number of possible input sequences, may be used a measure of how much we have narrowed down the identity of the $X$ sequence by observing the $Y$ sequence.
This fractional size is 
\[\frac{2^{NH(X|Y)}}{2^{NH(X)}}=\frac{1}{2^{NH(X:Y)}},\]
and the smaller this fraction---hence the greater $H(X:Y)$---the more one learns from learning the $Y$ sequence.

The most important interpretation of the mutual information does derive from the noisy coding theorem, however. Consider, as usual, sequences of length $N$, where $N$ is large; the input distribution to our channel is $p(x_{i})$. Roughly speaking, the noisy coding theorem tells us that it is possible to find $2^{NH(X:Y)}$ $X$ sequences of length $N$ (code words) such that on observation of the $Y$ sequence produced following preparation of one of these code words, it is possible to infer which X sequence was prepared, with a probability of error that tends to zero as $N$ tends to infinity \citep{shannon}. So if we were now to consider an information source $W$, producing messages with an information of $H(W)=H(X:Y)$, each output sequence of length $N$ from this source could be associated with an $X$ code word, and hence messages from $W$ be sent over the channel with arbitrarily small error as $N$ is increased\footnote{This result is particularly striking as it is not intuitively obvious that in the presence of noise, arbitrarily good transmission may be achieved without the per letter rate of information transmission also tending to zero. The noisy coding theorem assures us that it can be achieved.}.

The \textit{capacity}, $\mathcal{C}$, of a channel is defined as the supremum over all input distributions $p(x_{i})$ of $H(X:Y)$. The noiseless coding theorem states that given a channel with capacity $\mathcal{C}$ and an information source with an information of $H\leq \mathcal{C}$, there exists a coding system such that the output of the source can be transmitted over the channel with an arbitrarily small frequency of errors.

\subsection{Interlude: Abstract/concrete; technical, everyday}\label{abstract/concrete}

Part of my aim in this chapter has been to deflect the pressure of the question `What is information?' by following the lead of Austin (and, of course, Wittgenstein\footnote{`The questions ``What is length?", ``What is meaning?", ``What is the number one?" etc., produce in us a mental cramp. We feel that we can't point to anything in reply to them and yet ought to point to something. (We are up against one of the great sources of philosophical bewilderement: a substantive makes us look for a thing that corresponds to it.)' \citet[p.1]{witt:bluebook}.}) and pointing to the fact that `information' is an abstract noun: correspondingly we should not seek to illuminate the term by attempting fruitlessly to grasp for something that it corresponds or refers to, but by considering simple examples of its function and in particular considering its relations to grammatically simpler and less mystifying terms like `inform'.  

Now, when turning to information in the technical sense of Shannon's theory, we explicitly do \textit{not} seek to understand this noun by comparison with the verb `inform'. `Information' in the technical sense is evidently not derived from a nominalization of this verb. Nonetheless, `information' remains an abstract, rather than a concrete noun: it doesn't serve to refer to a material thing or substance. In this regard, note that the distinction `abstract/concrete' as applied to nouns does not map onto a distinction between physical concepts and concepts belonging to other categories. Thus the fact that `information', in the technical sense of Shannon's theory, may be included as a concept specifiable in physical terms does not entail that it stands for a concrete particular, entity or substance. For example, energy is a paradigmatic physical concept (to use another relevant term, energy is a physical quantity), yet `energy' is an abstract (mass) noun (akin to a property name). The interesting differences that exist between energy and the technical notion of information as examples of physical quantities deserve further analysis. See Chapter~\ref{study}, Sections~\ref{dissolving};~\ref{study concluding} for some remarks in this direction.  

Why my insistence that `information' in the technical sense remains an abstract noun? Well, consider that two strategies present themselves for providing an answer to the question `What is information' in the case of information theory. On the first the answer is: what is quantified by the Shannon information and mutual information. On the second it is: what is transmitted by information sources. 
These different strategies provide differing, but complementary answers. Under both, however, `information' is an abstract noun.

Taking the first strategy, one considers what is quantified by the Shannon information and mutual information. As we have seen, the Shannon information serves to quantify how much messages produced by a source can be compressed and the mutual information quantifies the capacity of a channel (for a particular input source distribution) to transmit messages. But this is evidently not to quantify an amount of stuff (even of some very diaphanous kind); and the amount that messages can be compressed and the capacity of a channel are no more concrete things than the size of my shoe is a concrete thing.

Now consider the second strategy. Recall our earlier quotation from Shannon. There he described the fundamental aim of communication theory as that of reproducing at one point a message that was selected at another point. Thus we might say (very roughly) that in the technical case, information is what it is the aim of a communication protocol to transmit: information (in the technical sense) is what is produced by an information source that is required to be reproduced if the transmission is to be counted a success\footnote{Note that this formulation is left deliberately open. What counts as successful transmission and therefore, indeed, as what one is trying to transmit, depends upon one's aims and interests in setting up a communication protocol.}.

However, the pertinent sense of `what is produced' is not the one pointing us towards the concrete systems that are produced by the source on a given occasion, but rather the one which points us towards the particular \textit{type} (sequence or structure) that these tokens instantiate. But a \textit{type} is not a concrete thing, hence `information', in this technical sense, remains an abstract noun. 

So, for example, if the source $X$ produces a string of letters like the following:\[x_{2}x_{1}x_{3}x_{1}x_{4}\ldots x_{2}x_{1}x_{7}x_{1}x_{4},\] say, then the type is the sequence `$x_{2}x_{1}x_{3}x_{1}x_{4}\ldots x_{2}x_{1}x_{7}x_{1}x_{4}$'; we might name this `sequence 17'. The aim is to produce at the receiving end of the communication channel another token of this type. What has been transmitted, though, the information transmitted on this run of the protocol, is sequence 17; and this is not a concrete thing.   

At this point we may draw an illustrative, albeit partial, analogy with information in the everyday sense. Imagine that I write down a message to a friend on a piece of paper (using declaritive sentences, to keep things simple); one will distinguish in the standard way between the sentence tokens inscribed and what is said by the sentences: the propositions expressed\footnote{Note, of course, that the propositions expressed are not to be identified with the sentence types of which the tokens I write are particular instances. (Consider, for example, indexicals.)}. It is the latter, \textit{what is said}, that is the information (everyday sense) I wish to convey. Similarly with information in the technical sense just described: one should distinguish between the concrete systems that the source outputs and the \textit{type} that this output instantiates. Again, it is the latter that is important; this is the information (technical sense) that one is seeking to transmit.

An important disanalogy between the technical and everyday notions of information now forcibly presents itself: the restatement of a by-now familiar point. In the everyday case, when I have written down my message to my friend, one not only has the sentence tokens and the sentence type they instantiate but also the propositions these sentences express; and again, it is these last that are the information I wish to convey. In the case we have just outlined for the information-theoretic notion of information, though, one only has the tokens produced by the source and the type they instantiate; it is this type that is transmitted, that constitutes the information in the technical sense we have just sketched. The further level, if any, of what various types might mean, or what instances of these types might convey, is not relevant to, or discussed by information theory: the point once more that information in the technical sense is not a semantic notion. Indeed, considered from the point of view of information theory, the output of an information source does not even have any syntactic structure.   

\section{Aspects of Quantum Information}\label{aspects}

Quantum information is a rich theory that seeks to describe and make use of the distinctive possibilities for information processing and communication that quantum systems provide. What draws the discipline together is the recognition that far from quantum behaviour presenting a potential \textit{nuisance} for computation and information transmission (in light of the trend towards increasing miniaturisation) the fact that the properties of quantum systems differ so markedly from those of classical objects actually provides \textit{opportunities} for interesting new communication protocols and forms of information processing. Entanglement and non-commutativity, two essentially quantum features, can be \textit{used}.

To give some examples: \citet{deutsch:1985} introduced the concept of the \textit{universal quantum computer}, and the evidence suggests that quantum computers are exponentially more powerful than classical computational models for the important task of factoring large numbers \citep{shor}; meanwhile quantum cryptography makes use of the fact that non-orthogonal quantum states cannot be perfectly distinguished in designing protocols for sharing secret random keys \citep[e.g.,][]{BB:84} thus holding out the promise of security of communication guaranteed by the laws of physics; entanglement may also be used in such protocols \citep{ekert:1991}.

Although the field of quantum information began to emerge in the mid-1980s, 
the \textit{concept} of quantum information itself was not truly available until the quantum analogue of Shannon's noiseless coding theorem---the \textit{quantum noiseless coding theorem}---was developed by Schumacher \citep{qcoding,schumacher:jozsa}\footnote{Historical note: Chris Fuchs has informed me that Ben Schumacher recollects first presenting the notion of quantum information at the IEEE meeting on the Physics of Computation in Dallas in October 1992. The germ of the idea and the term `qubit' arose in conversation between Schumacher and Wootters some months earlier.}. 


Quantum information theory may be considered as an extension of classical information theory which introduces new communication primitives, e.g., the \textit{qubit} (two-state quantum system) and shared entanglement, while providing quantum generalisations of the notions of sources, channels and codes. We will now review a selection of results (by no means comprehensive) that will be relevant to what follows. For systematic presentations of quantum information theory, see \citet{nielsen:chuang,physicsofqi,preskill,bennettshor:1998}. \citet{ekertjozsa:1996} also provides a nice review of quantum computation up to and including the development of Shor's algorithm.        

The first type of task one might consider consists of using quantum systems to transmit classical information. Whilst we are used to thinking that an $n$-dimensional quantum system possesses at most $n$ mutually distinguishable (i.e. orthogonal) states in which it might be prepared, one is also free to prepare such a system in one of any number of \textit{non}-orthogonal states. The price, of course, is that one will not then be able to determine perfectly \textit{which} state was prepared. This already makes things interesting.
It forces us to draw a distinction that is not needed in the classical case, that is, a distinction between the amount of information that is used to prepare, or is needed to specify, the state of a quantum system (the \textit{specification information}) and the information that has actually been encoded into a system (the \textit{accessible information}).

So, consider a classical information source, $A$, that has outputs $a_{i}, i=1\ldots k$, which occur with probabilities $p(a_{i})$. We will attempt to encode the output of this source into sequences of $n$-dimensional quantum systems, as follows. On receipt of output $a_{i}$ of $A$, a quantum system is prepared in the signal state $\rho_{a_{i}}$. These signal states may be pure or mixed, and may or may not be orthogonal. If the number of outputs, $k$, of the classical source is greater than $n$, though, the signal states \textit{will} have to be non-orthogonal.

We may consider sequences of length $N$ of signal states being prepared in this manner, where $N$ is very large. The amount of information needed to specify this sequence will be $NH(A)$ bits. The specification information, then, is the number of bits per system in the sequence needed to specify the whole sequence of states, and is given by the information of the classical source.

The quantum analogue of the Shannon information $H$ is the \textit{von Neumann entropy} \citep{wehrl}:
\begin{equation}\label{vn entropy}
S(\rho)\stackrel{\mathrm{def}}{=}-\mathrm{Tr}\rho\log \rho = -\sum_{i=1}^{n}\lambda_{i}\log \lambda_{i},
\end{equation}
where $\rho$ is a density operator on an $n$-dimensional Hilbert space and the $\lambda_{i}$ are its eigenvalues. For very large $N$, the sequence of quantum systems produced by our preparation procedure may be considered as an ensemble described by the density operator 
\begin{equation}\label{operator}
\rho=\sum_{i=1}^{k}p(a_{i})\rho_{a_{i}}.
\end{equation}
Equally, if one does not know the output of the classical source on a given run of the preparation procedure, then the state of the individual system prepared on that run may also be described by this density operator.      

The von Neumann entropy takes its maximum value, $\log n$, when $\rho$ is maximally mixed, and its minimum value, zero, if $\rho$ is pure. It also satisfies the inequality \citep{wehrl}:
\begin{equation}\label{mixing vn entropy}
S\bigl(\sum_{i=1}^{k}p(a_{i})\rho_{a_{i}}\bigr) \leq H(A) + \sum_{i}^{k}p(a_{i})S(\rho_{a_{i}}),
\end{equation}
which holds with equality \textit{iff} the $\rho_{a_{i}}$ are mutually orthogonal, i.e., $\rho_{a_{i}}\rho_{a_{j}}=0$,for $i\neq j$. Thus the specification information of the sequence, which is limited only by the number of outputs $k$ of the classical source, may be much greater than its von Neumann entropy, which is limited by the dimensionality of our quantum systems.

So, how much information have we actually managed to encode into these quantum systems? To answer this question we need to consider making measurements on the systems, and the resulting mutual information $H(A:B)$, where $B$ labels the observable measured, having outcomes $b_{j}$, with probabilities $p(b_{j})$, $j=1\ldots m$. Taking `encoded' to be a `success' word (something cannot be said to have been encoded if it cannot in principle be decoded), then the maximum amount of information encoded in a system is given by the accessible information \citep[cf.][]{qcoding}, that is, the maximum over all decoding observables of the mutual information. A well known result due to \citet{holevo} provides an upper bound on the mutual information resulting from the measurement of any observable, including positive operator valued (POV) measurements (which, recall, may have more outcomes than the dimensionality of the system being measured). This bound is:
\begin{equation}\label{holevo}
H(A:B)\leq S(\rho) -  \sum_{i}^{k}p(a_{i})S(\rho_{a_{i}}),
\end{equation}
with equality \textit{iff} the $\rho_{a_{i}}$ commute.

The Holevo bound~(\ref{holevo}) implies the weaker inequality 
\[H(A:B)\leq S(\rho)\leq \log n,\]
reinforcing our intuitive understanding that the maximum amount of information that may be encoded into a quantum system is limited by the number of orthogonal states available, i.e., by the dimension of the system's Hilbert space (even if we allow ourselves POV measurements to try to distinguish better non-orthogonal states). In particular, note that for a single qubit, the most that can be encoded is one bit of information. 

Again, from the Holevo bound and inequality~(\ref{mixing vn entropy}) it follows that
\[H(A:B)\leq S(\rho) -  \sum_{i}^{k}p(a_{i})S(\rho_{a_{i}}) \leq H(A).\]
The inequality on the right hand side will be strict if the encoding states $\rho_{a_{i}}$ are not orthogonal, implying that the accessible information will be strictly less than the specification information $H(A)$ in this case. This is a way of making precise the intuition that when encoding in non-orthogonal states, it is not possible to determine which states were prepared. If $H(A:B)<H(A)$ for any measurement $B$, then it is impossible to determine accurately what sequence of states was prepared by performing measurements on the sequence. 

Let us now look at quantum coding. Rather than beginning by considering a classical source, we could instead begin with a \textit{quantum} source. If a classical source is modelled by an ensemble $A$ from which letters $a_{i}$ are drawn with probabilities $p(a_{i})$, the quantum source will be modelled similarly by an ensemble of systems in states $\rho_{a_{i}}$, produced with probabilities $p(a_{i})$ \citep{qcoding}. We will assume these states to be pure, $\rho_{a_{i}}=\ket{a_{i}}{}\bra{}{a_{i}}$.  
Then, just as Shannon's noiseless coding theorem introduces the concept of the bit as a measure of information, the quantum noiseless coding theorem introduces the concept of the qubit as a measure of quantum information, characterising the quantum source.

By an ingenious argument, the quantum noiseless coding theorem runs parallel to Shannon's noiseless coding theorem, using much the same mathematical ideas. If we consider a long sequence of $N$ systems drawn from the quantum source, their joint state will be \[\rho^{\otimes N}=\rho^{1}\otimes\rho^{2}\otimes\ldots\otimes\rho^{N},\] where $\rho^{i}$ is the density operator for the $i$th system, given by eqn.~(\ref{operator}), with $\rho_{a_{i}}=\ket{a_{i}}{}\bra{}{a_{i}}$. In the classical case, for large enough $N$, we needed only to consider sending typical sequences of outcomes, of which there were $2^{NH(A)}$ for a source $A$, as only these had non-vanishing probability. Similarly in the quantum case, for large enough $N$, the joint state $\rho^{\otimes N}$ will have support on two orthogonal subspaces, one of which, the \textit{typical subspace}, will have dimension $2^{NS(\rho)}$ and will carry the vast majority of the weight of $\rho^{\otimes N}$, whilst the other subspace will have vanishingly small weight as $N\rightarrow\infty$ \citep{qcoding}\footnote{To see this, note that $\rho^{\otimes N}$ can be written as a weighted sum of $N$-fold tensor products of one dimensional eigenprojectors of $\rho$, with weights given by the products of the corresponding eigenvalues $\lambda_{i}$ of $\rho$. For large $N$ there will be $2^{NH(\vec{\lambda})}$, with $H(\vec{\lambda})=-\sum_{i=1}^{n}\lambda_{i}\log \lambda_{i}$, equiprobable typical sequences of eigenprojectors in this sum, i.e., sequences in which the relative frequency of occurrence of a given projector is equal to its associated eigenvalue, while all other sequences in the sum have very small weight. But $-\sum_{i=1}^{n}\lambda_{i}\log \lambda_{i}$ is just the von Neumann entropy $S(\rho)$.}. 
Because of this, the state $\rho^{\otimes N}$ may be transmitted with arbitrarily small error by being encoded onto a channel system of only $2^{NS(\rho)}$ dimensions \citep{qcoding,schumacher:jozsa}, for example, onto $NS(\rho)$ qubits. These channel systems may then be sent to the receiver and the original state recovered with near perfect fidelity. Thus, analogously to the classical case, we have a measure of the resources (now \textit{quantum} resources, mind) required to transmit what is produced by our quantum source. The von Neumann entropy provides a measure, in qubits, of the amount by which the output of our source may be compressed, hence provides a measure of the amount of quantum information the source produces\footnote{The converse to the quantum noiseless coding theorem, that $2^{NS(\rho)}$ qubits are \textit{necessary} for accurate transmission was proved in full generality by \citet{barnumfuchsjozsa:1996}.}.    

The use of entanglement as a communication resource is a centrally important feature of quantum information theory. The two paradigmatic examples of entanglement-assisted communication are dense coding and teleportation. In dense coding \citep{superdense} prior shared entanglement between two widely separated parties, Alice and Bob, allows Alice to transmit to Bob \textit{two} bits of information when she only sends him a \textit{single} qubit. This would be impossible if they did not share a maximally entangled state, e.g., the singlet state (one of the four \textit{Bell states}, see Table~\ref{bell states dense coding}) beforehand. The trick is that Alice may use a local unitary operation to change the global state of the entangled pair into one of four different orthogonal states. If she then sends Bob her half of the entangled pair he may perform a suitable joint measurement to determine which operation she applied; thence acquiring two bits of information.

\begin{table}
\begin{center}
\begin{minipage}{2.3in}
\[ \left. \begin{array}{c}
\ket{\phi^{+}}{}=1/\sqrt{2}(\ket{\up}{}\ket{\up}{}+\ket{\down}{}\ket{\down}{})\\
\ket{\phi^{-}}{}=1/\sqrt{2}(\ket{\up}{}\ket{\up}{}-\ket{\down}{}\ket{\down}{})\\
\ket{\psi^{+}}{}=1/\sqrt{2}(\ket{\up}{}\ket{\down}{}+\ket{\down}{}\ket{\up}{})\\
\ket{\psi^{-}}{}=1/\sqrt{2}(\ket{\up}{}\ket{\down}{}-\ket{\down}{}\ket{\up}{})
\end{array} \right\}\]
\end{minipage} \ =  \
\begin{minipage}{1.5in}
\[ \left\{\begin{array}{r}
-i\sigma_{y}\otimes \mathbf{1}\ket{\psi^{-}}{}\\
-\sigma_{x}\otimes \mathbf{1}\ket{\psi^{-}}{}\\
\sigma_{z}\otimes \mathbf{1}\ket{\psi^{-}}{}\\
\mathbf{1}\otimes\mathbf{1}\ket{\psi^{-}}{}
\end{array} \right. \]
\end{minipage}
\end{center}
\caption{The four Bell states, a maximally entangled basis for $2\otimes2$ dim. systems. A choice of one of four of the operations $\{{\mathbf 1},\sigma_{x},\sigma_{y},\sigma_{z}\}$ applied to her system by Alice may transform, for example, the singlet state to one of the other three states orthogonal to it.}\label{bell states dense coding}
\end{table}

In the teleportation protocol, by contrast \citep{teleportation} instead of being used to help send classical information, shared entanglement is used to transmit an unknown quantum state from Alice to Bob, with, remarkably, nothing that bears any relation to the identity of the state travelling between them. Furthermore, during the protocol, the state being teleported `disappears' from Alice's location before `reappearing' at Bob's a little while later, thus providing the inspiration for the science fiction title of the protocol. 
Also during the protocol, the intial shared entanglement is destroyed. One \textit{ebit} (the amount of entanglement in a maximally entangled state of a $2\otimes 2$ system) is used up in teleporting an unknown qubit state from one location to another.

Since teleportation is a linear process it may also be used for \textit{entanglement swapping}. Let's say that Alice and Bob, who are widely spatially separated, share a maximally entangled state of a pair of particles labelled 3 and 4. Alice may decide to perform the teleportation operation on a system, 2, which is half of an entangled pair, 1 and 2. Following the protocol, the entanglement between 1 and 2, and between 3 and 4 is destroyed, but 1 and 4 will end up entangled, whereas before they had been uncorrelated. The entanglement of 1 and 2 has been swapped onto entanglement of systems 1 and 4.

We shall be considering dense coding and teleportation in detail in later chapters. 

We should note, finally, a very important restricting principle for quantum protocols. This is the \textit{no cloning} theorem \citep{dieks,wootters:zurek}: It is impossible to make copies of an unknown quantum state. This marks a considerable difference with classical information processing protocols, as in the classical case, the value of a bit may be freely copied into numerous other systems, perhaps by measuring the original bit to see its value, and then preparing many other bits with this value. The same is not possible with quantum systems. As is well known, it is impossible to determine the state of a single quantum system by measurement (for a nice discussion see \citet{busch:observable}), so the measuring  approach would clearly be a non-starter\footnote{It is in the context of state determination and superluminal signalling that the question of cloning first arose. If it were possible to clone an unknown quantum state, then we could multiply up an individual system into a whole ensemble in the same state; and it would then be quite possible to determine what that state was (see, e.g., Section~\ref{Hilbert-Schmidt}). This, of course, would then give rise to the possibility of superluminal signalling using entanglement in an EPR-type setting: one would be able to distinguish between different preparations of the same density matrix, hence determine superluminally which measurement was performed on a distant half of an EPR pair.}.

To see that no more general scheme would be possible either, consider a device that makes a copy of an unknown state $\ket{\alpha}{}$. This would be implemented by a unitary evolution\footnote{Is it too restrictive to consider only unitary evolutions? One can always consider a non-unitary evolution, e.g. measurement, as a unitary evolution on a larger space. Introducing auxiliary systems, perhaps including the state of the apparatus, doesn't affect the argument.}  $U$ that takes the product $\ket{\alpha}{}\ket{\psi_{0}}{}$, where $\ket{\psi_{0}}{}$ is a standard state, to the product $\ket{\alpha}{}\ket{\alpha}{}$. Now consider another possible state $\ket{\beta}{}$. Suppose the device can copy this state too: $U\ket{\beta}{}\ket{\psi_{0}}{}=\ket{\beta}{}\ket{\beta}{}$. If it is to clone a general unknown state, however, it must be able to copy a superposition such as $\ket{\xi}{}=1/\sqrt{2}(\ket{\alpha}{}+\ket{\beta}{})$ also, but the effect of $U$ on $\ket{\xi}{}$ is to produce a fully entangled state  $1/\sqrt{2}(\ket{\alpha}{}\ket{\alpha}{} + \ket{\beta}{}\ket{\beta}{})$ rather than the required $\ket{\xi}{}\ket{\xi}{}$. It follows that no general cloning device is possible.

In fact it may be seen in the following way that if a device can clone more than one state, then these states must belong to an orthogonal set. We are supposing that $U\ket{\alpha}{}\ket{\psi_{0}}{}=\ket{\alpha}{}\ket{\alpha}{}$ and $U\ket{\beta}{}\ket{\psi_{0}}{}=\ket{\beta}{}\ket{\beta}{}$. Taking the inner product of the first equation with the second implies that 
$ \langle\alpha|\beta\rangle=\langle\alpha|\beta\rangle^{2},$
which is only satisfied if $\langle\alpha|\beta\rangle=0 \text{ or } 1$, i.e., only if $\ket{\alpha}{}$ and $\ket{\beta}{}$ are identical or orthogonal.          

\section{Information is Physical: The Dilemma}

A very striking claim runs through much of the literature in quantum information theory and quantum computation. This is the claim that `Information is Physical'. From the conceptual point of view, however, this statement is rather baffling; and it is perhaps somewhat obscure precisely what it might mean. Be that as it may, the slogan is often presented as the fundamental insight at the heart of quantum information theory; and it is frequently claimed to be entailed, or at least suggested, by the theoretical and practical advances of quantum information and computation\footnote{Perhaps the most vociferous proponent of the idea that information is physical was the late Rolf Landauer \citep[e.g.][]{landauer:1991,landauer:1996}.}.

However, it would seem that the slogan `Information is Physical' faces a difficult dilemma. If it is supposed to refer to information in the everyday sense then, whatever its precise meaning, it certainly implies a very strong reductionist claim. It would have to amount, amongst other things, to a claim that central semantic and mental attributes or concepts are reducible to physical ones. \textit{This}, however, is a purely philosophical claim, and a contentious one at that. As such it is hard to see how it could be supported by the claims and successes \textit{in physics} of quantum information theory.   

So is `information' in the slogan supposed to construed in the technical sense, then? Well, perhaps. But if so, then the claim is merely that some physically defined quantity is physical; and that is hardly an earth-shattering revelation. In particular it is now hard to see how it could represent an important new theoretical insight\footnote{Another possible reading of the slogan will be discussed briefly at the end of Chapter~\ref{comp}.}.

Of course, there is in philosophy a tradition occupied by those who hope, or expect, to achieve the reduction of semantic and related concepts to respectable physical ones. Representatives of this tradition we might term the \textit{semantic naturalizers}. We will discuss \citet{dretske:1981} as a well known example of such an approach briefly, below. The semantic naturalizer would not jib at the claim that information in the everyday sense is physical; indeed would undoubtedly endorse it. However, this does not affect the point that if `Information is Physical' adverts to information in the everyday sense, then what is at issue is a philosophical claim about the relations between different groups of concepts; and quantum information theory does not engage in this debate. Rather, as we have seen, this piece of \textit{physical} theory seeks to describe the distinctive ways in which quantum systems, with all their unusual properties, may be used for various tasks of information processing and transmission. It does not, therefore, adjudicate upon, nor provide evidence for or against a philosophical claim concerning the reduction of semantic properties to physical ones; and it is none the worse for that.       

In any case, it should be noted that the success, and even well-groundedness, of the project of naturalizing semantic properties can hardly be said to be a settled question. Certainly, a successful completion of the project has by no means been achieved. While \citet{adams} presents an up-beat account of progress, two recent sympathetic reviews \citep{loewer:1997,mclaughlin:rey} suggest that the project has yet to overcome important systematic difficulties. 

As noted in these reviews, proposals for naturalizing semantics typically face two sorts of problems, whose ancestry, in fact, may be traced back to difficulties that \citet{grice:1957} raised for the crude causal theory of meaning. These are what may be called the \textit{problem of error} and the \textit{problem of fine grain}.

In brief: it is an essential part of a proposal to naturalize semantics that an account be given of the content of beliefs (or of propositional attitudes in general). The problem of error relates to the feature of intentionality mentioned earlier: one might believe that $p$ when $p$ is not the case; and this is hard to accomodate in a naturalized account of content. (A very simple illustration: we might suggest that one has the belief that $p$ when one's belief is caused by the fact that $p$. But then one could only believe that $p$ if $p$ were the case; and this is false.) The problem of fine grain is in articulating the detailed structure of what is believed without using linguistic resources, as semantic relations have a finer grain than causal ones. (To use a hackneyed example, my belief that $x$ is a creature with a heart is distinct from my belief that $x$ is a creature with a kidney, yet the properties of having a heart and having a kidney are (nomologically) co-instantiated. Whatever is caused by a creature that has a heart is caused by a creature that has a kidney.) There is no consensus on whether these problems have been, or can be, satisfactorily addressed while an account still mantains its credentials as a fully naturalistic one.

Moreover, we should note that there are many who would be inclined to argue that there is system in our apparent failure to provide a satisfactory naturalized account of semantics thus far. The pertinent thought is that language, being a rule governed activity, has an essential normative component that cannot be captured by any naturalistic explanation. The impetus behind this line of thought derives from Wittgenstein's reflections on meaning and rule-following \citep{wittgenstein:investigations}.     

Returning to quantum information theory, the following quotation from a recent article in \textit{Reviews of Modern Physics} provides an apt illustration of the problematic claim that `Information is Physical'. 
\begin{quoting}   
What is...surprising is the fact that quantum physics may influence the field of information and computation in a new and profound way, getting at the very root of their foundations...

But why has this happened? It all began by realizing that information has a physical nature (Landauer, 1991;1996;1961).
It is printed on a physical support..., it cannot be transported faster than light in vacuum, and it abides by natural laws.
\textit{The statement that information is physical does not simply mean that a computer is a physical object, but in addition that information itself is a physical entity.}

In turn, this implies that the laws of information are restricted or governed by the laws of physics. In particular, those of quantum physics. \citep{galindo:martin-delgado}
\end{quoting}
Whilst illustrating the problem, this passage also invites a simple response, one indicating the lines of a solution.

Let's pick out three phrases:
\begin{enumerate}
\item `The statement that information is physical does not simply mean that a computer is a physical object'
\item `in addition...information itself is a physical entity'
\item `In turn, this implies that the laws of information are restricted or governed by the laws of physics.'
\end{enumerate}
Statement (2) is the one that purports to be presenting us with a novel ontological insight deriving from, or perhaps driving, quantum information theory. The difficulty is in understanding what this portentous sounding phrase might mean and, most especially, understanding what r\^ole it is supposed to play.  

For it is statement (3) (with `laws of information' understood as `laws governing information processing') that really seems to be the important proposition, if our interest is what information processing is possible using physical systems, as it is in quantum information theory. And (2) is entirely unnecessary to establish (3), despite their concatenation in the quotation above. All that we in fact require is part of statement (1): computers, or more generally, information processing devices, are physical objects. What one can do with them is necessarily restricted by the laws of physics.  

Quantum information theory and quantum computation are theories about what we can \textit{do} using physical systems, stemming from the recognition that the peculiar characteristics of quantum systems might provide opportunities rather than drawbacks. This project is evidently quite independent of any philosophical claim regarding the everyday concept of information. There is therefore no need for the quantum information scientist to take a stand on contentious questions such as whether semantic and mental concepts are reducible to physical ones. We have already noted that `Information is Physical', with `information' understood in the everyday sense, is not supported by the success of quantum information theory; no more, we now see, would such a claim be needed for it. 
All that is required is the obvious statement that the devices being used for information processing are physical devices. \textit{Contra} statement (1) and the suggestion of Galindo and Mart\'{\i}n-Delgado above, if anything more than this is  meant (literally) by `Information is Physical' then it is irrelevant to quantum information theory. 

It is perhaps helpful to note that part of the obscurity of statements like (2) may be the result of their seeming to incorporate a category mistake. Another example would be the following statement by Landauer:
\begin{quoting}
Information is not a disembodied abstract entity; it is always tied to a physical representation. \citep[p.188]{landauer:1996}
\end{quoting}
To see the nature of the mistake, let us return to the example I gave in Section~\ref{abstract/concrete} of writing down a message to my friend. As we noted, one distingushes between the sentence tokens inscribed (the collection of ink marks on the page) and what is written down: the propositions expressed.


Now if `information' refers to what is written down, inscribed, encoded, in or on various physical objects, or to what is conveyed by a message---as it might well be thought to do in statements like (2) and the Landauer quotation---then it makes no sense to say that information is physical. For \textit{what} is written down, as opposed to the collection of ink marks on the page, is not physical. This would be the category error. (The same argument applies if we consider information in the technical sense of what is produced by an information source, see Section~\ref{abstract/concrete}. The token is physical, but the type belongs to a different logical category.)

While it is true, as Landauer says, that information is not a disembodied abstract entity, this does not mean it is an embodied concrete entity: it is no sort of entity at all.
`Information' is functioning as an abstract noun and hence does not refer to an entity, nor indeed to some sort of substance.     
Talk of the necessity of a physical representation (cf. \citet[p.5]{steane:1997}: `no information without physical representation!') only amounts to (or need only amount to) the truism that if we are writing information down, or storing it in a computer memory then we need something to write it on, or store it in. But this doesn't make what is written down, or what is stored, physical\footnote{Note, furthermore, that it is by no means clear that with possessing information (as opposed to containing it) there is any useful sense in which information finds a \textit{representation} (a much over-used term); although, it may be the case that, as a matter of contingent fact, someone's possessing information \textit{supervenes} on facts about their brain, nervous system and, perhaps, unrestrictedly large regions of the universe.}.





\section{Alternative approaches: Dretske}\label{dretske}

So far, little mention has been made of other philosophical discussions of the nature of information. Instead, we have noted some features of the everyday concept of information and seen how, in particular, this concept is distinct from the concept of information due to Shannon. \citet{floridi} provides a useful summary of various other approaches to the concepts of information to be found in the philosophical literature.

However, there is one particular approach that we must look into in greater detail---that of Dretske in \textit{Knowledge and the Flow of Information} \citep{dretske:1981}. Dretske is a proponent of semantic naturalism; and in this book he articulates a position that is directly opposed to the view that I have advocated regarding the significance of the communication-theoretic notion of information. His distinctive claim is that a satisfactory semantic concept of information is indeed to be found in information theory and may be achieved with a simple extension of the Shannon theory: in his view there is not a significant distinction between the technical and everyday concepts of information. 

I shall suggest, however, that Dretske fails to establish this claim. Moreover, whether or not his proposed semantic concept of information is in fact a satisfactory one, it enjoys no licit connection with Shannon's theory. 

Whilst agreeing with Shannon that the semantic aspects of information are irrelevant to the engineering problem, Dretske also concurs with Weaver's assessment of the converse proposition: ``But this does not mean that the engineering aspects are necesssarily irrelevant to the semantic aspects" \citep[p.8]{shannon:weaver}.   
Of course, if the engineering aspects of mechanical communication systems \textit{are} relevant, though, it still needs to be demonstrated precisely what their relevance is.

Dretske begins by noting that one reason why the Shannon theory does not provide a semantic notion of information is that it does not ascribe an amount of information to individual messages, yet it is to individual messages that semantic properties would apply. To circumvent this difficulty, he introduces the following quantity as a measure of the amount of information that a \textit{single} event $y_{j}$, which may be a signal, carries about another event, or state of affairs, $x_{i}$:
\begin{definition}[Dretske's information measure]\label{dretske measure}
 \[I_{x_{i}}(y_{j}) = -\log p(x_{i}) - H\bigl(p(x_{i^{\prime}}|y_{j})\bigr), \]
\end{definition}   
where $x_{i} \in \{x_{i^{\prime}}\}, i^{\prime}=1,\dots, m; y_{j}\in \{y_{j^{\prime}}\}, j^{\prime}=1,\ldots,n$. 
That is, the amount of information that the occurrence of $y_{j}$ carries about the the occurrence (or obtaining) of $x_{i}$ is given by the surprise information of $x_{i}$, minus the uncertainty (as quantified by the Shannon measure) in the conditional probability distribution for the $x_{i^{\prime}}$ events (states of affairs) given that $y_{j}$ occurred.

From this definition of the amount of information that a single event carries, he moves to a definition of \textit{what} information is contained in a signal $S$:
\begin{definition}[Dretske's information {\em that}]\label{info dretske}  
\[\text{A signal $S$ contains the information {\em that} $q$}\;\; \stackrel{\mathrm{def}}{=}\;\; p\,(q|S)=\!1.\]
\end{definition}
The point of this definition is that there is to be a perfect correlation between the occurrence of the signal and what it is supposed to indicate: that $q$. 

Does this establish a link between the technical communication-theoretic notions of information and a semantic, everyday one? Not yet, at any rate. Whether definition~(\ref{info dretske}) supplies a satisfactory semantic notion of information isn't to be settled by stipulation, but would need to be established by the successful \textit{completion} of a programme of semantic naturalism demonstrating that Dretske's notion of information \textit{that} is indeed an adequate one. We have already noted that the question of whether such an objective might be achieved remains open.

However, perhaps more tellingly, there appear in any case to be major difficulties in the other direction---for the thought that Dretske's notion of information \textit{that} has any genuine ties to information theory. I shall mention two main sources of difficulty, either of which appears on its own sufficient to frustrate the claim that there are such ties.

In Dretske's proposal, the link to information theory is supposed to be mediated by definition~(\ref{dretske measure}) of the amount of information that an individual event carries about another event or state of affairs. He argues that if a signal is to carry the information that $q$ it must, amongst other things, carry as much information as is generated by the obtaining of the fact that $q$. 

Unfortunately, the quantity $I_{x_{i}}(y_{j})$  cannot play the r\^ole of a measure of the amount of information that $y_{j}$ carries about $x_{i}$. To see this we need merely note that the surpise information associated with $x_{i}$ is largely independent of the uncertainty in the conditional probability distribution for $x_{i^{\prime}}$ given $y_{j}$.
For example, our uncertainty in $x_{i^{\prime}}$ given $y_{j}$ might be very large, implying that we would learn little from $y_{j}$ about the value $x_{i^{\prime}}$, yet still the amount said to be carried by $y_{j}$ about $x_{i}$, under Dretske's definition, could be arbitrarily large, if the surprise information of $x_{i}$ dominates. Or again, the channel might be so noisy that we can learn nothing at all about $x_{i}$ from $y_{j}$--- the two are uncorrelated, no information can be transmitted---yet still $I_{x_{i}}(y_{j})$ could be strictly positive and very large (if the probability of $x_{i}$ is sufficiently small). This is sufficient to show that $I_{x_{i}}(y_{j})$ is unacceptable as a measure. The hoped-for link to information theory is snapped\footnote{One might try to finesse this difficulty by proposing different definitions for the amount of information that a single event carries about another, or more likely, adopt a direct criterion for when a signal carries `as much' information as is generated by the obtaining of the fact that $q$ (see below). None of the obvious approaches, though, suggest that the appeal to an \textit{amount} of information content (and hence a link to a quantitative theory of information) is really anything other than a free-wheel.}. 

The second main source of difficulty is that in most realistic situations it would appear very difficult to specify how much information should be associated with the fact that $q$. It would be straightforward enough, perhaps, if we always had a natural fixed range of options to choose between, as we are supposing the set $\{x_{i^{\prime}}\}$ provides, but how should  the different options in a realistic perceptual situation, say, be counted? The suspicion is that typically, there will be no well-defined range of distinct possibilities. Dretske himself notes this problem:
\begin{quoting}  \sloppypar
How, for example, do we calculate the amount of information generated by Edith's playing tennis?...[O]ne needs to know: (1) the alternative possibilities...(2) the associated probabilities...(3) the conditional probabilities...Obviously, in most ordinary communication setings one knows none of this. It is not even very clear whether one \textit{could} know it. What, after all, are the alternative possibilities to Edith's playing tennis? Presumably there are some things that are possible (e.g., Edith going to the hairdresser instead of playing tennis) and some things that are not possible (e.g., Edith turning into a tennis ball), but how does one begin to catalog these possibilities? If Edith might be jogging, shall we count this as \textit{one} alternative possibility? Or shall we count it as more than one, since she could be jogging almost anywhere, at a variety of different speeds, in almost any direction? \citep[p.53]{dretske:1981}
\end{quoting}   
His answer is that this spells trouble only for specifying \textit{absolute} amounts of information; and it is comparative amounts of information with which he is concerned, in particular, with whether a signal carries \textit{as much} information as is generated by the occurrence of a specified event, whatever the absolute values. But this response is surely too phlegmatic. If the ranges of possibilities aren't well-defined, then the associated measure of information is not well-defined; and the \textit{difference} between the two quantities will not then be well-defined: two wrongs don't make a right. Dretske's attempt to forge a link with a theory of quantity-of-information-carried seems highly doubtful. 

Of course, at this point Dretske could re-trench and argue that what he means by a signal carrying the same amount of information as is associated with the fact that $q$ is simply that signal and fact are perfectly correlated. This would be consistent, but would make it very plain that the digression via a quantitative theory of how much information a signal contains or an event generates is superfluous. It would now just be the concept of \textit{perfect correlation} that is operative \textit{ab initio}, not anything to do with measuring amounts of information that a signal can contain. This contrasts with Dretske's original hope that the requirement of perfect correlation between a signal and what it indicates could be motivated or derived from constraints on how much information a signal can carry. As is well known, in his later work Dretske did in fact move away from conditional probabilities in defining his concept of information \textit{that}, using the idea of perfect lawlike correlation instead (although for different reasons than the ones we have been dwelling on here \citep{dretske:1983,dretske:1988}), further emphasising that concepts from information theory really play no genuine r\^ole in his framework.

It thus seems that the appearance of a link between Dretske's 1981 semantic notion of information and information theory is illusory. No ideas that involve quantifying amounts of information transmitted truly play any substantive r\^{o}le in arriving at definition~(\ref{info dretske}). This means, first of all, that Dretske's notion of information \textit{that} gains no validation from the direction of information theory; and second, that his argument does not establish that there are closer ties between the communication-theoretic notion of information and the everyday notion than are usually admitted.       

It should be noted, finally, that care is required when considering Dretske's definition~(\ref{info dretske}) (and the later statements that do not involve conditional probabilities) as a possible primitive notion of information \textit{that}\footnote{By `primitive', I mean a notion of information that comes before the concepts of knowledge and cognitive agent and may be used to explain these latter concepts. Cf. Dretske: `In the beginning there was information. The word came later. The transition was achieved by the development of organisms with the capacity for selectively exploiting this information in order to survive and perpetuate their kind.' \citep[p.vii]{dretske:1981}.}. 
One must be aware that the definition may appear intuitively appealing for illegitimate reasons: as the result of the new notion it introduces being conflated with the idea of containing information inferentially, for example. With this latter notion of containing information, it is clear enough why perfect correlation can have a link to information: someone who knows of the correlation between signal and state of affairs may learn something about the state of affairs by observing the signal, in virtue of an inference. However, \textit{this} notion of information, containing information inferentially, is evidently not apt for the r\^ole of a primitive notion of information \textit{that}, as it relies upon the prior concept of a cognitive agent who may use their knowledge of the correlation to gain further knowledge. For Dretske, information is `that commodity capable of yielding knowledge' \citep[p.44]{dretske:1981}, but the obvious ways in which perfect correlation can yield knowledge---via an inference, or as part of a natural sign that may be understood or interpreted---are not available for picking out a primitive notion of information \textit{that}, on pain of the homunculus fallacy.

\section{Summary}

\sloppypar One of the main aims of this chapter has been to emphasise the distinction between the everyday concept of information---with its links to knowledge, language and meaning---and the technical notions of information that are developed in information theory\footnote{Warnings---more or less felicitous---that one should make this distinction abound. For example, Weaver:
`...\textit{information} must not be confused with meaning. In fact, two messages, one of which is heavily loaded with meaning and the other of which is pure nonsense, can be exactly equivalent from the present viewpoint as regards information.' \citep[p.8]{shannon:weaver}
Similarly Feynman:
`...``information" in our sense tells us nothing about the usefullness or otherwise of the message.' \citep[p.118]{feynman:1999}.}. 
It is not just that these technical ideas are introduced only to provide notions of an \textit{amount} of information, but that in most cases, information theory is silent on how much information in the everyday sense a message might contain or convey, if any. In general, what is quantified in information theory is emphatically \textit{not} information in the everyday sense of the word.



The following table provides a summary of some of the points that have been argued and of some of the positions that might be adopted. `SN' stands for `Semantic Naturalizer' and `$\neg$ SN' for `non-Semantic Naturalizer'.\vspace{2\baselineskip}

\begin{center}

\begin{tabular}{c|c|c}
 & \parbox{1in}{\begin{center}\large Yes\end{center}} & {\large No}  \\ \hline
\parbox{1.2in}{\begin{singlespacing}\large Information\\ is Physical?\\\end{singlespacing}} & SN & $\neg$SN \\ \hline
\parbox{1.3in}{\begin{singlespacing}\large Everyday distinct\\ from Technical?\\\end{singlespacing}} & $\neg$SN, SN & SN (early Dretske)\\ 
\end{tabular}

\end{center}\vspace{2\baselineskip}

Regarding the question of the physicality of information, if we are concerned with information in the everyday sense, then the dispute is between proponents of semantic naturalism and others, and concerns the reducibility (or otherwise) of semantic concepts. The semantic naturalizer will assert that information is, or may be, physical, while the non-semantic naturalizer will deny this. This philosophical debate remains unsettled, and I have suggested that it is a debate quantum information theory has no bearing upon. Equally, the outcome of the debate has no bearing on quantum information theory. 

Regarding the relationship between the everyday concept of information and information theoretic notions, both the semantic naturalizer and the non-semantic naturalizer can agree that these concepts are quite distinct. An attempt to naturalize semantics need not proceed by way of information theory; and given the very pronounced \textit{prima facie} divergences between information theoretic notions and the everyday concept, it does not look a terribly promising avenue to explore. The early Dretske did attempt such an approach, however; and would claim that the distinction between information theory and the everyday notion of information may be elided. I have suggested, though, that this attempt to build bridges between information theory and the everyday concept of information is not successful.


\end{doublespacing}



\chapter[Inadequacy of Shannon Information in QM?]{On a Supposed Conceptual Inadequacy of the Shannon Information in Quantum Mechanics}\label{supposed conceptual inadequacy}






\begin{doublespacing}


\section{Introduction}

In Part II of this thesis, we will be considering the implications of quantum information theory for the foundations of quantum mechanics. One of the topics we shall be investigating there is the approach of Zeilinger, who has put forward an information-theoretic principle which he suggests might serve as a foundational principle for quantum mechanics \citep{foundationalprinciple}.

As a part of this foundational project, \citet{conceptualinadequacy} have  
criticised Shannon's (1948) measure of information, the quantity fundamental to the discussion of information in both classical and quantum information theory. They claim that the Shannon information is not appropriate as a measure of information in the quantum context and have proposed in its stead their own preferred quantity and a notion of `total information content' associated with it, which latter is supposed to supplant the von Neumann entropy 
\citep{operationallyinvariant,encodingdecoding,reply}.

The main aim in \citet{conceptualinadequacy} is to establish that the Shannon information is intimately tied to classical notions, in particular, to the preconceptions of classical measurement, and that in consequence it cannot serve as a measure of information in the quantum context. They seek to establish this in two ways. First, by arguing that the Shannon measure only makes sense when we can take there to be a pre-existing sequence of bit values in a message we are decoding, which is not the case in general for measurements on quantum systems (consider measurements on qubits in a basis different from their eigenbasis); and second, by suggesting that Shannon's famous third postulate, the postulate that secures the uniqueness of the form of the Shannon information measure \citep{shannon} and has been seen by many as a necessary axiom for a measure of information, is motivated by classical preconceptions and does not apply in general in quantum mechanics where we must consider non-commuting observables.

These two arguments do not succeed in showing that the Shannon information is `intimately tied to the notion of systems carrying properties prior to and independent of observation' \citep[p.1]{reply}, however.
The first is based on too narrow a conception of the meaning of the Shannon information and the second, primarily, on a misreading of what is known as the `grouping axiom'. We shall see that the Shannon information is perfectly well defined and appropriate as a measure of information in the quantum context as well as in the classical (Section~\ref{arguments}). 


Brukner and Zeilinger have a further argument against the Shannon information (Section~\ref{finalargument}). They suggest it is inadequate because it cannot be used to define an acceptable notion of `total information content'. Equally, they insist, the von Neumann entropy cannot be a measure of information content for a quantum system because it has no general relation to information gain from the measurements that we might perform on a system, save in the case of measurement in the basis in which the density matrix is diagonal. By contrast, for a particular set of measurements, their preferred information measure sums to a unitarily invariant quantity that they interpret as `information content', this being one of their primary reasons for adopting this specific measure. This property will be seen to have a simple geometric explanation in the Hilbert-Schmidt representation of density operators however, rather than being of any great information theoretic significance; and this final argument found unpersuasive, as the proposed constraint on any information measure regarding the definition of `total information content' seems unreasonable. Part of the problem is that information content, total or otherwise, is not a univocal concept and we need to be careful to specify precisely what we might mean by it in any given context.

\section{Two arguments against the Shannon information\label{arguments}}

\subsection{Are pre-existing bit-values required?}\label{preexistingbits}
Since the quantity $-\sum_{i}{p_{i}}\log p_{i}$ is meaningful for any (discrete) probability distribution $\vec{p}=\{p_{1},\ldots,p_{n}\}$ (and can be generalised for continuous distributions), the point of Brukner and Zeilinger's first argument must be that when we have probabilities arising from measurements on quantum systems, $-\sum_{i}{p_{i}}\log p_{i}$ does not correspond to a concept of \textit{information}. Their argument concerns measurements on systems that are all prepared in a given state $|\psi\rangle$, where $|\psi\rangle$ may not be an eigenstate of the observable we are measuring. The probability distribution $\vec{p}$ for measurement outcomes will be given by $p_{i}={\mbox Tr}(|\psi\rangle\langle \psi |  P_{i})$, where $P_{i}$ are the operators corresponding to different measurement outcomes (projection operators in the spectral decomposition of the observable, for projective measurements). 

Brukner and Zeilinger suggest that the Shannon information has no meaning in the quantum case, because the concept lacks an `operational definition' in terms of the number of binary questions needed to specify an actual concrete sequence of outcomes. In general in a sequence of measurements on quantum systems, we cannot consider there to be a pre-existing sequence of possessed values, at least if we accept the orthodox eigenvalue-eigenstate link for the ascription of definite values (see e.g. \citet{bub:1997})\footnote{In a footnote, Brukner and Zeilinger suggest that the Kochen-Specker theorem in particular raises problems for the operational definition of the Shannon information. It is not clear, however, why the impossibility of assigning \textit{context independent} yes/no answers to questions asked of the system should be a problem if we are considering an \textit{operational} definition. Presumably such a definition would include a concrete specification of the experimental situation, i.e. refer to the context, and then we are not concerned with assigning a value to an \textit{operator} but to the outcome of a specified experimental procedure, and this can be done perfectly consistently, if we so wish. The de-Broglie Bohm theory, of course, provides a concrete example \citep{bell:1982}.\label{KS}}, and this rules out, they insist, interpreting the Shannon measure as an amount of information:
\begin{quoting}
The nonexistence of well-defined bit values prior to and independent of observation suggests that the Shannon measure, as defined by the number of binary questions needed to determine the particular \textit{observed} sequence 0's and 1's, becomes problematic and even untenable in defining our uncertainty as given \textit{before} the measurements are performed. \citep[p.1]{conceptualinadequacy}\vspace{\baselineskip}\\
...No definite outcomes exist before measurements are performed and therefore the number of different possible sequences of outcomes does not characterize our uncertainty about the individual system before measurements are performed. \citep[p.3]{conceptualinadequacy} 
\end{quoting}
These two statements should immediately worry us, however.
Recall the key points of the interpretation of the Shannon information (Section~\ref{shannon concepts}): given a long message (a long run of experiments), we \textit{know} that it will be one of the typical sequences that is instantiated. Given $\vec{p}$, we can say what the typical sequences will be, how many there are, and hence the number of bits ($NH(X)$) needed to specify them, 
independent of whether or not there is a pre-existing sequence of bit values.
It is irrelevant whether there already is some concrete sequence of bits or not; all possible sequences that will be produced will require the same number of bits to specify them as any sequence produced will always be one of the typical sequences. It clearly makes no difference to this whether the probability distribution is given classically or comes from the trace rule.
Also, the number of different possible sequences does indeed tell us about our uncertainty before measurement: what we know is that one of the typical sequences will be instantiated, what we are ignorant of is which one it will be, and we can put a measure on how ignorant we are simply by counting the number of different possibilities. Brukner and Zeilinger's attempted distinction between uncertainty before and after measurement is not to the point, the uncertainty is a function of the probability distribution and this is perfectly well defined before measurement\footnote{We may need to enter at this point the important note that the Shannon information is not supposed to describe our \textit{general} uncertainty when we know the \textit{state}, this is a job for a measure of mixedness such as the von Neumann entropy, see below.}. 

Brukner and Zeilinger have assumed that it is a necessary and sufficient condition to understand $H$ as a measure of information that there exists some concrete string of $N$ values, for then and only then can we talk of the minimum number of binary questions needed to specify the string.
But as we have now seen, it is \textit{not} a necessary condition that there exist such a sequence of outcomes.

We are not in any case forced to assume that $H$ is about the number of questions needed to specify a sequence in order to understand it as a measure of information; we also have the interpretations in terms of the maximum amount a message drawn from an ensemble described by the probability distribution $\vec{p}$ can be compressed, and as the expected information gain on measurement. (And as we have seen, one of these two interpretations must in fact be prior.)
Furthermore, the absence of a pre-existing string need not even be a problem for the minimum average questions interpretation --- we can ask about the minimum average number of questions that \textit{would} be required if we \textit{were} to have a sequence drawn from the ensemble. So again, the pre-existence of a definite string of values is not a necessary condition.

It is not a sufficient condition either, because, faced with a string of $N$ definite outcomes, in order to interpret $NH$ as the minimum \textit{average} number of questions needed to specify the sequence, we need to know that we in fact have a typical sequence, that is, we need to imagine an ensemble of such typical sequences and furthermore, to assume that the relative frequencies of each of the outcomes in our actual string is representative of the probabilities of each of the outcomes in the notional ensemble from which the sequence is drawn. If we do not make this assumption, then the minimum number of questions needed to specify the state of the sequence must be $N$ --- we cannot imagine that the statistical nature of the source from which the sequence is notionally drawn allows us to compress the message. So even in the classical case, the concrete sequence on its own is not enough and we need to consider an ensemble, either of typical sequences or an ensemble from which the concrete sequence is drawn. In this respect the quantum and classical cases are completely on a par. The same assumption needs to be made in both cases, namely, that the probability distribution $\vec{p}$, either known in advance, or derived from observed relative frequencies, correctly describes the probabilities of the different possible outcomes.
The fact that no determinate sequence of outcomes exists before measurement does not pose any problems for the Shannon information in the quantum context.

Reiterating their requirements for a satisfactory notion of information, Brukner and Zeilinger say:
\begin{quoting}
We require that the information gain be directly based on the observed probabilities, (and not, for example, on the precise sequence of individual outcomes observed on which Shannon's measure of information is based). \citep[p.1]{reply}
\end{quoting}
But as we have seen, it is false that the Shannon measure must be based on a precise sequence of outcomes (this is not a necessary condition) and the Shannon measure already \textit{is} and \textit{must be} based on the observed probabilities (a sequence of individual outcomes on its own is not sufficient).

There is, however, a difference between the quantum and classical cases that Brukner and Zeilinger may be attempting to capture. Suppose we have a sequence of  $N$ qubits that has actually been used to encode some information, that is, the sequence of qubits is a channel to which we have connected a classical information source. For simplicity, imagine that we have coded in orthogonal states. Then the state of the sequence of qubits will be a product of $|{0}\rangle$'s and $|1\rangle$'s and for measurements in the encoding basis, the sequence will have a Shannon information equal to $NH(A)$ where $H(A)$ is the information of the classical source. If we do not measure in the encoding basis, however, the sequence of 0's and 1's we get as our outcomes will differ from the values originally encoded and the Shannon information of the resulting sequence will be greater than that of the original\footnote{We may think of our initial sequence of qubits as forming an ensemble described by the density operator $\rho = p_{1}|0\rangle\langle 0| + p_{2}|1\rangle\langle 1|$, where $p_{1}, p_{2}$ are the probabilities for 0 and 1 in our original classical information source. Any (projective) measurement that does not project onto the eigenbasis of $\rho$ will result in a post-measurement ensemble that is more mixed than $\rho$ (see e.g. \citet{characterizingmixing,peres} and below) and hence will have a greater uncertainty, thus a greater Shannon information, or any other measure of information gain.}. We have introduced some `noise' by measuring in the wrong basis. As we have seen, however, the way we describe this sort of situation \citep[e.g.][]{qcoding}, is to use the Shannon mutual information $H(A:B)=H(A)-H(A|B)$, where $B$ denotes the outcome of measurement of the chosen observable (outcomes $b_{i}$ with probabilities $p(b_{i})$) and the `conditional entropy' $H(A|B)=\sum_{i=1}^{n}p(b_{i})H(p(a_{1}|b_{i}),\ldots,p(a_{m}|b_{i}))$, characterises the noise we have introduced by measuring in the wrong basis. $H(B)$ is the information (per letter) of the sequence that we are left with after measurement, $H(A:B)$ tells us the amount of information that we have actually managed to transmit down our channel, i.e. the amount (per letter) that can be decoded when we measure in the wrong basis.

\subsection{The grouping axiom}
The first argument has not revealed any difficulties for the Shannon information in the quantum context, so let us now turn to the second. 

In his original paper, Shannon put forward three properties as reasonable requirements on a measure of uncertainty and showed that the only function satisfying these requirements has the form $H=-K\sum_{i}p_{i}\log p_{i}$.\footnote{In contrast to some later writers, however, notably \citet{jaynes:1957}, he set little store by this derivation, seeing the justification of his measure as lying rather in its implications \citep{shannon}. Save the noiseless coding theorem, the most significant of the implications that Shannon goes on to draw are, as has been pointed out by Uffink, consequences of the property of Schur concavity and hence shared by the general class of measures of uncertainty derived in \citet{jos}.}

\begin{figure}[btp]
\begin{center}
\includegraphics*[178,499][321,575]{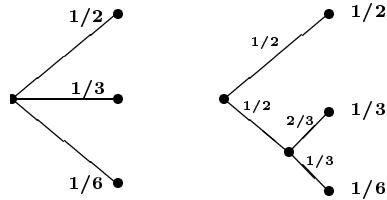}
\end{center}
\caption[Shannon's third requirement]{
The probability distribution $\vec{p}= \{\frac{1}{2},\frac{1}{3},\frac{1}{6}\}$ can be considered as given for the three outcomes directly, or we could consider first a choice of two equiprobable events, followed by a second choice of two events with probabilities $\frac{2}{3},\frac{1}{3}$, conditional on the second, say, of the first two events occurring, a `decomposition' of a single choice into two successive choices, the latter of which will only be made half the time. Shannon's third requirement says that the uncertainty in $\vec{p}$ will be given by $H(\frac{1}{2},\frac{1}{3},\frac{1}{6})= H(\frac{1}{2},\frac{1}{2}) + \frac{1}{2}H(\frac{2}{3},\frac{1}{3})$: the uncertainty of the overall choice is equal to the uncertainty of the first stage of the choice, plus the uncertainty of the second choice weighted by its probability of occurrence.}\label{decomposition}
\end{figure}

The first two requirements are that $H$ should be continuous in the $p_{i}$ and that for equiprobable events ($p_{i}=1/n$), $H$ should be a monotonic increasing function of $n$.
The third requirement is the strongest and the most important in the uniqueness proof. It states that if a choice is broken down into two successive choices, the original $H$ should be a weighted sum of the individual values of $H$. The meaning of this rather non-intuitive constraint is usually demonstrated with an example (see Fig. \ref{decomposition}).
A precise statement of Shannon's third requirement (one that includes also the second requirement as a special case) is due to \citet{faddeev} and is often known as the Faddeev grouping axiom:
\newtheorem{groupingaxiom}{Grouping Axiom}
\begin{groupingaxiom}[Faddeev] \label{faddeevgroupingaxiom}
For every $n\geq2$
\begin{equation}
H(p_{1},p_{2},\ldots,p_{n-1},q_{1},q_{2}) = H(p_{1},\ldots,p_{n-1},p_{n}) + p_{n}H(\frac{q_{1}}{p_{n}},\frac{q_2}{p_{n}})
\end{equation}
where $p_{n}=q_{1}+q_{2}$.
\end{groupingaxiom} 
The form of the Shannon information follows uniquely from requiring $H(p, 1-p)$ to be continuous for $0\leq p \leq 1$ and positive for at least one value of $p$, permutation invariance of $H$ with respect to relabelling of the $p_{i}$, and the grouping axiom.

`Grouping axiom' is an appropriate name. As it is standardly understood (see e.g. \citet{ash,jos,jaynes:1957}), we consider that instead of giving the probabilities $p_{1},\ldots,p_{n}$ of the outcomes $x_{1},\ldots,x_{n}$ of a probabilistic experiment directly, we may imagine grouping the outcomes into composite events (whose probabilities will be given by the sum of the probabilities of their respective component events), and then specifying the probabilities of the outcome events conditional on the occurrence of the composite events to which they belong; this way of specifying the probabilistic experiment being precisely equivalent to the first. So we might group the first $k$ events together into an event $A$, which would have a probability $p(A)=\sum_{i=1}^{k}p_{i}$, and the remaining $n-k$ into an event $B$ of probability $p(B)=\sum_{i=k+1}^{n}p_{i}$; and then give the conditional probabilities of the events $x_{1},\ldots,x_{k}$ conditional on composite event $A$ occurring, $(p_{1}/p(A)),\ldots,(p_{k}/p(A))$, and similarly the conditional probabilities for the events $x_{k+1},\ldots,x_{n}$ conditional on event $B$. The grouping axiom then concerns how the uncertainty measures should be related for these different descriptions of the same probabilistic experiment. It says that our uncertainty about which event will occur should be equal to our uncertainty about which group it will belong to plus the expected value of the uncertainty that would remain if we were to know which group it belonged to (this expected value being the weighted sum of the uncertainties of the conditional distributions, with weights given by the probability of the outcome lying within a given group).

So in particular, let us imagine an experiment with $n+1$ outcomes which we label $a_{1}, a_{2},\ldots, a_{n-1},b_{1},b_{2}$, having probabilities $p_{1},\ldots,p_{n-1},q_{1},q_{2}$ respectively. We can define an event $a_{n}=b_{1}\cup b_{2}, b_{1}\cap b_{2}=\emptyset$, which would have probability $p_{n}=q_{1}+q_{2}$ and the probabilities for $b_{1}$ and $b_{2}$ conditional on $a_{n}$ occurring will then be $\frac{q_{1}}{p_{n}}, \frac{q_{2}}{p_{n}}$ respectively. Grouping Axiom \ref{faddeevgroupingaxiom} says that the uncertainty in the occurrence of events $a_{1}, a_{2},\ldots, a_{n-1},b_{1},b_{2}$ is equal to the uncertainty for the occurrence of events $a_{1},\ldots,a_{n}$ plus the uncertainty for the occurrence of $b_{1},b_{2}$ conditional on $a_{n}$ occurring, weighted by the probability that $a_{n}$ should occur. 

Brukner and Zeilinger suggest that the grouping axiom, however, embodies certain classical presumptions that do not apply in quantum mechanics. This entails that the axiomatic derivation of the form of the Shannon measure does not go through and that the Shannon information ceases to be a measure of uncertainty in the quantum context.
The argument turns on their interpretation of the grouping axiom, which differs from the standard interpretation in that it refers to joint experiments.

\subsubsection{Brukner and Zeilinger's interpretation}
\begin{sloppypar}
If we take an experiment, $A$, with outcomes $a_{1},\ldots,a_{n}$ and probabilities $(p(a_{1}),\ldots,p(a_{n})) = (p_{1},\ldots,p_{n})$ and an experiment, $B$, with outcomes $b_{1},b_{2}$, then for the joint experiment $A \wedge B$, the event $a_{n}$ is the union of the two disjoint events $a_{n}\wedge b_{1}$ and $a_{n}\wedge b_{2}$. Let us assign to these two events the probabilities $q_{1}$ and $q_{2}$ respectively. Then $p(a_{n})=p(a_{n}\wedge b_{1})+p(a_{n}\wedge b_{2})=q_{1}+q_{2}=p_{n}$. On this interpretation, the left hand side of Grouping Axiom \ref{faddeevgroupingaxiom} is to be understood as denoting the uncertainty in the experiment with outcomes $a_{1},a_{2},\ldots,a_{n-1},a_{n}\wedge b_{1},a_{n}\wedge b_{2}$.
\end{sloppypar}

If $a_{n}$ occurs, the conditional probabilities for $b_{1},b_{2}$ will be $p(a_{n}\wedge b_{1})/p(a_{n})=q_{1}/p_{n}, p(a_{n}\wedge b_{2})/p(a_{n})=q_{2}/p_{n}$ respectively, and so $H(\frac{q_{1}}{p_{n}},\frac{q_{2}}{p_{n}})$ is the uncertainty in the value of $B$ given that $a_{n}$ occurs.

The grouping axiom can now be rewritten as:
\begin{groupingaxiom}[Brukner and Zeilinger]\label{BZ1} \mbox{ }
\begin{multline}
H \left(p(a_{1}),p(a_{2}),\ldots,p(a_{n-1}),p(a_{n}\wedge b_{1}),p(a_{n}\wedge b_{2})\right )\\
= H\left(p(a_{1}),p(a_{2}),\ldots,p(a_{n})\right) 
  + p(a_{n})H\left(p(b_{1}|a_{n}),p(b_{2}|a_{n})\right).
\end{multline}
\end{groupingaxiom}
Generalizing to the case in which we have $m$ outcomes for experiment $B$ and distinguish $B$ values for all $n$ $A$ outcomes, so that we have $mn$ outcomes $a_{i}\wedge b_{j}$, the grouping axiom becomes:
\begin{groupingaxiom}[Brukner and Zeilinger]\label{BZ2} \mbox{ }
\[H(A\wedge B)=H(A)+H(B|A)\]
\end{groupingaxiom}
From the point of view of Shannon's original presentation, this expression appears as a theorem rather than an axiom, being a consequence of the logarithmic form of the Shannon information and the definition of the conditional entropy.

\subsubsection{The inapplicability argument}

The classical assumptions made explicit, Brukner and Zeilinger suggest, in Grouping Axioms \ref{BZ1} and \ref{BZ2} are that attributes corresponding to all possible measurements can be assigned to a system simultaneously (in this case, $a_{i}, b_{j}$ and $a_{i}\wedge b_{j}$); and that measurements can be made ideally non-disturbing. Grouping Axiom \ref{BZ2}, for example, is supposed to express the fact that classically, the information we expect to gain from a joint experiment $A\wedge B$, is the same as the information we expect to gain from first performing $A$, then performing $B$ (where the uncertainty in $B$ is updated conditional on the $A$ outcome, but our ability to predict $B$ outcomes is not degraded by the $A$ measurement).

Their inapplicability argument is simply that as the grouping axiom requires us to consider joint experiments, the uniqueness proof for the Shannon information will fail in the quantum context, because we can consider measurements of non-commuting observables and the joint probabilities on the left hand side of Grouping Axiom \ref{BZ1} will not be defined for such observables; thus the grouping axiom will fail to hold. Furthermore, the grouping axiom shows that the Shannon information embodies classical assumptions, so the Shannon measure will not be justified as a measure of uncertainty because these assumptions do not hold in the quantum case. The result is that
\begin{quoting}
...only for the special case of commuting, i.e., simultaneously definite observables, is the Shannon measure of information applicable \textit{and the use of the Shannon information justified to define the uncertainty given before quantum measurements are performed}. \citep[p. 4]{conceptualinadequacy}, \small{my emphasis}. 
\end{quoting} 

This argument is problematic, however. Let us begin with the obvious point that a failure of the argument for uniqueness does not automatically rule out the Shannon information as a measure of uncertainty. In fact, the Shannon information can be seen as one of a general class of measures of uncertainty, characterised by a set of axioms in which the grouping axiom does not appear \citep{jos}, hence the grouping axiom is not necessary for the interpretation of the Shannon information as a measure of uncertainty.
(Uffink in fact has previously argued that the grouping axiom is not a natural constraint on a measure of information and should not be imposed as a necessary constraint, even in the classical case \citep[\S 1.6.3]{jos}.)
So from the fact that on the Brukner/Zeilinger reading, the grouping axiom seems to embody some classical assumptions that do not hold in the quantum case, it does not follow that the concept of the Shannon information as a measure of uncertainty involves those classical assumptions.

Furthermore, Brukner and Zeilinger's grouping axiom is not in fact equivalent to the standard form and the standard form is equally applicable in both the classical and quantum cases. Thus the Shannon information has not been shown to involve classical assumptions and the standard axiomatic derivation can indeed go through in the quantum context. The probabilities appearing in Grouping Axiom \ref{faddeevgroupingaxiom} are well defined in both the classical and quantum cases. 

\begin{figure}
\begin{center}
\includegraphics*[151,612][388,725]{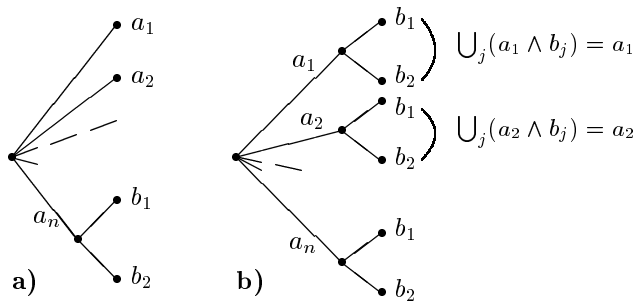}
\end{center}
\caption{{\bf a}) in Grouping Axiom \ref{faddeevgroupingaxiom} and eqn.~(\ref{conditional}), the `$B$' outcomes $b_{1}$ and $b_{2}$ cannot occur without $a_{n}\stackrel{\rm def}{=}b_{1}\cup b_{2}$ occurring. {\bf b}) in the joint experiment scenario, $b_{1}$ or $b_{2}$ can occur without $a_{n}$ occurring, but this is made to appear the same by coarse graining the joint experiment and only recording $B$ values when we get $A$ outcome $a_{n}$.}\label{difference}
\end{figure}

In Brukner and Zeilinger's notation, Grouping Axiom \ref{faddeevgroupingaxiom} would be written as
\begin{multline}\label{conditional}  
H(p(a_{1}),p(a_{2}),\ldots,p(a_{n-1}),p(b_{1}),p(b_{2}))     \\
 = H(p(a_{1}),\ldots,p(a_{n-1}),p(b_{1}\vee b_{2})) \\
 +p(b_{1}\vee b_{2})H(p(b_{1}|b_{1}\vee b_{2}),p(b_{2}|b_{1}\vee b_{2})) 
\end{multline}
This refers to an experiment with $n+1$ outcomes labelled by $a_{1},\ldots a_{n-1},b_{1},b_{2}$ and the grouping of two of these outcomes together, and is clearly different from Grouping Axiom \ref{BZ1} (see Fig. \ref{difference}).
Brukner and Zeilinger's Grouping Axiom \ref{BZ1} is the result of applying (\ref{conditional}) to a coarse grained joint experiment in which we only distinguish the $B$ outcomes of $A\wedge B$ for $A$ outcome $a_{n}$, and Grouping Axiom \ref{BZ2} is the result of applying the simple grouping axiom to the fine grained joint experiment $n$ times. These are not, then, expressions of the grouping axiom, but rather demonstrate its effect when actually applied to an already well defined joint probability distribution. The absence of certain joint probability distributions in quantum mechanics does not, however, affect the meaningfulness of the grouping axiom, because in its proper formulation it does not refer to joint experiments\footnote{To see that the standard case and the joint experiment case are mathematically distinct, note that the joint experiment formalism cannot express the situation in which $B$ events only happen if the $a_{n}$ event occurs. For that we would require that $p(a_{n})=p(b_{1})+p(b_{2})$, but then the marginal distribution for $B$ outcomes in the joint experiment does not sum to unity as is required for a well defined joint experiment, $\sum_{i} p(b_{i})=p(a_{n}) \neq 1$ in general.}
. (Note that if the outcomes of the experiment in eqn.~(\ref{conditional}) were represented by one dimensional projection operators, the event $b_{1}\vee b_{2}$ would be represented by a sum of orthogonal projectors which commutes with the remaining projectors; a similar relation (\textit{coexistence}) holds if the outcomes are represented by POV elements (\textit{effects}), \citep{busch:1996}.)

Thus we see that the explicit argument fails. There may remain, however, a certain intuitive one. Brukner and Zeilinger are perhaps suggesting that we miss something importantly quantum by using the Shannon information as its derivation is restricted to the case of commuting observables. Or rather (since we have seen that the grouping axiom does not explicitly refer to joint experiments and we know that it is not in fact necessary for the functioning of the Shannon information as a measure of uncertainty anyway), because it can only tell us about the uncertainty in joint experiments for mutually commuting observables rather than the full set of observables. However, when we recall that a measure of uncertainty is a measure of the spread of a probability distribution, we see that this simply amounts to the truism that one cannot have a measure of the spread of a joint probability distribution unless one actually has a joint probability distribution, and it in no way implies that there is anything un-quantum about the Shannon information itself.

The fact that joint probability distributions cannot be defined for all possible groupings of experiments that we might consider does not tell us anything about whether a certain quantity is a good or bad measure of uncertainty for probability distributions that \textit{can} be defined (e.g. any probability distribution derived from a quantum state by the trace rule, or conditional probabilities given by the L\"uders rule). We must be careful not to confuse the question of what makes a good measure of uncertainty with the question of when a joint probability distribution can be defined.

We already know that a function of a joint probability distribution cannot be a way of telling us how much we know, or how uncertain we are, \textit{in general} when we know the state of a system, because we know that a joint probability distribution for all possible measurements does not exist. It is for this reason in quantum mechanics that we introduce measures of mixedness such as the von Neumann entropy, which are functions of the \textit{state} rather than of a probability distribution. It is not a failing of the Shannon information as a measure of uncertainty or expected information gain that it does not fulfil the same role.
Part of Brukner and Zeilinger's worry about the Shannon information thus seems to arise because they are trying to treat it too much like a measure of mixedness, a measure of how uncertain we are in general when we know the state of a quantum system\footnote{This is illustrated for example in their reply to criticism of their grouping axiom argument by Hall \citep{reply,hall}.
Hall presents an interpretation of the grouping axiom concerning the increase in randomness on mixing of non-overlapping distributions, to which Brukner and Zeilinger's worries about joint experiments would not apply. Their reply, in essence, is that the density matrix cannot be simultaneously diagonal in non-commuting bases, therefore it cannot be thought to be composed of non-overlapping classical distributions, hence Hall's grouping axiom will not apply, further supporting their original claim that the Shannon measure is tied to the notion of classical properties. What this reply in fact establishes, however, is that Hall's axiom applied to mixtures of \textit{classical} distributions is not relevant to characterising the randomness of the \textit{density matrix}; but this is something with which everyone would agree, and this job certainly not one for which the Shannon information is intended. (If we did wish to use the grouping axiom in characterising the randomness of the density matrix, we would apply Hall's version to mixtures of density operators with orthogonal support; this would then pick out the von Neumann entropy up to a constant factor \citep{wehrl}.)}. 

\section{Brukner and Zeilinger's `Total information content'\label{finalargument}}

The final argument proposed against the Shannon information is that it is not appropriately related to a notion of `total information content' for a quantum system. It is also suggested that the von Neumann entropy, which has a natural relation to the Shannon information, is not a measure of information content as it makes no explicit reference to information gain from measurements in general \citep{conceptualinadequacy,reply}.

In place of the Shannon information, Brukner and Zeilinger propose the quantity
\begin{equation}
I(\vec{p})={\cal N}\sum_{i=1}^{n}\left (p_{i}-\frac{1}{n}\right )^{2}, \label{BZ info}
\end{equation}
from which they derive their notion of total information content as follows:

\begin{sloppypar}
A set of measurements is called \textit{mutually unbiased} or \textit{complementary} \citep{schwinger} if the sets of projectors $\{P\}, \{Q\}$ associated with any pair of measurement bases satisfy $\mbox{Tr}(PQ)=1/n$, where $n$ is the dimensionality of the system. There can exist at most $n+1$ such bases \citep{optimaldetermination}, constituting a \textit{complete} set,
and as was shown by \citet{ivanovic}, measurement of such a complete set on an ensemble of similarly prepared systems determines their density matrix $\rho$ completely. In analogy to acquiring the information associated with a (pointlike) classical system by learning its state (determining its position in phase space), Brukner and Zeilinger then suggest that the \textit{total information content} of a quantum system should be given by a sum of information measures for a complete set of mutually unbiased measurements. Taking $I(\vec{p})$ as the measure of information, the result is  a unitarily invariant quantity:
\begin{equation}\label{totalinformation}
I_{tot}=\sum_{j=1}^{n+1}I(\vec{p^{j}})=\sum_{ji}\left( p_{i}^{j}-\frac{1}{n} \right) ^{2}=\mbox{Tr}\left( \rho -\frac{\mathbf{1}}{n}\right)  ^2.
\end{equation}

This also provides their argument against the Shannon information. It is a necessary constraint on a measure of total information content, they argue, that it be unitarily invariant, but substituting $H(\vec{p})$ for $I(\vec{p})$ in (\ref{totalinformation}) does not result in a unitarily invariant quantity, that is, we do not have a sum to a `total information content'. Let us call the requirement that a measure sum to a unitarily invariant quantity for a complete set of mutually unbiased measurements the `total information constraint'. The suggestion is that the Shannon measure is inadequate as a measure of information gain because it does not satisfy the total information constraint and hence does not tell us how much of the total information content of a system we learn by performing measurements in a given basis. Similarly, the complaint against the von Neumann entropy is that it is merely a measure of mixedness, as unlike $I_{tot}$, it has no relation to the information gained in a measurement unless we happen to measure in the eigenbasis of $\rho$.
\end{sloppypar}

A few remarks are in order. First, $I(\vec{p})$ and $I_{tot}$ are not unfamiliar expressions. The quantity $\sum_{i}(p_{i}-1/n)^2$ is one of the class of measures of the concentration of a probability distribution given by \citet{jos}, and \citet{fano}, for example, remarks that $\mbox{Tr}(\rho^2)$ can serve as a good measure of information; furthermore, the relation expressed in eqn.~(\ref{totalinformation}) has previously been employed by \citet{larsen} in discussing exact uncertainty relations. Note also that $I(\vec{p})$ is an \textit{increasing} function of the concentration of a probability distribution, hence a measure of \textit{how much we know} given a probability distribution, rather than being a measure of uncertainty like $H(\vec{p})$; similarly $I_{tot}$ is an \textit{increasing} function of the purity of $\rho$.

More importantly, however, `information content' might mean several different things. It may not, then, be reasonable to require that every meaningful information measure sum to a unitarily invariant quantity that can be interpreted as an information content. Moreover, we may well ask why information measures for a complete set of mutually unbiased measurements should be expected to sum to \textit{any} particularly interesting quantity.
That the measure $I(\vec{p})$ happens to sum to a unitarily invariant quantity is, as we shall presently see, the consequence of a geometric property tangential to its role as a measure of information.

\subsection{Some Different Notions of Information Content}\label{differentnotions}

It is useful to distinguish between the information encoded in a system, the information required to specify the state of a system (more precisely, the information required to specify a \textit{sequence} of states drawn from a given ensemble) and states of complete and less than complete knowledge or information. Each of these can serve as a notion of information content in an appropriate context. 
In the classical case, their differences can be largely ignored, but in the quantum case there are important divergences. As we saw in the last chapter, it is necessary, for instance, to introduce the concept of the accessible information to characterize the difference between information encoded and specification information.

If we consider encoding the outputs $a_{i}$ of a classical information source $A$ into pure states $|a_{i}\rangle$ of an ensemble of quantum systems, then the state of the ensemble will be given by $\rho=\sum_{i}p(a_{i})|a_{i}\rangle\langle a_{i} |$. The von Neumann entropy, $S(\rho)=-\mbox{Tr}\rho\log\rho$, is a measure of how mixed this state is, giving us one sense of information content---the more mixed a state, the less information we have about what the outcome of measurements on systems described by the state will be\footnote{Mixed states are also sometimes said to be states of less than complete information due to a lack of information about the way a system was prepared, represented by a probability distibution over possible pure states. Our reading is to be preferred given the many-one relation of preparation procedures to density operators and the fact that density operators can also result from tracing out unwanted degrees of freedom.}.

If we are presented with a sequence of systems drawn from an ensemble prepared in this manner, each will be in one of the pure states, and the number of bits per system required to specify this sequence will be $H(A)$, the information of the classical source (which will be greater than $S(\rho)$ unless we have coded in orthogonal states). As we know, this is the specification information, also the amount of information required to prepare the sequence. For the amount of information that has actually been encoded into the systems, however, we need to consider measurements on the ensemble and the Shannon mutual information $H(A:B)$.

As already remarked (Section~\ref{preexistingbits}), if we encode using a certain basis (our $|a_{i}\rangle$ form an orthonormal set) and we measure in a different basis, then $H(A:B)< H(A)$; quantum `noise' reduces the amount we can decode. More significantly, if we have coded in non-orthogonal states then no measurement can distinguish these states perfectly and we cannot recover the complete classical information.   
The maximum amount of information encoded in a system is given by the accessible information (the maximum over all decoding observables of the mutual information) and using non-orthogonal coding states, the amount we can encode is less than the specification information. As we have seen, the Holevo bound \citep{holevo} provides an upper bound on the mutual information resulting from measurement of any observable (including POV measurements). For the case we are considering of pure encoding states, this reduces to
\[H(A:B) \leq S(\rho).\] 
This provides a very strong sense in which the von Neumann entropy does give us a notion of the total information content of a quantum system---it is the maximum amount that can actually be encoded in the system.      

Brukner and Zeilinger do not consider a quantum communication channel but are concerned rather with the information content of a single system considered in isolation. 
This information content is supposed to relate to how much we learn from learning the state, but if the system is being treated in isolation then by learning its state we are not acquiring a certain amount of information in virtue of the state being drawn from a given ensemble, as in the standard notion of information.
(Hence their analogy with gaining the information content of a classical system fails to hold.) 
In fact, their `total information content' seems best interpreted as a measure of mixedness analogous to the von Neumann entropy.

When introduced \citep{operationallyinvariant}, the information measure $I(\vec{p})$ is presented as a measure of how much we know about what the outcome of a particular experiment will be, given the state. The \textit{total} information of the state, then, would seem to be a measure of how much we know \textit{in general} about what the outcomes of experiments will be given the state; and this is precisely a question of the degree of mixedness of the state\footnote{Recently it has been noted that $I_{tot}$ is also related to the average distance of our estimate of the unknown state from the true state (measured in the Hilbert-Schmidt norm), given only a finite number $N$ of experiments in each mutually unbiased basis \citep{hradil}. This seems best understood as indicating that the mean error in our state estimation is inversely related to $N$, with a constant of proportionality that depends on the dimension of the system and the mixedness of the state. In any case, Brukner and Zeilinger are primarily interested in how much we know when the state has been determined to arbitrary accuracy.}. 

\subsubsection{Measures of mixedness}\label{mixedness}

The functioning of measures of mixedness can usefully be approached via the notions of majorization and Schur convexity (concavity).
The majorization relation $\prec$ imposes a pre-order on probability distributions \citep{jos,characterizingmixing}. A probability distribution $\vec{q}$ is majorized by $\vec{p}$, $\vec{q}\prec\vec{p}$, iff $q_{i}=\sum_{j}S_{ij}p_{j}$, where $S_{ij}$ is a doubly stochastic matrix. That is (via Birkhoff's theorem), if $\vec{q}$ is a mixture of permutations of $\vec{p}$. Thus if $\vec{q} \prec \vec{p}$, then $\vec{q}$ is a more mixed or disordered distribution than $\vec{p}$.

Schur convex (concave) functions respect the ordering of the majorization relation: a function $f$ is Schur convex if, if $\vec{q}\prec \vec{p}$ then $f(\vec{q})\leq f(\vec{p})$, and Schur concave if, if $\vec{q}\prec \vec{p}$ then $f(\vec{q})\geq f(\vec{p})$ (for strictly Schur convex(cave) functions, equality holds only if $\vec{q}$ and $\vec{p}$ are permutations of one another). This explains the utility of such functions as measures of the concentration and uncertainty of probability distributions, respectively.

The majorization relation will apply equally to the vectors of eigenvalues of density matrices. 
It can be shown that the vector of eigenvalues $\vec{\lambda^{\prime}}$ of the density matrix $\rho^{\prime}$ of the post measurement ensemble for a (non-selective) projective measurement is majorized by the vector of eigenvalues $\vec{\lambda}$ of the pre-measurement state $\rho$ \citep{characterizingmixing}. (If we measure in the eigenbasis of $\rho$, then there is, of course, no change in the eigenvalues). The $\lambda^{\prime}_{i}$ are just the probabilities of the different outcomes of the measurement in question, thus the probability distribution for the outcomes of any given measurement will be more disordered or spread than the eigenvalues of $\rho$.

If we take any Schur concave function we know to be a measure of uncertainty, for instance the Shannon information $H(\vec{p})$, and $\vec{p}$ is the probability distribution for measurement outcomes, we then know that $H(\vec{p})\geq H(\vec{\lambda})$, for any projective measurement we might perform. This explains why $H(\vec{\lambda})=S(\rho)$, a measure of mixedness, is a measure of how much we know given the state: the more mixed a state, the more uncertain we must be about the outcome of any given measurement. Similarly, if we take a measure of the concentration of a probability distribution, a Schur convex function such as $I(\vec{p})$, then we know that for any measurement with outcome probability distribution $\vec{p}$, $I(\vec{p})\leq I(\vec{\lambda})=I_{tot}$; and this explains why $I_{tot}$ is a measure of how much we know given $\rho$: the less the value of $I_{tot}$, the less able we are to predict the outcome of any given experiment. Note, however, that it is the structure that is imposed by the majorization relation that is of underlying importance. Choosing a particular measure $H(\vec{p})$, $I(\vec{p})$, or any other member $U_{r}(\vec{p})$ of Uffink's general class of measures of uncertainty, is simply a matter of choosing by convention a numerical measure to lay on top of this structure, for convenience.       

Brukner and Zeilinger would of course deny that their total information content is merely a measure of mixedness. The argument that it is more than this rests on the satisfaction of the total information constraint, the relation between the measure of information $I(\vec{p})$ and $I_{tot}$ for a complete set of mutually unbiased measurements as expressed in eqn.~(\ref{totalinformation}). We shall now see that this relation can be given a simple geometric explanation using the Hilbert-Schmidt representation of density operators.

\subsection{The Relation between Total Information Content and $I(\vec{p})$}\label{Hilbert-Schmidt}

The set of complex $n\times n$ Hermitian matrices forms an $n^{2}$-dimensional real Hilbert space $V_{h}(\mathbb{C}^n)$ on which we have defined an inner product $(A,B)=\mbox{Tr}(AB); A,B \in V_{h}(\mathbb{C}^n)$ and a norm $\|A\|=\sqrt{\mbox{Tr}(A^2)}$ \citep{fano,wichmann}.
The density matrix $\rho$ of an $n$ dimensional quantum system can be represented as a vector in this space. The requirements on $\rho$ of unit trace and positivity imply that the tip of any such vector must lie in the $n^2 -1$ dimensional hyperplane $\bm{T}$ a distance $1/\sqrt{n}$ from the origin and perpendicular to the unit operator $\mathbf{1}$, and on or within a hypersphere of radius one centred on the origin.

It is useful to introduce a set of basis operators on our space; we require $n^{2}$ linearly independent operators $\Gamma_{i}\in V_{h}(\mathbb{C}^n)$ and it may be useful to require orthogonality and a fixed norm: $\mathrm{Tr}(\Gamma_{i}\Gamma_{j})=\mbox{const.}\times \delta_{ij}$.
Any operator on the system can then be expanded in terms of this basis and in particular, $\rho$ can be written as a vector 
\[ \boldsymbol{\varrho}=\frac{\mathbf{1}}{n} + \sum_{i=1}^{n^{2}-1}\mbox{Tr}(\rho \Gamma_{i})\Gamma_{i},\]
where we have chosen $\Gamma_{0}=\mathbf{1}$ to take care of the trace condition. A very familiar example of this formalism is the Bloch sphere for two-dimensional quantum systems. Here the basis operators $\Gamma_{i}$ are given by the three Pauli matrices $\sigma_{i}$.

Evidently, $\rho$ may be determined experimentally by finding the expectation values of the $n^{2}-1$ operators $\Gamma_{i}$ in the state $\rho$. If we include the operator $\mathbf{1}$ in our basis set, then the one-dimensional projectors associated with measurement of any maximal (non-degenerate) observable will provide a maximum of a further $n-1$ linearly independent operators. Obtaining the probability distribution for a given maximal observable will thus provide $n-1$ of the parameters required to determine the state, and the minimum number of measurements of maximal observables that will be needed in total is $n+1$, if each observable provides a full complement of linearly independent projectors. 

Each such set of projectors spans an $n-1$ dimensional hyperplane in $V_{h}(\mathbb{C}^n)$ and their expectation values specify the projection of the state $\rho$ into this hyperplane. Ivanovic (1981) noted that projectors $P,Q$ belonging to any two different mutually unbiased bases will be orthogonal in $\bm{T}$, hence the hyperplanes associated with measurement of mutually unbiased observables are orthogonal in the space in which density operators are constrained to lie in virtue of the trace condition. If $n+1$ mutually unbiased observables can be found, then, $V_{h}(\mathbb{C}^n)$ can be decomposed into orthogonal subspaces given by the one dimensional subspace spanned by $\mathbf{1}$ and the $n+1$ subspaces associated with the mutually unbiased observables. The state $\rho$ can then be expressed as:
\begin{equation}
\boldsymbol{\varrho}= \frac{\mathbf{1}}{n} +\sum_{j=1}^{n+1}\sum_{i=1}^{n}q_{i}^{j}\bar{P}_{i}^{j}, \label{rho}
\end{equation}
where $\bar{P}_{i}^{j}=P_{i}^{j}-\mathbf{1}/n$ is the projection onto $\bm{T}$ of the $i$th one-dimensional projector in the $j$th mutually unbiased basis set, and $q_{i}^{j}=(p_{i}^{j}-1/n)$ is the expectation value of this operator in the state $\rho$.
 For a given value of $j$, the vectors $\bar{P}_{i}$ span an $(n-1)$ dimensional orthogonal subspace and the square of the length of a vector expressed in the form (\ref{rho}) lying in subspace $j$ will be given by $\sum_{i=1}^{n}(q_{i}^{j})^2 = I(\vec{p^{j}})$.       

The geometrical explanation of $I_{tot}$ is then simply as follows. $\mbox{Tr}(\rho^{2})$ is the square of the length of $\boldsymbol{\varrho}$ in $V_{h}(\mathbb{C}^n)$ and $\mbox{Tr}(\rho - \mathbf{1}/n)^{2}$ is the square of the distance of $\rho$ from the maximally mixed state (the length squared of $\rho$ in $\bm{T}$). This squared length will just be the sum of the squares of the lengths of the components of the vector $\rho-\mathbf{1}/n$ in the orthogonal subspaces into which we have decomposed $V_{h}(\mathbb{C}^n)$, i.e. it will be given by $\sum_{ji}(q_{i}^{j})^2$. This is what eqn.~(\ref{totalinformation}) reports and it explains how $I_{tot}$ and $I(\vec{p})$ satisfy the total information constraint.

Thus we see that if $I_{tot}$ differs from being a simple measure of mixedness, then that is because it is a measure of length also; and this explains why it can be given by a sum of quantities $I(\vec{p})$ for a complete set of mutually unbiased measurements. As measures of \textit{how much we know given the state}, however, $I_{tot}$ and $S(\rho)$ bear the same relation to their appropriate measures of information, as we saw in the previous section. Equally, as measures of information, $H(\vec{p})$ stands to $S(\rho)$ in the same relation as $I(\vec{p})$ to $I_{tot}$.  
$I_{tot}$ is the \textit{upper} bound on the amount we can \textit{know} about the outcome of a measurement as measured by $I(\vec{p})$; $S(\rho)$ is the \textit{lower} bound on our \textit{uncertainty} about what the outcome of a measurement will be, as measured by $H(\vec{p})$.

The complaint against the Shannon information was supposed to be that as $H(\vec{p})$ fails to satisfy the total information constraint, it does not tell us the information gained from a particular measurement; the complaint against the von Neumann entropy that as $S(\rho)$ is not given by a sum of measures for a complete set of mutually unbiased measurements, it is not suitably related to the information gained from a measurement.
However, we can now see that insisting on the total information constraint in this way is tantamount to insisting that only a function which measures the length of the component of $\rho$ lying in a given hyperplane can be a measure of information, and correlatively, that the only viable notion of total information content is a measure of the length of $\rho$ in $V_{h}(\mathbb{C}^{n})$.
But $H(\vec{p})$ can be a perfectly good measure of information without having to be a measure of the length of the projection of $\rho$ into the subspace associated with an observable; and as we have just seen, $S(\rho)$ does have an explicit relation to the information gain from measurement that justifies its interpretation as a total information content. A relation, moreover, that $I_{tot}$ also possesses and which serves to justify \textit{its} interpretation as a measure of how much we know given the state.

Hence our conclusion must be that the total information constraint is not a reasonable requirement on measures of information.   

Of course, $H(\vec{p})$ does not tell us the information gain on measurement it we take, as Brukner and Zeilinger seem to, `the information encoded in a basis' simply to \textit{mean} the length squared of the component of the state lying in the measurement hyperplane; but this is a non-standard usage. $H(\vec{p})$ certainly remains a measure of our expected information gain from performing a particular measurement (how much the outcome will surprise us, on average, given that we have the probability distribution); and if we are interested in the amount of information encoded, in the usual sense of the word, that can be decoded using a particular measurement, i.e., if we have a string of systems into which information has actually been encoded, then we may always just consider the Shannon mutual information associated with that measurement. (The `total information' associated with this quantity will then be given, via the Holevo bound, as the von Neumann entropy, for pure encoding states.)  

Perhaps the fundamental error in this final argument of Brukner and Zeilinger is their failure to appreciate that the choice of measure of information one should adopt is a largely conventional matter, depending on what one's aims are and accordingly, which measure is most convenient. As such, trying to use the total information constraint to rule certain measures out as incorrect is simply mistaken. The geometrical property that the measures $I(\vec{p})$ and $I_{tot}$ possess is indeed a nice one that will be useful in certain contexts, for example, if one wishes to provide certain exact uncertainty relations instead of inequalitites \citep{larsen}. But this just serves to make the point that it should be horses for courses. As we have seen, the Shannon information and von Neumann entropy have multiple important and central uses as measures of information in the quantum context.

\section{Conclusion}

Of the three arguments that Brukner and Zeilinger have presented against the Shannon information, we have seen that the first two fail outright. These arguments sought to establish that the notion of the Shannon information is undermined in the quantum context due to a reliance on classical concepts. With regard to the first we saw that, contrary to Brukner and Zeilinger, the existence of a pre-determined string is neither necessary nor sufficient for the interpretation of $H(\vec{p})$ as a measure of information, hence the absence of such a string would not cause any problems for the Shannon information in quantum mechanics.

The objective of their second argument was to highlight classical assumptions in the grouping axiom that would prevent the axiomatic derivation of the Shannon information going through in the quantum case. This argument turned out to be based on an erroneous reading of the grouping axiom that appeals to joint experiments. The grouping axiom is in fact perfectly well defined in the quantum case and the standard axiomatic derivation of the form of $H(\vec{p})$ can indeed go through. 
The grouping axiom does not reveal any problematic classical assumptions implicit in the Shannon information.

In their final argument, Brukner and Zeilinger suggest that defining the notion of the total information content of a quantum system in terms of the Shannon information would lead to a quantity with the unnatural property of unitary non-invariance. But this is not a compelling argument against the Shannon quantity as a measure of information. We have seen that it is not a necessary requirement on every meaningful measure of information that it sum to a unitarily invariant quantity for a complete set of mutually unbiased measurements; nor, conversely, is it necessary that every viable notion of total information content be given by such a sum of individual measures of information. 

Brukner and Zeilinger's arguments thus fail to establish that the Shannon information involves any particularly classical assumptions or that there is any difficulty in the application of the Shannon measure to measurements on quantum systems. The Shannon information is perfectly well defined and appropriate as a measure of information in the quantum context as well as in the classical.


\end{doublespacing}


\chapter{Case Study: Teleportation}\label{study}





\begin{doublespacing}

\section{Introduction}

The phenomenon of teleportation \citep{teleportation} is perhaps the most striking example of entanglement assisted communication. It illustrates several distinctive features associated with quantum information protocols, most notably the fact that entanglement (a characteristically quantum property) serves as an important resource, and that unknown quantum states cannot be cloned. Our main concern in this chapter, though, will be to consider teleportation as an example of how conceptual puzzles can arise if one thinks of information in the wrong way. That is, if one neglects the fact that `information' is an abstract noun. For teleportation has certainly often been seen as a conceptually puzzling process. I will suggest that these puzzles generally arise as a consequence of a familiar philosophical error---in fact the one that Strawson warns of in the epigraph to this Part---that is, the error of assuming that every grammatical substantive, in this instance the word `information', is a referring term. Let us begin with a brief review of the teleportation protocol\footnote{Helpful discussions of further conceptual aspects of teleporation, in particular concerning the relation of teleportation to nonlocality, may be found in \citet{hardy:disentangling}, \citet{jon1} and \citet{cliftonpope}. \citet{mermin:teleportation} also provides an interesting perspective.}.

\section{The quantum teleportation protocol}

In the teleportation protocol we consider two parties, Alice and Bob, who are widely separated, but each of whom possess one member of a pair of particles in a maximally entangled state. Alice is presented with a system in some unknown quantum state, and her aim is to transmit this state to Bob. In the standard example, Alice and Bob share one of the four Bell states  and she is presented with a spin-1/2 system in the unknown state $\ket{\chi}{}=\alpha\ket{\up}{}+\beta\ket{\down}{}$.

\begin{figure}
\includegraphics{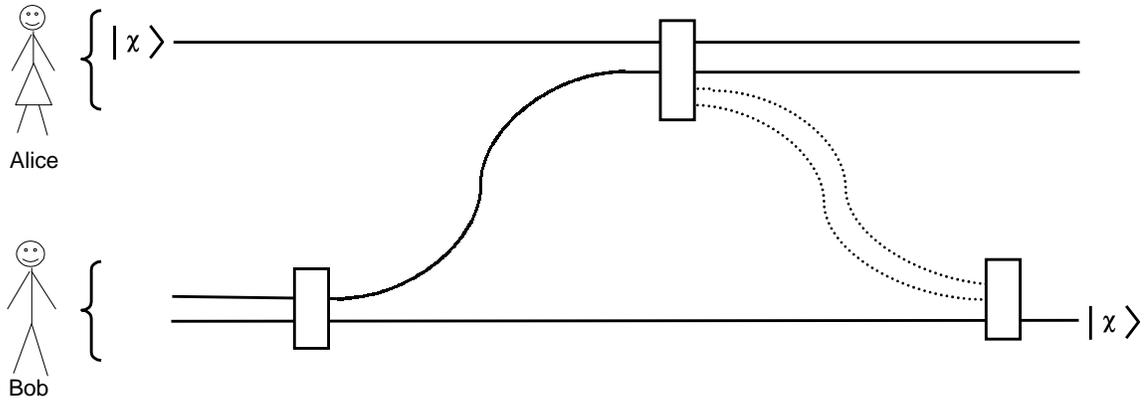}
\caption{\small Teleportation. A pair of systems is first prepared in an entangled state and shared between Alice and Bob, who are widely spatially separated. Alice also possesses a system in an unknown state $\ket{\chi}{}$. Once Alice performs her Bell-basis measurement, two classical bits recording the outcome are sent to Bob, who may then perform the required conditional operation to obtain a system in the unknown state $\ket{\chi}{}$. (Continuous black lines represent qubits, dotted lines represent classical bits. Time runs along the horizontal axis.)\label{telep1}}
\end{figure}

By performing a suitable joint measurement on her half of the entangled pair and the system whose state she is trying to transmit (in this example, a measurement in the Bell basis), Alice can flip the state of Bob's half of the entangled pair into a state that differs from $\ket{\chi}{}$ by one of four unitary transformations, depending on what the outcome of her measurement was. If a record of the outcome of Alice's measurement is then sent to Bob, he may perform the required operation to obtain a system in the state Alice was trying to send (Fig.~{\ref{telep1}}). 

The result of the protocol is that Bob has obtained a system in the state $\ket{\chi}{}$, with nothing that bears any relation to the identity of this state having traversed the space between him and Alice. Only two classical bits recording the outcome of Alice's measurement were sent between them; and the values of these bits are completely random, with no dependence on the parameters $\alpha$ and $\beta$. Meanwhile, no trace of the identity of the unknown state remains in Alice's region, as is required in accordance with the no-cloning theorem (the state of her original system will usually now be maximally mixed). The state has `disappeared' from Alice's region and `reappeared' in Bob's, hence the use of the term \textit{teleportation} for this phenomenon.

Of course, this quantum mechanical process differs from science fiction versions of teleportation in at least two ways. First, it is not \textit{matter} that is transported, but simply the quantum state $\ket{\chi}{}$; and second, the protocol is not instantaneous, but must attend for its completion on the arrival of the classical bits sent from Alice to Bob. Whether or not the quantum protocol approximates to the science fiction ideal, however, it remains a very remarkable phenomenon from the information-theoretic point of view\footnote{Interestingly, it can be argued that quantum teleporation is perhaps not so far from the sci-fi ideal as one might initially think. \citet{vaidman} suggests that if all physical objects are made from elementary particles, then what is distinctive about them is their form (i.e. their particular state) rather than the matter from which they are made. Thus it seems one could argue that \textit{objects} really are teleported in the protocol.}. For consider what has been achieved. An unknown quantum state has been sent to Bob; and how else could this have been done? Only by Alice sending a quantum system \textit{in} the state $\ket{\chi}{}$ to Bob\footnote{Or by her sending Bob a system in a state explicitly related to $\ket{\chi}{}$ \citep[cf.][]{park:1970}.}, for she cannot determine the state of the system and send a description of it instead. (Recall, it is impossible to determine an unknown state of an individual quantum system.)

If, however, Alice did \textit{per impossibile} somehow learn the state and send a description to Bob, then systems encoding that description would have to be sent between them. In this case something that \textit{does} bear a relation to the identity of the state is transmitted from Alice to Bob, unlike in teleportation. Moreover, sending such a description would require a \textit{very great deal} of classical information, as in order to specify a general state of a two dimensional quantum system, two \textit{continuous} parameters need to be specified.

The picture we are left with, then, is that in teleportation there has been a transmission of something that is inaccessible at the classical level (often loosely described as a transmission of \textit{quantum} information); in the transmission this information has been in some sense disembodied; and finally, the transmission has been very efficient---requiring, apart from prior shared entanglement, the transfer of only two classical bits.

\subsection{Some information-theoretic aspects of teleportation}      

There are two information-theoretic aspects of the teleporation protocol it is helpful to go into in somewhat more detail. 
The first concerns our reason for saying that a very large amount of information is required to specify the state that is teleported. 

As we just noted, in order to describe an arbitrary (pure) state of a two dimensional quantum system, it is necessary to specify two continuous parameters. Recalling the Bloch sphere representation (cf. Section~\ref{Hilbert-Schmidt}), we may specify two real numbers (angles) to determine a point on the sphere. Why should doing this have associated with it an amount of information? If it is to do so we will need to imagine a classical information source that is selecting these pairs of angles with various probabilities; then a certain Shannon information may be ascribed to the process. Given a particular output of this information source, a quantum system is prepared in the state corresponding to the two angles selected. The quantum states prepared in this manner will then have associated with them a \textit{specification} information (cf. Section~\ref{aspects}) given by the information of the source. Once a system has been prepared in some state in this way, it is presented to Alice, who may proceed to teleport the state to Bob.

Rather than the pairs of angles being selected from their full, continuous, range of possible values, the surface of the sphere might be coarse-grained evenly to give a finite number of choices. One might pick the angles specifying the mid-point, say, of each small element of surface area to provide the finite set of pairs of angles to choose between. Loosely speaking this coarse-graining corresponds to considering angles only to a certain degree of accuracy. As this accuracy is increased (the choices made more finely grained), the number of bits required to specify the choice increases without bound. If our information source is selecting states to an arbitrarily high accuracy then, the specification information is unboundedly large. (On the other hand, if the information source is only selecting between a small number of distinct states, then the specification information is correspondingly small. From now on we will assume that unless otherwise stated, the unknown states to be teleported are selected from a suitable coarse-graining of the whole range of possible angles.) It is essential to note, however, that even if the specification information associated with the state that has been teleported to Bob is exceedingly large, the majority of this information is not accessible to him. This leads on to the second point.

As will be recalled from the earlier discussion (Sections~\ref{aspects}; \ref{differentnotions}) when one considers encoding classical information in quantum systems, it is necessary to distinguish between specification information and accessible information. The specification information refers to the information of the classical source that selects sequences of quantum states, the accessible information to the maximum amount of information that is available following measurements on the systems prepared in these states. In teleportation, of course, the systems are prepared near Alice before teleportation of their states to Bob. He may then perform various measurements to try and learn something. Call the information of the source selecting the states to be teleported by Alice $H(A)$; the mutual information $H(A:B)$ will determine the amount of classical information per system that Bob is able to extract by performing some measurement, $B$, following successful teleportation of the unknown state. The accessible information is given by the maximum over all decoding measurements of $H(A:B)$. As we know, the Holevo bound restricts the amount of information that Bob may acquire to a maximum of one bit of information per qubit, that is, to a maximum of one bit of information per successful run of the teleportation protocol.

So this gives us the sense in which the very large amount of information that may be associated with the unknown state being teleported to Bob is largely inaccessible to him. Note that the amount of information that Bob may acquire from the teleported state is less than the amount of classical information---two bits---that Alice had to send to him during the protocol. This fact is of the utmost importance, for if the Holevo bound did not guarantee this, and Bob were able to extract more than two bits of information from his system, then teleportation would give rise to paradox (when embedded in a relativistic theory) as superluminal signalling would be possible\footnote{The argument parallels the one given by \citet{teleportation} to the effect that two full classical bits are required in teleportation. In essence, if Bob were able to gain more than two bits of information in the protocol, then even if he were not to wait for Alice to send him the pair of bits each time and simply guessed their values instead, then some information would still get across.}.


So the Holevo bound ensures that teleportation is not paradoxical, but it also means that teleportation, when considered as a mode of ordinary \textit{classical} information transfer, is pretty inefficient, requiring two classical bits to be sent for every bit of information that Bob can extract at his end.

\section{The puzzles of teleportation}

Let us return to the picture of teleportation that was sketched earlier. An unknown quantum state is teleported from Alice to Bob with nothing that bears any relation to the identity of the state having travelled between them. The two classical bits sent are quite insufficient to specify the state teleported; and in any case, their values are independent of the parameters describing the unknown state. The unboundedly large specification information characterizing the state---information that is inaccessible at the classical level---has somehow been disembodied, and then reincarnated at Bob's location, as the quantum state first `disappears' from Alice's system and then `reappears' with Bob.

The conceptual puzzles that this process presents seem to cluster around two essential questions. First, how is \textit{so much} information transported? And second, most pressingly, just \textit{how} does the information get from Alice to Bob?

Perhaps the prevailing view on how these questions are to be answered is the one that has been expressed by \citet{jozsa:1998,jozsa:2003} and \citet{penrose:1998}. In their view, the classical bits used in the protocol evidently can't be carrying the information, for the reasons we have just rehearsed; therefore the entanglement shared between Alice and Bob must be providing the channel down which the information travels. They conclude that in teleportation, an indefinitely large, or even infinite amount of information travels backwards in time from Alice's measurement to the time at which the entangled pair was created, before propagating forward in time from that event to Bob's performance of his unitary operation and the attaining by his system of the correct state. Teleportation seems to reveal that entanglement has a remarkable capacity to provide a hitherto unsuspected type of information channel, which allows information to travel backwards in time; and a very great deal of it at that. Further, since it is a purely quantum link that is providing the channel, it must be purely \textit{quantum} information that flows down it. It seems that we have made the discovery that quantum information is a new \textit{type} of information with the striking, and non-classical, property that it may flow backwards in time.   

The position is summarized succinctly by Penrose:

\begin{quoting}
How is it that the \textit{continuous} `information' of the spin direction of the state that she [Alice] wishes to transmit...can be transmitted to Bob when she actually sends him only two bits of discrete information? The only other link between Alice and Bob is the quantum link that the entangled pair provides. In spacetime terms this link extends back into the past from Alice to the event at which the entangled pair was produced, and then it extends forward into the future to the event where Bob performs his [operation].

Only \textit{discrete} classical information passes from Alice to Bob, so the complex number ratio which determines the specific state being `teleported' must be transmitted by the \textit{quantum} link. This link has a channel which `proceeds into the past' from Alice to the source of the EPR pair, in addition to the remaining channel which we regard as `proceeding into the future' in the normal way from the EPR source to Bob.There is no other physical connection.
\citep[p.1928]{penrose:1998}
\end{quoting}

But one might feel, with good reason, that this explanation of the nature of information flow in teleportation is simply too outlandish. This is the view of \citet{dh}, who conclude instead that with suitable analysis, the message sent from Alice to Bob can, after all, be seen to carry the information characterizing the unknown state. The information flows from Alice to Bob hidden away, unexpectedly, in Alice's message. This approach, and the question of what light it may shed on the notion of quantum information, is considered in detail in the next chapter. Suffice it to say at present that Deutsch and Hayden disagree with Jozsa and Penrose over the nature of quantum information and how it may flow in teleportation.

One might adopt yet a third, and perhaps more prosaic response to the puzzles that teleportation poses. This is to adopt the attitude of the \textit{conservative classical quantity surveyor}\footnote{A resolution along these lines, tied also to an ensemble view of the quantum state (\textit{vide infra}) has been suggested by \citet{jon1} and \citet{petermorgan}.\label{ccqs ensemble}}. According to this view, an amount of information cannot be said to have been transmitted to Bob unless it is accessible to him. But of course, as we noted above, the specification information associated with the state teleported to Bob is \textit{not} accessible to him: he cannot determine the identity of the unknown state. On this view, then, the information associated with selecting some unknown state $\ket{\chi}{}$ will not have been transmitted to Bob until an entire ensemble of systems in the state $\ket{\chi}{}$ has been teleported to him, for it is only then that he may determine the identity of the state\footnote{Note that we will need to adjust our scenario slightly to incorporate this view. In our initial set-up, the source $A$ selected a sequence of states which were then teleported one by one to Bob. Now we imagine instead that following some particular output of $A$, an entire ensemble of systems is prepared in the pure state associated with that output; then this ensemble of systems --- all in the same unknown pure state --- is teleported. This adjustment is required because in our initial set-up for the teleportation procedure, the only way in which an ensemble of systems all in the same state could be teleported to Bob would be by setting the information of the source $A$ to zero, with the tiresomely paradoxical result that Bob could now determine the state all right, but would gain no information by doing so.}. To teleport a whole ensemble of systems, though, Alice will need to send Bob an infinite number of classical bits; and now there isn't a significant disparity between the amount of information that has been explicitly sent by Alice and the amount that Bob ends up with. One needs to send a very large number of classical bits to have transmitted by teleportation the very large amount of information associated with selecting the unknown state. 

This approach does not seem to solve all our problems, however. Someone sympathetic to the line of thought espoused by Jozsa and Penrose can point out in reply that there still remains a mystery about \textit{how} the information characterizing the unknown state got from Alice to Bob --- the bits sent between them, recall, have no dependence on the identity of the unknown state. So while the approach of the conservative classical quantity surveyor may mitigate our worry to some extent over the first question, it does not seem to help with the second.


\section{Resolving (dissolving) the problem}\label{dissolving}

Dwelling on the question of how the information characterizing the unknown state is transmitted from Alice to Bob has given rise to some conundrums. Should we side with Jozsa and Penrose and admit that quantum information may flow backwards in time down a channel constituted by shared entanglement? Or perhaps with Deutsch and Hayden, and agree that information should flow in a less outlandish fashion, but that quantum information may be squirrelled away in seemingly classical bits? Counting conservatively the amounts of information available after teleportation may make us less anxious about the load carried in a single run of the protocol, but the question still remains: how did the information, in the end, get to Bob? Should we just conclude that it is transported nonlocally in some way? But what might that mean?

If the question `How does the information get from Alice to Bob?' is causing us these difficulties, however, perhaps it might pay to look at the question itself rather more closely. In particular, let's focus on the crucial phrase `the information'.

Our troubles arise when we take this phrase to be referring to a particular, to some sort of substance or entity whose behaviour in teleportation it is our task to describe. The assumption common to the approaches of Deutsch and Hayden on the one hand, and Jozsa and Penrose on the other, is that we need to provide a story about how some \textit{thing} denoted by `the information' travels from Alice to Bob. Moreover, it is assumed that this supposed thing should be shown to take a spatio-temporally continuous path.

But recall that `information' is an abstract noun. This means that `the information' certainly does \textit{not} refer to a substance or to an entity. The shared assumption is thus a mistaken one, and is based on the error of hypostatizing an abstract noun. (We shall return to this issue in the context of the Deutsch-Hayden approach once again in the following chapter). If `the information' doesn't introduce a particular, then the question `How does the information get from Alice to Bob?' cannot be a request for a description of how some thing travels. It follows that the locus of our confusion is dissolved.

But if it is a mistake to take `How does the information get from Alice to Bob?' as a question about how some thing is transmitted, then what is its legitimate meaning, if any? It seems that the only legitimate use that can remain for this question is as a flowery way of asking: what are the physical processes involved in the transmission? Now \textit{this} question is a perfectly straightforward one, even if, as we shall see (Section~\ref{interpretations}), the answer one actually gives will depend on the interpretation of quantum mechanics one adopts. But there is no longer a \textit{conceptual} puzzle over teleportation. Once it is recognised that `information' is an abstract noun, then it is clear that there is no further question to be answered regarding how information is transmitted in teleportation that goes beyond providing a description of the physical processes involved in achieving the aim of the protocol. That is all that `How is the information transmitted?' can intelligibly mean; for there is not a question of information being a substance or entity that is transported, nor of `the information' being a referring term. Thus, one does not face a double task consisting of a) describing the physical processes \textit{by which} information is transmitted, followed by b) tracing the path of a ghostly particular, information. There is only task (a). 

The point should not be misunderstood:  I am not claiming that there is no such thing as the transmission of information,
but simply that one should not understand the transmission of information on the model of transporting potatoes, or butter, say, or piping water\footnote{Note that we do sometimes talk of a flow of information; and we do say of many physical quantities that are not entities or substances --- for example, energy, heat --- that they flow. But there is no analogy between the two cases, for what this latter description means is that the quantities in question obey a local conservation equation. It is not clear that it is at all intelligible to suggest that information should obey a local conservation equation. Certainly, the concept of quantity of information that is provided by the Shannon theory does not give us a concept of a quantity it makes sense to suggest might obey such an equation. (On this, see Section~\ref{study concluding} below.)}. 

\subsection{The simulation fallacy}\label{simulation fallacy}

Whilst paying due attention to the status of `information' as an abstract noun provides the primary resolution of the problems that teleportation can sometimes seem to present us with, there is a secondary possible source of confusion that should be noted. This is what may be termed the \textit{simulation fallacy}.

Imagine that there is some physical process $\cal{P}$ (for example, some quantum-mechanical process) that would require a certain amount of communication or computational resources to be simulated classically. Call the classical simulation using these resources $\cal{S}$. The simulation fallacy is to assume that because it requires these classical resources to simulate $\cal{P}$ using $\cal{S}$, there are processes going on when $\cal{P}$ occurs that are physically equivalent to (are instantiations of) the processes that are involved in the simulation $\cal{S}$ itself (although these processes may be being instantiated using different properties in $\cal{P}$). In particular, when $\cal{P}$ is going on, the thought is that there must be, at some level, physical processes involved in $\cal{P}$ which correspond concretely to the evolution of the classical resources in the simulation $\cal{S}$.
The fallacy is to read off features of the simulation as real features of the thing simulated\footnote{Note that it will not always be fallacious to take features of a simulation to correspond to features of the simulated --- if the features in question are explicitly \textit{analogues} of features of the system or process being simulated. One should thus distinguish between i) simulations that involve analogues and ii) functional `black-box', or input-output simulations.}.

A familiar example of the simulation fallacy is provided by Deutsch's argument that Shor's factoring algorithm supports an Everettian view of quantum mechanics \citep[p.217]{FoR}. The argument is that if factoring very large numbers would require greater computational resources than are contained in the visible universe, then how could such a process be possible unless one admits the existence of a very large number of (superposed) computations in Everettian parallel universes? A computation that would require a very large amount of resources if it were to be performed classically is explained \textit{as} a process that consists of a very large number of classical computations. But of course, considered as an argument, this is fallacious. The fact that a very large amount of classical computation might be required to produce the same result as a quantum computation does not entail that the quantum computation consists of a large number of parallel classical computations\footnote{For further discussion of Deutsch's conception of quantum `parallel processing', see \citet{steane} and \citet{horsman}.}.   

The simulation fallacy is also evident in the common claim that Bell's theorem shows us that quantum mechanics is nonlocal, or the claim that the experimental violation of Bell inequalities means that the world must be nonlocal. Of course, what is in fact shown by these well-known results is that no local hidden variable model can simulate the predictions of quantum mechanics, nor provide a model for the experimentally observed correlations. But these facts about simulation don't lead directly to facts about the simulated: the fact that any adequate hidden variable model must be nonlocal does not show that quantum mechanics is nonlocal (this, of course, is an interpretation dependent property), nor show the world to be nonlocal.

While the question of what classical resources would be required to simulate a given quantum process is an indispensible guide in the search for interesting quantum information protocols and is vitally important for that reason, the simulation fallacy indicates that it is by no means a sure guide to ontology. 

With regard to teleportation, it is important to recognise the simulation fallacy in order to assuage any worries that might remain over the question `How does so much information get from Alice to Bob?', and to undermine further the thought that teleportation must be understood as a flow of information. 

For the fact that it would take a very large number of classical bits to transmit the identity of an unknown state from Alice to Bob does not entail that
in teleportation there is a real corresponding transmission of information, some physical process going on that instantiates, albeit in a different medium, the transport of this large amount of information\footnote{Nor, for example, does the fact that there are protocols in which the state of a qubit can be substituted for an arbitrarily large amount of classical information \citep{lucien:substituting} imply that this large amount of information is really there in the qubit. }. (Note that the flow of the hypostatized `quantum information' of Jozsa and Penrose plays precisely this r\^ole: the analogue, in a different medium, of the transport of the large amount of classical information.) Equivalence from the point of view of information processing does not imply physical equivalence. 

Awareness of the simulation fallacy is particularly relevant when we consider the approach of the conservative classical quantity surveyor. Recall that the point of this approach is to deny that a large amount of information can be said to have been transported to Bob in teleportation until that information is actually available to him. However, it might be objected to this that after a single run of the teleportation protocol, the information characterizing the state is certainly present at Bob's location, even if inaccessible to him, as a system \textit{in} the unknown state is present\footnote{It is for this reason that it is natural to marry conservative classical quantity surveying with an ensemble view of the quantum state (see footnote~\ref{ccqs ensemble}), for then this objection would not go through --- when the two positions are conjoined, not only is the information characterizing the state not available until the whole ensemble is teleported, but neither has the \textit{state} been teleported until the whole ensemble has been teleported to Bob.}.

This contention would seem to rest on an argument of the following form:
The only way the unknown state can appear at Bob's location is if the information characterizing the state has actually been transported to Bob, hence on appearance of the state, the specification information associated with the state has indeed been transported to Bob's location. (Crudely, if a system in the given state is present, then the information is present, as it takes this information to specify the state.) But such an argument needs to be treated with care, for the main premise appears to rest on the simulation fallacy. Just because it would take a large amount of information to specify a state doesn't mean that we should conclude that this amount of information has been physically transported in teleportation when Bob's system acquires the state.

In any event, the simplest way to remain clear on whether or not, or in what way, information can be said to be present at Bob's location following a single run of the teleportation protocol is to respect the distinction between the specification information associated with a system and the amount of information that may be said to be encoded or contained in the system. Once Bob's system has acquired the state $\ket{\chi}{}$ teleported by Alice, then his system has associated with it the same specification information, $H(A)$: \textit{if} one were now \textit{asked} to specify the state of Bob's system, then this number of bits would be required, on average. This quantity of information is not encoded or contained in the system however. The mutual information $H(A:B)$ and the accessible information provide the relevant measures of how much information Bob's system can be said to contain, for they govern the amount that may be decoded. But of course, as `information' is an abstract noun, containing information is not containing some \textit{thing}, however aethereal.

\section{The teleportation process under different interpretations}\label{interpretations}

By reflecting on the logico-grammatical status of the term `information' we have been able to replace the (needlessly) conceptually puzzling question of how the information gets from Alice to Bob in teleportation, with the simple, genuine, question of what the physical processes involved in teleportation are. While this may not, perhaps, be quite enough to still all the controversy that trying to understand teleportation has evoked, the controversy is now of a very familiar kind: it concerns what interpretation of quantum mechanics one adopts. For the detailed story one tells about the physical processes involved in teleporation will of course depend upon one's interpretive stance. Two questions in particular will find different answers under different interpretations: first, is nonlocality involved in teleportation? and second, has anything interesting happened before Alice's classical bits are sent to Bob and he performs the correct unitary operation? 

We will now see how some of these differences play out in the following familiar interpretations (the list of approaches considered is by no means exhaustive).   

\subsection{Collapse interpretations: Dirac/von Neumann, GRW}

The natural place to begin is with the orthodox approach of \citet{dirac} and \citet{vN} in which there is a genuine process of collapse on measurement\footnote{One of the defining features of what I here term `orthodoxy' is the adoption of the standard eigenstate-eigenvalue link for the ascription of definite values to quantum systems. See e.g. \citet{bub:1997}.}. (The vagueness over where, when, why and how this collapse takes place might be alleviated along lines suggested by \citet{grw}, perhaps.) If one has a genuine process of collapse then as noted long ago by \citet*{EPR}\footnote{See \citet{erpart1} for a recent discussion.}, one has action-at-a-distance. In the presence of entanglement, a measurement on one system can result in a real change to the possessed properties of another system, even when the two systems are widely separated. (Although, as is well known, these changes do not allow one to send signals superluminally --- this is known as the \textit{no-signalling theorem}\footnote{An early version of the no-signalling theorem, specialised to the case of spin 1/2 EPR-type experiments appears in \citet{bohm}. Later, more general versions are given by \citet{tausk,eberhard,grw:no-signalling}. See also \citet{shimony}, \citet[Chpt. 4.6]{redheadbook}.}.) 

In teleportation, then, under a collapse interpretation, the effect of Alice's Bell-basis measurement will be to prepare Bob's system, at a distance, in one of four pure states which depend on the unknown state $\ket{\chi}{}$, by using the nonlocal effect of collapse. It then only remains for Alice to send her two bits to Bob to tell him which (type of) state he now has in his possession. Under this interpretation, teleportation explicitly involves nonlocality, or action-at-a-distance; and it is precisely because of the nonlocal effect of collapse, preparing Bob's system in a state that differs in one of only four ways from $\ket{\chi}{}$, that a mere two classical bits need be sent by Alice in order for Bob's system to acquire a state parameterised by two continuous values.

It is enlightening to compare the effect of collapse in this scenario to that of a rigid rod held by two parties. Imagine that Alice wanted to let Bob know the value of a parameter that could take on values in the interval $[0,1]$. If they were each holding one end of a long rigid rod, then Alice could let Bob know the value she has in mind simply by moving her end of the rod along in Bob's direction by a suitable distance. Bob, seeing how far his end of the rod moves, may infer the value Alice is thinking of\footnote{Of course, in a relativistic setting, rigid bodies would not be permissible, although they are in non-relativistic quantum mechanics. This does not in any case affect the point of the analogy.}. There is no mystery here about how the value of the continuous parameter is transmitted from Alice to Bob. Alice, by moving her end of the rod, moves Bob's by a corresponding amount. In teleportation, the effect of collapse is somewhat analogous: Bob's system is prepared, by the nonlocal effect of collapse, in a state that depends on the two continuous parameters characterizing $\ket{\chi}{}$. As we have said, collapse allows a real change in the physical properties that a distant system possesses, if there was prior entanglement. Compare: pushing one end of a rigid rod axially leads to a change in the position of the far end. The nonlocal effect of collapse, which is here understood as a real physical process, is providing the main physical mechanism behind teleportation; and recall that once the physical mechanisms have been described (I have argued) there is no further question to be asked about how information is transmitted in the protocol.       

In a collapse interpretation, teleportation thus involves nonlocality, in the sense of action-at-a-distance, crucially. Also, something interesting certainly has happened once Alice performs her measurement and before she sends the two classical bits to Bob. There has been a real change in the physical properties of Bob's system, as it acquires one of four pure states. (Although note that at this stage the probability distributions for measurements on Bob's system will nonetheless not display any dependence on the parameters characterizing $\ket{\chi}{}$, in virtue of the no-signalling theorem. It is only once the bits from Alice have arrived and Bob has performed the correct operation that measurements on his system will display a dependence on the parameters $\alpha$ and $\beta$.)

\subsection{No collapse and no extra values: Everett}

It is possible to retain the idea that the wavefunction provides a complete description of reality while rejecting the notion of collapse; this way lies the Everrett interpretation \citep{everett}\footnote{It should be noted that there have been a number of different attempts to develop Everett's original ideas into a full-blown interpretation of quantum theory. The most satisfactory of these would appear to be an approach on the lines of Saunders and Wallace \citep{simon1,simon2,simon3,simonrelativism,wallace:worlds,wallace:structure} which resolves the preferred basis problem and has made considerable progress on the question of the meaning of probability in Everett (on this, see in particular \citet{deutsch:decision,wallace:rationality}).}. The characteristic feature of the Everett interpretation is that the dynamics is always unitary; and no extra values are added to the description provided by the wavefunction in order to account for definite measurement outcomes. Instead, measurements are simply unitary interactions which have been chosen so as to correlate states of the system being measured to states of a measuring apparatus. Obtaining a definite value on measurement is then understood as the measured system coming to have a definite state (eigenstate of the measured observable) \textit{relative to} the indicator states of the measuring apparatus and ultimately, relative to an observer\footnote{This is the case for ideal first-kind (non-disturbing) measurements. The situation becomes more complicated when we consider the more physically realistic case of measurements which are not of the first kind; in some cases, for example, the object system may even be destroyed in the process of measurement. What is important for a measurement to have taken place is that measuring apparatus and object system were coupled together in such a way that if the object system had been in an eigenstate of the observable being measured prior to measurement, then the subsequent state of the measuring apparatus would allow us to infer what that eigenstate was. In this more general framework the importance is not so much that the object system is left in a eigenstate of the observable relative to the indicator state of the measuring apparatus, but that we have definite indicator states relative to macroscopic observables.}. 
A treatment of teleportation in the Everettian context was given by \citet{vaidman}. \citet{braunstein:irreversible} provides a detailed discussion of the teleportation protocol within unitary quantum mechanics without collapse.

With teleportation in an Everettian setting, and unlike teleportation under the orthodox account, it is clear that there will be no action-at-distance in virtue of collapse when Alice performs her measurement, for the simple reason that there is no process of collapse. Instead, the result of Alice's measurement will be that Bob's system comes to have definite relative states related to the unknown state $\ket{\chi}{}$, with respect to the indicator states of the systems recording the outcome of Alice's measurement. (It will be argued in the next chapter that this does not amount to a new form of nonlocality.) Note, though, that at this stage of the protocol, the \textit{reduced} state of every system involved will now be maximally mixed\footnote{This would not in general be the case if the initial entangled state were not maximally entangled, or if Alice's measurement were not an ideal measurement; with these eventualities, the teleportation would be imperfect (fidelity less than 1).}. As \citet{braunstein:irreversible} notes, this feature corresponds to the `disembodiment' of the information characterizing the unknown state in the orthodox account of teleportation: following Alice's measurement, all the systems involved in the protocol will have become fully entangled. Dependence on the parameters characterizing the unknown state will only be observable with a suitable \textit{global} measurement, not for any local measurements. In particular, one can consider the correlations that now exist between the systems recording the outcome of Alice's measurement and Bob's system. Certain of the joint (and irreducible) properties of these spatially separated systems will depend on the identity of the unknown state. In this sense, the information characterizing $\ket{\chi}{}$ might now be said to be `in the correlations' between these systems. (This is the terminology Braunstein adopts.) 

Once Bob has been sent the systems recording the outcome of Alice's measurement, however, he is able to disentangle his system from the other systems involved in the protocol. Its state will now factorise from the joint state of the other systems; and will in fact be the pure state $\ket{\chi}{}$. Dependence on the parameters $\alpha$ and $\beta$ \textit{will} finally be observable for local measurements once more, but this time, only at Bob's location. 

In collapse versions of quantum mechanics, the nonlocal effect of collapse was the main physical mechanism underlying teleportation. In the no-collapse Everettian setting, the fundamental mechanism is provided by the fact that in the presence of entanglement, local unitary operations---in this case, Alice's measurement---can have a non-trivial effect on the global state of the joint system.

So, has anything significant happened at Bob's location before Alice sends him the result of her measurement and he performs his conditional unitary operation? Well, arguably not: nothing has happened other than all of the systems involved in the protocol having become entangled, as a result of the various local unitary operations.

\subsection{No collapse, but extra values: Bohm}

The Bohm theory account provides us with an interesting intermediary view of teleportation, in which there is no collapse of the wavefunction, but nonlocality plays an interesting r\^ole. We shall follow the analysis of \citet{maroney:hiley}.

The Bohm theory \citep{bohm:1952} is a nonlocal, contextual, deterministic hidden variable theory, in which the wavefunction $\Psi(\mathbf{x}_{1},\mathbf{x}_{2}\ldots\mathbf{x}_{n},t)$ of an $n$-body system evolves unitarily according to the Schr\"odinger dynamics, but is supplemented with definite values for the positions $\mathbf{x}_{1}(t),\mathbf{x}_{2}(t)\ldots\mathbf{x}_{n}(t)$ of the particles. Momenta are also defined according to $\mathbf{p}_{i}=\nabla_{i}S$, where $S$ is the phase of $\Psi$, hence a definite trajectory may be associated with a system, where this trajectory will depend on the many-body wavefunction (and thus, in general, on the positions and behaviour of all the other systems, however far away). If the initial probability distribution for particle positions is assumed to be given by $|\Psi|^{2}$, then the same predictions for measurement outcomes will be made as in ordinary quantum mechanics. For detailed presentations of the Bohm theory, see \citet{bohmhiley} and \citet{holland:1995}.

The guiding effect of the wavefunction on the particle positions may also be understood in terms of a new \textit{quantum potential} that acts on particles in addition to the familiar classical potentials. The quantum potential is given by 
\[Q(\mathbf{x}_{1},\mathbf{x}_{2},\ldots,\mathbf{x}_{n})=-\hbar^{2}\sum_{i=1}^{n}\frac{\nabla^{2}_{i}R}{2m_{i}R},\]
where $R$ is the amplitude of $\Psi$ and $m_{i}$ is the mass of the $i$-th particle. Among the ways in which this quantity differs from a classical potential is that it will in general give rise to a nonlocal dynamics (that is, in the presence of entanglement, the force on a given system will depend on the instantaneous positions of the other particles, no matter how far away); and it may be large even when the amplitude from which it is derived is small. \citet[\S 3.2]{bohmhiley} suggest that the quantum potential should be understood as an `information potential' rather than a mechanical potential, as a way of accounting for its peculiar properties.

The determinate values for position in the Bohm theory are usually understood as providing the definite outcomes of measurement\footnote{Note, though, that measurement may not usually be understood as revealing pre-existing values in the Bohm theory.} that would appear to be lacking in a no-collapse version of quantum mechanics, in the absence of an Everett-style relativization. Following a measurement interaction, the wavefunction of the joint object-system and apparatus will have separated out (in the ideal case) into a superposition of non-overlapping wavepackets (on configuration space) corresponding to the different possible outcomes of measurement.
The determinate values for the positions of the object-system and apparatus pointer variable will pick out a point in configuration space; and the outcome that is observed, or is made definite, is the one corresponding to the wavepacket whose support contains this point. The wavefunction for the total system remains as a superposition of all of the non-overlapping waverpackets, however. \citet{bohmhiley} introduce the notions of \textit{active}, \textit{passive} and \textit{inactive} information to describe this feature of the theory. If $\Psi$ may be written as a superposition of non-overlapping wavepackets, then they suggest that the definite configuration point of the total system picks out one of these wavepackets (the one whose support contains the point) as active. The evolution of the point is determined solely by the wavepacket containing it; and in keeping with their conception of $Q$ as an information potential, the information associated with this wavepacket is said to be active. The information associated with the other wavepackets is termed either `passive', or `inactive'. `Passive', if the different wavepackets may in the future be made to overlap and interfere, `inactive' if such interference would be a practical impossibility (as for example, if environmental decoherence has occurred in a measurement-type situation --- this corresponds to the case of `effective collapse' of the wavefunction). 

In their discussion of the teleportation protocol, Maroney and Hiley adopt the approach in which a definite spin vector is also associated with each spin 1/2 particle, in addition to its definite position.
The idea is that with each system is associated an orthogonal set of axes (body axes) whose orientation is specified by a real three dimensional spin vector, $\mathbf{s}$, along with an angle of rotation about this vector; where these quantities are determined by the wavefunction\footnote{This is the approach to spin of \citet{bst:1955}. For a systematic presentation see \citet[\S 10.2-10.3]{bohmhiley} or \citet[Chpt. 9]{holland:1995}. Other approaches to spin are possible, e.g., \citet[\S 10.4-10.5]{bohmhiley}, \citet[Chpt. 10]{holland:1995}, or the `minimalism' of \citet{bell:1966,bell:1981}, in which no spin values are added.}.  

The analysis of teleportation then proceeds much as in the Everett interpretation, save that we may also consider the evolution of the determinate spin vectors associated with the various systems. Initially, the system in the unknown state $\ket{\chi}{}$ will have some definite spin vector that depends on $\alpha$ and $\beta$, $\mathbf{s}(\alpha,\beta)$, while it turns out that if Alice and Bob share a singlet state, the spin vectors for their two systems will be zero \citep[\S10.6]{bohmhiley}. Now Alice performs her Bell-basis measurement. As in the Everettian picture, the effect of measurement is to entangle the systems being measured with systems recording the outcome of the measurement. But this is not the only effect, in the Bohm theory. The total wavefunction is now a superposition of four terms corresponding to the four possible outcomes of Alice's measurement; and one of these four terms will be picked out by the definite position value of the measuring apparatus pointer variable. For each of these four terms taken individually, Bob's system will be in a definite state related to the state $\ket{\chi}{}$, thus with each will be associated a definite spin vector $\mathbf{s}^{j}(\alpha,\beta)$, $j=1,\ldots,4$, pointing in some direction. When one of the four terms is picked out as active, and the others rendered passive (or inactive), following Alice's measurement, the spin vector for Bob's system will change instantaneously from zero to one of the four $\mathbf{s}^{j}(\alpha,\beta)$ \citep{maroney:hiley}.        

Thus in the Bohm theory, teleportation certainly involves nonlocality; and moreover, something very interesting does happen as soon as Alice has made her measurement. Bob's system acquires a definite spin vector that depends on the parameters characterizing the unknown state, as a result of a nonlocal quantum torque \citep{maroney:hiley}. Furthermore, there is a one in four chance that this spin vector will be the same as the original $\mathbf{s}(\alpha,\beta)$; and all this while the total state of the system remains uncollapsed, with all the particles entangled. 

Finally, as we have seen before, once Alice sends Bob systems recording the outcome of her measurement, he may perform the conditional unitary operation necessary to disentangle his system from the others, and leave his system in the state $\ket{\chi}{}$. The spin vector of his system will now be $\mathbf{s}(\alpha,\beta)$ with certainty.

\subsubsection{A note on active information}\label{active information}

The conclusion of \citet{maroney:hiley} and \citet{hiley:1999} is that according to the Bohm theory, what is transferred from Alice's region to Bob's region in the teleportation protocol is the active information that is contained in the quantum state of the initial system. However questions may be raised about how apposite this description is.

For ease of reference, let us introduce labels for some of the systems involved in teleportation. Call the system whose unknown state is to be teleported, system 1; Alice's half of the entangled pair, system 2; and Bob's half, system 3. Also let us label the pointer degree of freedom of the measuring apparatus by $x_{0}$.
At the beginning of the teleportation protocol, the state of system 1 factorises from the entangled joint state of 2 and 3; and the state of the measuring apparatus will also factorise. Accordingly, the quantum potential will be given by a sum of separate terms:
\begin{equation}\label{qpotential before}
Q(\mathbf{x}_{1},\mathbf{x}_{2},\mathbf{x}_{3},x_{0}) = Q(\mathbf{x}_{1},\alpha,\beta) + Q(\mathbf{x}_{2},\mathbf{x}_{3}) + Q(x_{0}),
\end{equation}
where it has been noted that the first term, the one that will determine the motion of system 1, depends on the parameters characterizing the unknown state\footnote{The component of the force on the $i$-th system due to the quantum potential is given by $m_{i}\ddot{\mathbf{x}}_{i}=-\nabla_{i}Q$ \citep[cf.][\S 7.1.2]{holland:1995}; therefore, only terms in the sum which depend on $\mathbf{x}_{i}$ will contribute to the motion of the $i$-th system.}.

Once Alice performs her Bell basis measurement, however, all the systems become entangled; and the potential will be of the form:
\begin{equation}\label{qpotential middle}
Q(\mathbf{x}_{1},\mathbf{x}_{2},\mathbf{x}_{3},x_{0}) = Q(\mathbf{x}_{1},\mathbf{x}_{2},x_{0}) + Q(\mathbf{x}_{3},x_{0},\alpha,\beta)
\end{equation}
The part of the quantum potential that will affect system 3 now depends on $\alpha$ and $\beta$.

Finally, at the end of the protocol, systems 1, 2 and the measuring apparatus are left entangled; and system 3, in the pure state $\ket{\chi}{3},$ factorises. The quantum potential then takes the form:
\begin{equation}\label{qpotential end}
Q(\mathbf{x}_{1},\mathbf{x}_{2},\mathbf{x}_{3},x_{0}) = Q(\mathbf{x}_{1},\mathbf{x}_{2},x_{0}) + Q(\mathbf{x}_{3},\alpha,\beta)
\end{equation}

Maroney and Hiley say:
\begin{quoting}
What we see clearly emerging here is that it is active information that has been transferred from particle 1 to particle 3 and that this transfer has been mediated by the nonlocal quantum potential.\citep[p.1413]{maroney:hiley}

\noindent ...it is the objective active information contained in the wavefunction that is transferred from particle 1 to particle 3. \citep[p.1414]{maroney:hiley}
\end{quoting}

Note that the part of the potential that is active on system 3 will already have acquired a dependence on $\alpha$ and $\beta$ before the end of the protocol; that is, as soon as Alice has performed her measurement. So if active information depending on these parameters is transferred at all, it will have been transferred before the end of the protocol. However it is not until Alice has sent her message to Bob and he performs his conditional operation that the term  $Q(\mathbf{x}_{3},\alpha,\beta)$ in eqn.~\ref{qpotential end} will take the same form as the initial $Q(\mathbf{x}_{1},\alpha,\beta)$.

The difficulties for the stated conclusion arise when we consider more closely what is meant by `active information'. In \citet{maroney:hiley,hiley:1999}, the connection is made with a different sense of the word `information' than the ones with which we have so far been concerned in this thesis. This is a sense that derives from the verb `inform' under its branch I and II senses (Oxford English Dictionary), \textit{viz.} to give form to, or, to give formative principle to (this latter, a Scholastic Latin offshoot).  

Thus `information' as it appears in `active information' and company, means the action of giving form to\footnote{Cf. OED `information', sense 7.}. `The information of $x$' (read: The \textit{in}-formation of $x$) means the action of giving form to $x$.

Now, while we may understand what is meant by $Q$ being said to be an information potential --- it is a potential that gives form to something, presumably the possible trajectories associated with particles (although note that the distinction with mechanical potentials is now blurred, as these give form to the possible trajectories too) --- and may understand the term `active' as picking out the part of the quantum potential that is shaping the actual trajectory in configuration space of the total system, it does not make sense to say that active information is transferred in teleportation. Because `information' here refers to a particular action --- the giving of a form to something --- and an action is not a \textit{thing} that can be moved\footnote{On some accounts, an action is the bringing about of some event or state of affairs by an agent \citep{hyman:alvarez}; on others, an action is an event \citep{davidson:actionsandevents}. On no account is an action something which can intelligibly be said to be moved about.}. The same \textit{type} of action may be taking place at two different places, or at two different times, but an action may not be moved from $A$ to $B$.   

Thus with `active information' understood in the advertised way,
all that can be said is that an action of the same \textit{type} is being performed (by the quantum potential) on system 3 at the end of the teleportation protocol as was being performed on system 1 at the beginning, not that something has been transferred between the two. We may not, then, understand `transfer' literally.  When all is said and done, it is perhaps clearer simply to adopt the standard description and say that the quantum \textit{state} of particle 1 has been `transferred' in teleportation; that is (as a quantum state is a mathematical object and therefore cannot literally be moved about either), that system 3 has been made to acquire (is left in) the unknown state $\ket{\chi}{}$.

To sum up: it perhaps looked as if the Bohmian notion of active information might provide us with a sense of what is transported in teleportation if we insist that \textit{information}, `the information in the wavefunction', is, in a literal sense, transported. But this proves not to be the case.

\subsection{Ensemble and statistical viewpoints}

So far, in all the interpretations we have considered, the quantum state may describe individual systems. Let us close this section by looking briefly at approaches in which the state is taken only to describe \textit{ensembles} of systems.

We may broadly distinguish two such approaches. The first I will term an \textit{ensemble} viewpoint. In this approach, the state is taken to represent a real physical property, but only of an ensemble. Following a measurement, the ensemble must be left in a \textit{proper} mixture\footnote{See \citet{d'Espagnat} for this terminology, also \citet{impsep}.}, in order for there to be definite outcomes, i.e., the ensemble is left in an appropriate mixture of sub-ensembles, each described by a pure state (eigenstate of the measured observable). Thus there will be a real process of collapse, but only at the level of the ensemble, not for individual systems (which are not being described by a quantum state, if at all). 

The second approach I call a \textit{statistical} interpretation. (This is the interpretation that would be adopted by instrumentalists, for example.) On this view, the quantum formalism merely describes the probabilities for measurement outcomes for ensembles, there is no description of individual systems and collapse does not correspond to any real physical process. 

On both these approaches, as the state is only associated with an ensemble, it is not until an entire ensemble has been teleported to Bob (that is, Alice has run the teleportation protocol on every member of an ensemble in the unknown state $\ket{\chi}{}$) that he acquires something in the state $\ket{\chi}{}$. An ensemble or statistical viewpoint thus makes a natural partner to conservative classical quantity surveying in teleportation.  

Under the statistical interpretation, there is clearly no nonlocality involved in teleportation, as there is no real process of collapse; and nothing of any interest has happened before the required classical bits are sent to Bob. (The no-signalling theorem entails that Alice's measurement won't affect the probability distributions for distant measurements.) The end result of the completed teleportation process is that Bob's ensemble is ascribed the state $\ket{\chi}{}$; where this merely means that the statistics one will expect for measurements on Bob's ensemble are now the same as those one would have expected for measurements on the initial ensemble presented to Alice.   
 
The ensemble viewpoint presents a rather different picture, as it does involve a real process of collapse, even if only at the ensemble level. Let us suppose that Alice has performed the Bell basis measurement on her ensembles, but has not yet sent the ensemble of classical bits to Bob. The effect of this measurement will have been to leave Bob's ensemble in a proper mixture composed of sub-ensembles in the four possible states a fixed rotation away from $\ket{\chi}{}$. Thus there has been a nonlocal effect: that of preparing what was an improper mixture into a particular proper mixture, whose components depend on the parameters characterizing the unknown state. The use of the flock of classical bits that Alice sends to Bob is to allow him to separate out the ensemble he now has into four distinct sub-ensembles, on each of which he performs the relevant unitary operation, ending up with all four being described by the state $\ket{\chi}{}$.

\section{Concluding remarks}\label{study concluding}

The aim of this chapter has been to show how substantial conceptual difficulties can arise if one neglects the fact that `information' is an abstract noun. This oversight seems to lie at the root of much confusion over the process of teleportation; and this gives us very good reason to pay attention to the logical status of the term. A few closing remarks should be made.

Schematically, a central part of the argument has been of the following form:   

Puzzles arise when we feel the need to tell a story about how something travels from Alice to Bob in teleportation. 
In particular, it might be felt that this something needs to travel in a spatio-temporally continuous fashion; and one might accordingly feel pushed towards adopting something like the Jozsa/Penrose view.

But if `the information' doesn't pick out a particular, then there is no thing to take a path, continuous or not, therefore the problem is not a genuine one, but an illusion.

We can imagine a number of objections. A very simple one might take the following form: You have said that information is not a particular or thing, therefore it does not make sense to inquire how \textit{it} flows (but only inquire about the means by which it is transmitted). But don't we have a theory that quantifies information (\textit{viz.} communication theory); and if we can say how much of something there is, isn't that enough to say that we have a thing, or a quantity that can be located?

This objection is dealt with quickly. Note that this form of argument will not work in general---one can say how much a picture might be worth in pounds and pence, for example, but this is not quantifying an amount of stuff, nor describing a quantity with a location---and it does not work in this particular case either (cf. Section~\ref{shannon concepts}). The Shannon information doesn't quantify an amount of stuff that is present in a message, say, nor the amount of a certain quantity that is present at some spatial location. The Shannon information $H(X)$ describes a specific property of a \textit{source} (not a message), namely, the amount of channel resources that would be required to transmit the messages the source produces. This is evidently not to quantify an amount of stuff, nor to characterize a quantity that has a spatial location. (The source certainly has a spatial location, but its information does not.) Or consider the mutual information. Loosely speaking, this quantity tells us about the amount we may be able to infer about some event or state of affairs from the obtaining of another event or state of affairs. But how much we may infer is not a quantity it makes sense to ascribe a spatial location to.

Another objection might be as follows: You have suggested that it is a mistake to hypostatize information, to talk of it as a thing that moves about. How is this to be reconciled with some of the ways we often talk about information in physics, especially the example in relativity, where the most natural way of stating an important constraint is to say that relativity rules out the propagation of \textit{information} faster than the speed of light?

The response is that one can admit this mode of talking without it entailing a hypostatized conception of information. The constraint is that superluminal signalling is ruled out on pain of temporal loop paradoxes \citep[e.g.][\S 7.ix]{rindler}. What this means is that no \textit{physical process} is permissible that would allow a signal to be sent superluminally and thus allow information to be transmitted superluminally. What are ruled out are certain types of physical processes, not, save as a metaphor, certain types of motion of information\footnote{The types of processes in question might not be identifiable without recourse to concepts of what would count as successful transmission of information, but this does not mean that one has to conceive of information as an entity or substance, just that one needs a concept of what it means to receive a signal from which one can learn something.}. 

A final objection that might be raised to support the line of thought that inclines one towards the Jozsa and Penrose conception of teleportation is just this: 
Well, don't we after all require that information be propagated in a spatio-temporally continuous way? Even if this is not to be construed as a flow of stuff, or the passage of an entity?

The response illustrates part of the value of noting the features of the term `information' that have been emphasized in this chapter. 

The genuine question we face is: what are the physical processes that may be used to transmit information? Not the (obscure) question `How does information behave?'. Once we see what the question is clearly, then the answer, surely, is to be given by our best physical theory describing the protocol in question. To be sure, many of the most familiar classical examples we are used-to use spatio-temporally continuous changes in physical properties to transmit information (a prime example might be the use of radio waves), but it is up to physical theory to tell us about the nature of the processes we are using to transmit information in any given situation. And the examples we have found in entanglement assisted communication seem precisely to be examples in which \textit{global} rather than local properties are being used to carry information; and there seems not to be a useful sense in which information is being carried in a spatio-temporally continuous way (although, see Chapter~\ref{dh} for further discussion of Deutsch and Hayden's opposing view).

It is not the nature of information that is at issue, but the nature of the physical objects and the physical properties we may use to transmit information.

The value of getting clear on the real nature of the question one faces about information transmission in teleportation will become evident again in the following chapter.

On a final note, the deflationary approach that has been adopted towards teleportation in this chapter should be compared with what may be termed the `nihilist' approach of \citet{duwell:2003}. While I am in broad sympathy with much of what Duwell has to say, we differ on some important points. Duwell also advocates the view that quantum information is not a substance, but reaches from this the strong conclusion that quantum information does not exist. From the current point of view this conclusion is unwarranted. Certainly, quantum information is not a substance or entity, but this does not mean that it doesn't exist, it is just a reflection of the fact that `information' is an abstract noun. `Beauty' for example, is an abstract noun, but no one would want to conclude that there is no beauty in the world. Moreover, Duwell's conclusion could only possibly be hyperbolical, for if classical information can be said to exist, then so too can quantum information; and contrapositively, if quantum information does not exist, then no more does classical information. The concept of classical information is given by Shannon's noiseless coding theorem, the concept of quantum information, by the quantum noiseless coding theorem. As we are by now vividly aware, these are not concepts of material quantities or things. But rejecting the concept of quantum information would be akin to cutting off one's nose to spite one's face; and is by no means necessary in order to get a proper understanding of teleportation.

Teleportation is not rendered unproblematic by trying to do without the notion of quantum information and facing the protocol equipped only with Shannon's concept, but simply by resisting the temptation to hypostatize an abstract noun; and, having  recognised the status of `information' as an abstract noun, by realising that the only genuine question one faces is the relatively straightforward one of describing the physical processes by which information is transmitted.

\end{doublespacing}



\chapter[The Deutsch-Hayden Approach]{The Deutsch-Hayden Approach: Nonlocality and Information Flow}\label{dh}







\begin{doublespacing}


\section{Introduction}

The existence of entanglement, and the associated questions concerning nonlocality, are of perennial interest in the foundations of quantum mechanics \citep{EPR,schrodinger1,schrodinger:1936,bell:1964,redheadbook,Maudlin:non-loc}. As we have seen, following the development of quantum information theory, 
entanglement assisted communication \citep{superdense,teleportation} has presented a new sphere in which puzzles may arise. In this context, an important development has been the claim of \citet{dh} to provide an especially local story about quantum mechanics, by making use of the Heisenberg picture. They claim, moreover, finally to have clarified the nature of information flow in entangled quantum systems, reaching the conclusion that information is a local quantity, even in the presence of entanglement. The approach of Deutsch and Hayden was mentioned in passing in the previous chapter. The aim of this chapter is to assess their claims in detail.

Their discussion takes place within the context of unitary quantum mechanics without collapse, and without the addition of determinate values; and they proceed to make two claims to locality. First, they suggest, even in the presence of entanglement, the state of the global system can in fact be seen to be completely determined by the states of the individual subsystems, when these states are properly construed (a conclusion not available in the usual Schr\"odinger picture and one supposed to chime with Einstein's well-known demand for a \textit{real state} for spatially separated systems \citep[pp.77-83]{einsteinauto}). Second, the effects of local unitary operations, again, even in the presence of entanglement, are \textit{explicitly} seen to be local in their picture.   

However, before the implications of their formalism may be assessed, something needs to be said about how it is to be interpreted. Deutsch and Hayden are not explicit on this point and do not offer any interpretation. This proves problematic as two different modes of interpretation of their formalism may be discerned---what may be called the \textit{conservative} and the \textit{ontological} interpretations---and quite different conclusions follow concerning the questions of locality 
and information flow within these interpretations.

The conservative interpretation, perhaps the most natural way of reading the Deutsch-Hayden paper, takes the formalism at face value, simply as a re-writing of standard unitary quantum mechanics. In this case, we shall see, there are no novel gains with respect to locality and Deutsch and Hayden's claims about information flow prove at best misleading. Under the ontological interpretation, though, a dramatic departure from our usual ways of understanding quantum mechanics is made and a wholly new range of intrinsic properties of subsystems introduced. These would substantiate Deutsch and Hayden's claims, but at a certain cost of plausibility. We should note too that the ontological interpretation of the Deutsch-Hayden formalism is best seen as the postulation of a new type of theory, rather than being a new way of interpreting familiar quantum mechanics.   

The discussion will begin in Section~\ref{formalism}, where the machinery of the Deutsch-Hayden approach is outlined, in particular, the mathematics that lies behind the two claims to locality. These claims are then assessed (Section~{\ref{assess}}), for the conservative and ontological interpretations in turn. 

Note that in Deutsch-Hayden we have a formalism without collapse and without the addition of determinate values. If we are to consider the question of the locality of their approach, the appropriate comparisons are therefore with other approaches that are consistent with this assumption. On the one hand we should compare with a realist approach of the Everett stripe \citep{everett,simonrelativism,wallace:worlds}, while on the other we should compare with a form of statistical interpretation, by which, recall, I mean an interpretation in which quantum mechanics merely describes probabilities for measurement outcomes for ensembles, there is no description of individual systems and collapse does not correspond to any real physical process for individual systems. The question to be  answered, then, is: do Deutsch and Hayden present us with advantages with respect to locality that are not also shared by these other approaches? We shall see that under the conservative interpretation, they do not. 

In Section~\ref{info}, attention finally turns to the question of information flow in entangled systems. In Section~\ref{whereabouts} the nature of the question at issue is clarified, before  Deutsch and Hayden's explanation of quantum teleportation and their introduction of the concept of locally inaccessible information is considered (Section~\ref{explaining}). Their claims regarding the nature of information flow are then evaluated for the conservative and ontological interpretations in turn (Section~\ref{assessinfo}), along axes provided by three questions: i) Have Deutsch and Hayden finally given the correct account of teleportation, as compared to related accounts such as that of \citet{braunstein:irreversible}? ii) Is the concept of locally inaccessible information useful? iii) Have they provided us with a new concept of information, or quantum information? We close with a brief summary.

\section{The Deutsch-Hayden Picture}\label{formalism}

Deutsch and Hayden consider a network of $n$ interacting qubits as their model of a general quantum system. They take as the object describing the state of the $i$th qubit at time $t$ a triple \[\mathbf{q}_{i}(t)=(q_{i,x}(t),q_{i,y}(t),q_{i,z}(t))\] of $2^{n}\times2^{n}$ Heisenberg picture operators satisfying the familiar commutation and anti-commutation relations of the Pauli spin operators. This object they term the `descriptor' of a system. To see how this representation works, let us first recall the basics of the Heisenberg picture. 

\begin{sloppypar}
As expressed in the equations
\begin{equation}
\bra{}{\psi(t)}A\ket{\psi(t)}{}=\bra{}{\psi}U^{\dagger}AU\ket{\psi}{}=\bra{}{\psi}A(t)\ket{\psi}{},
\end{equation}
time dependence in quantum mechanics can either be associated with the vector (\textit{ket}) representing the state, or with the operator representing the observable. In the Schr\"odinger picture, the state ket undergoes unitary evolution ($\ket{\psi}{} \mapsto U\ket{\psi}{}$); in the Heisenberg picture, the state ket remains unchanged and the basis kets $\{\ket{\alpha_{i}}{}\}$ of the Hilbert space are evolved ($\ket{\alpha_{i}}{} \mapsto U^{\dagger}\ket{\alpha_{i}}{}$).
Another useful way of representing these facts is given by the Hilbert-Schmidt representation.
\end{sloppypar}

As we have seen (Section~\ref{Hilbert-Schmidt}), the set of $N\times N$ complex Hermitian matrices forms an $N^{2}$ dimensional real vector space, $V_{h}(\mathbb{C}^{N})$, on which we may define an inner product $(A,B)=\mathrm{Tr}(AB),\; A,B\in V_{h}(\mathbb{C}^{N})$ and norm $||A||=\sqrt{\mathrm{Tr}A^2}$; and just as in our familiar examples of vector spaces, e.g. Euclidean ${\mathbb R}^3$, it is useful to define a set of basis vectors for the space. We require $N^{2}$ linearly independent operators $\Gamma_{j} \in V_{h}(C^N)$, and we may require orthogonality and a fixed normalisation: $\mathrm{Tr}(\Gamma_{j}\Gamma_{j^{\prime}})=\mathrm{const.}\delta_{jj^{\prime}}$.

Recall that any observable can then be represented in this space in the form:
\begin{equation}
\mathbf{A}=\sum_{j=0}^{N^2-1}\mathrm{Tr}(A\Gamma_{j})\Gamma_{j}=\sum_{j=0}^{N^2-1}a_{j}\Gamma_{j},
\end{equation}
where the $\mathrm{Tr}(A\Gamma_{j})=a_{j}$ are the components of the vector $\mathbf{A}$ representing the observable $A$. In particular the density matrix $\rho$ can also be written as a vector:
\begin{equation}\label{density matrix}
{\boldsymbol{\varrho}}=\frac{\mathbf{1}}{N} + \sum_{j=1}^{N^{2}-1}\mathrm{Tr}(\rho \Gamma_{j})\Gamma_{j}=\sum_{j=0}^{N^2-1}\rho_{j}\Gamma_{j},
\end{equation}
where $\Gamma_{0}$ has been chosen as $\mathbf{1}$, the identity.
In this representation, the expectation value of $A$ is just the projection of the vector $\boldsymbol{\varrho}$ onto the vector $\mathbf{A}$: $\langle A \rangle_{\rho}= \mathrm{Tr}(A\rho)=(\mathbf{A}.\boldsymbol{\varrho}$).
The equivalence between the Schr\"odinger and Heisenberg pictures now takes on a very graphic form. We can either picture leaving the basis vectors (operators) as they are and rotating the vector $\boldsymbol{\varrho}$ under time evolution, or we can picture rotating the basis vectors (and hence any observable $A$) in the opposite sense, and leaving $\boldsymbol{\varrho}$ unchanged. In either case, the angle between the two resulting vectors and hence the expectation value is clearly the same: $\mathbf{A}(t).\boldsymbol{\varrho}=\mathbf{A}.\boldsymbol{\varrho}(t)$.

Writing the time dependence out explicitly, we will have, in the Heisenberg picture:
\begin{equation}
\mathbf{A}(t)=\sum_{j}a_{j}U^{\dag}(t)\Gamma_{j}U(t), \label{A(t)}
\end{equation}
while in the Schr\"odinger picture,
\begin{equation}
\boldsymbol{\varrho}(t)=\sum_{j}\mathrm{Tr}(\rho U^{\dag}(t)\Gamma_{j}U(t))\Gamma_{j} = \sum_{j}\langle\Gamma_{j}(t)\rangle_{\rho}\Gamma_{j}. \label{rho(t)}
\end{equation}
The expectation value of observable $A$ at time $t$ is simply $\sum_{j}a_{j}\langle\Gamma_{j}(t)\rangle_{\rho}$. 

Notice that in both expressions (\ref{A(t)}) and (\ref{rho(t)}), the time evolved operators $\Gamma_{j}(t)= U^{\dag}(t)\Gamma_{j}U(t)$ feature. These operators, along with their expectation values $\langle\Gamma_{j}(t)\rangle_{\rho}$, will be our main objects of interest.

What should we choose as basis vectors? For $N=2$, the set of Pauli operators forms an orthogonal basis set, $\mathrm{Tr}(\sigma_{i}\sigma_{j})=2\delta_{ij}$, (we adopt the convention that $\sigma_{0}$ denotes the identity) thus we can choose $\sqrt{2}\,\Gamma_{j}\in\{\mathbf{1},\sigma_{x},\sigma_{y},\sigma_{z}\}$ to provide an orthonormal basis $\{\Gamma_{j}\}$.\footnote{The choice of the Pauli operators as a basis set gives us the familiar Bloch sphere representation of the density matrix of a two-state system.} We are then interested in the behaviour of the set $\{U^{\dag}(t)\,(\sigma_{i}/\sqrt{2})\,U(t)\}$.

So far, all we have done is translate some very familiar results into the language of the space $V_{h}(\mathbb{C}^{N})$. We now make the all-important move that provides the core result of the Deutsch-Hayden picture (following \citet{gottesman}). That is, we note that unitary transformations of operators have the property of being a multiplicative group homomorphism\footnote{A map $f:\mathcal{A}\mapsto \mathcal{B}$ is a \textit{group homomorphism} if $\forall a_{1},a_{2}\in \mathcal{A}, f(a_{1}a_{2})=f(a_{1})f(a_{2})$.}:
\begin{equation} U^{\dag}ABU=(U^{\dag}AU)(U^{\dag}BU). \end{equation}
In other words, the time evolution of a product will be given by the product of the time evolution of the individual operators. Thus we do not need to follow the evolution of the whole basis set of operators, but only of a generating set. For example, in the $N=2$ case, noting that $\sigma_{x}\sigma_{y}=i\sigma_{z}$, we see that $\sigma_{z}(t)=-i\sigma_{x}(t)\sigma_{y}(t)$ and that we need only follow the evolution of the generating set $\{\sigma_{x},\sigma_{y}\}$ to capture the time evolution of the whole system. (For completeness, note that $\sigma_{i}^{2}=\mathbf{1}$; the time evolution of the identity is of course trivial.)  

For $N=2^{n}$, $n$-fold tensor products of Pauli matrices will provide us with an orthogonal set, thus our basis operators will be
\begin{equation}\label{gammaj}
\Gamma_{j}=\frac{1}{\sqrt{2^{n}}}\:\sigma^{1}_{m_{1}}\otimes\sigma^{2}_{m_{2}}\otimes\ldots\otimes\sigma^{n}_{m_{n}};
\end{equation}
where the index $j$ runs from 0 to $(4^{n}-1)$ and labels an ordered $n$-tuple $<\!\!m_{1},m_{2},\ldots,m_{n}\!\!>$, $m_{i}\in\{0,1,2,3\}$. We are interested in the behaviour of the $4^{n}$ $\Gamma_{j}(t)$; again, however, we need only track the evolution of objects of the form
\[ \mathbf{1}\otimes\mathbf{1}\otimes\ldots\otimes\sigma^{i}_{m_{i}}\otimes\ldots\otimes\mathbf{1}, \]
which we denote $q_{i,m_{i}}$; the $\Gamma_{j}$ are given by ordinary matrix multiplication of these objects:
\begin{equation}
\Gamma_{j}=\prod_{i=1}^{n}\frac{1}{\sqrt{2}}\:q_{i,m_{i}}\,.
\end{equation}
The behaviour of the $\Gamma_{j}(t)$ is thus determined by following the time evolution of a minimum of $2n$ of the $q_{i,m_{i}}$ and taking appropriate products. 

The $q_{i,m_{i}}$ with $m_{i}$ running from 1 to 3 are, of course, the components of the Deutsch-Hayden descriptor $\mathbf{q}_{i}$. This choice of three operators per system as the basic objects whose time evolution we are to follow is more than is strictly necessary for a generating set, but it leads to a very simple description of an individual system, as we shall shortly see. First, however note that
the density matrix at time $t$ can now be written as
\begin{equation}
\boldsymbol{\varrho}(t)=\frac{1}{2^n}\sum_{m_{1}m_{2}\ldots m_{n}}\biggl\langle \prod_{i}q_{i,m_{i}}(t)\biggr\rangle_{\!\!\!\rho}\prod_{i}q_{i,m_{i}}\,. \label{density}
\end{equation} 
That is, the $4^{n}$ components $\rho_j(t)$ of the vector representing the density matrix at time $t$ are given by the expectation values of products of the $q_{i,m_{i}}(t)$. The state of the joint system at time $t$ is thus completely determined by the time evolution of the $2n$ or $3n$ chosen $q_{i,m_{i}}$ and the initial state $\rho$. To see the significance of the triple $\mathbf{q}_{i}$, note that any observable $A^{i}$ on the $i$th system alone will have the form:
\begin{equation}
A^{i}=\sum_{m_{i}=0}^{3}a_{m_{i}}(\mathbf{1}\otimes\mathbf{1}\otimes\ldots\otimes\sigma^{i}_{m_{i}}\otimes\ldots\otimes\mathbf{1})=a_{0}\mathbf{1}^{\otimes n}+\sum_{m_{i}=1}^{3}a_{m_{i}}q_{i,m_{i}},
\end{equation}
thus $\mathbf{q}_{i}(t)$ tells us about observables on the $i$th system at time $t$ and $\langle \mathbf{q}_{i}(t) \rangle_{\rho}$ determines their expectation values. Equivalently, the three components of $\langle \mathbf{q}_{i}(t)\rangle_{\rho}$ give us the interesting components of the vector $\boldsymbol{\varrho}(t)$ lying in the subspace spanned by observables pertaining to the $i$th system alone; and with renormalisation, the components, in our vector representation, of the reduced density matrix of the $i$th system.

Explicitly, this reduced density matrix is:
\begin{equation}
\rho^{i}(t)=\frac{1}{2}\sum_{m_{i}}\langle q_{i,m_{i}}(t)\rangle_{\rho}\,\sigma^{i}_{m_{i}}.\label{rdmi}
\end{equation}
It is also easy to write down the reduced density matrix for any \textit{grouping} of subsystems. If we were interested in the systems $i$, $j$ and $k$, say, taking the partial trace of (\ref{density}) over the other systems will give us a reduced state of the form:
\begin{equation} 
\rho^{ijk}(t)=\frac{1}{8}\sum_{m_{i}m_{j}m_{k}}\langle q_{i,m_{i}}(t)q_{j,m_{j}}(t)q_{k,m_{k}}(t)\rangle_{\rho}\,\sigma^{i}_{m_{i}}\otimes\sigma^{j}_{m_{j}}\otimes\sigma^{k}_{m_{k}}.\label{rdmijk}
\end{equation}

So we have now seen the basis for the first claim to locality: given just the descriptors $\mathbf{q}_{i}(t)$ for each individual system, and the initial state $\rho$, we may calculate the reduced density matrix for each subsystem, \textit{and} the density matrix for successively larger groups of subsystems, up to and including the density matrix for the system as a whole.

We may note in passing another interesting feature of the Deutsch-Hayden formalism. A question that often arises, particularly in discussion of quantum correlations, is whether different preparations of the same density matrix really correspond to physically distinct situations, as all observable properties of systems having the same density matrix are identical. A pleasing aspect of the Deutsch-Hayden set-up is that it provides a representation in which differences in the way systems are prepared may find direct expression in the formalism\footnote{Although, it must be noted that as we are in the context of no-collapse quantum mechanics, the possibility does not obtain of preparing a distant system in a particular way via collapse, \textit{\`a la} EPR.}. For example, it may be the case that $\langle \mathbf{q}_{i}(t)\rangle_{\rho}=\langle \mathbf{q}_{j}(t)\rangle_{\rho}$ i.e., the two systems have the same reduced density matrix, but that $\mathbf{q}_{i}(t)$ and $\mathbf{q}_{j}(t)$ differ, representing differences in their histories.

\subsection{Locality claim (2): Contiguity}

Let us now consider the second claim to locality. This, recall, was the claim that it can be seen explicitly in the Deutsch-Hayden formalism that local unitary operations have only a local effect. 
As \citet{jozsa:private} has emphasized, this aspect of the Deutsch-Hayden picture is in fact a re-expression of the no-signalling theorem.

\begin{figure}
\scalebox{0.75}{\includegraphics{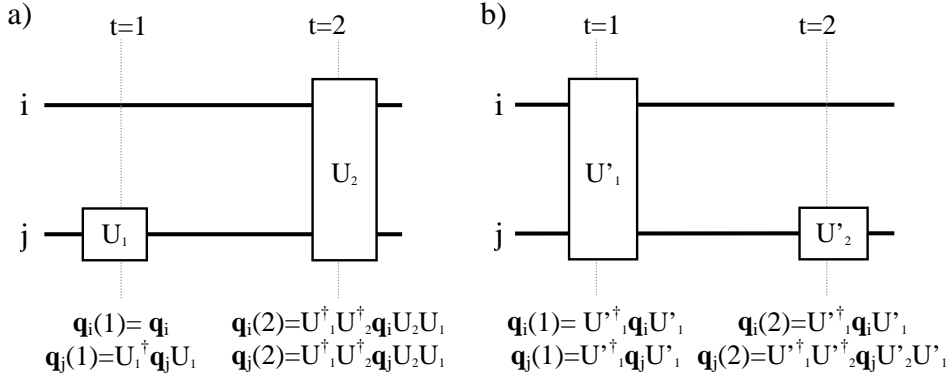}}
\caption{a) At $t=1$, a unitary operation, $U_{1}$, which acts only on system $j$ is applied; the descriptor of system $i$, $\mathbf{q}_{i}(1)$, is unaffected. After $i$ and $j$ interact via $U_{2}$ at $t=2$, however, $\mathbf{q}_{i}(2)$ will depend on the operation $U_{1}$. In (b) systems $i$ and $j$ initially interact via $U^{\prime}_{1}$. At $t=2$, $U^{\prime}_{2}$, acting on $j$ alone, is applied; $\mathbf{q}_{i}(2)$ is unaffected.
}\label{contiguity}
\end{figure}

In the Heisenberg picture, a sketch of a simple version of the theorem would be as follows: let us write an observable acting on subsystem $i$ alone as $A^{i}=\mathbf{1}\otimes A$; at time $t$, $A^{i}(t)=U^{\dag}(t)(\mathbf{1}\otimes A)U(t)$. Suppose $U(t)$ does not act on $i$, then $A^{i}(t)=(U^{\dag}\otimes\mathbf{1})(\mathbf{1}\otimes A)(U\otimes\mathbf{1})=\mathbf{1}\otimes A$, i.e. an observable is unaffected by unitary operations on systems it does not pertain to. Now consider our $q_{i,m_{i}}$; the foregoing clearly applies to them---a unitary operation on a system $j$ does not affect $q_{i,m_{i}}$. More generally, if our network of $n$ systems were divided up into two subsets of systems, $M$ and $N$, whose members interact amongst themselves but not with systems from the other subset, then the unitary operator describing the time evolution of the network will factorise: $U^{M}\otimes U^{N}$. Then the $q_{i,m_{i}}$ for $i\in M$ will not be affected by $U^{N}$, nor those for $i \in N$ by $U^{M}$. We can do more than merely note that the descriptors of a set of interacting systems do not depend on unitary operations on a disjoint set, however. In fact we can see that the descriptor at time $t$ of a given system will depend, apart from the history of operations applied to it alone, only on its previous interactions and on the histories and past interactions of the systems it has interacted with. This property may be called \textit{contiguity}; and is best seen with a simple example (Fig~\ref{contiguity}). 

Imagine we have two systems, $i$ and $j$ and that we are going to perform two unitary operations. First, at $t=1$, we perform $U_{1}$, which acts on $j$ alone; clearly, after this operation, $q_{i,m_{i}}(1)=U^{\dag}_{1}q_{i,m_{i}}U_{1}=q_{i,m_{i}}$. Next we allow $i$ and $j$ to interact via $U_2$; now, however, $q_{i,m_{i}}(2)=U^{\dag}_{1}U^{\dag}_{2}q_{i,m_{i}}U_{2}U_{1}$. Because $U_{2}$ acts on both $i$ and $j$, $U_{1}$ no longer factors out; interaction causes the $q_{i,m_{i}}$ to lose the form of a product of a single Pauli operator with the identity and they can pick up a dependence on what has happened to the system that $i$ has interacted with. We can say that all this remains happily local, however, as this dependence on the history of $j$ only arises following an entangling interaction between the two systems. The reasoning extends in the obvious way to more complicated chains; if $j$ had previously interacted with $k$, then once $i$ and $j$ interact, the $q_{i,m_{i}}(t)$ pick up what they would not previously have had, a dependence on what has happened to $k$; and so on.

To re-emphasize that the Deutsch-Hayden descriptor of a system at time $t$ will not, however, depend on what happens at $t$ to a system with which it has interacted in the past, we take the following simple example (Fig~\ref{contiguity}). Again consider two systems $i$ and $j$; this time, however, we begin by allowing them to interact via a unitary operation $U^{\prime}_{1}$, then 
\begin{eqnarray}
q_{i,m_{i}}(1) & = & U^{\prime\dag}_{1}q_{i,m_{i}}U^{\prime}_{1}\neq q_{i,m_{i}},\;\; \mathrm{and} \nonumber\\
q_{j,m_{j}}(1) & = & U^{\prime\dag}_{1}q_{j,m_{j}}U^{\prime}_{1}\neq q_{j,m_{j}}. 
\end{eqnarray}
\sloppy Now we perform $U^{\prime}_{2}$, which acts on $j$ alone.
Whilst $q_{j,m_{j}}(2)= U^{\prime\dag}_{1}U^{\prime\dag}_{2}q_{j,m_{j}}U^{\prime}_{2} U^{\prime}_{1}$, for the descriptor of $i$ we have
\begin{equation} q_{i,m_{i}}(2)= U^{\prime\dag}_{1}U^{\prime\dag}_{2}q_{i,m_{i}}U^{\prime}_{2} U^{\prime}_{1}=U^{\prime\dag}_{1}q_{i,m_{i}}U^{\prime}_{1},\end{equation} 
$U^{\prime}_{2}$ factors out; there is no immediate dependence on what happens at the present only to $j$, even when $i$ and $j$ have interacted in the past.

\begin{figure}\begin{center}
\scalebox{0.5}{\includegraphics{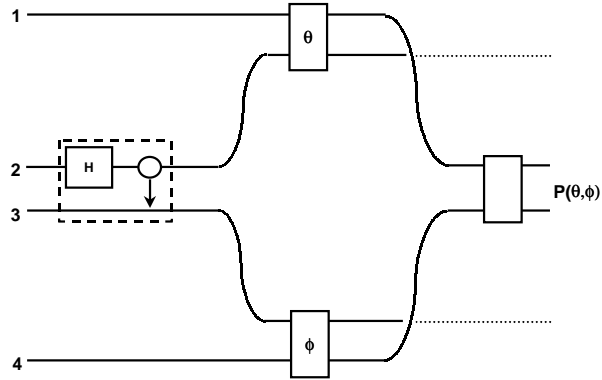}}\end{center}
\caption{A Bell experiment. An entangled state of systems 2 and 3 is prepared (here by the action of a Hadamard gate, H, which performs a rotation by $\pi$ around an axis at an angle of $\pi/4$ in the $z$-$x$ plane; followed by a controlled-NOT operation---the circle indicates the control qubit, the point of the arrow, the target, to which $\sigma_{x}$ is applied if the control is in the $0$ computational state) and the entangled pair is shared between two distant locations. A measurement at an angle $\theta$ is performed on 2 and the outcome recorded in system 1; a measurement at an angle $\phi$ made on 3 and recorded in 4. Time runs along the horizontal axis. Note that in no-collapse quantum mechanics without added values, correlations do not obtain until they are displayed by a suitable joint measurement. }\label{Bell}
\end{figure}

The picture, then, is that following an interaction, the descriptor of a system $i$ picks up a backwards looking (and hence what we might call a local, or contiguous) dependence on what has happened to the system that $i$ has interacted with, and on the previous interactions of that system. 
As an illustration, let us consider how the non-factorisable probability distributions for Bell-type experiments come about in this formalism (Fig.~\ref{Bell}).

As usual, we begin by preparing a pair of systems (2 and 3) in an entangled state. These systems are spatially separated and two local measurements performed, at an angle $\theta$ on system 2 and an angle $\phi$ on system 3. The outcomes are recorded into systems 1 and 4 respectively. Immediately following the measurement, the descriptor of system 1 will depend on $\theta$, but not on the parameter characterizing the distant measurement, $\phi$. However, as system 1 has interacted with system 2, its descriptor will \textit{also} depend on what has happened to 2 in the past; which was, in this case, an entangling interaction between 2 and 3. Similarly, the descriptor of 4 following the local measurement will depend on $\phi$ and not on $\theta$, but will depend too on what happened to system 3---that is, on 3's initial entangling with 2. Because the descriptors of 1 and 4 depend, following the pair of local measurements, on the initial entangling interaction between 2 and 3, their product can give rise to the familiar non-factorisable probability distribution when 1 and 4 are subsequently brought together and joint measurements performed.      

It is tempting to think of the contiguity property of the Deutsch-Hayden descriptors 
as depicting a causal chain in which dependence on the parameters characterising the history of a system is passed on during interactions, or even more metaphorically, in terms of information about the relevant history of a system being transmitted via local interactions. More soberly, we see that if the $\mathbf{q}_{i}$ are taken to be the primary objects of interest then the effects of local unitary operations on these are indeed explicitly seen to be local, as the descriptor of a system cannot come to depend on a parameter characterising a unitary operation selected in a distant region without the system having undergone an appropriate chain of \textit{local} interactions.  
As we have said, however, this is just the no-signalling theorem writ large.

\section{Assessing the Claims to Locality}\label{assess}

Having outlined the machinery of the Deutsch-Hayden approach, we may now consider the status of its claim to provide a particularly local picture of quantum mechanics. As remarked in the introduction, it is necessary to distinguish two modes of interpretation of the formalism.

\subsection{The Conservative Interpretation}

The \textit{conservative interpretation} is to take the formalism at face-value, simply as a re-writing of standard (unitary) quantum mechanics, in which we fix the initial state $\rho$ and track time evolution via the $\mathbf{q}_{i}(t)$. If we want to talk in terms of properties, we may see the $\mathbf{q}_{i}(t)$, against the background of a chosen $\rho$, as denoting propensities for the display of certain individual and joint probability distributions for measurement outcomes, via equations (\ref{rdmi}) and (\ref{rdmijk}).    

\subsubsection{Locality Claim (1)}\label{loc1}

The first claim to locality was that the global state can be seen to be determined by the states of individual subsystems. What is certainly true is that given the $n$ $\mathbf{q}_{i}(t)$, the $4^{n}$ $\Gamma_{j}(t)$ are determined and hence we can keep track of the changes to the joint system over time. Note, however, that the initial global state $\rho$ still has to be specified and plays a very important role. It is needed to determine the experimentally accessible properties of individual and joint systems; both the $\Gamma_{j}(t)$ \textit{and} $\rho$ are required to determine expectation values of measurements. That it is the \textit{global} state is crucial, as in general in the presence of entanglement, $\langle q_{i,m_{i}}\!(t)\,q_{j,m_{j}}\!(t)\rangle_{\rho}\neq \langle q_{i,m_{i}}(t)\rangle_{\rho}\langle q_{j,m_{j}}(t)\rangle_{\rho}$. 

With the global state of the system still playing such an important role, however, it is not clear that we have yet gained much in the way of locality by considering the Deutsch-Hayden construction under the conservative interpretation.
Taking the simplest picture of a time evolving density operator, products of the $\mathbf{q}_{i}(t)$ determine how \textit{any} given initial state will evolve; it is no surprise if the initial state of the joint system is specified and we have kept track of the changes to the system (albeit that these are fixed by the individual $\mathbf{q}_{i}(t)$) that we then know what the final state will be.

In reply it is open to Deutsch and Hayden to argue that appeal to the global state is in fact innocuous, as a standard initial state can always be chosen and the $\mathbf{q}_{i}(0)$ adjusted accordingly. To be sustained, however, this line of argument commits one to the ontological interpretation, which we shall consider in due course. For now, let us consider the status of the second locality claim under the conservative interpretation.

\subsubsection{Locality Claim (2)}

We begin by asking why it might seem important to show explicitly that local unitary operations have only a local effect. (We recall, of course, that the standard no-signalling theorem already assures us that local unitaries will not have any effect on the probability distributions for distant measurements). It is clear that if we were only to consider the question of nonlocality as it is usually raised in the context of Bell-type experiments, then the Deutsch-Hayden approach would not offer us any distinctive advantages. For, as has been mentioned, their point of departure is to assume no-collapse quantum mechanics with no determinate values added, thus the appropriate comparisons must either be with an Everettian or a statistical interpretation. But it is well known that the Everett interpretation does not suffer from the familiar difficulties with nonlocality in the Bell or EPR setting that accrue to theories involving collapse or additional variables (indeed, this is often presented as one of the selling-points of the approach); while for a statistical interpretation, the familiar no-signalling theorem does all that could be required to ensure that nonlocality does not arise (see \citet{erpart1} for further discussion and references). Thus if one is considering the question of locality in this context, the crucial factor is the assumption of quantum mechanics without a real process of collapse, and without additional variables, rather than anything distinctive about the Deutsch-Hayden approach.

However, things may look rather less clear-cut when one considers the phenomena of entanglement assisted communication such as superdense coding \citep{superdense} and teleportation \citep{teleportation}. These phenomena vividly illustrate the fact that in the presence of entanglement, local unitary operations can have a very significant effect on the \textit{global} state of the system.
And might this not indicate a novel sort of nonlocality of which even the Everett interpretation would be guilty? 
If so, the Deutsch-Hayden approach would seem to offer a clear advantage, with its explicit locality regarding the effects of local unitary operations.

\begin{figure}
\begin{center}\scalebox{0.7}{\includegraphics{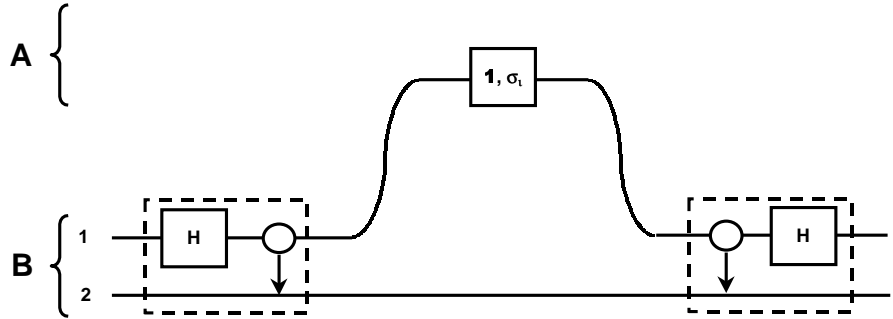}}\end{center}
\caption{Superdense coding. A maximally entangled state of systems 1 and 2 is prepared by Bob (B). System 1 is sent to Alice (A) who may do nothing, or perform one of the Pauli operations. On return of system 1, Bob performs a measurement in the Bell basis, here by applying a controlled-NOT operation, followed by the Hadamard gate. This allows him to infer which operation was performed by Alice.}\label{sdcoding}
\end{figure}

Consider the example of superdense coding in more detail (Fig.~\ref{sdcoding}).  
In this protocol, Alice is able to send Bob two bits of information with the transmission of a \textit{single} qubit, by making use of the global effect of a local operation. 

The two parties begin by sharing a maximally entangled state; let us say the singlet state. Then, simply by applying one of the Pauli operators to her half of the shared system, Alice may flip the joint state into one of the others of the four orthogonal, maximally entangled Bell states: a local operation has resulted in a change in the global state that is as great as could be---from the initial state to one orthogonal to it. Now, if Alice sends her half of the shared system to Bob, he just needs to measure in the Bell basis to determine which of the four operations Alice performed, arriving at two bits of information. In this protocol, the possibility of changing the global state by a local operation has been used to send information in a very unexpected way. The phenomenon of teleportation may also be viewed as arising from the fact that the set of maximally entangled states may be spanned by local unitary operations \citep{braunstein:twist}.

So, does the example of entanglement assisted communication indicate an important sphere in which Deutsch-Hayden presents benefits of locality? Note that these examples do not affect the question of locality for the statistical interpretation, as on this interpretation the quantum state does not correspond to anything real. But one might be interested in a more realist approach. Thus we should ask how the Everett interpretation fares with locality in entanglement assisted communiction.   

It can in fact be argued that the examples of superdense coding and teleportation do not demonstrate a new form of nonlocality in Everett. Our worry is about the effect on the global state of local operations; however, even if we are being robustly realist about it, the global state is not itself a spatio-temporal entity. Thus changes in the global state do not correspond to local \textit{or} to non-local changes. It is better to think in terms of changes to properties of the systems; but it is clear that unlike the sort of change that would be associated with collapse, the effects of local unitary operations that we are considering do not give rise to any changes in local and non-relational properties of the separated systems (i.e., locally observable probability distributions are unchanged). 
Thus, although certainly striking, and non-classical, the potential global effects of a local unitary operation in the presence of entanglement are not appropriately construed as non-local. 

The case is clear enough for superdense coding; teleportation invites a further brief comment. When this protocol is analysed from the Everett perspective, the significant feature is that
immediately following Alice's measurement and before she sends a record of her outcome to Bob, Bob's system will already have acquired a definite state related to the state Alice is sending, relative to the outcome of Alice's measurement. And this may look like a form of non-locality: the pertinent relative state of Bob's system has come to depend on the parameters characterising the state being sent by Alice, merely as a result of a local operation (measurement) carried out at a distance by Alice, and without any direct interaction between the two sides of the experiment.    

It seems that this appearance of non-locality is again not genuine, however. What have changed as a result of Alice's measurement are the relative states of Bob's system; that is, roughly, relational properties of his system. It is no mystery that relational properties can be affected unilaterally by operations on one of the \textit{relata} and it certainly does not connote non-locality\footnote{Consider the following classical example: We have two heaps of sand, $x$ and $y$, piled on the ground, some distance apart. Let us say $x$ is heavier than $y$. By adding a few more shovel-fulls to $y$, we may make this statement false; but this does not imply a non-local effect on $x$.}. 
The effect of Alice's measurement has been to entangle further systems with the initial entangled pair, namely, the system whose state was to be transmitted and systems recording the outcome of the Bell measurement. The trick is that the type of measurement interaction Alice performs has been chosen such that the way in which the systems recording the outcome of her measurement are allowed to become related to Bob's system (in virtue of the initial entanglement) entails that relative to their outcome recording states, Bob's system will have the required states. That is, the genuine change is in fact all on Alice's side. 
(\citet{vaidman} has also argued to the effect that teleportation does not involve nonlocality, when understood in Everettian terms.)

The conclusion is that when considered under the conservative interpretation, the explicit locality in the effect of local unitary operations that the Deutsch-Hayden formalism provides in the contiguity of changes in the $\mathbf{q}_{i}(t)$ does not vouchsafe an important sense of locality that would be lacking in an Everettian or statistical interpretation. 
Indeed we can see that it would necessarily be quite misleading to suggest that the contiguity property points to a novel feature of locality in the Deutsch-Hayden formalism interpreted conservatively. As we have noted, the novelty must be supposed to concern the absence of any effect on the global state from local unitary operations, even in the presence of entanglement; and this indeed follows, in a trivial sense, if we \textit{fix} the initial state $\rho$ and track time evolution via the $\mathbf{q}_{i}(t)$, adopting the Heisenberg viewpoint. But what we described in the Schr\"odinger picture as a change in the global state following a local operation now merely becomes, in the Heisenberg picture, a change in the expectation values for some joint observables that can't be understood in terms of changes in expectation values for observables pertaining to subsystems. But why, if we were supposed to be worried at all, should we be less worried by changes in these joint expectation values as a result of local unitary operations, than in changes to the global state?

\subsection{The Ontological Interpretation}

Maudlin, in the course of his careful discussion of the question of holism in quantum mechanics, arrives at the following dialectical position:
\begin{quoting}
We now have a reasonably clear question: according to the quantum theory, can the physical state of a system be completely specified by the attribution of physical states to the spatial parts of the system, together with facts about how those parts are spatiotemporally related? \citep[p.50]{maudlin}
\end{quoting}
In standard quantum theory, the answer, of course, is \textit{no}. The point of the Deutsch-Hayden approach under the ontological interpretation is to answer instead `yes'.

To see how this might be achieved, recall why the conservative interpretation must fail to give an affirmative answer to Maudlin's question.

In the conservative interpretation, the assignment of properties at a given time is necessarily a joint venture between the global state $\rho$ and the descriptors; and as we noted (Section~\ref{loc1}), appeal \textit{has} to be made to global properties of the state. The $\mathbf{q}_{i}(t)$ cannot themselves be said to denote properties of the subsystems, rather, they determine what the effects of dynamical evolution would be for any possible initial state of the whole system. It is only when some particular initial state is specified that we may begin to talk about the properties of subsystems and of the whole; denoted by expectation values of the $\mathbf{q}_{i}(t)$ and products of the $q_{i,m_{i}}(t)$, respectively. And we have already noted a crucial feature several times: in general, the properties that are assigned to joint systems (expectation values for joint observables, or propensities for the display of certain joint probability distributions on measurement), will not be reducible to properties assigned to subsystems (individual expectation values and propensities). 

The ontological interpretation departs from this in two ways. First, the status of the global quantum state is fundamentally revised. A fixed standard state is adopted by convention (for example, the computational basis state $\ket{0}{}\ket{0}{}\ldots\ket{0}{}$) and it is delegated to playing a purely mathematical r\^ole in the machinery of the theory, rather than representing any physical contingency. Its status is now simply that of a \textit{rule} for reading off the observable properties of systems. Secondly, the $\mathbf{q}_{i}(t)$ are taken to represent intrinsic (i.e., non-relational) and occurrent (i.e., non-dispositional) properties of individual subsystems. The first feature is required of these properties if the global properties of the total system are to be reduced to the properties currently possessed by its subsystems; the second feature is a natural requirement in this context. A change in the descriptor of a system now represents a change in the actually possessed, intrinsic properties of the system. These intrinsic properties are clearly of a new sort; and they do not receive any further characterisation or explanation than is provided by their r\^ole in the formalism. Thus on the ontological interpretation, the content of the first claim to locality is that the global properties of the joint system are reducible to local, intrinsic properties of subsystems, while the content of the second is that \textit{changes} in the global properties are reducible to changes in the currently possessed properties of subsystems. Under the ontological intepretation, then, we certainly have an interesting thesis. Note that now, as adumbrated earlier, changes in the initial conditions of a system may be reflected in changes in the $\mathbf{q}_{i}(0)$, whereas under the conservative interpretation they would be represented by changes in the time-zero density matrix, $\rho(0)$\footnote{A half-way house is unsatisfactory. One might adopt a conventional fixed initial state in the conservative interpretation and adjust the $\mathbf{q}_{i}(0)$ accordingly, but this would not eliminate the global r\^ole of the state in determining joint properties, i.e. we do not have reducibility to individual properties, as in this interpretation the $\mathbf{q}_{i}(t)$ do not represent intrinsic properties.}.

It can hardly be emphasized enough that the approach of the ontological interpretation marks a considerable departure from our usual ways of thinking about quantum mechanics. Indeed it might best be thought of as the proposal of a new theory, in which the behaviour of the intrinsic properties denoted by the $\mathbf{q}_{i}(t)$ is fundamental\footnote{Note, however, that the ontological interpretation of Deutsch-Hayden lacks a measurement theory. Although we have a prescription for what the probability distributions associated with various measurements will be, we do not yet have a description of the measurement process itself, or of the obtaining of various outcomes, in terms internal to the theory. It might be thought that some sort of Everettian approach could be adopted, but as the relative state finds no place in the Deutsch-Hayden framework, it appears, at least \textit{prima facie}, to be resistant to standard Everettian analysis.}.  

In gaining with respect to reducibility, however, the ontological interpretation acquires what might be felt to be some rather objectionable features. The first is a problem of underdetermination.

The central, distinctive, claim of the ontological interpretation is that the intrinsic properties of a subsystem, denoted by the descriptor $\mathbf{q}_{i}(t)$ are fundamental. This means that there is a fact about which properties a given system actually possesses at any stage; and thus also, a fact about what the true descriptor of the system is. However the interpretation also involves a strict distinction between observable and unobservable properties. The observable properties are those that are given by expectation values. But this means that we can never in fact know the true descriptor of a system. We only have empirical access to expectation values and to the density matrices of systems, but continuously many different $\mathbf{q}_{i}(t)$ will be compatible with this data. The true descriptor of a system could be any one of the many that would provide consistency with both the density matrix of the subsystem (eqn.~(\ref{rdmi})) and that of the total system (eqn.~(\ref{density})). Thus the facts about the true descriptors; and hence about the intrinsic properties that systems actually possess, although supposedly the fundamental reality, are empirically inaccessible. According to the ontological interpretation, there is an important fact about what the correct descriptors of a set of systems are, but any assignment of descriptors to such a set will necessarily be underdetermined by the accessible data.

As a corollary of this point, it is worth remarking that the analogy Deutsch and Hayden suggest between their descriptors and Einstein's desired `real state' for separated systems might be overstated. While it may be the case that under the ontological interpretation, subsystems do indeed possess independent real states, we would still face the epistemological problem that this real state could never be determined by local measurements---we could at most only ever learn the $\langle \mathbf{q}_{i}(t)\rangle_{\rho}$ for a system, when presented with a sufficient number of identically prepared systems.    
 
The second difficulty for the ontological interpretation, and one closely related to the underdetermination problem, is that the shift in meaning of the $\mathbf{q}_{i}(t)$, from determining time evolution for any given initial state, to denoting intrinsic properties of subsystems, induces a worrisome redundancy. In the normal quantum mechanical picture one can think of the $\mathbf{q}_{i}(t)$ in the following way.

Take some fixed sequence of unitary operations performed on a group of systems. This sequence will correspond to some particular evolution of the set of $\mathbf{q}_{i}(t)$. Now we could consider different initial quantum states for the set of systems; these states would evolve variously under the sequence of unitary operations whose effect is captured in the evolving $\mathbf{q}_{i}(t)$. 
At any given time, the actual quantum state of our group of systems could be one from a whole range, depending on which initial state was in fact chosen. The evolution of some particular initial state from time 0 to time $t$ may therefore be said to depict one history from the range of possible ones. To use the term favoured by philosophers, the evolution of this state represents the history of one possible world. A choice of different initial state is a choice of different possible world. 

Now the $\mathbf{q}_{i}(t)$ capture the effects of our sequence of unitary operations for all initial states. Thus their time evolution can be said to depict the histories of the \textit{entire set} of possible worlds; whilst the world from amongst these that is realised is determined by which initial state is chosen. However, when we move to the ontological view, the \textit{very same structure} (the sequence of time evolving $\mathbf{q}_{i}(t)$) only represents a \textit{single} world, as the choice of initial state is a fixed part of the formalism. What seems like it can represent a range of possible worlds, we are to suppose, can only represent a single one; and conversely, the structure being used to describe a single world in the ontological Deutsch-Hayden picture is one we know in fact to be adequate to describe a whole set of possible worlds in quantum mechanics. Thus the Deutsch-Hayden picture, taken ontologically, would seem to be extremely, perhaps implausibly, extravagant in the structure it uses to depict a single world. This difficulty, whilst certainly not a knock-down objection to the ontological intrepretation, nonetheless seves to highlight some of its unpalatable features.

\section{Information and Information Flow}\label{info}

We have seen that under the conservative interpretation, the Deutsch-Hayden formalism does not confer any benefits with respect to locality that do not follow directly from adopting no-collapse, unitary, quantum mechanics as a basic theory, and hence would be equally available with an Everettian interpretation, or, if one were perhaps to allow a formal collapse, but deny that it corresponded to any real process, on a statistical interpretation. With the ontological interpretation, by contrast, we do find something new, but this is better characterized as concerning the reducibility of global properties to local intrinsic properties of subsystems, rather than being a question of locality or nonlocality.

One of the most important aspects of the Deutsch-Hayden approach, however, is the claim that their formalism finally clarifies the nature of information flow in quantum systems; indeed, that it reveals that information can be seen to be transported locally in quantum systems, the phenomena of entanglement assisted communication notwithstanding. It is to this question that we now turn. Again, the matter must be assessed independently for the two different modes of interpretation of the formalism. We shall begin, however, with a few general remarks about the topic of information flow.   

\subsection{Whereabouts of information}\label{whereabouts}

As we saw in the previous chapter, the puzzle that seems to be posed by the examples of teleportation and the like is over the question `How does the information get from $A$ to $B$?'. This is a perfectly legitimate question if it is understood as a question about what the causal processes involved in the transmission of the information are, but recall that it would be a mistake to take it as a question concerning how information, construed as a particular, or as some pseudo-substance, travels. `Information' is an abstract noun and doesn't serve to refer to an entity or substance. Thus when considering an information transmission process, one that involves entanglement or otherwise, we should not feel it incumbent upon ourselves to provide a story about how some thing, denoted by `the information', travels from $A$ to $B$; nor, \textit{a fortiori}, worry about whether this supposed thing took a spatio-temporally continuous path or not. By contrast, we might very well be interested in the behaviour of the \textit{physical systems} involved in the transmission process and which may or may not usefully be said to be information carriers during the process.    

A second general point concerns what it might mean to ask whether or not information is a `non-local quantity' \citep[p.1759]{dh}. 
Note that for the reason just stated, information is not something that can be said to have a spatio-temporal character, but nonetheless one can, in certain contexts, intelligibly ask `Where is the information?' This question is a fairly specialised one, though: it presupposes that we have some specific piece, or type, of information in mind and asks where this may be found, in the sense of asking where one might learn, or learn about, the fact, or facts, it pertains to. (And, of course, to specify where something may be learnt is not to say that \textit{what is learnt} has to be located there.) Sometimes no very precise answer to this question in terms of a designated spatio-temporal region will be possible, or particularly helpful. 

As a particular example of the latter case, and one that will figure again later, consider the following scenario of encrypting a message. Let us say that Alice and Bob are spatially separated but share a secret random bit string, the key. Alice also has in her possession a message she wishes to send to Bob, a string of bits denoting something; this is the information we are interested in. At this stage, we can say that Alice's notebook, in which the message is written, contains the information. If she then encrypts the message by adding (mod 2) the message string to the key, writes the result down (producing the cyphertext); and destroys both the original message and her copy of the key, then the question `Where is the information now?' leaves us without a straightforward answer. We can't answer by gesturing to Alice's side, or to Bob's side, or to the cyphertext, since from none of these, taken individually, may we learn what the message was; although if we had access both to Bob's key and the cyphertext then we should be able to learn it. A simple request for a location doesn't have a useful answer in this scenario. For this reason, we introduce further vocabulary and talk instead of the message being \textit{encrypted} in the cyphertext. It is not to be found wherever the cyphertext is located, rather, it may be learnt whenever cyphertext and key are brought together, and not otherwise; the asymmetry in the r\^oles of the cyphertext and key is captured by the fact that it is the cyphertext and not the key in which the message is said to be encrypted (although not located).
The bald question `where is the information throughout this protocol?' does not, in this case, invite answers with sufficient articulation for a perspicuous description of what is going on.

Deutsch and Hayden, however, have something specific in mind when they raise the question of whether in quantum systems, information is a local or non-local quantity. If it is the case that a joint quantum system can have global properties that are not reducible to local properties of subsystems, then these global properties might be used to encode and transmit information in a way that cannot be understood as subsystems individually carrying the information. This is what they would mean by information being a non-local quantity. 
The issue is whether we can, in general, always understand an information transmission process involving quantum systems in terms of the properties of subsystems being used to carry the information.
The examples of entanglement assisted communication, as usually understood, would strongly suggest otherwise.

We shall focus on teleportation as the most interesting case; and one which displays the characteristic features at issue.

\subsection{Explaining information flow in teleportation: Locally accessible and inaccessible information}\label{explaining}
 
\begin{figure}
\scalebox{0.7}{\includegraphics{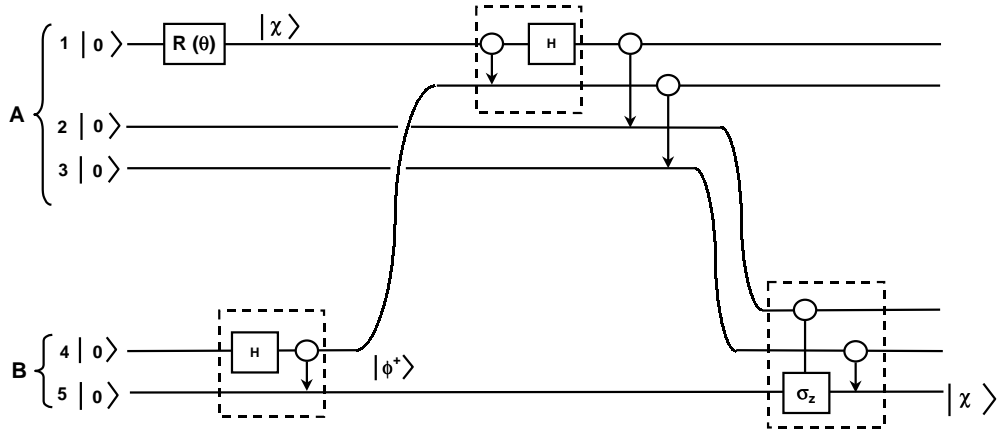}}
\caption{Teleportation. All systems begin in the 0 computational basis state. Bob (B) creates a maximally entangled state of systems 4 and 5. System 1 is prepared in some unknown state $\ket{\chi}{}$, by a rotation depending on the parameter $\theta$. When system 4 is sent to Alice (A), she performs a measurement in the Bell basis, recording the outcome in systems 2 and 3. Systems 2 and 3 are transported to Bob, who performs a controlled-$\sigma_{z}$ operation on 2 and 5,and a controlled-NOT on 3 and 5. System 5 is left in the original unknown state $\ket{\chi}{}$.}\label{teleportation}
\end{figure}

Let us recall once more what the teleportation protocol looks like in the absence of collapse (Fig.~\ref{teleportation}).
Sharing a maximally entangled state with Bob, Alice performs a joint measurement on her half of the entangled pair (4) and on a system (1) prepared in some unknown state, with the result that the state of Bob's system (5), relative to the outcomes of her measurement, is changed in a way that relates systematically to the unknown state to be teleported. At this stage of the protocol, every system involved is now in a maximally mixed state, i.e., the information that characterises the unknown state will not be available to local measurements.
As we have seen, the protocol continues with the sending of the systems (2 and 3) recording the outcome of Alice's measurement to Bob, who can now perform the conditional unitary operations required to disentangle his system (5) from the others, in such a way that it ends up in the original, unknown, state. The information characterising the unknown state is now available again to local measurements, but this time, only at Bob's location.  

The crucial feature in this protocol is the change in the relative states that is allowed by the \textit{global} property of entanglement. Subsystems, therefore, do not seem to be playing the r\^ole of information carriers in teleportation, and this conclusion is further supported by the fact that the only systems that are sent from Alice to Bob during the protocol are both maximally mixed.

Deutsch and Hayden, though, wish to give an account of teleportation in which information flow is local; that is, in which subsystems can indeed be seen to carry information from Alice to Bob. In particular, they are concerned to rebut claims such as that of \citet{braunstein:irreversible}, who suggests that the information characterising the unknown state is contained in the global system rather than in subsystems during the protocol; or the---by now very familiar---approach of \citet{penrose:1998}, who suggests that the information must flow along a channel constituted by the initial shared entanglement between Alice and Bob, first backwards, and then forwards again in time.

Clearly, a good starting point for the debate would be an appropriate criterion for when a system may be said to contain information. 
Deutsch and Hayden would seem to have one of two slightly different necessary and sufficient conditions in mind, although they are not explicit.

They begin by introducing a fairly familiar \textit{sufficient} condition for a system $S$ to contain information about a parameter $\theta$: If a suitable measurement on $S$ would display a probabilistic dependence on $\theta$, then $S$ may be said to contain information about $\theta$. Then a \textit{necessary} condition for containing information is presented: $S$ can be said to contain information about $\theta$ only if its descriptor depends on $\theta$. These definitions motivate an informal argument of roughly the following form: Let us say we have a group of systems that includes $S$; denote this group by $S \cup S^{\perp}$. Assume that the descriptor of $S$ alone depends on $\theta$. If we know that the group $S \cup S^{\perp}$ as a whole contains information about $\theta$, because global measurements would display suitable probabilistic dependence, but $S^{\perp}$ does not (as the descriptors of the systems in $S^{\perp}$ do not depend on $\theta$), then the information must be in $S$, in virtue of $S$'s descriptor depending on $\theta$. Therefore from the fact that the descriptor of $S$ depends on $\theta$, we may infer that it contains information about $\theta$.

This conclusion would be underwritten by either one of the following two definitions\footnote{The two statements that follow must be understood as proposed definitions, as they are not entailed by Deutsch and Hayden's argument, just sketched. The argument uses the necessary and the sufficient condition for containing information, and the rule of inference: if a group of systems contains information about $\theta$, and a subgroup does not, then the complement of that subgroup contains the information about $\theta$. However if we have more than one system whose descriptor depends on $\theta$, then all that the argument based on these principles allows us to conclude is that their union contains the information, not each system individually, which is the desired conclusion.}:

\begin{definition}
$S$ contains information about $\theta$ $\leftrightarrow$ its descriptor depends on $\theta$
\end{definition}
\begin{definition}
$S$ contains information about $\theta$ $\leftrightarrow$ its descriptor depends on $\theta$ and measurements on the global system $S \cup S^{\perp}$ would display a probabilistic dependence on $\theta$.   
\end{definition}
These two definitions differ as it is possible for the $\mathbf{q}_{i}(t)$ to depend on $\theta$, but for $\rho(t)$ not to (recall the problem of underdetermination). The second is rather more natural, particularly if we are to tie the notion of information being used to the context of definite communication-theoretic procedures.

With one of these definitions of containing information in hand, Deutsch and Hayden's claim for the locality of information flow follows directly from the contiguity property of the changes in the $\mathbf{q}_{i}(t)$. The proposal is that teleportation should now be understood in the following way. System 1 is prepared in some state characterised by the parameter $\theta$; its descriptor now depends on $\theta$. Following Alice's Bell-basis measurement, the descriptors of the `message qubits' 2 and 3 also come to depend on $\theta$. These two systems, as they are transported, carry the information about $\theta$ to Bob's location, where, following a suitable local interaction, the descriptor of his system (5) also comes to depend on $\theta$. We must note the further, crucial, point, however, that the systems 2 and 3 carry the information to Bob in a \textit{locally inaccessible} manner. Although their descriptors depend on $\theta$, and hence the systems may be said to carry information under the Deutsch-Hayden definition, this dependence may not be revealed by measurements on the systems individually---their reduced density matrices are maximally mixed.

Deutsch and Hayden define locally inaccessible information as information that is present in a system, but that may not be revealed by individual measurements on the system. The explanation of teleportation, then, is that the message qubits do actually carry the information characterising the unknown state to Bob, but they do so locally inaccesibly. The general conclusion is that subsystems can always be thought to carry information in entanglement assisted communication protocols (hence `information is a local quantity'), it is just that these protocols involve locally inaccessible information.  

\subsection{Assessing the claims for information flow}\label{assessinfo}

How satisfactory is this account as an explanation of teleportation, and, indeed as a general picture for information transmission in quantum systems? We shall consider three questions: First, have Deutsch and Hayden finally given the correct account of teleportation, as opposed, say, to Braunstein? Second, is the concept of locally inaccessible information useful? Third, do Deutsch and Hayden provide us with a new concept of information, or quantum information? 
We must consider the answers to these questions for the two modes of interpretation of the formalism in turn.

Before that, a preliminary remark. Recall that as properly understood, the question `How does information get from Alice to Bob?' is a question about the causal processes involved in the transmission. It is clear that simply answering: `the information is carried in the message qubits' would not be enough to explain teleportation on its own, as this information might never be made accessible again at Bob's location, or it might be made locally accessible, perhaps, but not in such a way that Bob's system is actually to be found in the original unknown state. Obviously, the explanation has also to refer to the r\^ole of the initial entanglement and the changes in the global properties of the system that this entanglement allows, and which the teleportation protocol exploits. This suggests a moderate way of understanding the application of the Deutsch-Hayden formalism in teleportation that would not involve commitment to their claims about locality or information flow.

On this view, the advantage their formalism presents is simply in highlighting the difference in r\^oles played by the initial entanglement and the message qubits in teleportation. The asymmetry in these r\^oles is, as Deutsch and Hayden point out, analogous to the asymmetry in the r\^oles of the key and cyphertext in classical encryption based on a shared secret random string\footnote{The analogies and, importantly, disanalogies, between entanglement and shared secret bits are developed in detail in \citet{collins:popescu}.}. Before the final stage of the protocol, it is the message qubits, and not Bob's qubit, that have had the direct dynamical coupling to the system whose state is to be teleported (reflected in the fact that their descriptors depend on $\theta$)---compare with the classical cyphertext, which is generated from the message. But it is the correlations that are established between the relative states of the message systems and Bob's qubit, in virtue of the initial entanglement, that allow the unknown state to be recovered by Bob. (Similarly, the classical correlations between the key and cyphertext allow the encrypted message to be recovered). This suggests that it may well be useful to distinguish between the question of whether an analysis in terms of the $\mathbf{q}_{i}(t)$ helps us understand an aspect of teleportation; and whether the account in terms of information flow does so.

Returning to our three questions. The adjective `correct' in the first question might be understood in one of two ways; either correct \textit{simpliciter}, or correct \textit{given} the background assumptions. In order to be correct \textit{simpliciter}, the account of teleportation would clearly have to be, first of all, correct given the background assumptions, while these background assumptions themselves also have to be correct. The relevant background assumption when we consider the conservative interpretation is that unitary (no-collapse) quantum mechanics is our setting; this is the setting also for \citet{braunstein:irreversible}, hence the point of the comparison.  

\subsubsection{Conservative interpretation}

From the previous remarks on the conservative interpretation, we know that the assignment of properties to systems involves both the global state and the $\mathbf{q}_{i}(t)$: we do not have reducibility of global properties to properties of subsystems and therefore subsystems cannot, after all, always be thought to carry information in entanglement assisted communication. It makes no odds whether one adopts the Heisenberg or the Schr\"odinger viewpoint, it is still the case that joint (and irreducible) properties of subsystems are being used to carry information in the protocols. In Braunstein's account of teleportation, after Alice's Bell-basis measurement, the information characterising the unknown state is said to be in the correlations between the message qubits and Bob's qubit, i.e., it is carried by certain joint properties of these systems. The same is true in the Deutsch-Hayden setting, understood conservatively; so we are not in fact being offered a substantially different account of teleportation. This entails part of the answer to the second question.

Under the conservative interpretation, there is an important sense in which there is no difference between saying that a system contains locally inaccessible information and saying that the information is in the correlations. In both cases this would translate into: the information is carried by joint, and not individual, properties of subsystems. One can frequently make perfectly good sense of a system being said to contain information about a parameter if a suitable measurement on the system would display a probabilistic dependence on the parameter, for then one can learn something about the parameter by performing the measurement. But if the information is locally inaccesible, then this means either i) for some \textit{different} initial state of the global system then there will be a probabilistic dependence for the local measurement---but this would be physically irrelevant to the situation actually being considered; or ii) for some measurement on the \textit{global} system, a probabilistic dependence on the parameter will be displayed---and this is no different from what one would say on Braunstein's account.

So where, if anywhere, does a difference lie? In marking an asymmetry. But note that the pertinent aysmmetry may also be understood in a Schr\"odinger picture account such as Braunstein's. In teleportation, the point being emphasized is that it is the message qubits, and not Bob's qubit, that have had the direct dynamical coupling to the system that was prepared in the state characterized by the parameter $\theta$; and this is clear enough without invoking locally inaccessible information. (The significance, of course, is that we know from the no-signalling theorem that dependence on a parameter chosen in one region may not be displayed in another unless there has been a direct, or indirect, dynamical coupling between systems from the two regions.) Another way to mark the asymmetry would begin by pointing out that the initial entanglement, the sending of the message qubits to Bob, and the correct sequence of unitary operations being performed by Alice and Bob, are individually necessary, and jointly sufficient conditions for a successful teleportation protocol. If we were to miss any one of these out, then the protocol would fail, but evidently, for different reasons in each case.         

The preceding discussion indicates that under the conservative interpretation, the concept of locally inaccessible information is not playing a very useful explanatory r\^ole. It is misleading to suggest that the message qubits really carry anything---at best this is a roundabout way of saying that joint properties do\footnote{Recall from the comments in Section~\ref{whereabouts} and the previous chapter that we are not \textit{forced} to say that the information must be located in one system rather than another, or that it is carried by one system rather than another. The assumption that we \textit{must} is predicated upon the misleading picture of information as a particular or substance.}. 
This conclusion in turn casts doubt on the value of adopting either of the proposed definitions of containing information in the context of the conservative interpretation.

However, it would be precipitate to conclude from this that we may in fact learn nothing from the analysis of teleportation in the Deutsch-Hayden formalism. As suggested earlier, one can distinguish between the description using the $\mathbf{q}_{i}(t)$ being useful and the concept of locally inaccessible information being so. Deutsch and Hayden are certainly right that an analysis in terms of their descriptors does help emphasize the important asymmetry between the r\^oles in the protocol of sending the message qubits and the existence of the initial entanglement; and due consideration of this asymmetry contributes, for example, towards undermining the plausibility of a Penrose-type explanation. The analogy with the cyphertext and key is also enlightening in this regard. But as we have just noted, it is quite possible to mark this asymmetry without needing to invoke talk of containing information, which has potential to mislead.

The answer to the third question under the conservative interpretation is perhaps the most intriguing. We have seen that locally inaccessible information does not figure successfully in an attempt to retain subsystems as information carriers in the presence of entanglement, but have Deutsch and Hayden nonetheless succeeded in shedding light on the---sometimes obscure seeming---concept of quantum information? They say, for example:
\begin{quoting}\vspace*{-.2in} 
...it is impossible to characterize quantum information at a given instant using the state vector alone. To investigate where information is located, one must also take into account how the state came about. In the Heisenberg picture this is taken care of automatically, precisely because the Heisenberg picture gives a description that is both complete and local. \citep[p.1773]{dh}
\end{quoting}
It seems, though, that this suggestion would incorporate a number of confusions.

While it is true that the $\mathbf{q}_{i}(t)$ provide more information than simply following the time evolved state would, this is not information about the time evolution of particular systems that the latter description lacks. The $\mathbf{q}_{i}(t)$ look more informative because they capture time evolution for any given initial state, thus they say more about the dynamics a system has been subject to; but in the conservative interpretation, this is not to say more about the system, but rather about the \textit{unitary operators}.
This extra information that one gets is not then `complete', i.e., information that would be lacking in the description of a given network of systems in the Schr\"odinger picture, but is given one in Heisenberg. Instead, it is information about something else; about how other systems, prepared in a different way would react, or information about, for example, the fields that have driven the systems' evolution.

Furthermore, one can readily accept that one has more information if one knows how the state came about, but deny that this information is a property that has to be located. So again, one can, in fact should, deny that there is information located with systems that is lacking from the state vector picture. The `extra' information represented in the $\mathbf{q}_{i}(t)$ consists of facts about the unitary operations undergone; and this information cannot be said to be here, there, or anywhere, as it makes no sense to ask where these facts are. Facts are of the wrong logical category to possess a location (cf.~\citet{strawson}). The underlying thought seems to be that the description in terms of the $\mathbf{q}_{i}(t)$ allows us to `determine where the information about a given parameter is located at a given instant' \citep[p.1771]{dh}. But note that the question `Where is the dependence on the parameter?' could be a bad question; one inviting us to confuse the description of a thing with the thing itself. It is \textit{what} depends on the parameter that is important; and in entanglement assisted communication, under the conservative interpretation, this will often only be \textit{joint}, and not individual properties.   


\subsubsection{Ontological interpretation}

The discussion of our three questions for the ontological interpretation may be somewhat more brief. As to the first: on the ontological interpretation, global properties are reduced to intrinsic properties of subsystems, therefore, the properties of subsystems may indeed be thought to be carrying the information in entanglement assisted communication protocols. Thus, adopting the Deutsch-Hayden formalism understood in the ontological way, we would have an explanation of teleportation in which
the information that the system carries as a whole can be thought a consequence of information being carried by subsystems; in which information is genuinely carried between Alice and Bob in the message qubits during teleportation. (Of course, this explanation may not be reflected back onto our more usual ways of understanding quantum mechanics, but relies on the ontological interpretation. As such it has no power to confute opposing views, such as Braunstein's, that derive from a different set of assumptions.) 


Why does it now seem acceptable to say that information is carried in subsystems, despite the fact that it may not be possible to learn anything by performing measurements on an individual system? Because in the ontological interpretation, the explanation of the physical processes by which information is transmitted from $A$ to $B$ (answering `How does the information get from $A$ to $B$' in the legitimate way,) involves the intrinsic properties of subsystems denoted by the $\mathbf{q}_{i}(t)$.
In contrast to the conservative interpretation, we are now able to answer the question `What depends on the parameter?' with: the intrinsic properties of subsystems. 
As the intrinsic properties of subsystems are being used as the information bearing properties under the ontological interpretation, the definitions given above of containing information would have a point\footnote{Although it is not clear that they are wholly trouble-free. Under definition (1), for example, there will be cases in which a system is said to contain information locally inaccessibly, but where it could never be made accessible, i.e. could never be displayed even under global measurements. This would tend to undermine the plausibility of the claim that the system does in fact contain information, which casts doubt on the acceptability of the definition. So again, definition (2) would seem preferable. But  
it might be beneficial to restrict talk of containing information still further, to cases in which some particular information transmission protocol is envisaged, or in which an agent would stand to learn something by performing measurements on a group of systems. 
}.
 
Regarding the usefulness of the concept of locally inaccessible information, the purpose of the introduction of this category is to recognise that there are two ways in which a system may be said to carry information in the ontological interpretation; either in its observable, or in its unobservable, empirically inaccessible, properties. This distinction is necessary for the explanation of entanglement assisted communication in the ontological interpretation, thus the introduction of the category is useful.

In answer to our third question, however, it is important to recognise that the ontological interpretation of Deutsch-Hayden is not providing us with an account of a new type of information, but of new properties, new ways in which information may be carried. Again, because this turns on the details of the ontological interpretation, it cannot be taken to provide us with a new understanding of information, or quantum information, that could be transferred back to more familiar quantum mechanical settings.

\section{Conclusion}

Deutsch and Hayden present their formalism as an avowedly local account of quantum mechanics, which finally clarifies the nature of information transmission in entangled quantum systems. To what extent is this successful? We have seen that in order to assess the claims of locality, and the claims regarding the nature of information flow, it is essential to distinguish between a conservative and an ontological interpretation of the formalism, as very different conclusions follow. To summarise:

On the conservative interpretation, there are no benefits with respect to locality that do not follow immediately from adopting a version of quantum mechanics in which there is no genuine process of collapse and no additional properties added (and which, consequently, would be shared by an Everettian or a statistical interpretation); thus no distinctive feature of the Deutsch-Hayden approach is in play. As far as information transmission is concerned, the formalism does not show that information is after all, a local quantity (in Deutsch and Hayden's sense), as it remains the case that joint, rather than individual, properties are used to carry information in entanglement assisted communication protocols. The explanation proffered of teleportation does not differ in substance from that which would be given by an account sharing the same initial assumptions, such as that of Braunstein. Furthermore, we have seen that it would be confused to think that the description in terms of the $\mathbf{q}_{i}(t)$ fills-in an account of information, and where it is located in quantum systems, that is missing in the usual Schr\"odinger picture. The additional information the $\mathbf{q}_{i}(t)$ provide (when they do so) consists of certain facts about the unitary operations undergone (not information carried by systems); and it makes no sense to propose that these facts have a location. 

With the ontological interpretation, on the other hand, we have an interesting result; although one better characterized as regarding the reducibility of global properties of quantum systems to individual properties, rather than as a question of locality or nonlocality. With this reducibility, the claim about the locality of information transmission, even in the presence of entanglement, follows. However, as the ontological interpretation provides a picture which differs so markedly from our usual ways of understanding quantum mechanics, these results clearly cannot be taken to shed light on the nature of information flow in entangled quantum systems when we have \textit{not} taken the dramatic step of introducing an entirely new range of intrinsic properties of systems. And reducibility does not come free: one is confronted with an unpleasant form of underdetermination and the bogey of redundancy.   

Unfortunately, Deutsch and Hayden do not distinguish the two different modes of interpretation of their formalism; indeed they are arguably conflated, to deleterious effect. The reason to believe that they must have something along the lines of the ontological interpretation in mind is that their main claims would not be true in any interesting way otherwise; but at certain points they would seem to suggest clearly that the conservative reading is correct: when they imply that it is merely the move to the Heisenberg picture which does the work (p.1759); when suggesting that they have simply provided a reformulation of Schr\"odinger picture quantum mechanics (p.1773).
As we have seen, however, if there is equivocation between the conservative and the ontological interpretation, then it is impossible to draw any conclusion regarding information flow and locality.

So, having drawn this all-important distinction, the conclusion of our discussion is that in the ontological interpretation, we have a bold thesis which might be adopted, despite its objectionable features, in order to obtain reducibility of global properties to local properties, if this was thought particularly desirable for some reason. Retaining the conservative approach, on the other hand, we would have a formalism with some occasionally useful features, but not one which provides a novel sense of locality, nor, indeed, of information flow. \textit{En route}, the discussion should have shed some more light on the puzzles that so often seem to surround the question of information transmission in entangled quantum systems.

\end{doublespacing}



\chapter[Entanglement in Deutsch-Hayden]{Characterizing Entanglement in the Deutsch-Hayden Formalism}\label{entdh}






\begin{doublespacing}


In the previous chapter, frequent reference was made to the fact that in general in quantum mechanics, the properties assigned to joint systems are not reducible to properties possessed by individual systems: this is the result of entanglement. (The ontological approach to Deutsch-Hayden is then seen as a theory with new types of intrinsic properties of subsystems which allow such reduction, even in the presence of entanglement.)

A natural question to consider next, therefore, is how one may characterize entanglement within the Deutsch-Hayden formalism. 
Based as it is on the Hilbert-Schmidt representation, one may also hope to to gain some geometric insight into entanglement, given the pleasing geometrical picture of quantum states that the formalism provides.   

In this chapter we will be concerned with bipartite entanglement, pure and mixed, in the Deutsch-Hayden formalism. (Related investigations of the entanglement of $2\otimes 2$ systems are \citet{horodeckisPRA:1996,kummer1,kummer2}.)
As we have seen, one of the primary benefits of the formalism is in tracking the history of the evolution of systems. However, for the bare question of whether or not certain systems are entangled, the details of the dynamical history are irrelevant---we need only consider whether the density matrix is entangled or not. The relevant form of an $n$-system density matrix will therefore be (cf. eqn.~(\ref{density}) Chapter~\ref{dh}): 
\begin{equation}
{\boldsymbol{\varrho}}(t)=\frac{1}{2^n}\sum_{m_{1}m_{2}\ldots m_{n}}\biggl\langle \prod_{k=1}^{n}q_{k,m_{k}}\biggr\rangle_{\!\!\!\rho(t)}\prod_{k=1}^{n}q_{k,m_{k}}\,. 
\end{equation} 
For the case of $2\otimes 2$ systems (now dropping the explicit time parameter) the density matrix may be written in the convenient form:
\begin{equation}
\boldsymbol{\varrho}=\frac{1}{4}\Bigl( \mathbf{1}\otimes\mathbf{1} + \mathbf{a}.\boldsymbol{\sigma}\otimes\mathbf{1} + \mathbf{1}\otimes\mathbf{b}.\boldsymbol{\sigma} +\sum_{ij}c_{ij}\sigma_{i}\otimes\sigma_{j}\Bigr),\label{ab form}
\end{equation}      
where $a_{i}$, $b_{i}$ and $c_{ij}$ ($i$,$j=1\ldots 3$) are the expectation values of the operators $\sigma_{i}\otimes\mathbf{1}$, $\mathbf{1}\otimes\sigma_{i}$ and $\sigma_{i}\otimes\sigma_{j}$, respectively, i.e., the values $\langle q_{1,i}\rangle_{\rho}$, $\langle q_{2,i}\rangle_{\rho}$, $\langle q_{1,i}q_{2,j}\rangle_{\rho}$ in the previous notation. 

In this form, the 3-vectors $\mathbf{a}$, $\mathbf{b}$ are the Bloch vectors for the reduced density matrices of systems 1 and 2 respectively, hence determine the expectation values for individual experiments; while the 3 by 3 matrix with components $c_{ij}$---which we will term the \textit{correlation} matrix---determines the results of joint experiments. It is helpful to bear in mind that $\boldsymbol{\varrho}$ in eqn.~(\ref{ab form}) is itself a unit vector (in the 16 dimensional real space of Hermitian operators on $\mathbb{C}^{2}\otimes\mathbb{C}^2$) which may be expressed in column form as:

\begin{singlespacing}
\[\boldsymbol{\varrho}=\frac{1}{2}\begin{pmatrix} 1 \\ \mathbf{a} \\ \mathbf{b} \\ c_{ij}\end{pmatrix}.\]\end{singlespacing}

Let us review some terminology. A state is called entangled if it is not separable, that is, for bipartite systems, if it cannot be written in the form:
\[ |\Psi\rangle_{12}=|\phi\rangle_{1}|\psi\rangle_{2}, \textrm{ for pure, or  } \rho^{12}=\sum_{i}\lambda_{i}\rho^{1}_{i}\otimes \rho^{2}_{i},\textrm{  for mixed states,}\]
where $\lambda_{i} \geq 0, \sum_{i}\lambda_{i}=1$ and 1, 2 label the two distinct subsystems. The case of pure states of bipartite systems is made particularly simple by the existence of the Schmidt decomposition---such states can always be written in the form:
\begin{equation}\label{schmidt}
|\Psi\rangle_{12}=\sum_{i}\sqrt{p_{i}}\,|\bar{\phi_{i}}\rangle_{1}|\bar{\psi_{i}}\rangle_{2},
\end{equation}
where $\{|\bar{\phi_{i}}\rangle\},\{|\bar{\psi_{i}}\rangle\}$ are orthonormal bases for systems 1 and 2 respectively, and $p_{i}$ are the (non-zero) eigenvalues of the reduced density matrices of the subsystems. The number of coefficients in any decomposition of the form (\ref{schmidt}) is fixed for a given state $|\Psi\rangle_{12}$, hence if the state is separable (unentangled), there is only one term in the Schmidt decomposition, and conversely. Pure state entanglement also has a simple relation to Bell inequality violation. Entanglement is a necessary condition for Bell inequality violation; and for bipartite and $n$-partite pure state entanglement, it is sufficient too \citep{gisinperes,popescurohrlich}. That is, all pure entangled states violate some Bell inequality.

Things are more complex for mixed state entanglement. The simple Schmidt decomposition test does not exist; and it was shown by \citet{werner} that some mixed entangled states do not violate any Bell inequality: remarkably, entanglement is not sufficient for Bell inequality violation\footnote{Although it was later shown by \citet{popescu:hidden} that Werner states for dimensions greater than or equal to five could be made to violate a Bell inequality if sequential measurements are allowed---this is termed \textit{hidden nonlocality}. See \citet{barrett:2002} for further discussion.}. The provision of a simple necessary and sufficient condition---the \textit{positive partial transpose} condition---for mixed state entanglement in $2\otimes 2$ and $2 \otimes 3$ dimensional systems, was finally achieved by the Horodeckis in \citep{horodeckisPLA:1996}. One of our aims in this chapter will be to gain some understanding of the positive partial transpose condition in the Deutsch-Hayden formalism. Throughout we will move freely from positive to contrapositive forms of statement, so if a property $P$ is a necessary condition for separability, say, then $\neg P$ is a sufficient condition for entanglement; and so on.

\section{Background}

It will be useful to review some of the pertinent results, beginning with a summary of the Horodecki's positive partial transpose condition.

\subsection{Entanglement witnesses and the Horodecki's PPT condition}\label{entanglement witnesses and the PPT condition}

We will begin with the concept of a positive map. A linear map is said to be positive if it maps positive operators to positive operators. So if $A$ is a positive operator\footnote{By a positive operator I mean what is sometimes called more precisely a positive semi-definite operator. An operator $A$ acting on a Hilbert space $\mathcal{H}$ is positive semi-definite if $\forall \ket{\psi}{} \in \mathcal{H}, \bra{}{\psi} A \ket{\psi}{}\geq 0$. If the operator $A$ acts on a complex vector space, as in quantum mechanics, then positivity of $A$ implies that it is Hermitian.}, $A\geq 0$, and $\Lambda$ is a positive map, then $\Lambda(A)\geq 0$. A positive map $\Lambda$ is termed \textit{completely positive} if, when we consider adding a further system and so enlarging the Hilbert space, the extended map $\mathbf{1}\otimes\Lambda$ is a positive map for operators on the larger space. (Completely positive maps correspond to the most general form of quantum dynamics, see, e.g., \citet[Chpt. 8]{nielsen:chuang}.)  

The transpose operator $T$ is an example of a map which is positive, but not completely positive. The effect of $T$ on an operator is defined in terms of its effect on the operator when written in a particular basis: $(TA)_{ij}=A^{\mathrm{T}}_{ij}=A_{ji}$, i.e. it is just the familiar process of matrix transposition.
The \textit{partial} transpose is when $T$ is applied to a subsystem only. 
So consider an operator $A$ on a tensor product Hilbert space $\mathcal{H}^{1}\otimes\mathcal{H}^{2}$, expressed in terms of a product basis $\{\ket{i}{1}\ket{k}{2}\}$. Its matrix components will be $A_{ik,jl}$, where the indices $i$,$j$ refer to the first subsystem and $k,l$ to the second.  
The effect of taking the partial transpose on the second system, $(\mathbf{1}\otimes T)A$, will be:
\[A^{\mathrm{T}_{2}}_{ik,jl} = A_{il,jk}.\]
Thus the indices referring to system 2 are permuted.

\citet{peresseparability} noticed that the partial transpose could be used to provide a necessary condition for separability. Since $T$ is a positive map and doesn't affect the trace of an operator, its effect, $T\rho$, on a density operator will be to produce another valid density operator, $\rho^{\mathrm{T}}$. Thus the effect of the partial transpose on a \textit{separable} density operator is necessarily to produce another density operator:
\begin{equation}\label{ptseparable}
(\mathbf{1}\otimes T)\rho_{\mathrm{sep}}=(\mathbf{1}\otimes T)\sum_{i}\lambda_{i}\rho^{1}_{i}\otimes \rho^{2}_{i}
=\sum_{i}\lambda_{i}\rho^{1}_{i}\otimes (\rho^{2})^{\mathrm{T}}_{i};
\end{equation}
hence if a state is separable, it will remain positive under partial transposition. As the transpose is not completely positive, however, there exist states which do not have a positive partial transpose; and these must therefore be entangled. The positive partial transpose condition proved better than Bell inequality and entropic (\textit{vide infra}) criteria for distinguishing the entanglement of Werner states, thus it was natural to conjecture that having a positive partial transpose (PPT) is also a sufficient condition for entanglement.

This conjecture was proven to hold for $2\otimes 2$ and $2\otimes 3$ dimensional systems by the Horodeckis; but also to fail in higher dimensions---then PPT is only a necessary condition.

\citet{horodeckisPLA:1996} begin by introducing the concept of an \textit{entanglement witness}\footnote{The name, though, is due to Terhal, e.g. \citep{terhal}.}. An entanglement witness is an Hermitian operator $W$ which has a positive expectation value for all separable states \samepage($\forall \rho (\rho\; \text{is separable}\rightarrow \mathrm{Tr}(W\rho)\geq 0$)), but negative expectation value for at least one entangled state. Thinking in terms of the vector space of Hermitian operators, $\mathrm{Tr}(W\rho)=0$ defines a plane (normal to the vector $\mathbf{W}$ and containing the origin) on one side of which lie all the separable states, $\mathrm{Tr}(W\rho)\geq 0$, while on the other, lies at least one entangled state.

Now, it is clearly true by definition that
\begin{proposition}
\[\rho \;\text{\em is separable} \rightarrow \forall A \Bigl(\forall \rho^{\prime}\bigl(\rho^{\prime}\text{\em is separable}\rightarrow \mathrm{Tr}(A\rho^{\prime})\geq 0\bigl)\rightarrow \mathrm{Tr}(A\rho)\geq 0 \Bigr),\]
\end{proposition}
where $A$ ranges over Hermitian operators and $\rho^{\prime}$ over density matrices. The Horodeckis point out that it is a consequence of the Hahn-Banach theorem that the converse holds too \citep{horodeckisPLA:1996}. The set $S$ of separable states is a convex set, an entangled state $\tilde{\rho}$ lies beyond this set; and for any entangled state there exists a plane separating it from $S$. Thus for any entangled state there is an entanglement witness, namely, the operator $W$ defining this plane \citep[Lemma 1]{horodeckisPLA:1996}. 

The contrapositive statement is that if for a state $\rho$, $\mathrm{Tr}A\rho\geq 0$ for all $A$ which have positive expectation values on all separable states, then $\rho$ is separable (because if it weren't, there would be some $A$---the entanglement witness---for which $\mathrm{Tr}(A\rho)$ would be less than 0). We thus have 
\begin{proposition}\label{lemma1+}
\[\rho \;\text{\em is separable} \leftrightarrow \forall A \Bigl(\forall \rho^{\prime}\bigl(\rho^{\prime}\,\text{\em is separable}\rightarrow \mathrm{Tr}(A\rho^{\prime})\geq 0\bigl)\rightarrow \mathrm{Tr}(A\rho)\geq 0 \Bigr).\]
\end{proposition}
Noting that 
$\forall \rho (\rho \;\text{is separable}\rightarrow \mathrm{Tr}(A\rho)\geq 0)$ is logically equivalent to \[\forall P\forall Q (\mathrm{Tr}(AP\otimes Q) \geq 0 ),\] 
where $P$ and $Q$ are projectors on $\mathcal{H}^{1}$ and $\mathcal{H}^{2}$ respectively\footnote{For the forward implication, if $\mathrm{Tr}A\rho\geq 0$ for all separable states, it is so for all tensor products of 1-d projectors; for higher dimensional projectors, just take sums of 1-d cases. For the converse, if $\forall P\forall Q (\mathrm{Tr}(AP\otimes Q) \geq 0 )$, then $\mathrm{Tr}(A P_{i}\otimes Q_{i})\geq 0$, where $P_{i},Q_{i}$ are 1-d projectors; and it follows that $\mathrm{Tr}(A\sum_{i}\lambda_{i}P_{i}\otimes Q_{i})\geq 0$. But any separable $\rho$ may be written as $\rho_\mathrm{sep} = \sum_{i}\lambda_{i}P_{i}\otimes Q_{i}$, therefore $\mathrm{Tr}(A\rho_{\mathrm{sep}})\geq 0$. QED.}, we reach

\begin{theorem}[Horodeckis, 1996]
\[\rho \;\text{\em is separable} \leftrightarrow \forall A \Bigl(\forall P\forall Q \bigl(\mathrm{Tr}(AP\otimes Q) \geq 0 \bigr)\rightarrow \mathrm{Tr}(A\rho)\geq 0 \Bigr).\]
\end{theorem}
Again, the forward implication is true trivially; it is the converse, based on the lemma concerning entanglement witnesses, that is profound. The next step is to restate Theorem 1 in terms of positive maps\footnote{In \citep{horodeckisPLA:1996}, positive maps between spaces of operators with differing dimensionality are considered, to allow for systems where dim$\mathcal{H}^{1}\neq\text{dim}\mathcal{H}^{2}$. This complication will be suppressed in the following.}.

Here use is made of the isomorphism between positive maps and operators on $\mathcal{H}^{1}\otimes\mathcal{H}^{2}$ \citep{jamiolkowski}. \citet{horodeckisPLA:1996} note that a map $\Lambda$ will be positive \textit{iff} the associated operator $\mathcal{S}(\Lambda)$ is Hermitian and $\mathrm{Tr}(\mathcal{S}(\Lambda)P\otimes Q)\geq 0$, for all $P$, $Q$ on $\mathcal{H}^{1}$, $\mathcal{H}^{2}$ respectively. This allows us to replace the quantification over Hermitian operators $A$ in Theorem 1 with a quantification over maps $\Lambda$. The isomorphism $\mathcal{S}(\Lambda)$ is chosen as $\mathcal{S}(\Lambda)=(\mathbf{1}\otimes\Lambda)P_{0}$, where $P_{0}$ is the projector onto a maximally entangled state; and we reach:
\begin{proposition}\label{original SLambda statement}
$\rho\; \text{\em is separable} \leftrightarrow \forall \Lambda (\Lambda \; \text{$+$ve} \rightarrow \mathrm{Tr}(\mathbf{1}\otimes\Lambda P_{0}\rho)\geq 0)$.
\end{proposition}
It is then argued that the condition on the trace in the consequent of the RHS of prop. \ref{original SLambda statement} is equivalent to the condition $\mathrm{Tr}(\mathbf{1}\otimes\Lambda \rho P_{0})\geq 0$, i.e., a positive map is now considered acting on $\rho$, rather than $P_{0}$. Note that
\begin{proposition}\label{operator to trace}
$\forall \Lambda \bigl(\Lambda\; \text{$+$ve}\rightarrow \mathbf{1}\otimes\Lambda\rho\geq 0\bigr)\rightarrow \forall \Lambda \bigl(\Lambda\; \text{$+$ve}\rightarrow\mathrm{Tr}(\mathbf{1}\otimes\Lambda\rho P_{0})\geq 0\bigr),$
\end{proposition}
as if $\mathbf{1}\otimes\Lambda\rho$ is a positive operator it will have positive expectation value with \textit{all} projectors, including $P_{0}$.
This, together with prop. \ref{original SLambda statement} entails the backwards implication in the following theorem:
\begin{theorem}[Horodeckis, 1996]
$\rho\; \text{\em is separable} \leftrightarrow \forall \Lambda (\Lambda \text{$+$ve}\rightarrow \mathbf{1}\otimes\Lambda\rho\geq 0)$.
\end{theorem}
Again, the forward implication is straightforward, while it is the converse that rests, via props. \ref{operator to trace},\ref{original SLambda statement} on the lemma concerning entanglement witnesses (prop. \ref{lemma1+}). 

Theorem 2 succeeds in characterizing fully the property of separability in terms of the requirement of remaining positive under all positive maps; however it does involve a quantification over \textit{all} positive maps. To reach a manageable operational characterization, \citet{horodeckisPLA:1996} note that for $2\otimes 2$ and $2\otimes 3$ dimensional systems, the positive maps are decomposable in the form: $\Lambda= \tilde{\Lambda}_{1} + \tilde{\Lambda}_{2}T$, where $\tilde{\Lambda}$ are completely positive maps \citep{stromer,woronowicz}. Thus $\mathbf{1}\otimes\Lambda\rho\geq 0$ for all positive maps $\Lambda$ is true if (and only if) $\mathbf{1}\otimes T\rho\geq 0$, as the only way in which $\mathbf{1}\otimes\Lambda\rho$ could fail to be positive is if $\mathbf{1}\otimes T\rho$ fails to be; and we finally reach the conclusion
\begin{theorem}[Horodeckis 1996]
A state $\rho$ acting on $\mathbb{C}^{2}\otimes\mathbb{C}^{2}$ or $\mathbb{C}^{2}\otimes\mathbb{C}^{3}$ is separable {\em if and only if} its partial transpose is a positive operator.
\end{theorem}

\subsection{The majorization condition}\label{entanglement majorization condition}

It is a remarkable feature of entanglement that the state of a joint system may be pure while the states of the individual subsystems are mixed. It is this aspect of entanglement that Schr\"{o}dinger had in mind in his well-known statement that  
\begin{quoting}
Maximal knowledge of a total system does not necessarily include total knowledge of all its parts, not even when these are fully separated from each other and at the moment are not influencing each other at all. \citep[\S 10]{schrodinger:cats}\footnote{Note, however, that this statement is not the most felicitous, as it is ambiguous between the thought that we lack total knowledge of the subsystems because there are facts to know about the individual susbsystems of which we are ignorant; and the---perhaps happier---thought that there simply is no further knowledge to be had regarding the properties of subsytems individually than is given by their reduced density matrix, which in the case being considered, won't be pure.}     
\end{quoting}

For example, with a pair of qubits in the singlet state, the joint state is pure, while the reduced states of the subsystems are maximally mixed. If we look at the von Neumann entropy as a measure of mixedness of these states, the entropy of the singlet state will be zero, while the entropies of each of the subsystems will be 1. This phenomenon couldn't obtain with the Shannon information of a pair of classical random variables, as $H(X\wedge Y) \geq H(X), H(Y)$; and this line of thought has led to the investigation of various entropic inequalities as criteria for entanglement \citep{horodeckis:alphaentropy,cerfadami:1999,tsallis:nonextensiveentropy}.  

This aspect of entanglement achieved its definitive characterization in the majorization criterion of \citet{separabledisorder}.
We have already met the majorization relation, $\prec$, as the underlying notion of disorder on which measures of uncertainty provide a conventionally chosen numerical scale; and we saw that the relation applied both to probability distributions and to the vectors of eigenvalues of density matrices. For a bipartite system, therefore, we may consider how the vector of eigenvalues $\vec{\lambda}(\rho^{12})$ of the joint system compares to the vectors of eigenvalues of the subsystems, $\vec{\lambda}(\rho^{1}), \vec{\lambda}(\rho^{2})$.

\citet{separabledisorder} showed that if the state $\rho^{12}$ is separable, then
\begin{equation}\label{majorization condition}
\vec{\lambda}(\rho^{12}) \prec \vec{\lambda}(\rho^{1})\;\; \text{and}\;\;   \vec{\lambda}(\rho^{12}) \prec \vec{\lambda}(\rho^{2}).    
\end{equation}
That is, in words: if a state is separable, then it is more disordered globally than it is locally---the vector of eigenvalues of the joint state is majorized by the vectors of eigenvalues of the reduced states of the subsystems. If we then have any Schur concave function $U$ that may serve as a measure of uncertainty, we will have the inequalities:
\begin{equation}
U(\vec{\lambda}(\rho^{12})) \geq U(\vec{\lambda}(\rho^{1})), U(\vec{\lambda}(\rho^{2})).
\end{equation}
For a separable state there is more uncertainty associated with the global state than with the states of subsystems (for all measures of uncertainty). Contrapositively, if there is less uncertainty associated with the global state than there is with the states of the subsystems, then the global state must be entangled.

Importantly, \citet{separabledisorder} also proved that the majorization condition is only a necessary condition for separability and not a sufficient one, as there exist entangled states with the same global and local spectra as separable ones---in this case, eqn.~(\ref{majorization condition}) will not be able to distinguish between entangled and separable states. This demonstrates the inherent limitation of the thought expressed in the quotation of Schr\"{o}dinger above as a characterisation of entanglement\footnote{It was perhaps not widely appreciated immediately that the Nielsen and Kempe result brings to an end at a stroke the programme of finding entropic and related criteria for entanglement e.g. using Renyi and Tsallis entropies. This is evident following Uffink's characterisation of uncertainty measures based on the majorization relation---which includes quantities of this type---as all such criteria will be implied by the condition~(\ref{majorization condition}). (Latterly there is some appreciation of this, see e.g. \citet{rossignoli:canosa}.) Furthermore in light of the Nielsen and Kempe result, we know without further ado that criteria of this form can only be sufficient and not necessary for entanglement.}.

\subsection{The tetrahedron of Bell-diagonal states}

As our final piece of background it will be helpful to note 
that the representation of density operators in the form (\ref{ab form}) becomes particularly simple if the matrix $c_{ij}$ happens to be diagonal; then we may represent the correlation matrix in the easily visualisable form of a vector, $\mathbf{c}$, in 3 dimensional real space, given by the components $c_{ii}$. This fact is made use of in \citet{horodeckisPRA:1996}, for example.

In particular, if it is also the case that $\mathbf{a}=\mathbf{b}=0$, i.e., the states of the subsystems are maximally mixed, then the density operator is completely characterized by the vector $\mathbf{c}$. This corresponds to the class of \textit{Bell-diagonal} states, \textit{viz.} the class of states that results from taking convex combinations of Bell state projectors.

\begin{figure}
\scalebox{1.5}{\includegraphics{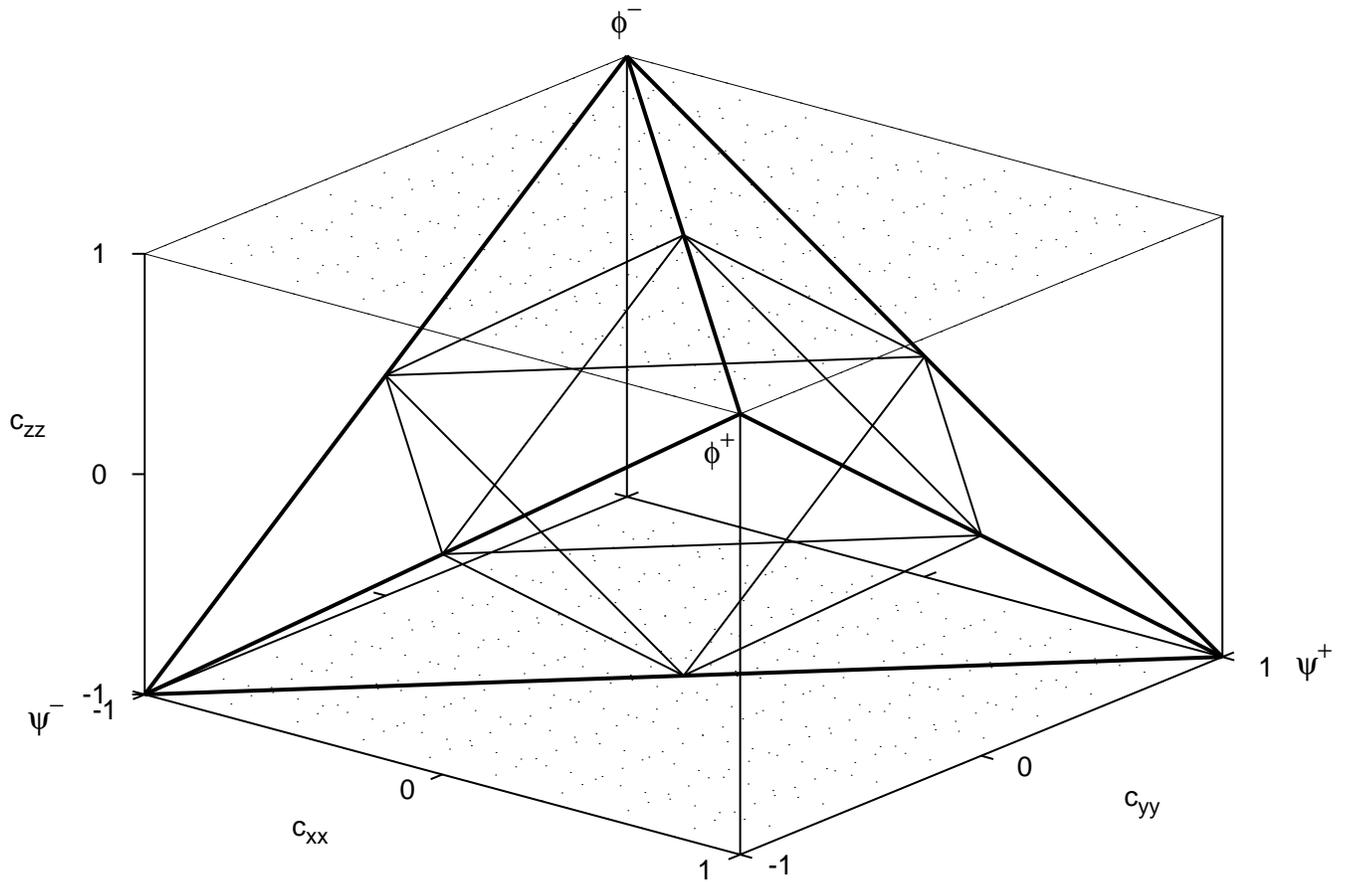}}
\caption{\small Bell-diagonal states may be represented by the diagonal components of the correlation matrix $c_{ij}$. The vertices of the tetrahedron $\cal{T}$ correspond to the four Bell states $\ket{\phi^{+}}{},\ket{\phi^{-}}{},\ket{\psi^{+}}{},\ket{\psi^{-}}{}$. A Bell-diagonal state is separable iff it corresponds to a point belonging to the central octohedron $\cal{T}\cap-\cal{T}$ \citep{horodeckisPRA:1996}.\label{tetrahedron}}
\end{figure}

For the four projectors onto the Bell states, $\mathbf{a}$ and $\mathbf{b}$ are indeed zero, as these are maximally entangled states; and as is well known, the $\mathbf{c}$ vectors corresponding to the Bell state projectors are:

\begin{singlespacing}
\begin{equation}
\mathbf{c}_{\phi +}= \begin{pmatrix} 1 \\ -1 \\ 1\end{pmatrix}, \mathbf{c}_{\phi -}= \begin{pmatrix} -1 \\ 1 \\ 1\end{pmatrix}, \mathbf{c}_{\psi +}= \begin{pmatrix} 1 \\ 1 \\ -1\end{pmatrix}, \mathbf{c}_{\psi -}= \begin{pmatrix} -1 \\ -1 \\ -1\end{pmatrix}.  
\end{equation} \end{singlespacing} 
These four vectors correspond to the vertices of a tetrahedron $\cal{T}$ centred on the origin (see Figure~\ref{tetrahedron}); and the Bell-diagonal states will thus correspond to vectors lying on or within the surfaces of $\cal{T}$. (Furthermore, every vector $\mathbf{c}$ whose endpoint lies on or within $\mathcal{T}$ will correspond to a Bell-diagonal state.)

Now consider density matrices $\boldsymbol{\varrho}$ with $\mathbf{a},\mathbf{b} \neq 0$, but still with $c_{ij}$ diagonal. The tetrahedron $\mathcal{T}$ remains important for these states, as no vector whose end point lies beyond $\mathcal{T}$ can correspond to a valid density matrix, as \citet{horodeckisPRA:1996} note.
If we were to take such a vector, $\mathbf{c}^{\prime}$, then it would give rise to an operator that is not positive, for it would have a negative expectation value with one of the Bell state projectors. 

To see this, note that the expectation value of the projector $P_{\psi^{-}}$ onto the singlet state, say, with a valid density operator, will take the form:

\begin{singlespacing}
\begin{equation} 
P_{\psi^{-}}.\boldsymbol{\varrho} = \frac{1}{2}\begin{pmatrix} 1 \\ 0 \\ 0 \\ \mathbf{c}_{\psi_{-}}\end{pmatrix}.\frac{1}{2}\begin{pmatrix} 1 \\ \mathbf{a} \\ \mathbf{b} \\ \mathbf{c}\end{pmatrix} = \frac{1}{4}\bigl(1 + \mathbf{c}_{\psi^{-}}.\mathbf{c}\bigr)
\end{equation}\end{singlespacing}
This expectation value has to be greater than or equal to zero if $\boldsymbol{\varrho}$ is to be a positive operator; and now note that $1+ \mathbf{c}_{\psi^{-}}.\mathbf{c} = 0$ defines a plane normal to $\mathbf{c}_{\psi^{-}}$ (and a distance of $-1/\sqrt{3}$ from the origin), on the positive side of which the vector $\mathbf{c}$ has to lie, if it is to belong to a positive operator. This plane, of course, is the face of $\mathcal{T}$ normal to $\mathbf{c}_{\psi_{-}}$. The other faces of the tetrahedron arise in the same way, as planes beyond which a vector $\mathbf{c}^{\prime}$ would belong to an operator having a negative expectation value with the corresponding Bell state projector. Thus in order for a diagonal matrix $c_{ij}$ to belong to a positive operator, the vector composed of its diagonal elements must lie within the tetrahedron $\mathcal{T}$.

It was shown by \citet{horodeckisPRA:1996} (using entanglement witnesses based on Werner's `swap' operation \citep{werner}) that a necessary condition for a density operator with diagonal correlation matrix to be separable is that the end point of its $\mathbf{c}$ vector lie on or within the octohedron given by the intersection of $\cal{T}$ with its reflection through the origin, $-\cal{T}$ (Figure~\ref{tetrahedron}). Furthermore, for  Bell-diagonal states, this condition is a sufficient one too, for every vector lying within the octohedron corresponds to a positive operator (this property in general fails for $\mathbf{a},\mathbf{b}\neq 0$); in particular, note that for $\mathbf{a}=\mathbf{b}=0$ the vertices of the octohedron correspond to separable states, hence points within the octohedron correspond to convex combinations of separable states, which themselves will be separable.

Finally, note that the entanglement properties of a state are invariant under $U_{1}\otimes U_{2}$ rotation. \citet{horodeckisPRA:1996} point out that with such unitary operations on subsystems, the correlation matrix of a joint system can always be brought into diagonal form\footnote{These unitary operations give rise to 3-d rotation matrices $R_{1}$ and $R_{2}$, whose effect on the correlation matrix $(C)_{ij}=c_{ij}$ will be: $R_{1}CR^{\mathrm{T}}_{2}$.}, hence the simplified 3 dimensional representation with tetrahedron and octohedron may be used to study the entanglement properties of all $2\otimes 2$ states. For all states with maximally mixed reduced states for subsystems, therefore, the condition of their corresponding $\mathbf{c}$ vector belonging to the octohedron is both necessary and sufficient for separability \citep{horodeckisPRA:1996}.

\section{Characterizations in the Deutsch-Hayden representation}

With this background material behind us, let's begin with some very simple properties of the Deutsch-Hayden representation in the form~(\ref{ab form}).

First, recall that the requirement of unit trace and positivity imply that for a density operator $\rho$, $\mathrm{Tr}(\rho^{2})\leq 1$. This further implies that 

\begin{singlespacing}
\begin{equation}\label{normalisation} 
\boldsymbol{\varrho}.\boldsymbol{\varrho} =\frac{1}{2}\begin{pmatrix} 1 \\ \mathbf{a} \\ \mathbf{b} \\ c_{ij}\end{pmatrix}.\frac{1}{2}\begin{pmatrix} 1 \\ \mathbf{a} \\ \mathbf{b} \\ c_{ij}\end{pmatrix} = \frac{1}{4}( 1 + a^{2} + b^{2} + \sum_{ij}c_{ij}^{2}) \leq 1,\end{equation}\end{singlespacing}

\noindent hence for valid density matrices, $a^{2} + b^{2} + \sum_{ij}c_{ij}^{2} \leq 3$, with equality being achieved for pure states.

For product states ($\rho^{12}=\rho^{1}\otimes\rho^{2}$), $\boldsymbol{\varrho}$ will clearly take the form:
\begin{equation}\label{product states}
\boldsymbol{\varrho}_{\mathrm{prod}} = \frac{1}{4}\bigl(\mathbf{1}\otimes\mathbf{1} + \mathbf{a}.\boldsymbol{\sigma}\otimes\mathbf{1} + \mathbf{1}\otimes\mathbf{b}.\boldsymbol{\sigma} +\sum_{ij}a_{i}b_{j}\sigma_{i}\otimes\sigma_{j}\bigr),
\end{equation} 
while separable states will be of the form $\rho_{\mathrm{sep}}=\sum_{k}\lambda_{k}\rho^{1}_{k}\otimes\rho^{2}_{k}$, thence:
\begin{equation}\label{separable states}
\boldsymbol{\varrho}_{\mathrm{sep}} = \frac{1}{4}\sum_{k} \lambda_{k}\bigl(\mathbf{1}\otimes\mathbf{1} + \mathbf{a}^{k}.\boldsymbol{\sigma}\otimes\mathbf{1} + \mathbf{1}\otimes\mathbf{b}^{k}.\boldsymbol{\sigma} +\sum_{ij}a_{i}^{k}b_{j}^{k}\sigma_{i}\otimes\sigma_{j}\bigr),
\end{equation} 
where the $\lambda_{k}$ are convex coefficients. Here the vector $\mathbf{a}$ is given by $\sum_{k}\lambda_{k}\mathbf{a}^{k}$, $\mathbf{b}$ by $\sum_{k}\lambda_{k}\mathbf{b}^{k}$, and similarly, $c_{ij}$ by $\sum_{k}\lambda_{k}a_{i}^{k}b_{j}^{k}$. 

The conditions for pure state entanglement are straightforward. The joint state must be pure, hence $a^{2}+b^{2}+\sum_{ij}c_{ij}^2=3$, while the expectation values for joint observables do not factorise: $c_{ij}\neq a_{i}b_{j}$. If $c_{ij}= a_{i}b_{j}$ then the state is always unentangled, but the converse does not hold in general unless the joint state is pure\footnote{For mixed states, factorisation might fail because of classical correlations, as in eqn. (\ref{separable states}).}.

Another way to state the necessary and sufficient condition for pure state entanglement that makes more use of the geometric properties of the representation is to note that if $\boldsymbol{\varrho}$ is pure, it will be entangled \textit{iff} the reduced states are mixed. In this representation, the simplest way to investigate purity is just to consider the length of vectors---pure states will be of unit length, mixed states of less than unit length. Hence if $\boldsymbol{\varrho}.\boldsymbol{\varrho}=1$, then $\boldsymbol{\varrho}$ is entangled \textit{iff} $a^{2},b^{2}<1$.

\subsection{Some sufficient conditions for entanglement}

Let us now consider some sufficient conditions for entanglement when the joint state is mixed. 

A direct route is to consider a constraint that is implied by the correlation matrix for separable states being given by $c_{ij}=\sum_{k}\lambda_{k}a^{k}_{i}b^{k}_{j}$. Note that $f(x)=x^{2}$ is a convex function. It follows that
\[c_{ij}^{2}=\bigl(\sum_{k}\lambda_{k}a^{k}_{i}b^{k}_{j}\bigr)^{2} \leq \sum_{k} \lambda_{k}\bigl(a_{i}^{k}b^{k}_{j}\bigr)^{2};\]  
and thence
\[ \sum_{ij}c_{ij}^{2} \leq \sum_{\substack{ k\\ ij}}\lambda_{k}\bigl(a_{i}^{k}b^{k}_{j}\bigr)^{2}=\sum_{k}\lambda_{k}(a^{k})^{2}(b^{k})^2.\]
But $(a^{k})^{2}$ and $(b^{k})^2$ are just the lengths squared of the Bloch vectors of the reduced density matrices of systems 1 and 2 respectively, for the $k$th state making up the mixture; and therefore $(a^{k})^{2},(b^{k})^2\leq 1$. Therefore we obtain the constraint:
\begin{proposition}\label{c_{ij}^{2} leq one} If $\boldsymbol{\varrho}$ is separable then $\sum_{ij}c_{ij}^{2}\leq 1$.
\end{proposition}

$\sum_{ij}c_{ij}^{2}$ is the length squared of the component of the vector $\boldsymbol{\varrho}$ in the 9 dimensional subspace of $V_{h}(\mathbb{C}^{N})$ spanned by joint observables. We learn from prop.~\ref{c_{ij}^{2} leq one} that if the length of this component of the vector is greater than 1, then the state is entangled.

This squared length is invariant under unitary operations of the form $U_{1}\otimes U_{2}$, hence the constraint is unchanged if such a transformation of $\boldsymbol{\varrho}$ is applied in order to diagonalize the correlation matrix. Following diagonalization, the constraint will be that $\mathbf{c}.\mathbf{c} \leq 1$ for separable states. The equation  $\mathbf{c}.\mathbf{c} = 1$ is the unit 3-sphere enclosing the octohedron of separable states, making it clear that prop.~\ref{c_{ij}^{2} leq one} is not a sufficient condition for separability: there exist entangled states with $\sum_{ij}c_{ij}^{2} \leq 1$.     

A stronger condition may be obtained by making use of the majorization criterion. In the Deutsch-Hayden representation, the simplest measure of disorder will be the length-squared measure. With this measure of disorder, the majorization criterion (\ref{majorization condition}) implies that for separable states, the length squared of $\boldsymbol{\varrho}$ is less than or equal to the length squared of the reduced state of system 1; and similarly, less than or equal to the length squared of the reduced state of system 2. In terms of components we will have:
\begin{eqnarray}
\frac{1}{4}\bigl( 1 + a^{2} + b^{2} + \sum_{ij}c_{ij}^{2}\bigr) & \leq & \frac{1}{2}(1 + a^{2}), \;\text{and} \\
 & \leq & \frac{1}{2}(1 + b^{2}), 
\end{eqnarray}
which rearrange to give the pair of constraints:
\begin{proposition}\label{spheres} If $\boldsymbol{\varrho}$ is separable then $\sum_{ij}c_{ij}^2 \leq 1 \pm |a^{2}-b^{2}|$.\end{proposition}

Together with prop.~\ref{c_{ij}^{2} leq one}, then, this gives us a nested trio of hyperspheres in the 9 dimensional subspace of joint observables. On the inside is the sphere given by $\sum_{ij}c_{ij}^{2} = 1-|a^{2}-b^{2}|$, then there is the sphere $\sum_{ij}c_{ij}^{2} = 1$; while the outermost sphere is $\sum_{ij}c_{ij}^{2} = 1+|a^{2}-b^{2}|$. If a state is separable then the component of $\boldsymbol{\varrho}$ in the subspace of joint observables must lie on or within the innermost sphere; if it lies beyond it, the state is entangled.

Of course, we know that the majorization criterion is only a necessary, and not a sufficient condition for separability; and the application made of the criterion here is further weakend as a result of being specialised to a particular choice of measure of disorder. Thus we know that props.~\ref{c_{ij}^{2} leq one} and \ref{spheres} are in general only sufficient conditions for entanglement, but nevertheless they provide us with a pleasing geometrical picture of certain conditions under which a state will be entangled. 

\subsection{The PPT and reduction criteria}

Let us now consider the positive partial transpose condition in the Deutsch-Hayden formalism. The effect of the partial transpose is straightforward. The Pauli matrices $\sigma_{x}$ and $\sigma_{z}$ are invariant under transposition, so the only changes will be in the $\sigma_{y}$ components of the state: $\sigma_{y}^{\mathrm{T}}=-\sigma_{y}$, thus:
\begin{equation}\label{pt effect}
\mathbf{1}\otimes T \boldsymbol{\varrho} = \frac{1}{4}\bigl( \mathbf{1}\otimes \mathbf{1} + \mathbf{a}.\boldsymbol{\sigma}\otimes\mathbf{1} + \mathbf{1}\otimes \mathbf{b}^{\prime}.\boldsymbol{\sigma} +\sum_{ij}c_{ij}^{\prime}\sigma_{i}\otimes\sigma_{j}\bigr),
\end{equation} 
where \begin{singlespacing}
\begin{equation}\label{b prime c prime}
\mathbf{b}^{\prime}=\begin{pmatrix} b_{1}\\-b_{2}\\b_{3}\end{pmatrix}\;\text{and}\; c^{\prime}_{ij} = \begin{pmatrix} c_{11} & -c_{12} & c_{13}\\c_{21} & -c_{22} & c_{23}\\ c_{31} & -c_{32} & c_{33}\end{pmatrix}.
\end{equation}
\end{singlespacing}

Now it is clear enough why positive partial transpose is a \textit{necessary} condition for separability. If we consider first a product state, the effect of the partial transpose is to reflect the Bloch vector for system 2 in the $z$-$x$ plane, thus we end up with a perfectly good state again; if our initial state is a separable state, each of the $\mathbf{b}^{k}$ vectors over which we take a convex sum is similarly reflected, each again represents a perfectly good state, so their convex sum represents a perfectly good state. 

It would be interesting to see, however, whether studying the representation of states in the Deutsch-Hayden form could provide us with insight into why, if a state is entangled, it will become non-positive under the action of the partial transpose. By investigating eqn.~(\ref{pt effect}), can we learn something about why the replacement $\mathbf{b}\mapsto \mathbf{b}^{\prime}, c_{ij}\mapsto c_{ij}^{\prime}$ leads to non-positive operators when we have entanglement? We have already seen why this transformation is unproblematic for separable states: then $c_{ij}\mapsto c_{ij}^{\prime}$ can simply be understood as 
\[\sum_{k}\lambda_{k}a_{i}^{k}b^{k}_{j}\mapsto\sum_{k}\lambda_{k}a_{i}^{k}(b^{k}_{j})^{\prime},\]
which is clearly consistent with the change $\mathbf{b}\mapsto \mathbf{b}^{\prime}$ and will give rise to a valid state. What is it about entangled states that changes this?

Of course, in an important sense, the result of \citet{horodeckisPLA:1996} gives a complete answer to this question: if a state were not to become non-positive under partial transposition it would not be entangled---it would be separable instead. As we shall see however, using the Deutsch-Hayden representation it is also possible, in some cases, to get a more direct answer to our question, in addition.

We will also consider another condition, equivalent to the positive partial transpose condition for $2\otimes 2$ and $2\otimes 3$ dimensions, known as the \textit{reduction} criterion \citep{horodeckis:reduction}, which is based on the positive map $\Lambda_{\mathrm{red}}(A) = \mathbf{1}\mathrm{Tr}(A) - A$. The effect of this map applied to subsystem 2 of a joint state $\rho^{12}$ will be: 
\[\mathbf{1}\otimes\Lambda_{\mathrm{red}}\,\rho^{12} = \rho^{1}\otimes \mathbf{1} - \rho^{12},\]
where $\rho^{1}$ is the reduced density operator of system 1. This gives rise to the following criterion for $2\otimes 2$ and $2\otimes 3$ dimensional systems:
\begin{proposition}[Reduction criterion]\label{reduction criterion} $\rho \;\text{\em is separable} \leftrightarrow  \rho^{1}\otimes \mathbf{1} - \rho^{12} \geq 0.$\end{proposition} 
(As for the positive partial transpose, this condition will only be a necessary one for higher dimensions.) Clearly a similar condition may be generated with the application of $\Lambda_{\mathrm{red}}$ to the first system instead.
For qubits, $\Lambda_{\mathrm{red}}$ is equivalent to a $\pi$ rotation about the $y$-axis, followed by the transpose operation.

In the Deutsch-Hayden representation, the effect of the reduction map will be
\begin{equation}
\mathbf{1}\otimes\Lambda_{\mathrm{red}}\, \boldsymbol{\varrho} = \frac{1}{4}\bigl(\mathbf{1}\otimes\mathbf{1} + \mathbf{a}.\boldsymbol{\sigma} - \mathbf{1}\otimes\mathbf{b}.\boldsymbol{\sigma} - \sum_{ij}c_{ij}\sigma_{i}\otimes\sigma_{j}\bigr).
\end{equation}
The Bloch vector for system 2 has been reflected through the origin, while $c_{ij}\mapsto -c_{ij}$. Again, can we gain any insight into why a change of this sort will lead to non-positive operators when we have entanglement?

Let us begin with the simplest case. Can we explain why a pure entangled state $\boldsymbol{\varrho}$ will become non-positive under partial transposition? We can, in at least two ways.

Consider first the effect of a $\pi$ rotation about the $y$-axis applied to the second of our pair of systems in the joint state $\boldsymbol{\varrho}$. The result will be:

\begin{singlespacing}
\[ \mathbf{b}\mapsto \mathbf{b}^{\prime\prime} = \begin{pmatrix}-b_{1}\\b_{2}\\-b_{3}\end{pmatrix}=-\mathbf{b}^{\prime},\;\; c_{ij}\mapsto c_{ij}^{\prime\prime} = \begin{pmatrix} -c_{11} & c_{12} & -c_{13}\\-c_{21} & c_{22} & -c_{23}\\ -c_{31} & c_{32} & -c_{33}\end{pmatrix} = - c_{ij}^{\prime};\]
\end{singlespacing}

\noindent where, as before, $\mathbf{b}^{\prime}$ and $c_{ij}^{\prime}$ are the system 2 Bloch vector and the correlation matrix for $\boldsymbol{\varrho}$ under partial transposition respectively (cf. eqn.~(\ref{b prime c prime})). The expectation value of the partially transposed $\boldsymbol{\varrho}$ with this rotated state will be:

\begin{singlespacing}
\begin{equation} 
\frac{1}{2}\begin{pmatrix} 1 \\ \mathbf{a} \\ \mathbf{b}^{\prime} \\ c_{ij}^{\prime}\end{pmatrix}.
\frac{1}{2}\begin{pmatrix} 1 \\ \mathbf{a} \\ -\mathbf{b}^{\prime} \\ -c_{ij}^{\prime}\end{pmatrix}=\frac{1}{4}(1 + a^{2} - b^{2} -\sum_{ij}c_{ij}^{2}),\end{equation}
\end{singlespacing}
since $\mathbf{b}^{\prime}.\mathbf{b}^{\prime}= b^{2}$ and $c_{ij}^{\prime}c_{ij}^{\prime} = c_{ij}^{2}.$ 

It follows that $\mathbf{1}\otimes T \boldsymbol{\varrho}$ will have a negative expectation value with the rotated state if
\begin{equation}\label{negative expectation condition}
(b^{2}-a^{2}) + \sum_{ij}c_{ij}^{2} > 1.
\end{equation}
We know from the Schmidt decomposition (eqn.~\ref{schmidt}), however, that for a pure joint state, the reduced states of the subsystems will be equally mixed, therefore in this case $a^{2}=b^{2}$, and the first term in (\ref{negative expectation condition}) will be zero. Furthermore, since $\boldsymbol{\varrho}$ is pure, we know (eqn.~(\ref{normalisation})) that $a^{2}+b^{2}+\sum_{ij}c_{ij}^{2}=3$, while as it is entangled, $a^{2},b^{2}<1$; therefore, $\sum_{ij}c_{ij}^{2}$ must be strictly greater than 1. 

It follows that for any pure entangled state $\boldsymbol{\varrho}$, its partial transpose will have a negative expectation value with a canonically chosen one dimensional projector---in fact, the one attained from $\boldsymbol{\varrho}$ by the $\pi$ rotation about the $y$-axis on system 2---and for this reason the partial transpose of such a state will not be positive. Similarly, the effect of the reduction map on a pure entangled state $\boldsymbol{\varrho}$ will be to produce an operator which has negative expectation value with a canonically chosen projector---in this case, $\boldsymbol{\varrho}$ itself.    

So we may see why, for pure states to begin with, if $\sum_{ij}c_{ij}^{2}>1$, the effect of partial transposition will be to give rise to a non-positive state; and we have seen furthermore, that any pure entangled state will certainly have $\sum_{ij}c_{ij}^{2}>1$.

This condition can be related to some of our earlier discussions. Let's consider again the tetrahedron and octohedron of Figure~\ref{tetrahedron}. Assume that the correlation matrix of the state we are interested in has been diagonalized by a suitable $U_{1}\otimes U_{2}$ unitary operation, so that we may consider the vector $\mathbf{c}$ of the diagonal components alone. The effect of partial transposition on $\mathbf{c}$ will be to reflect it in the $z$-$x$ plane; and we may now see why the end point of a $\mathbf{c}$ vector has to be within the central octohedron in order to be associated with a separable state. 

Recall that if the end point of a $\mathbf{c}$ vector lies beyond the tetrahedron $\mathcal{T}$ then this implies a negative expectation value with the Bell projector defining the appropriate face. The effect of the partial transposition will be to reflect all of the vectors picking out points lying within $\mathcal{T}$ in the $z$-$x$ plane; the image of $\mathcal{T}$ under this transformation will coincide with $-\mathcal{T}$. If a vector $\mathbf{c}$ does not, after the partial transpose transformation, still lie within $\mathcal{T}$, it will now give rise to an operator that has a negative expectation value with one of the Bell projectors. Therefore, it is only vectors picking out points within the intersection of $\mathcal{T}$ with its reflection in the $z$-$x$ plane, that is, in the central octohedron, that correspond to separable states. For $\mathbf{c}$ vectors whose endpoints lie in the `arms' of the tetrahedron, beyond the octohedron, the effect of the partial transpose will be to reflect them beyond $\mathcal{T}$; and we see that the operators associated with these vectors will consequently be non-positive after partial transposition as they will have a negative expectation value with one of the Bell projectors.

For diagonalized $c_{ij}$, the condition $\sum_{ij}c_{ij}^{2}>1$ translates into $\mathbf{c}.\mathbf{c}>1$. The sphere $\mathbf{c}.\mathbf{c}=1$, is recall, the sphere enclosing the central octohedron, so we see that if for a state $\boldsymbol{\varrho}$, $\sum_{ij}c_{ij}^{2}>1$, then the partial transpose of this state will be an operator which has a negative expectation value with some maximally entangled state and hence must be non-positive. The maximally entangled state in question will be a Bell projector rotated by the $U_{1}\otimes U_{2}$ operation that diagonalized $c_{ij}$ for the given $\boldsymbol{\varrho}$.  

We have now seen two reasons why a pure entangled state $\boldsymbol{\varrho}$ will give rise a non-positive operator on partial transpose. From the requirements of normalisation and positivity (eqn.~(\ref{normalisation})), we know that for a pure state to be entangled, $\sum_{ij}c_{ij}^{2}>1$, i.e., the length of the component of the state in the joint observable susbspace is greater than one. This means i) that the partial transpose of $\boldsymbol{\varrho}$ will have a negative expectation value with the $\mathbf{1}\otimes \sigma_{y}$ rotated $\boldsymbol{\varrho}$\,; and ii) that the partial transpose of $\boldsymbol{\varrho}$ will also have a negative expectation value with a maximally entangled state. These give us direct reasons why these entangled states will have a non-positive partial transpose. 

In fact the considerations that lead to property (ii) do not depend essentially on $\boldsymbol{\varrho}$ being pure. We know from prop.~\ref{c_{ij}^{2} leq one} that if $\sum_{ij}c_{ij}^{2}>1$, then the associated state, whether pure or mixed, is entangled; and from the reasoning we have just rehearsed concerning the tetrahedron, we may infer that any such state will certainly give rise to a non-positive operator under partial transposition, as it will have a negative expectation value with a maximally entangled state.     

The reasoning that led to property (i) may also be generalised to cover a large class of mixed entangled states. In eqn.~(\ref{negative expectation condition}), assuming the purity of $\boldsymbol{\varrho}$ meant that the first term became equal to zero; this will not be so in general with mixed states. The other r\^{o}le of the purity assumption was in allowing us to pick out a \textit{projector} with which the partially transposed operator would have a negative expectation value. If $\boldsymbol{\varrho}$ is not a pure state then the $\mathbf{1}\otimes \sigma_{y}$ rotated $\boldsymbol{\varrho}$ will obviously not be a projector, though.

We may still use eqn.~(\ref{negative expectation condition}) to understand why some entangled states will give rise to non-positive operators on partial transposition, or following application of the reduction map, however. When $\boldsymbol{\varrho}$ is mixed, (\ref{negative expectation condition}) is indicating a simple condition under which the partially transposed $\boldsymbol{\varrho}$ becomes an \textit{entanglement witness}. If for a given $\boldsymbol{\varrho}$,
\[\sum_{ij}c_{ij}^2>1-(b^{2}-a^{2}),\]
then $\mathbf{1}\otimes T \boldsymbol{\varrho}$ is an entanglement witness for the state resulting from applying the unitary operation $\mathbf{1}\otimes\sigma_{y}$ to $\boldsymbol{\varrho}$. The partially transposed operator will have positive expectation value for all separable states (this property of the original density operator is not changed under partial transpose) and a negative expectation value for the stated entangled state. Similarly, when (\ref{negative expectation condition}) holds, $\mathbf{1}\otimes\Lambda_{\mathrm{red}}\,\boldsymbol{\varrho}$ will be an entanglement witness for $\boldsymbol{\varrho}$ itself. We know, finally, from the majorization criterion applied with the length squared measure of disorder, prop. \ref{spheres}, that whenever (\ref{negative expectation condition}) holds, then $\boldsymbol{\varrho}$ will be entangled.     

It is an interesting fact that the sufficient condition for entanglement derived from the majorization condition should coincide with the condition under which the partially transposed $\boldsymbol{\varrho}$ can be seen immediately to be an entanglement witness for a particular entangled state. To re-cap, the insight provided by this into why some entangled states become non-positive under partial transposition is that any $\boldsymbol{\varrho}$ for which (\ref{negative expectation condition}) holds will be entangled; and we can see it will certainly end up not being positive under partial transposition, as it will be transformed into an entanglement witness for a particular known state, hence will be an operator with a negative expectation value for that state, hence will not be a positive operator.

\section{Summary}

In this chapter we have gained some insight into when a state will be entangled in the Deutsch-Hayden formalism. For pure states, the necessary and sufficient condition will be that $\sum_{ij}c_{ij}^{2}>1$; that is, the component of the vector $\boldsymbol{\varrho}$ in the 9 dimensional subspace of joint observables is of length greater than 1. For mixed states, this weakens to a sufficient condition. By applying the majorization condition, a generally wider sufficient condition is found: if the component of $\boldsymbol{\varrho}$ in the joint observable subspace lies beyond the hypersphere given by $\sum_{ij}c_{ij}^{2}=1-|a^{2}-b^{2}|$ then the state is entangled. 
We were also able to gain some further understanding of why certain classes of entangled states will become non-positive under the effect of the partial transpose operation. A pure entangled state, we have said, will have $\sum_{ij}c_{ij}^{2}>1$, but then such a state can be seen to become non-positive under partial transpose in at least two direct ways: it will have a negative expectation value with a certain canonically chosen one-dimensional projector; and it will have a negative expectation value with a maximally entangled state attained from a Bell state by the $U_{1}\otimes U_2$ rotation that diagonalizes the correlation matrix. This analysis can be extended to some mixed states. Those mixed states with $\sum_{ij} c_{ij}^{2}>1$ will be entangled and can also be seen to become non-positive under partial transposition due to a negative expectation value with a maximally entangled state. Finally, those mixed states with $\sum_{ij}c_{ij}^2>1-(b^{2}-a^{2})$ will be entangled, and they become non-positive as they are transformed into entanglement witnesses.

We have been concerned in this chapter exclusively with bipartite entanglement. However it would be interesting to extend the discussion to multipartite cases. As Hewitt-Horsman has noted\footnote{Talk at Oxford Philosophy of Physics Discussion Group, 12 February 2003.}, the Deutsch-Hayden formalism is useful in the context of tripartite pure state entanglement for determining the presence of genuine three-party entanglement, as opposed to two-party.
One begins by checking conditions of the form $\langle q_{1,i}q_{2,j}q_{3,k}\rangle \neq\langle q_{1,i}\rangle\langle q_{2,j}\rangle\langle q_{3,k}\rangle$. If these are satisfied, then as the joint state is pure, there must be some entanglement present. The next stage is to check the following three sets of conditions:
\begin{eqnarray*}
\langle q_{1,i}q_{2,j}q_{3,k}\rangle & \neq & \langle q_{1,i}\rangle\langle q_{2,j}q_{3,k}\rangle, \\
\langle q_{1,i}q_{2,j}q_{3,k}\rangle & \neq & \langle q_{2,j}\rangle\langle q_{1,i}q_{3,i}\rangle, \\
\langle q_{1,i}q_{2,j}q_{3,k}\rangle & \neq & \langle q_{1,i}q_{2,j}\rangle\langle q_{3,k}\rangle. 
\end{eqnarray*}
There are then two possibilities. Either all three (sets of) conditions are satisfied, in which case we have genuine 3 party entaglement, or two are satisfied and one is not, in which case we have 2-party entanglement (and depending on which two are satisfied, this will tell us which two systems are entangled.) 

Within the class of genuine three-party entangled states, there is a distinction between GHZ-type entanglement \citep{GHZ}, states of the form:
\[|\psi\rangle=\frac{1}{\sqrt{2}}(|000\rangle+|111\rangle);\]
and W-type entanglement \citep{wtype}, states of the form:
\[|\phi\rangle=\frac{1}{\sqrt{3}}(|001\rangle+|010\rangle+|100\rangle).\]
It is plausible to conjecture that an extension of the approach used in this chapter will prove useful for distinguishing between states of these two different classes, as tracing out one system of a GHZ state leaves a separable (classically correlated) state, while tracing out one system of a W state leaves a component of maximally entangled two-party entanglement. This fact should find expression in the Deutsch-Hayden formalism in terms of differing lengths of components in the subspace belonging to two-party joint observables. \enlargethispage*{1000pt}

\end{doublespacing}



\chapter[Quantum Computation and the C-T Hypothesis]{Quantum Computation and the Church-Turing Hypothesis}\label{comp}





\begin{doublespacing}

\section{Introduction}\label{intro}

In this chapter we will be considering some of the philosophical questions raised by the theory of quantum computing. First, and briefly, whether the efficiency of quantum computing gives us an argument for a substantive notion of quantum information (Section~\ref{containing}); and second, in more detail, we shall consider some questions regarding the status of the Church-Turing hypothesis (Sections \ref{tpvct} and \ref{constraint?}). 

The advent of quantum computers\index{computer!quantum} has raised a question concerning the relationship between the classical theory of computation, based on the Church-Turing hypothesis\index{Turing!Church-Turing hypothesis}, and the quantum theory. It is quite common to find the claim that the quantum theory of computation is the more fundamental. However, one sometimes also encounters a much stronger claim to the effect that the quantum computer has succeeded in finally making sense of Turing's theory of computation, or that Turing's machines were really quantum mechanical all along. 
We shall be considering some of the issues that have arisen around this question of the relation between the classical and quantum theories of computation.

Richard Feynman\index{Feynman, R} was the prophet of quantum computation. He pointed out that it seems that one cannot simulate the evolution of a quantum mechanical system efficiently on a classical computer\index{simulation!of q.m. systems}. He took this to imply that there might be computational benefits to be gained if computations are carried out using quantum systems themselves rather than classical systems; and he went on to describe a universal quantum simulator \citep{feynman:1982}\index{quantum!simulator, universal}. However, it is with Deutsch's\index{Deutsch, D} introduction of the concept of the universal quantum computer\index{universal!quantum computer}\index{computer!quantum} in his 1985 paper that the field really begins~\citep{deutsch:1985}. 

Deutsch's\index{Deutsch, D} paper is the seed from which the riches of quantum computation\index{computer!quantum} theory have grown, but in it are to be found roots of philosophical confusion over the notion of computation, in particular, in the claim that a physical principle, the Turing Principle\index{Turing!Principle}, underlies the Church-Turing hypothesis\index{Turing!Church-Turing hypothesis|(}. 

The Turing Principle\index{Turing!Principle|(} is stated as follows:
\begin{quoting}
Every finitely realizable physical system can be perfectly simulated by a universal model computing machine operating by finite means. \citep{deutsch:1985}
\end{quoting}
It is the claim that the Turing Principle underlies the Church-Turing hypothesis that is primarily responsible for the thought that quantum computers are necessary to make proper sense of Turing's theory. For the Turing Principle is not satisfied in classical physics, owing to the continuity of states and dynamics in the classical case, yet it is, Deutsch argues \citep[\S3]{deutsch:1985} in the case of quantum mechanics. If the Turing Principle really were the heart of the theory of computation, prior to the development of the notion of quantum computers we would have been faced with a considerable difficulty, as this supposedly fundamental Principle is false under classical mechanics. I shall be arguing, however, that it is a mistake to see the Turing Principle\index{Turing!Principle|)} as underlying the Church-Turing hypothesis\index{Turing!Church-Turing hypothesis|)} (Section~\ref{tpvct}), hence this issue does not arise. In Section \ref{constraint?} we will consider whether the Church-Turing hypothesis might play a r\^ole as a constraint on physical laws, as suggested in the quantum case by \citet{nielsen:ct}, for example.

\section{Quantum computation and containing information}\label{containing}

Before moving on to discuss the Church-Turing hypothesis and the Turing Principle, let us pause to consider briefly an argument suggesting that quantum systems should be seen to contain information in a more literal, or substantive, sense than I have so far allowed. This argument is based on the gains in efficiency over the best known classical algorithms that can be achieved for certain important computational tasks using quantum computers.
The argument is suggested by the presentation of \citet{jozsa:2000}.

It is very natural (although not wholly uncontroversial) to view the property of entanglement as the main source of the exponential speed-up given by quantum algorithms such as that of \citet{shor} \citep{jozsa:1998,ekertjozsa:1998,jozsa:2000,jozsa?}. This view can be motivated in the following way. If we consider specifying the state of a system composed of $n$ two state classical systems, then $n$ bits are needed. By contrast, in order to specify a general state of an $n$ qubit system, we will need to specify $2^{n}$ coefficients for the $2^{n}$ basis vectors of the system (because of the tensor product structure of the state space); the order of the number of bits needed will be exponential in $n$. It is often therefore said that `...a quantum system can embody exponentially more information than its classical counterpart' \citep[p.108]{jozsa:2000}.

Now when we consider information processing, i.e., evolving the quantum state in a particular way, then even the simple case of a single 1-qubit operation (a single computational step for a quantum computer) is equivalent to an exponentially large amount of classical computation, when the initial state is entangled. The effect of the unitary operation on the state would need to be calculated classically as a ($2\times 2$) matrix multiplication for \textit{each} of the $2^{n}$ coefficients specifying the state. The quantum evolution corresponds to exponentially much classical computation, in the presence of entanglement:\enlargethispage*{40pt}
\begin{quoting}
Natural quantum physical evolution may be thought of as the processing of quantum information....[T]o perform natural quantum physical evolution, Nature must process vast amounts of information at a rate that cannot be matched in real time by any classical means...\citep[p.109]{jozsa:2000}
\end{quoting}  

There is a strong suggestion that quantum evolution is doing a great deal of work---a great deal of work in processing something---and therefore, there is something a great deal of which is being processed: we should allow a more substantive notion of quantum information. 

This conclusion can be resisted by noting that we have here a further example of what I have termed the simulation fallacy (cf. Section~\ref{simulation fallacy}). The fact that quantum evolution corresponds to an exponentially large amount of classical computation implies that we can use quantum systems to do something that corresponds to a very great deal of work in classical terms. But we cannot infer from this that the quantum computer is \textit{doing} this amount of work, rather than merely causing, in a different way, a result which could only be brought about with a lot of effort, classically.  

\section{The Turing Principle versus the Church-Turing Hypothesis}
\label{tpvct}
Let us now turn to consider the Turing Principle. In his landmark 1985 paper, Deutsch\index{Deutsch, D} argues that underlying the Church-Turing hypothesis\index{Turing!Church-Turing hypothesis}, the basis for the classical theory of computation, there is an implicit physical assumption, namely, the Turing Principle\index{Turing!Principle}, which is, recall:
\begin{quoting}
Every finitely realizable physical system can be perfectly simulated by a universal model computing machine operating by finite means.\footnote{A computing machine $M$ is said to \textit{perfectly simulate}\index{simulation!perfect} a physical system $S$, under a given labelling of their inputs and outputs, if their exists a program $\pi(S)$ for $M$ that renders $M$ computationally equivalent to $S$ under that labelling.} \citep{deutsch:1985}
\end{quoting}
The Church-Turing hypothesis\index{Turing!Church-Turing hypothesis}, by contrast, he states as follows:
\begin{quoting}
Every `function which would naturally be regarded as computable' can be computed by the universal Turing machine. \citep{deutsch:1985}
\end{quoting}
The two main ways in which these statements differ are, first, that Turing's\index{Turing, A} `functions which would naturally be regarded as computable' has, in effect, been replaced by `functions which may in principle be computed by a physical system' \citep[p.99]{deutsch:1985}, the result of the stipulation that the universal computing machine\index{computer!universal} perfectly simulates every finite physical system; and second, that the reference to a specific form of universal computer---the universal Turing machine\index{Turing!machine}---has been replaced by an unspecified universal computing machine, with the requirement only that it operate by finite means.

The heuristic value of the move to the Turing Principle\index{Turing!Principle} is undoubted, for it led Deutsch\index{Deutsch, D} to define the universal quantum computer\index{universal!quantum computer} and hence spark a vigorous new field of physics. The \textit{psychological} liberalisation involved in this move from the Church-Turing hypothesis\index{Turing!Church-Turing hypothesis} was thus invaluable, but, I shall suggest, it is mistaken to argue that the Turing Principle underlies the Church-Turing hypothesis, or that this physical principle should be thought of as the real basis for the theory of computation.

To begin with, it is important to recognise that in his famous paper `On Computable Numbers', \citet{turing} \index{Turing, A} was concerned with what is computable by \textit{humans}, not with describing the ultimate limits of what we now mean by `computer'. Deutsch\index{Deutsch, D} is well aware of this fact, e.g. \citet[p.2]{deutsch:etal:1999}, but by glossing over it here, we would miss several important things. First, the purely mathematical element of Turing's thesis; second, the chance to separate out the precursors of the computational analogy\index{computational!analogy} from the foundations of the theory of computation\footnote{\citet{shanker:1987}\index{Shanker, S} investigates this area and undertakes this separation in detail.}; and third, the distinction between the task of characterizing the effectively calculable\index{effective calculability}, which had become so urgent by the mid 1930's and to which the Church-Turing hypothesis\index{Turing!Church-Turing hypothesis} was directed, and the rather different project of considering what classes of functions can be calculated by machines or physical processes most widely construed (a distinction which Copeland\index{Copeland, B.J.}, in particular, has emphasized e.g. \citet{copeland}). To see something of the significance of these points, let us make the comparison with Church's\index{Church, A} position in his 1936 paper.

Church\index{Church, A} proposed that the intuitive notion of effective calculability\index{effective calculability} be made precise by identifying effectively calculable functions with the recursive functions~\citep[\S 7]{church:1936}\index{recursive functions}. Again, calculability here means calculable by \textit{humans}. By contrast, Turing\index{Turing, A} presented the mathematical insight that if certain functions could be encoded in, for example, binary terms, then a \textit{machine} could be made to compute analogues of those functions. The machine was the Turing machine\index{Turing!machine} and it turned out, the functions the recursive functions\index{recursive functions}. The second part of his argument, \S 9 of the paper, was then to relate this to human calculation; an argument for why computability defined in terms of Turing machines should capture all that would `naturally be regarded as computable' by humans.
 
\begin{sloppypar}
As Shanker\index{Shanker, S}, for example, recounts \citep[\S 2]{shanker:1987}, the differences between Church's\index{Church, A|(} and Turing's\index{Turing, A|(} presentations was all important for G\"odel\index{G\"odel, K|(}. G\"odel did not accept what is best seen as Church's \textit{stipulation} that the effectively calculable functions\index{effective calculability|(} are the recursive functions\index{recursive functions} until Turing's argument in `On Computable Numbers' became known. His objection was that Church had not shown \textit{why} the properties associated with our intuitive notion of effective calculability would be captured by the class of recursive functions (see also \citet{davis:1982,soare}). That he came to accept Church's\index{Church, A|)} convention after `On Computable Numbers' shows that he took Turing to have solved this problem. Presumably, what was important about this solution was not Turing's demonstration of the capabilities of the Turing machine\index{Turing!machine}, but rather, the argument in \S 9 that Turing machine computability\index{Turing!computability} captures that which would `naturally be regarded as computable'. Thus G\"odel\index{G\"odel, K|)} was convinced of the adequacy of Turing's account of what it is for a human to calculate in a formal system; and that this was no different from the operation of a Turing machine\index{Turing!machine}. In this way Turing was supposed to have explicated the intuitive notion of effective calculability\index{effective calculability|)}.

However, we should note that it is precisely this step back to the notion of calculable-by-human from calculable-by-machine and attempting to explain the former in terms of the latter that gives rise to the computational analogy\index{computational!analogy}, which may well be seen as philosophically problematic\footnote{\citet{shanker:1987}\index{Shanker, S} locates the ultimate source of the pressures that lead here to the computational analogy\index{computational!analogy} (as he calls it, the Mechanist Thesis), with Hilbert\index{Hilbert,D}.}; and note further that this final `step back' is an entirely logically separable part of the argument. The class of functions that may be calculated algorithmically by a human computer may be co-extensive with the class of Turing machine computable functions, without one having to explain human computation in mechanical terms.
\end{sloppypar}

To continue: If we were to follow Deutsch\index{Deutsch, D} and reinterpret Turing's `functions which would naturally be regarded as computable' as the functions which may in principle be computed by a real physical system, then we are neglecting the fact that Turing meant computable by humans. This is no mere historical point. The most obvious consequence would be that we ignore the possibility of making the useful distinction between computing by human and computing by machine---a physical system considered as a computer. But perhaps more importantly, we miss the significance of Turing's purely mathematical thesis, his recognition that certain functions can be encoded and machines thus made to compute them for us. Deutsch's\index{Deutsch, D} argument for his reinterpretation is that
\begin{quoting}
...it would surely be hard to regard a function `naturally' as computable if it could not be computed in Nature, and conversely. \citep[p.99]{deutsch:1985}
\end{quoting}
In the first part of this, `computed in Nature' suffers from the suggestive ambiguity between computable by human and computable by physical object, so let us take it to mean computable by machine, or more widely, physical object considered as a computer. More important for the present is the converse, which would read:
\begin{quoting}
It would be hard to regard a function computable in Nature as not `naturally' computable.
\end{quoting} 
But this is rather a teasing play on words. Part of the point at issue is what it means for a function to be computable in Nature, for a function to be computed by a machine, a meaning that Turing had to provide \textit{en route} to determining what the relation between functions computable in Nature and the `naturally computable' might be. If we just claim that the `naturally computable' functions are all and only those functions that can be computed in physical reality, we not only, perforce, miss the original point of trying to capture the effectively calculable\index{effective calculability}, but more importantly for present purposes, we miss out the key \textit{mathematical} component at the heart of the theory of computation. For we have not provided, as Turing did, a specification of what it is for a physical object to compute, to give a mathematical meaning to the possible evolutions of physical states.\index{Turing, A|)} 

What can be computed in physical reality has two sorts of determinant, mathematical\index{mathematical!determinants} and physical\index{physical!determinants}. The mathematical determines what the evolution of given physical states into others in a certain way would mean, what would have been computed by such a process; and the physical determines whether such a process can occur. Identifying the `naturally computable' functions with those that can be computed by physical systems, we emphasize the physical determinant to the exclusion of the mathematical one---we say that what can be computed is \textit{whatever} can be computed by \textit{any} physical system, but we have not said what, if anything, these various physical processes amount to in mathematical terms.

When Deutsch\index{Deutsch, D} says that behind the Church-Turing hypothesis\index{Turing!Church-Turing hypothesis} is really an assertion of the Turing Principle\index{Turing!Principle} (\citet[p.99]{deutsch:1985},\citet[p.3]{deutsch:etal:1999}), what he is trying to capture is the imperious nature of the hypothesis: you can't find any computation that can be done that \textit{can't} be done by the universal Turing machine\index{Turing!machine}\index{computer!universal}. He takes this imperious claim to require the possible existence of a physical object that could actually perform every (physical) computation. For `...the computing power of \textit{abstract} machines has no bearing on what is computable in reality' \citep[p.134]{FoR}, what is important is whether the computational processes that the machine describes can actually occur. The essence of the universal computing machine\index{computer!universal} is supposed to be that the physical properties it possesses are the most general computational properties that any object can possess. It follows that if the universal machine is to be an interesting object of study, it must be physically possible for it to exist (although supplies of energy and memory may remain a little idealised), otherwise studying it could tell us nothing about what can be computed in reality.

The significance of the Turing machine\index{Turing!machine|(} is thus supposed to lie in the fact that its description is so general that it has been pared down to the bare essentials of computing, with the result that any computation by any object can be described in terms of the operation of a Turing machine\footnote{This is perhaps a common view of the significance of Turing's machine.}. Deutsch\index{Deutsch, D} considers Turing's machine to be a very good, but ultimately inadequate attempt to give a description of the most general computing machine possible \citep[p.252]{FoR}. He would suggest that Turing\index{Turing, A} had made himself hostage to fortune by offering such a concrete characterisation of what is supposed to be the most general computing machine, in particular by explicitly describing the machine in classical (mechanical) terms and not allowing for the possible implications of quantum mechanics\index{quantum!mechanics} or some other successor theory\footnote{Deutsch\index{Deutsch, D} cites, for example, Feynman's\index{Feynman, R} remark \textit{a propos} Turing\index{Turing, A}: `He thought that he understood paper.' \citep[p.252]{FoR}\label{feynman:paper}}. Taking Turing's\index{Turing, A} \textit{intention} to refer to the most general machine as the important thing and erasing the unnecessary physical details of the Turing machine\footnote{The essence of the Turing machine is retained in the requirement that the universal computing machine operate by \textit{finite means}, defined in \citet[p.100]{deutsch:1985}.}, the content of the Church-Turing\index{Turing!Church-Turing hypothesis|(} hypothesis becomes the assertion that this most general machine can exist. The hypothesis has become the physical Principle\index{Turing!Principle}---it is now just an empirical question whether the universal computing machine can exist.

But this misrepresents the import of the Church-Turing hypothesis\index{Turing!Church-Turing hypothesis|)}, for we have missed the mathematical component, the definitional r\^ole of the Turing\index{Turing!machine|)} machine in the theory of computation. Put baldly, the reason why there is no computation that cannot be performed by the universal computing machine\index{computer!universal|(} is not that it just so happens that this object can actually exist in physical reality, but rather that nothing could count as such a computation. A computation is \textit{defined} by reference to the abstract universal computing machine, the possible evolution of physical states given a mathematical meaning by reference to that model. What we call a computation is determined by the abstract model, hence there can be no such thing as a computation that cannot be performed by the universal computing machine.\index{computer!universal|)} 

Of course, it is conceivable that there could be physical processes that are not covered by our abstract model and which we decide we might want to call computations, but these processes still need to gain a mathematical meaning from somewhere; and once we have given them such meaning, we will have extended our definition of computing to cover these cases as well.\footnote{Note that this question differs from the question of whether the definition of machines computing captures all that would `naturally be regarded as computable' by humans. What is currently at issue is the mathematical meaning that can be given to various physical processes, not whether the definition of computing offered would include all and only that which falls under the intuitive notion of the effectively calculable\index{effective calculability}.} This does not, however, affect the point that by definition there can be no computation that cannot be performed by the universal computer\index{computer!universal}, as a corollary of these physical processes having mathematical meaning. (Until these processes are accepted under the definition, they are not yet computations. Compare p.~\pageref{piggyback} for an example of a specific type.)

Having noted that from two computing machines we can form a composite machine, whose set of computable functions contains the union of the sets of functions computable by its components, Deutsch\index{Deutsch, D} suggests that:
\begin{quoting}
There is no purely logical reason why we could not go on \textit{ad infinitum} building more powerful computing machines, nor why there should exist any function that is outside the computable set of every physically possible machine. \citep[p.98]{deutsch:1985}  
\end{quoting}
He goes on to suggest that it is physics rather than logic that provides the constraint (presumably the contingent physical fact that there can exist a universal computing machine\index{computer!universal} exhausting the possibilities). But this seems wrong. Our immediate response is to ask why might there not simply come a point after which no new functions are added and we would just keep adding ones we already have?  What might determine this? It is \textit{precisely} logic, logic and mathematics, that determine this question. Once we have defined our computational states in a certain way, it is mathematics that determines\index{mathematical!determinants} the set of functions that can be computed by all possible evolutions of those states. Deutsch\index{Deutsch, D} is correct that physics has a r\^ole to play in determining what is computable, but it can only get in on the act \textit{after} mathematics.

Another example of Deutsch\index{Deutsch, D} seeming to over-emphasize the r\^ole of physics at the expense of mathematics is the following passage:
\begin{quoting}
Computers are physical objects, and computations are physical processes. What computers can or cannot compute is determined by the laws of physics alone and not by pure mathematics. \citep[p.98]{FoR}
\end{quoting}
Computations, remembering that we are speaking strictly of mechanical, not human computers, are indeed physical processes, but what makes them a computation is not physical. The processes going on in a computer are governed by the laws of physics, but it would be wrong to say that the \textit{computation} is entirely governed by physics, for mathematics determines what the transitions from physical state to physical state mean. Physics determines\index{physical!determinants|(} what physical state can follow from what physical state, but mathematics\index{mathematical!determinants|(} determines whether or not this is a computation and what it is a computation of.

Deutsch\index{Deutsch, D} is quite right to emphasize that the physical determinants of computing should not be ignored in the theory of computation, but he has taken this insight too far by entirely neglecting the mathematical determinants. We must recognise that their place is \textit{prior} to that of the physical determinants. If our theory of computation is asking what the ultimate limits of computation are (again, computation by machines), the answer must involve two sorts of consideration. We are asking what is possible with physical computational states defined in a given way, so our first consideration is what can these states evolve to, whilst the second is: what does such an evolution mean? The first part is a physical question and the second part mathematical. Maths will determine what $\alpha_{i} \ldots \alpha_{f}$ means ({\boldmath $\alpha$} being physical states under some description), but it won't say if it is a possible evolution of states---that is for the laws of physics to decide.

We might want to say, then, that mathematics provides the ultimate bound on what is computable (most obviously, nothing could count as computing a contradiction); and it determines what progressions of physical states are computations and what they are computations of. But what progressions of physical states there can be is determined by physics.\index{mathematical!determinants|)}\index{physical!determinants|)}

In their admirably clear 1999 paper, Deutsch, Ekert and Lupacchini~\citep{deutsch:etal:1999}\index{Deutsch, D}\index{Ekert, A}\index{Lupacchini, R} admit that there are both logical and physical limits to the computations that can be performed by computing machines. They present the halting problem\index{halting problem|(} as an example in which logical and physical constraints are intimately linked; but their discussion still seems to betray confusion between the mathematical and the physical nature of computing.

From the halting problem, we learn that there are some computational problems, in particular, determining whether a specified universal Turing machine\index{Turing!machine} given a specified input will halt, that cannot be solved by any Turing machine\index{Turing!machine}; and it is logic that tells us this. Deutsch, Ekert and Lupacchini\index{Deutsch, D}\index{Ekert, A}\index{Lupacchini, R} go on to say that:
\begin{quoting}
In physical terms, this statement says that machines with certain properties cannot be physically built, and as such can be viewed as a statement about physical reality or equivalently, about the laws of physics. \citep[p.4]{deutsch:etal:1999}
\end{quoting}
But the halting problem tells us nothing of the sort. The halting problem lies primarily on the mathematical side of computing and teaches us nothing directly about the laws of physics. Given the specification of computing states we are dealing with, mathematics and logic tell us that nothing could count as providing the solution to this problem; no possible state is the solution. Thus the `certain properties' that the machines may not possess are mathematical properties, not physical ones. It is not that the machines are forbidden to possess these properties, that some force prevents it, it is that \textit{nothing would count as building a machine with these properties}. The halting problem,\index{halting problem|)} then, tells us nothing about what can be built; it tells us the mathematical constraints on what can be computed given the way we have defined computing. Failing to recognise this means failure to understand the way in which the definitional r\^ole of the abstract universal computer\index{computer!universal} gives mathematical meaning to the evolution of physical states. This in turn can be traced back to a failure to recognise Turing's\index{Turing, A} purely mathematical achievement in `On Computable Numbers', quite separate from the concern there with epistemological issues surrounding effective calculability\index{effective calculability}.\footnote{I am indebted to \citet{shanker:1987} for the emphasis on this separation.}

We have seen that Deutsch's emphasis on the possible physical existence of the universal computing machine\index{computer!universal|(} misrepresents its significance; missing entirely its essential r\^ole determining the mathematical meaning of the evolution of physical states. From this it is clear that insisting on the physical nature of the Turing Principle\index{Turing!Principle} debars it from playing the central r\^ole in the theory of computation. For it is not a contingent, empirical fact that there exists a universal computing machine, it is a necessary fact that arises from the way the abstract model determines the mathematical meaning of certain physical processes, making them computations. It is not that the universal machine covers all the possibilities, the universal machine \textit{determines} the possibilities.\index{computer!universal|)}

Where Deutsch\index{Deutsch, D} is correct, however, is that there is a clear sense in which we should be interested in the physical realization of the abstract computing machine. The importance of being able to build the machine, if only in principle, is that we want the progressions of states it describes to actually be \mbox{do-able!} This would clearly determine whether we have an interesting definition of computation and one worth pursuing. (We should emphasize here, just to be clear, that this issue is distinct from the issue of the definition of computation providing mathematical meaning for physical processes and distinct again from the task of characterizing the effectively calculable\index{effective calculability}.) However, we do not require that the universal computing machine\index{computer!universal} be a possible physical existent; all that is required is that physical analogues of the computational processes of the universal machine are physically possible processes.

\subsection{Non-Turing computability? The example of Malament-Hogarth spacetimes}

As a particularly striking example of where these concerns would be relevant, let us consider Hogarth's\index{Hogarth, M} presentation of non-Turing computability\index{Turing!non-computability|(} in certain relativistic spacetimes \citep{hogarth}. The idea is that in these spacetimes, dubbed \textit{Malament-Hogarth} spacetimes\index{Malament-Hogarth spacetime|(}, it appears possible to perform \textit{supertasks}\index{supertasks}---an infinite number of steps in a finite length of time. These spacetimes $(M,g)$ are such that they contain a path $\lambda$ that starts from a point $p$ and has infinite length, but that on this path it is always possible to signal to a point $q$ that can be reached from $p$ in a finite span of proper time.\footnote{That is, all points on $\lambda$ are contained in the chronological past of $q$. The chronological past of a point $q$ is the set of all points $p$ for which there is a nontrivial future directed timelike curve from p to q \citep[p.24, n.1]{earman:norton}.}

\begin{sloppypar}
A toy example of such a spacetime is given by \citet{earman:norton}. Starting with a Minkowski spacetime $(R^{4},\eta)$ we choose a scalar field $\Omega$ on $M$ such that $\Omega=1$ outside a compact set $C\subset M$ and $\Omega$ tends rapidly to infinity as we approach a point $r\in C$. The spacetime $(R^{4}-r,\Omega^{2}\eta)$ is then a Malament-Hogarth spacetime and the path $\lambda$ will start at p and go towards $r$. What we are supposed to do is project a given Turing machine\index{Turing!machine} down the path $\lambda$ and then travel to $q$, by which time the machine will have signalled to us if it has halted. Using this technique, we might, for example, solve the Goldbach conjecture\index{Goldbach's conjecture} by programming our Turing machine\index{Turing!machine} to check each even number in turn to determine whether it is the sum of two primes, and halt if it finds a counterexample. We then send it off down $\lambda$ and travel to $q$. If we have received a signal, the conjecture is false, if not, it is true. Generalising this approach, we appear able to solve Turing unsolvable problems in these spacetimes.
\end{sloppypar}

The \textit{decision problem}\index{decision problem} for a property $P$ is said to be solvable if there is a mechanical test (effective procedure) which will tell us whether or not any object (of the appropriate category) possesses $P$ in a finite number of steps \citep[p.115]{boolos:jeffrey}. Thus the decision problem\index{decision problem} for $P$ is Turing solvable if there is both a Turing machine\index{Turing!machine|(} that will halt after a finite number of steps if and only if $P$ holds and a Turing machine that will halt after a finite number of steps if and only if $P$ does not hold. If only one of these exist, the problem is \textit{partially Turing solvable}\index{partial solvability}. The halting problem\index{halting problem} and the decision problem for first order logic\index{decision problem!for first order logic|(} are partially Turing solvable\index{partial solvability}, but the full decision problem can be solved for them in a Malament-Hogarth spacetime. For the halting problem\index{halting problem}, all we need do is project the Turing machine in question down $\lambda$, set to signal if it halts. We travel to $q$ and if we have received a signal, we know the machine halts and if not, we know it never halts. Similarly for the decision problem for first order logic, noting that there exists a Turing machine\index{Turing!machine|)} that will halt after a finite number of steps if a given sentence $S$ is valid \citep[p.145]{boolos:jeffrey}, we adopt the same procedure---if we have a received a signal at $q$, the sentence is valid, if we have not, it is not\index{decision problem!for first order logic|)}. It is clear that the decision problem for any partially Turing solvable problem\index{partial solvability} is solvable in a Malament-Hogarth spacetime (we will have to vary our interpretation of signal/no-signal appropriately, of course).

Hogarth\index{Hogarth, M|(} goes on to describe more complicated computational processes that would seem to solve the decision problem for arithmetic\index{decision problem!for arithmetic}, but the simple case serves for our purposes. We have here a clear example of the question of the physical realizability of the processes described being all-important. If the processes Hogarth describes are physically possible, then we have a whole new class of computability distinct from Turing computability and we extend our notion of computability accordingly. Note that the mathematical meaning of the processes Hogarth describes \label{piggyback}piggy-backs on our current definition of computability---we think we can see clearly what these processes would mean if they were physically possible. Given the meaning we have already given to computational processes in terms of the universal Turing machine\index{computer!universal} and what it can compute, these meanings seem to follow.\footnote{I say `think' and `seem' here, for we may believe that these mathematical meanings unfold from, since they are already contained in, the mathematical concepts we have. But we may believe that the mathematical meaning of these processes ultimately rests on our \textit{decision} to accept the conclusions set out as following from our present stock of mathematical propositions. This allows for the positions of those who believe there is a fact about, for example, whether Golbach's conjecture is true independent of whether a proof or disproof has been or ever will be found; and those who believe there is no such fact \textit{until} a proof or disproof has been found.} 
The reason why the claim that it is a conceptual truth that our particular universal computing machine\index{computer!universal} can perform all possible computations is not undermined by the Hogarth example and others like it is that we have, as it were, recognised new possibilities in our (abstract) universal computing machine\index{computer!universal}, not discovered that it could not in fact perform all possible computations, which would be logically impossible. Or rather, to be more precise, by generalising or slightly adjusting the sets of physical states and their evolution for our \textit{definitional} universal machine (in the Hogarth case, by including evolutions in these unusual spacetimes), we change the class of computations and computable functions at the same time.

Returning to the question of the physical realizability of these Hogarthian processes, we need to recognise that the computational process extends from the initial launch of the Turing machine\index{Turing!machine} to the possible reception of the signal by the receiver. Thus whether these are physically possible computations will depend on whether a suitable Turing machine\index{Turing!machine} can exist in the spacetime in question (in particular we will be worried about what happens to it as it approaches $r$), whether a signal from the Turing machine can reach the observer intact, and of course, whether Malament-Hogarth spacetimes are physically possible.\index{Malament-Hogarth spacetime|)}\footnote{These questions should be approached with an open mind, see \citet{earman:norton} for an interesting discussion, and compare \citet[\S 6]{hogarth}} If it turns out that these processes are physically possible, then we must extend our notion of what can be computed to include these striking non-Turing computations. If they are not, then a definition of computability that included Hogarth's\index{Hogarth, M|)} computations would not be an interesting one for practical purposes---it would be no more than a mathematical toy. We cannot learn any maths from the conceivability of peculiar computational processes, for our knowledge of the relevant maths is already explicit in our conceiving them; that it might be an open question whether these processes are physically possible is only relevant to the question of what we can make machines (or physical objects in general since `machine' implies manufacturing), do for us.
\index{Turing!non-computability|)}

\subsection{Lessons}

We have seen, then, that Deutsch\index{Deutsch, D} has over-emphasized the physical determinants\index{physical!determinants} of computing to the exclusion of the mathematical\index{mathematical!determinants}. The Turing Principle\index{Turing!Principle} should not be seen to underlie the Church-Turing hypothesis\index{Turing!Church-Turing hypothesis}, for that misrepresents the mathematical significance of the concept of the universal computing machine. The universal machine\index{computer!universal} defines the mathematical meaning of the possible evolution of physical states and hence it is a necessary fact that the universal computing machine can perform every possible computation. It is certainly interesting that the Turing Principle\index{Turing!Principle} happens to be true in quantum mechanics\footnote{Intuitively, the state of any finite quantum system is just a vector in Hilbert space and can be represented to arbitrary precision by a finite number of qubits; and any evolution of the system is just a unitary transformation\index{unitary operator} of this vector and can be simulated by the universal quantum computer\index{universal!quantum computer}, which by definition can generate any unitary transformation with arbitrary precision. Deutsch\index{Deutsch, D} offers a more rigorous proof taking into account the fact that any sub-system must always be coupled to the environment \citep[\S 3]{deutsch:1985}}, but we should hesitate to draw any far-reaching conclusions from this. Certainly, the claim adumbrated in Section~\ref{intro} that the advent of the quantum computer\index{computer!quantum} makes sense of Turing's\index{Turing, A} theory of computation, that his machines were quantum mechanical\index{quantum!mechanics} after all,
is false. 

The discussion might be summarized in the following way. 

It is useful to distinguish between three different tasks with which the Church-Turing hypothesis\index{Turing!Church-Turing hypothesis} is associated: characterizing the effectively calculable, providing the evolution of physical states with mathematical meaning and fixing upon a useful definition of physical computability. The Turing Principle\index{Turing!Principle} could not replace or underlie the Church-Turing hypothesis\index{Turing!Church-Turing hypothesis} for any of these tasks. Not the first, because the Turing Principle\index{Turing!Principle} is supposed to concern all functions computable by physical systems, rather than what is computable by a human; and not the second or third because an empirical principle cannot play the crucial definitional mathematical r\^ole that I have emphasized. It is perhaps worth noting that the Turing Principle\index{Turing!Principle} is undoubtedly most closely tied \textit{in intention} to the third of these tasks rather than to the first. However, although it is true that Turing\index{Turing, A} did not consider the possibility of computations using explicitly quantum objects, this can hardly be said to be to the detriment of the Church-Turing hypothesis\index{Turing!Church-Turing hypothesis}. The third of the tasks I have mentioned, delimiting the bounds of physical computability, is not really, after all, the object of the Church-Turing hypothesis\index{Turing!Church-Turing hypothesis}.

As has been emphasized at various points, we have been talking in this section only of computation by machine or by physical object considered as a computer, as opposed to human computing or calculating. This is an important clarifying step that allows us to distinguish clearly the mathematical and physical sides of the theory of computation.
Having mentioned this convenient separation of human from machine, however, one's thoughts seem naturally drawn to the further, notoriously vexed, question of the relation between human cognition and machine computation. Rather than delve into this question here\footnote{See \citet[\S 4]{timpson:2004} for discussion of this question. One point that it is perhaps helpful to note is that the debate about the nature of human cognition and of thinking machines might generate less heat and confusion if the question of whether it might be possible to build a machine which we could appropriately ascribe mental conduct terms to were always clearly distinguished from the question of whether it is possible to analyse cognition and conation in computational terms.}, it suffices to note that even if it is thought that human calculation is no more than physical calculation with a cherry on top, this separation remains important, for it emphasizes the different types of r\^ole the mathematical and physical determinants\index{mathematical!determinants}\index{physical!determinants} of computation play; and this distinction in r\^ole is one which, I suggest, should be retained independently of any judgement on the value of the computational analogy. 

\section{The Church-Turing Hypothesis as a constraint on physics?}\label{constraint?}

In the preceding section we saw the necessity of distinguishing between a number of different ideas with which the Church-Turing hypothesis is often loosely associated; and it was emphasized in several places that the task of characterizing the effectively calculable functions should be distinguished from the task of delimiting the bounds of the physically computable, while it is the former task to which the Church-Turing hypothesis is directed. This important point has been ably expounded by \citet{copeland,copeland:encyclopedia} (see also \citet{pitowsky:2002})\footnote{For an intemperate reply to Copeland, in defence of the `orthodoxy' which conflates these and other ideas, see \citet{hodges:2004}. \citet[\S 5]{copeland:proudfoot:2004} reply.}.

On this topic, a telling observation concerns the nature of the evidence that is cited as endowing the Church-Turing hypothesis with the very high degree of entrenchment that it deservedly enjoys. This evidence generally centres on the fact that the large number of different attempts to make precise the intuitive notion of effective calculability all give rise to the very same class of computable functions, along with the fact that all the functions we intuitively take to be effectively calculable fall into this class. A representative textbook statement \citep{cutland} is the following (N.B. the basic computational model in this book, equivalent to the universal Turing machine, is the universal register machine (URM)):    
\begin{quoting}
The evidence for Church's thesis, which we summarise below, is impressive.
\begin{enumerate}
\item The Fundamental result: many independent proposals for a precise formulation of the intuitive idea led to the same class of functions, which we have called $\mathcal{C}$.
\item A vast collection of effectively computable functions has been shown explicitly to belong to $\mathcal{C}$ [...]
\item The implementation of a program P on the URM to compute a function is clearly an example of an algorithm; thus...we see that all the functions in $\mathcal{C}$ are computable in the informal sense. Similarly with all the other equivalent classes, the very definitions are such as to demonstrate that the functions involved are effectively computable.
\item No one has ever found a function that would be accepted as computable in the informal sense, that does not belong to $\mathcal{C}$.\\ \citep[p.67]{cutland}
\end{enumerate}
\end{quoting}
The point is that all this evidence, while certainly telling us something important, \textit{has no implications at all} for the question of what the bounds of physical computability are---on the question of what we can get physical systems to do for us. It simply points to the fact that Church, Turing and others did indeed succeed (amazingly well) in making precise the intuitive notion of effective calculability. And note that the facts cited are not really \textit{evidence} for a \textit{hypothesis}, but rather emphasize that the Church-Turing \textit{definition}, or stipulation, does not lead to conflict with any pre-theoretic notions of effective calculability. These facts are not evidence, then, but are \textit{reasons} why this definition is both a very good and a remarkably powerful one. 

The unimpeachable status that the Church-Turing hypothesis enjoys does not, therefore, impugn (nor could it be impugned by) the possibility of physical computational models that go beyond Turing computability (the example of Malament-Hogarth computability gave us a concrete example); the areas of concern are quite distinct. It follows that one shouldn't seek to use the Church-Turing hypothesis as a restricting principle on physical laws.

However a prominent example of this unfortunate mode of reasoning in the context of quantum mechanics is to be found in \citet{nielsen:ct}, for example. The abstract to this paper states:
\begin{quoting}   
We construct quantum mechanical observables and unitary operators which, if implemented in physical systems as measurements and dynamical evolutions, would contradict the Church-Turing thesis which lies at the heart of computer science. We conclude that either the Church-Turing thesis needs revision, or that only restricted classes of observables may be realized, in principle, as measurements, and that only a restricted class of unitary operators may be realized, in principle, as dynamics. \citep{nielsen:ct}
\end{quoting}

To give a flavour of the approach: the author begins by considering an observable defined by
\[ \hat{h}=\sum_{x=0}^{\infty}h(x)\ket{x}{}\bra{}{x},\]  
where $\{\ket{x}{}\}$ is an orthonormal basis for some physical system with a countably infinite dimensional Hilbert space (e.g. the number states of a particular mode of the e-m field), and $h(x)$ is the characteristic function for the halting problem. We may suppose that the various $\ket{x}{}$ states can reliably be prepared. Measurement of this observable on systems prepared in these states will then evaluate the halting function for us. Nielsen concludes that this would conflict with the Church-Turing hypothesis, therefore, we must either revise the hypothesis, or conclude that this type of measurement is not in fact physically possible. Given the entrenchment of the Church-Turing hypothesis, Nielsen opts for the latter. But the conclusion is misplaced, as the Church-Turing hypothesis does not rule out the possibility of non-Turing computability using physical systems; and the entrenchment of the hypothesis does not rest on empirical evidence about what can be computed by physical systems\footnote{In fact, one can raise a further problem for this example of Nielsen's---it is not clear that it \textit{would} constitute an example of non-Turing computability. In order to perform the measurement corresponding to the operator $\hat{h}$, we need to be able to pick out the correct piece of equipment in the lab. But in order to do this one would already have to have evaluated the halting function (imagine a shelf in the lab with a series of apparatuses all of which measure in the $\{\ket{x}{}\}$ basis, but have different eigenvalue spectra associated with them). Thus the outlined procedure would not count as an \textit{effective} procedure, as one can't pick out the desired piece of apparatus by an effective procedure. In essence, the solution to the halting function has been hardwired into the apparatus, but we can't get at it unless we already have the solution.}.

It is sometimes suggested that part of the meaning of the slogan `Information is Physical' for the quantum information scientist is to encapsulate the recognition of the need to go beyond the Church-Turing hypothesis in the theory of computation. Our reflections in this chapter give the lie to this conception, however. It is based on an equivocation between the task of characterizing the effectively calculable functions---the task of the Church-Turing hypothesis---and the distinct task of investigating the bounds of physical computability.  


\end{doublespacing}



\chapter{Morals}\label{morals}
\begin{doublespacing}

It is time to draw together some morals from the preceding chapters. 

The first is the simple statement that the everyday and the information-theoretic notions of information are indeed distinct and are to be kept apart. When this is done, I believe, we have a significant counter-weight to the rather breathless sorts of claims remarked on in the introduction, to the effect that the advent of quantum information theory heralds a revolution for our world picture in which the material will be replaced as fundamental by the immaterial: information. 

When we recognise that quantum information theory is not concerned with, and has no implications for, information in the everyday sense of the word, that is, for a notion of information involved with semantic and epistemic concepts, it can be seen that these claims for the implications of quantum information theory, which \textit{are} concerned with our epistemological position and the meaning of our discourse, seem groundless.   

The further point that it is important to keep the distinction between the technical and everyday notions of information clear when approaching foundational questions in quantum mechanics will be remarked on again, briefly, in the next chapter.

The second moral is a very general one regarding the nature of quantum information theory. It has been maintained throughout Part~I that `information' is an abstract noun. We saw, in particular, the value of being clear on this fact in the quantum case when considering the topic of teleportation. The conceptual troubles this process had seemed to present were revealed to be the result of the old philosophical error of hypostatizing an abstract noun.
But where, we might ask, does this leave quantum information theory? Without a subject matter? Indeed not.

The correct conclusion, it seems to me, is this:
Quantum information theory is best seen not as a theory of some strange new thing---quantum information---but rather as a new type of information theory. That is, a theory of communication and computation which identifies new types of physical resource---qubits, shared entanglement, etc.---and correspondingly identifies new sorts of task that can be achieved using them, along with some different (and hopefully, in at least some cases, better) ways of managing old tasks. The subject matter of the theory, if it is demanded that such be found, can simply be said to be these new types of physical resources and the tasks that can be achieved using them. 

One might say that it is a question of bracketing. Quantum information theory is not a
\begin{center}\large
(quantum information) theory \\
{\normalsize but a}\\
quantum (information theory).
\end{center}

Our final moral concerns the slogan `Information is Physical'. It might be recalled that having discussed the dilemma this proposition faces, I rather left the conclusion that should be drawn hanging. Let me make that good now. Recall that the dilemma was produced by the fact that if `information' in the slogan was supposed to advert to information in the everyday sense of the word, then the claim would seem to involve a commitment to the project of semantic naturalism. But it was argued that the success and practice of quantum information theory has no implications for this purely philosophical question; and no more would the outcome of the philosophical debate have any implications for quantum information theory. If, though, the slogan is supposed to refer to information in the technical sense of Shannon theory, then it is hard to see how the claim that some physically defined quantity is physical is in the least enlightening. (Alternatively, I suggested that the whole thing could simply be seen as a category mistake.) 

The resolution of the dilemma is this: the claim `Information is Physical' as made by a quantum information scientist should not, I suggest, be construed as an ontological claim, but rather as a \textit{methodological one}. It does not represent a claim about how the world is, or represent an insight into the nature of information, but it should be seen, rather, to express a commitment characteristic of the discipline. Roughly speaking, the view that it is a very interesting and fruitful business to study the information carrying, storing and processing capabilities of physical systems as described by our most fundamental physical theories. \textit{A priori} this need not have been so, but the vibrant health of quantum information science assures us, emphatically, that in the case of quantum systems, it most certainly is.

\end{doublespacing}


\part{Information and the Foundations of Quantum Mechanics}

\renewcommand{\bra}[2]{\,_{#1}\langle \, #2\!|}

\begin{quoting}
Information theory has, in the last few years, become something of a scientific bandwagon...

Although this wave of popularity is certainly pleasant and exciting for those of us working in the field, it carries at the same time an element of danger. While we feel that information theory is indeed a valuable tool...it is certainly no panacea for the communication engineer or, \textit{a fortiori}, for anyone else. Seldom do more than a few of nature's secrets give way at one time. It will be all to easy for our somewhat artificial prosperity to collapse overnight when it is realised that the use of a few exciting words like \textit{information}, \textit{entropy}, \textit{redundancy}, do not solve all our problems.\\ \citep{shannon:1956}
\end{quoting}


\chapter{Preliminaries}\label{talk}

\begin{doublespacing}

\section{Information Talk in Quantum Mechanics}

Shannon's words above represent a salutory warning for those of us interested in the question of whether quantum information theory has implications for the foundational problems of quantum mechanics. Is it, perhaps, that we have become overly excited by the appearance of a few trigger words (information, uncertainty, entropy...) in books, journals and pre-print servers dedicated to quantum theory? Compare on the other hand, Fuchs:
\begin{quoting}
...no tool appears better calibrated for a direct assault [on quantum foundations] than quantum information theory. Far from a strained application of the latest fad to a time-honored problem, this method holds promise precisely because a large part---but not all---of the structure of quantum theory has always concerned information. It is just that the physics community needs reminding. \citep{fuchs:only}
\end{quoting}

In this brief chapter I shall set out a few preliminaries; some points that are rather basic, but essential when trying to see what can be made of information talk in quantum mechanics.

Appeal in some form to the notion of information as a way of addressing the conceptual problems presented by quantum mechanics has been a recurrent feature of many discussions of the quantum foundations, particularly for those in the Copenhagen tradition; and this trend has been reinvigorated following the growth of quantum information theory. For a selection of more recent statements see, for example, \citet{fuchs:peres,mermin:whose,mermin:copenhagencomputation,peierls:ghost,peierls:defence,peres,wheeler:ghost,wheeler,foundationalprinciple}.

Very often, the suggestion proceeds along the lines that the traditional problems of measurement, nonlocality and so on, are resolved when one recognises that the quantum state should simply be viewed as representing one's \textit{knowledge} or \textit{information} rather than any objective property of the world. A representative formulation is the following, due to Hartle:
\begin{quoting}
The state is not an objective property of an individual system but is that information, obtained from a knowledge of how a system was prepared, which can be used for making predictions about future measurements.

...A quantum mechanical state being a summary of the observers' information about an individual physical system changes both by dynamical laws, and whenever the observer acquires new information about the system through the process of measurement. The existence of two laws for the evolution of the state vector...becomes problematical only if it is believed that the state vector is an objective property of the system...The ``reduction of the wavepacket" does take place in the consciousness of the observer, not because of any unique physical process which takes place there, but only because the state is a construct of the observer and not an objective property of the physical system. \citep[p.709]{hartle:1968}\footnote{It may be noted in passing that Hartle's \textit{argument} for these propositions is by no means entirely persuasive. While there is not room to go into details here, suffice it to say that his argument for construing the state of a system as information trades on an ambiguity between specifying what a state \textit{is} (e.g., an assignment of truth values to experimental propositions) and specifying what state something is \textit{in}; moreover, a realist opponent can always insist that the quantum state only allows us to make predictions about the behviour of a system precisely \textit{because} it corresponds to a system's possessing certain objective properties.}   
\end{quoting} 

As so often in the foundations of quantum mechanics, however, it is instructive to turn to the writings of John Bell; and there we find a warning. For `information' is on Bell's famous list of \textsc{bad words} that 
`have no place in a formulation with any pretence to physical precision' \citep[p.34]{bell:against}\footnote{To illustrate Bell's use of the term `formulation': ``Surely, after 62 years, we should have an exact formulation of some serious part of quantum mechanics? By `exact' I do not of course mean `exactly true'. I mean only that the theory should be formulated in mathematical terms, with nothing left to the discretion of the theoretical physicist...until workable approximations are needed in applications." \citep[p.33]{bell:against}.}. Bell indicates the pertinent sources of disquiet with two rhetorical questions: \textit{Information about What?}; \textit{Whose information?} 

These are indeed good questions, and the first most especially. For it presents a fundamental dilemma: the Scylla and Charybdis facing proponents of information talk in quantum mechanics.  

If the quantum state is to be construed in terms of representing one's information then it seems that there are two possible sorts of answer that could be given to the question `Information about what?':
\begin{enumerate}
\item Information about what the outcomes of experiments will be;
\item Information about how things are with a system prior to measurement, i.e., about hidden variables.
\end{enumerate}

Now the latter option is unlikely to be attractive to anyone who is trying to appeal to information as a way of avoiding the problems caused by the seemingly odd behaviour of the quantum state. The aim, roughly speaking, was to circumvent the problems associated with collapse or nonlocality by arguments of the form: there's not really any \textit{physical} collapse, just a change in our knowledge; there's not really any \textit{nonlocality}, it's only Alice's knowledge of (information about) Bob's system that changes when she performs a measurement on her half of an EPR pair. But we all know that if we are to have hidden variables lurking around then these are going to be \textit{very badly behaved indeed} in quantum mechanics (nonlocality, contextuality). So it surely can't be this second answer that our would-be informationist is really after.

But now consider the first answer. If the information that the state represents is information about what the results of experiments will be, then the difficulty is now to say anything interesting that doesn't simply slide into instrumentalism. Instrumentalism, of course, is the general view that scientific theories do not seek to describe the laws governing unobservable things, but merely function as devices for predicting the outcomes of experiments. An instrumentalist view of the quantum state understands the state merely as a device for calculating statistics for measurement outcomes: this is very close to the view that the state merely represents information about what the results of measurements will be. But if all that appeal to information were ultimately to amount to is a form of instrumentalism, then we would not have a particularly interesting---and certainly not a novel---intrepretational doctrine. It should be noted that merely presenting an old doctrine such as instrumentalism in the currently popular idiom of information does not make it any more (or any less, admittedly,) of an attractive doctrine. (Here Shannon's warning is very pertinent.)

Thus the dilemma. To present a distinctive and hence an interesting doctrine, it seems that the proponent of information has somehow to steer a course that avoids hidden variables, yet does not merely amount to instrumentalism; but it is not clear that this is easily done.   

One option might be the following: one could emphasise that in contrast to standard instrumentalism, the focus of one's interest is individual systems rather than the statistics of measurements for ensembles. But this approach suffers from a decisive objection.

For there is a very important further subtlety that needs to be highlighted if one is interested in viewing the quantum state as representing an agent's knowledge or information. This is the point that---to use the philosophical jargon---both the terms `knowledge' and `information' are factive. That is, one can't know that $p$ unless $p$ is the case; one can't have the information that $p$ unless it is true that $p$. 
The major difficulty this presents for those who may have hoped to avoid the conceptual problems of quantum mechanics by understanding the quantum state of an individual system in terms of information is that the factivity of knowledge and information entails precisely the sort of objectivity that the invocation of information was originally intended to bypass. 

The straightforward instrumentalist seeks to avoid the problems associated with measurement and nonlocality by remaining at the level of statistics only: individual systems are not described and collapse doesn't correspond to any real process. So far as it goes, this strategy is reasonably successful\footnote{See \citet{saunders:measurement} for some criticisms of instrumentalism as a solution to the problem of measurement, though.}. For someone taking the information route and associating a quantum state with individual systems, however, the essence of \textit{their} approach is that different agents can ascribe different states to a given quantum system, because they have different information regarding it. Thus in the Wigner's friend scenario, for example, \citep{wigner:1961} rather than there having to be a mysterious collapse at some point, both agents involved simply ascribe different states to the system being measured. There is not supposed to be one correct state which is in some sense an objective property of a system, rather, each agent will ascribe a different state based on their differing information (whether they are inside the lab doing the measurement, or waiting patiently outside for their friend). Similarly in the EPR case, Alice's measurement is understood not to change any real properties of Bob's system; her measurement merely provides her with some particular information about it, in virtue of the correlations involved in the inital entanglement. Post measurement, she will ascribe a new state to Bob's system---which is located at a distance---but since the state does not correspond to an objective property of the system, this does not connote nonlocality. Indeed, Bob continues, all the while, to ascribe the same old state (density operator) to his system as ever, until he performs a measurement of his own, or gets in touch with Alice.  

But the factivity of information and knowledge puts paid to these forms of argument: if the quantum state represents what one knows, or what information one has, then things have to be as they are known to be. For example, if I know what the probability distributions for the outcomes of various measurements on a system are, then the probabilities must indeed be thus and so. We have a matter of right or wrong determined by what the properties of a system actually are. If Alice performs a measurement on her half of an entangled pair in the singlet state and subsequently knows the pure state of Bob's system, then his system objectively has to be in that state. Alice now knows that a particular experiment will have some outcome as a certainty, whereas before it didn't; and this is a determinate matter of fact. 
Thus we end up, in this approach, having to talk again about objective properties of a system, and objective properties that can be changed at a distance, even after making our appeal to knowledge and information talk. No progress is thus made with the conceptual problems in this direction; the approach is a blind alley.   

I have so far emphasised only one of Bell's questions. The point of the second, `Whose information?' is presumably to highlight what Bell felt would be an unnacceptable level of vagueness associated with use of the term `information', if it were to occur in a putative formulation of fundamental theory. This vagueness could be seen to come from two different directions: first, a vagueness of anthropocentrism (how are we to specify with any precision what counts as a \textit{bona fide} cognitive agent?); second a vagueness associated with subjectivity (different agents might occupy different perspectives, perhaps) although this sort of worry is to some degree mollified by realising the factivity of information 
\footnote{
Interestingly, \citet{mermin:whose}, developing an idea due to \citet{peierls:defence}, has sought to respond to the challenge presented by the `Whose information?' question, by deriving conditions under which different density matrices can be thought to represent different knowledge that various agents might have about one and the same system (see also \citet{mermin:compatibility}, \citet{bfm}). This approach has rightly been criticised by Fuchs, however (see \citet[esp. pp.19-25;42-51]{fuchs:what}; and also \citet{fuchs:compatibility}) on the grounds that any approach in this vein, that involves assessing whether an agent's ascription of a state to a system is correct, or admissible, or what-not, amounts to giving up on the original desire for non-objectivity of the state that was supposed to be doing the distinctive conceptual work. If there is a question, ultimately, of being right or wrong, then one might as well openly admit that the quantum state is objective after all. In essence, the point here may be put in terms of factivity again: if we imagine different knowledge that people might have about a system and the different states they may assign on the basis of that knowledge, then there must exist determinate facts about the system that each of them is, to a greater or lesser degree, aware of. Although he does not himself put it in these terms, Fuchs' awareness of the factivity of the terms `knowledge' and `information' and his related criticism of Mermin, mark the change from the objective Bayesianism of \citet{fuchs:2001} to the more consistent subjectively Bayesian position of \citet{fuchs:only}.}.  

The dangers of amounting to no more than a form of instrumentalism and the factivity of the terms `knowledge' and `information' are the first two preliminary considerations that need to be borne in mind when assessing information based approaches to quantum mechanics. The third and final one is as follows. It was emphasised in Part~I that the everyday notion of information---with its links to knowledge, language and meaning---is to be firmly distinguished from the technical notion of information that arises in information theory. The latter is not a semantic or an epistemic concept; and \textit{pace} Dretske, considerations of mechanical communication systems would seem to have precious little to do with explaining semantic and epistemic properties. Now, keeping the distinction between the everyday and the technical notions of information clearly in mind is crucial when considering the r\^ole that quantum information theory might have to play in the foundations of quantum mechanics, for otherwise one may easily fall prey to some serious misconceptions.

For example, it might well be thought that it is simply obvious that quantum information theory will shed light on the interpretive problems of quantum mechanics. After all, the key conceptual problem in quantum mechanics is the problem of measurement; but what is measurement other than a transfer of information, an attempt to gain knowledge? As we are now equipped with a theory of information in the quantum domain, enlightenment is sure to follow! 

This line of thought rests, of course, on a flagrant confusion between information in its everyday and its technical senses; between an epistemic and an information-theoretic sense of information. The following is a concrete example of just such a confusion:
\begin{quoting}
Quantum measurements are usually analyzed in abstract terms of wavefunctions and Hamiltonians. Only very few discussions of the measurement problem in quantum theory make an explicit effort to consider the crucial issue---the transfer of information. Yet obtaining knowledge is the very reason for making a measurement. \citep[p.viii]{zurek:1990}
\end{quoting} 
However, if any link is to be established between the techniques and applications of quantum information theory and the conceptual puzzles of quantum mechanics, it is not to be achieved by a facile equation of radically different senses of the term `information'. Here, more than anywhere, we need to be vividly aware of Shannon's warnings about getting over-excited by a few heavily-loaded terms; and we need to be on the look-out to make sure no-one is being misled by an implicit or explicit slide between different senses of the term `information'. 

With these preliminary reflections behind us, we shall turn, in the next chapter, to consider some specific proposals for the application of information-theoretic ideas to the foundational problems of quantum mechanics. 

\end{doublespacing}

\chapter{Some Information-Theoretic Approaches}\label{approaches}

\begin{doublespacing}


If one of the \textit{prima facie} difficulties faced by attempts to appeal to notions of information in approaching foundational questions in quantum mechanics is that of avoiding an unedifying descent into instrumentalism, then where else may we hope to make progress with the project? One obvious avenue for attack is to investigate whether ideas from quantum information theory might help provide a perspicuous conceptual basis for quantum mechanics, perhaps by leading us towards an enlightening axiomatisation of the theory. Certainly, strikingly different possibilities for information transfer and computation are to be found in quantum mechanics when compared with the classical case, and might these facts not help us characterize how and why quantum theory has to differ from classical physics? 

The thought that ideas from quantum information might lead us towards a transparent conceptual basis for quantum mechanics has been expressed perhaps most powerfully by Fuchs and co-workers \citep[cf.][]{fuchs:paulian}. In this chapter, we shall investigate two particular approaches in this vein, the Foundational Principle of Zeilinger; and the information-theoretic characterization theorem of Clifton, Bub and Halvorson.

\section{Zeilinger's Foundational Principle}

In Chapter~\ref{supposed conceptual inadequacy}, Brukner and Zeilinger's attempted criticism of the Shannon information in quantum mechanics was discussed, and as remarked there, what provides the background to this criticism is 
Zeilinger's (1999) proposal of an information-theoretic foundational principle for quantum mechanics.

This foundational principle, Zeilinger suggests, is to play a role in quantum mechanics similar to that of the Principle of Relativity in Special Relativity, or to the Principle of Equivalence in General Relativity. Like these, the Foundational Principle is to be an intuitively understandable principle which plays a key r\^{o}le in deriving the structure of the theory. In particular, he suggests that the Foundational Principle provides an explanation for the irreducible randomness in quantum measurement and for the phenomenon of entanglement. We will begin by considering whether the Principle can indeed be successful as a foundational principle for quantum mechanics, before suggesting why advocacy of the Principle might be thought to require or 
motivate arguments against the Shannon information.

Before stating the Foundational Principle, it is helpful to identify two philosophical assumptions that  Zeilinger's position incorporates. The first is a form of phenomenalism: physical objects are taken not to exist in and of themselves, but to be mere constructs relating sense impressions \citep[p.633]{foundationalprinciple}\footnote{Here I take phenomenalism to be the doctrine that the subject matter of all conceivable propositions are one's own actual or possible experiences, or the actual and possible experiences of another.}; the second assumption is an explicit instrumentalism about the quantum state:
\begin{quoting}
The initial state...represents all our information as obtained by earlier observation...[the time evolved] state is just a short-hand way of representing the outcomes of all possible future observations. \citep[p.634]{foundationalprinciple}
\end{quoting}
With these assumptions noted, let us consider the two distinct formulations of the Principle presented in Zeilinger (1999):
\newtheorem{FP}{FP}
\begin{FP}\label{FP1}
An elementary system represents the truth value of one proposition.
\end{FP}
\begin{FP}\label{FP2}
An elementary system carries one bit of information.
\end{FP}

At first glance, these two statements appear most naturally to be concerned with the amount of information that can be encoded into a physical system. However, this interpretation is at odds with the passage in which Zeilinger motivates the Foundational Principle. In this passage, his concern is with the number of propositions required to \textit{describe} a system. He considers the analysis of a composite system into constituent parts and remarks that it is natural to assume that each constituent system will require fewer propositions for its description than the composite does. The end point of the analysis will be reached when we have systems described by a single proposition only; and it is these systems that are termed `elementary'. 

The apparent tension between these different ideas of how FP1 and 2 should be read is relieved when Zeilinger goes on to explain what he means by an elementary system carrying or representing some information:
\begin{quoting}
...that a system ``represents" the truth value of a proposition or that it ``carries" one bit of information only implies a statement concerning what can be said about possible measurement results. \citep[p.635]{foundationalprinciple}
\end{quoting}
Thus the Foundational Principle is not a constraint on how much information can be encoded into a physical system. It is a constraint on how much the state of an elementary system can say about the results of measurement. This interpretation is rendered consistent with the discussion in terms of the propositions required to describe a system, as from Zeilinger's instrumentalist point of view, describing (the state of) a quantum system can only be to make a claim about future possible measurement results. Furthermore, we can understand the peculiar idiom of a system `representing' some information, where this is taken not to refer to the encoding of some information into a system, when we recall that from the point of view of Zeilinger's phenomenalism, a physical system is not an actual thing. On his view, a system represents a quantity of information about measurement results because a physical system literally \textit{is} nothing more than an agglomeration of actual and possible sense impressions arising from observations.

In short, however, it seems that a clearer, and perhaps more philosophically neutral, statement of the Foundational Principle would be the following:
\begin{FP}\label{FP3}
The state of an elementary system specifies the answer to a single yes/no experimental question,
\end{FP}
where we have used the fact that by `proposition' Zeilinger means something that represents an experimental question. With this relatively clear statement of the Foundational Principle in hand, let us now consider its claims as a foundational principle for quantum mechanics. 

To begin with, we should note the limitations implied by Zeilinger's conception of the description of a system. It might not always be the case that the state of an individual system can be characterised appropriately as a list of experimental questions to which answers are specified; and in such a case, the terms of the Foundational Principle cannot be set up. Consider the de Broglie-Bohm theory, for example, with its elements of holism and contextuality---even though the theory is deterministic, the results of measurements are in general not determined by the properties of the object system alone but are the result of interaction between object system and measuring device. It would seem that this theory could neither be supported nor ruled out by the Foundational Principle, as we can neither identify something that would count as an elementary system in this theory, given the way `elementary system' has been defined, nor, \textit{a fortiori}, begin to enumerate how many experimental questions such an entity might specify. However, for present purposes, let us put this sort of worry to one side. 

Another concern arises when considering the distinction we have drawn between describing a system and encoding information into it. Unlike encoding, the notion of describing a system presupposes a certain language in which the description is made, and the description of a given system could be longer or shorter depending on the conceptual resources of the language used. If we are to make a claim about the number of propositions required to describe a system, then, as we must when identifying an elementary system to figure in the Foundational Principle, we must already have made a choice of the set of concepts with which to describe the system. But this is worrying if the purpose of the Foundational Principle is to serve as a basis from which the structure of our theory is to be derived. If we already have to make substantial assumptions about the correct terms in which the objects of the theory are to be described, then it may be that the Foundational Principle will be debarred from serving its foundational purpose.
With this worry in mind, let us now consider the first of the concrete claims for the Foundational Principle, that it explains the irreducible randomness of quantum measurements. 

Zeilinger's suggestion is that we have randomness in quantum mechanics because:
\begin{quoting}
...an elementary system cannot carry enough information to provide definite answers to all questions that could be asked experimentally \citep[p.636]{foundationalprinciple},
\end{quoting} 
and this randomness must be irreducible, because if it were reduced to hidden properties, then the system would carry more than one bit of information. Unfortunately, this does not constitute an explanation of randomness, even if we have granted the existence of elementary systems and adopted the Foundational Principle. For the following question remains: why is it that experimental questions exist whose outcome is not already determined by a specification of the finest grained state description we can offer? How is it that any space for randomness remains? Or again, why isn't one bit enough?

The point is, it has not been explained \textit{why} the state of an elementary system cannot specify an answer to all experimental questions: this does not in fact appear to follow from the Foundational Principle. 
The Foundational Principle says nothing about the structure of the set of experimental questions, yet this turns out to be all-important.

Consider the case of a classical Ising model spin, which has only two possible states, `up' or `down'; here one bit, the specification of an answer to a single experimental question (`Is it up?') \textit{is} enough to specify an answer to all questions that could be asked. There is no space for randomness here, yet this classical case is perfectly consistent with the Foundational Principle. Thus it seems that no explanation of randomness is forthcoming from the Foundational Principle and furthermore, it is far from clear that the Principle, on its own, in fact allows us to distinguish between quantum and classical. 

Of course, if one assumes that experimental questions are represented in the quantum way, as projectors on a complex Hilbert space, then even for the simplest non-trivial state space, there will be non-equivalent experimental questions, the answer to one of which will not provide an answer to another; but we cannot assume this structure if it is the very structure that we are trying to derive. It appears from the way in which the Foundational Principle is supposed to be functioning in the attempted explanation of randomness, that something like the quantum structure of propositions is being assumed. But this is clearly fatal to the prospects of the Foundational Principle as a foundational principle.\footnote{In a sense, we could say that Zeilinger's explanation of randomness is problematic because it fails to explain why the state space of quantum mechanics is so gratuitously large from the point of view of storing information \citep{fuchs:howmuch}. It is then striking that this attempted information-theoretic foundational approach to quantum mechanics has not allowed for one of the significant insights vouchsafed by quantum information theory.} 

Does the Principle fare any better with the proposed explanation of entanglement? The idea here is to consider $N$ elementary systems, which, following from the Foundational Principle, will have $N$ bits of information associated with them. The suggestion is that entanglement results when all $N$ bits are exhausted in specifying joint properties of the system, leaving none for individual subsystems \citep{foundationalprinciple}, or more generally, when more information is used up in specifying joint properties than would be possible classically. The underlying thought is that this approach captures the intuitive idea that when we have an entangled system, we know more about the joint system (which may be in a pure state) than we do about the individual sub-systems (which must be mixed states). 
The proposal is further developed in \citet{essence}, where Brukner and Zeilinger's information measure is used to provide a quantitative condition for $N$ qubits to be unentangled, which is then related to a condition for the violation of a certain $N$-party Bell inequality. 

To give a basic example of how the idea is supposed to work, consider the case of two qubits. Recall that the maximally entangled Bell states are joint eigenstates of the observables $\sigma_{x}\otimes\sigma_{x}$ and $\sigma_{y}\otimes\sigma_{y}$. From the Foundational Principle, only two bits of information are associated with our two systems, i.e., the states of these systems can specify the answer to two experimental questions only. If the two questions whose answers are specified are `Are both spins in the same direction along $x$?' ($1/2(\mathbf{1}\otimes\mathbf{1} + \sigma_{x}\otimes\sigma_{x})$) and `Are both spins in the same direction along $y$?' ($1/2(\mathbf{1}\otimes\mathbf{1} + \sigma_{y}\otimes\sigma_{y})$), then we end up with a maximally entangled state. If, by contrast, the two questions had been `Are both spins in the same direction along $x$?' and `Is the spin of particle 1 up along $x$?', the information would not have all been used up specifying \textit{joint} properties and we would have instead a product state (joint eigenstate of $\sigma_{x}\otimes\sigma_{x}$ and $\sigma_{x}\otimes \mathbf{1}$).    

Now, although this idea may have its attractions when used as a criterion for entanglement within quantum mechanics, it does not succeed in providing an explanation for the phenomenon of entanglement, which was the original claim.

If we return to the starting point and consider our $N$ elementary systems, all that the Foundational Principle tells us  regarding these systems is that their individual states specify the answer to a single yes/no question concerning each system individually. There is, as yet, no suggestion of how this relates to joint properties of the combined system. Some assumption needs to be made before we can go further. For instance, we need to enquire whether there are supposed to be experimental questions regarding the joint system which can be posed and answered that are not equivalent to questions and answers for the systems taken individually. (We know that this will be the case, given the structure of quantum mechanics, but we are not allowed to \textit{assume} this structure, if we are engaged in a foundational project.\footnote{To illustrate, a simultaneous truth value assignment for the experiments $\sigma_{x}\otimes\sigma_{x}$ and $\sigma_{y}\otimes\sigma_{y}$ cannot be reduced to one for experiments of the form $\mathbf{1}\otimes\mathbf{a}.\boldsymbol{\sigma}, \mathbf{b}.\boldsymbol{\sigma}\otimes\mathbf{1}$.}) If this \textit{is} the case then there can be a difference in the information associated with correlations (i.e., regarding answers to questions about joint properties) and the information regarding individual properties. But then we need to ask: why is it that there exist sets of experimental questions to which the assignment of truth values is not equivalent to an assignment of truth values to experimental questions regarding individual systems? 

Because such sets of questions exist, more information can be `in the correlations' than in individual properties. Stating that there is more information in correlations than in individual properties is then to report that such sets of non-equivalent questions exist, \textit{but it does not explain why they do so}. However, it is surely this that demands explanation---why is it not simply the case that all truth value assignments to experimental questions are reducible to truth value assignments to experimental questions regarding individual properties, as they are in the classical case? That is, why does entanglement exist? In the absence of an answer to the question when posed in this manner, the suggested explanation following from the Foundational Principle seems dangerously close to the vacuous claim that entanglement results when the quantum state of the joint system is not a separable state. 

Of course, if we are in the business of looking within quantum mechanics and asking how product and entangled states differ, then it is indeed legitimate to consider something like the condition \citet{essence} propose; and we can then consider how good this condition is as a criterion for entanglement\footnote{At this point it is worth noting that there have been other discussions of entanglement which develop the intuitive idea that when faced with entangled states, we know more about joint properties than individual properties. As we saw in Section~\ref{entanglement majorization condition}, a very general framework is presented by \citet{separabledisorder}, who use the majorization relation to compare the spectra of the global and reduced states of the system; a necessary (but not sufficient) condition for a state to be separable is then that it be more disordered globally than locally.}. But as mentioned before, if we are trying to explain the existence of entanglement then we cannot simply assume the quantum mechanical structure of experimental questions. 

Let us close by considering a final striking passage. Zeilinger suggests that the Foundational Principle might provide an answer to Wheeler's question `Why the quantum?' \citep{wheeler} in a way congenial to the Bohrian intuition that the structure of quantum theory is a consequence of limitations on what can be said about the world:
\begin{quoting}
The most fundamental viewpoint here is that the quantum is a consequence of what can be said about the world. Since what can be said has to be expressed in propositions and since the most elementary statement is a single proposition, quantization follows if the most elementary system represents just a single proposition. \citep[p.642]{foundationalprinciple}
\end{quoting}
But this passage contains a crucial non-sequitur. Quantization only follows if the propositions are projection operators on a complex Hilbert space. And why is it that the world has to be described that way? \textit{That} is the question that would need to be answered in answering Wheeler's question; and it is a question which, I have suggested, the Foundational Principle goes no way towards answering. 

\subsection{Word and world: Semantic ascent}\label{ascent}

At this juncture let us pause to consider the following paranthetical, but perhaps illuminating, remarks. 

The sentiment expressed in the last quotation of Zeilinger is evidently very close to that captured by the famous (or infamous) statement attributed to Bohr by Petersen:
\begin{quoting}
There is no quantum world. There is only an abstract quantum physical description. It is wrong to think that the task of physics is to find out how nature \textit{is}. Physics concerns what we can say about nature. \citep[p.12]{petersen:1963}
\end{quoting}
The last sentence is particularly pertinent: `Physics concerns what we can say about nature.' Compare again, another statement of Zeilinger's, `...what can be said about Nature has a constitutive contribution on what can be ``real".' (Reported in \citet[p.615]{fuchs:paulian}).

These views clearly pick out one strand of thought that can be seen to contribute to the wider speculative thesis that information may, in some sense, provide a new way in physics. If quantum mechanics reveals that the true subject matter of physics is what can be said, rather than how things are, it seems but a small step from there to the view that what is fundamental is the play of information.     

However, there is a very obvious difficulty with the thought that what can be said provides a consitutive contribution to what can be real and that physics correspondingly concerns what we can say about nature. Simply reflect that some explanation needs to be given of where the relevant constraints on what can be said come from. Surely there could be no other source for these constraints than the way the world actually is---it can't \textit{merely} be a matter of language\footnote{Of course, what statements can be made depends on what concepts we possess; and, trivially, in order to succeed in making a statement, one needs to obey the appropriate linguistic rules. But the point at issue is what can make one set of concepts more fit for our scientific theorising than another? For example, why do we have to replace commuting classical physical quantities with non-commuting quantum observables? As Quine perspicuously notes `...truth in general depends on both language and extra-linguistic fact. The statement ``Brutus killed Caesar" would be false if the world had been different in certain ways, but it would also be false if the word ``killed" happened to have the sense of ``begat".'\citep[p.36]{quine:1953}. The world is required to provide the extra-linguistic component that will make one set of concepts more useful than another; furthermore, without an extra-lingustic component to truth, we could only ever have analytic truths---and that would no longer be physics.}. It is because of the unbending nature of the world that we find the need to move, for example, from classical to quantum physics; that we find the need to revise our theories in the face of recalcitrant experience.
Zeilinger and Bohr (in the quotation above) would thus seem to be putting the cart before the horse, to at least some degree. Schematically, it's the way the world is (independently of our attempted description or systematisation of it) that determines what can usefully be said about it, and that ultimately determines what sets of concepts will prove most appropriate in our scientific theorising.

Another point can be drawn from the Petersen quotation. With its focus on the level of physical \textit{description} and what can be \textit{said} about nature (as opposed to how nature is) this passage can be seen to provide us with an example of what is often known as \textit{semantic ascent}. 

Semantic ascent is the move from what Carnap called the material mode to the formal mode, that is, roughly speaking, from talking about things to talking about words. As Quine says, `\textit{semantic ascent}...is the shift from talking in certain terms to talking about them' \citep[p.271]{quine:1960}. 
Bohr, it would seem, would have us ascend from the level of using words within our theory, to the level of describing our descriptions. This, the suggestion is, is the true task of physics.

What would such an ascent achieve? As Quine is quick to note, semantic ascent doesn't bake much ontological bread:
\begin{quoting}
Semantic ascent...applies anywhere. `There are wombats in Tasmania' might be paraphrased as ` ``Wombat" is true of some creatures in Tasmania', if there were any point in it. \citep[p.272]{quine:1960}
\end{quoting}

The point is this. It's true, but trivial, that if we ascend to a level at which we are describing what we say about nature, that is, take the physical description as our focus of interest, then our subject matter will no longer be the world, for we have moved from talking \textit{in} various terms to talking \textit{about} them. At this level there will, in a sense, be no quantum world, for we are talking about \textit{words} and not the world.

But the fact that we have ascended doesn't mean that the level we have ascended from goes away. The world doesn't disappear because we may be talking about the terms in which we describe it. It follows that one can't shirk the difficulties and mysteries of interpreting quantum mechanics by simply saying: `Physics concerns what we can say about nature,' for, crucially, we can always ask---well, what is said? (descent after our semantic ascent), as well as---how do we say it? (remaining at the ascended level).

The fact that one can always make a semantic ascent does not mean that one can do without the level from which ascent has been made\footnote{It might be felt, perhaps, that this is the real import of the Bohr quote, and serves to distinguish the quantum from the classical case: in the quantum case, we might be supposed to imagine that one \textit{can} intelligibly kick away the lower level, having made the semantic ascent. Such a suggestion (`vertiginous semantic ascent', as it might be called) is incoherent however. It would amount to the claim that the `descent' question `So: what was said?' becomes unintelligible, but this would entail that the terms under discussion have to become entirely devoid of meaning, and as such they would have no role whatsoever in physics.}. Indeed, the interesting interpretational questions concern why one should take one stance rather than another to claims made using terms within a theory, and the usual ranges of options (various forms of realism, instrumentalism and hybrids thereof) will remain open irrespective of ascent. It is important to realise that the semantic ascent of the Bohrian quote doesn't succeed in highlighting any differences between the classical world view and quantum mechanics. In so far as `there is no quantum world' is true in the Petersen quotation, it would be true of the classical world too: it is a universal and entirely innocuous observation that if we ascend to the level at which we are describing our physical-theory discourse, then our subject matter will be words rather than world.


The `There is no quantum world' passage is apt to induce apoplexy in the realist-minded, but there seems after all no call for raised blood-pressures. When analysed as an example of semantic ascent, it seems that the passage is---so far as it is intelligible---somewhat innocuous in import.


\subsection{Shannon information and the Foundational Principle}

To finish the story, let us consider how Zeilinger's approach based on the Foundational Principle interplays with the discussion of measures of information in quantum mechanics.

As we have noted, Zeilinger adopts an instrumentalist view of the quantum state, and such instrumentalist sentiments are common. Where Brukner and Zeilinger depart from the norm, however, is in adopting a very literal construal of the information taken to constitute the state, by adopting, at least in a simple case, the Hilbert-Schmidt representation of states:
\begin{quoting}
We describe a photon by a catalog of information (``information vector") $\vec{i}=(i_{1},i_{2})$ about mutually complementary propositions $\{{\cal P}_{1},{\cal P}_{2}\}$. Such propositions are, for example, ${\cal P}_{1}$: ``the polarization of the photon is vertical (horizontal)" \citep{malus}
\end{quoting} 
The component $i_{1}$ is defined as $(p-q)$, where $p$ and $q$ are the probabilities for vertical and horizontal polarization respectively\footnote{For this two-dimensional quantum system, we have here, essentially, the Bloch sphere representation.}. Thus, the components of the information vector $\vec{i}$ correspond, effectively, to the coefficients $q_{i}^{j}$ in eqn.~(\ref{rho}), and the propositions ${\cal P}$ to the operators $\bar{P}_{i}^{j}$.

On this conception, an amount of information in the form of probabilities has been associated to propositions representing the outcomes of mutually unbiased measurements; the information and the experimental propositions it is about can be read off directly from the Hilbert-Schmidt representation of the state, given some choice of basis operators (choice of complete set of mutually unbiased measurements). Illustrating the general idea, if probability 1 is associated to some proposition, then the state says the maximum possible about the outcome of the measurement with which that proposition is associated; if there is a flat distribution for outcomes of a measurement, the state contains no information about it. In general the state will contain \textit{partial} information about a number of mutually unbiased observables. Endorsing the instrumentalist line, \textit{all} that the state is is an amount of information in this way about mutually complementary observables. 

Now the statements FP1 and FP2 refer to an elementary system carrying or representing an amount of information.
As we have seen, rather than being a putative restriction on how much information might be encoded into, or read from, a physical system, the Foundational Principle is intended by Zeilinger to capture a restriction on how much can be said about measurement outcomes, and hence, in particular, is a restriction on how much the \textit{state} can say about measurement outcomes. 

For Zeilinger, the state will in general be constituted by amounts of partial information about measurement outcomes. The Foundational Principle requires that the state can only contain a limited amount of information, namely one bit; hence it follows that the amounts of partial information contained in the state, although how these are to be quantified has not yet been specified in detail, must add up to one bit's worth in total.

This, however, rules out the Shannon information as the measure of the amount `carried' by the state about a given measurement; we know that in general we will not have a sum to unity for amounts of partial information conceived in the way outlined (Section~\ref{finalargument}). (As $H(\vec{p})$ does not sum to a unitarily invariant quantity for a complete set of mutually unbiased measurements, we cannot guarantee that we will attain the value of one for any given pure state.)

Thus the conjunction of the Foundational Principle with Brukner and Zeilinger's brand of literal instrumentalism about the quantum state is inconsistent with adopting the Shannon information to measure the amount of information `carried' about a measurement in Zeilinger's sense. I suggest that it is this fact that tempts Brukner and Zeilinger to argue, unsuccessfully as it turns out, that the Shannon information is not the correct measure of information and cannot be applied in quantum mechanics.

We may close this discussion with two final comments. First, consider what someone rather more realist about the quantum state might make of the Foundational Principle. Here the information idiom would no longer be particularly enticing and a more precise statement of what is being expressed by the Foundational Principle in quantum mechanics would be natural:
\begin{quote}
\textbf{`R'\,FP)} \textit{Any projective measurement other than in the eigenbasis of $\rho$ results in a shorter vector in $V_{h}(\mathbb{C}^{n})$.}
\end{quote}
(`R'\,FP for `realist' Foundational Principle.) That is, any such measurement would result in a more spread probability distribution; if we began with a pure state then post- (non-selective) measurement, the ensemble will no longer be represented by a one-dimensional projector. Given this statement of the Principle, we see that it is a matter of choice whether or not, or with which quantities, we chose to discuss the uncertainties associated with the probability distributions generated by the state. (Recall that in the Hilbert-Schmidt representation of a density operator, the probability for a measurement outcome is simply the projection of the vector representing the state onto the vector representing the measurement outcome.)

Second, we might wonder whether the foregoing argument indicates that for the instrumentalist at least, $I(\vec{p})$ does after all represent the `correct' measure of information in quantum mechanics. Such a choice would appear very artificial given the close relation between the functioning of $I(\vec{p})$ and $H(\vec{p})$ discussed in Section~\ref{Hilbert-Schmidt}. Note, however, that one could still be an instrumentalist about the quantum state while adopting `R'FP as more genuinely informative than FP1 and FP2. The instrumentalist is not, then, forced to accept $I(\vec{p})$ as the only correct measure of information in quantum mechanics. This is reassuring, for as pointed out earlier, the choice of measure of information is a matter of convention and convenience, a mere matter of whether the quantity in question suits the task to hand. 

We saw in Chapter~\ref{supposed conceptual inadequacy} that Brukner and Zeilinger's arguments against the applicability of the Shannon information in quantum mechanics are unsuccessful; and we have now seen how these arguments would seem to be motivated by the conjunction of Zeilinger's Foundational Principle with a particular form of instrumentalism about the quantum state. Even if one has instrumentalist leanings, however, this does not imply that the Brukner-Zeilinger measure can be the only correct measure of information in quantum mechanics. 

\section{The Clifton-Bub-Halvorson characterization theorem}\label{cbh}

I have argued that Zeilinger's Foundational Principle does not constitute a principle from which we may derive the structure of quantum mechanics, nor which allows us to understand the origins of entanglement and quantum randomness. In essence, it is silent about the structure of the set of experimental questions, yet it is this that turns out to be crucial. The next approach we shall consider, that of  Clifton, Bub and Halvorson \citep{cbh}, provides a far happier conclusion. Their project of characterizing quantum mechanics in terms of three information-theoretic constraints is indeed successful, although it may be questioned whether all three are striclty necessary. I shall outline the approach here, before moving on to raise some questions concerning the initial assumption of a $C^{*}$-algebraic starting point; and then consider in what sense their axiomatic approach may be said to provide an information-theoretic \textit{interpretation} of quantum mechanics, or to motivate such an interpretation.

\subsection{The setting}

Proceeding within a $C^{*}$-algebraic framework, Clifton, Bub and Halvorson \citep{cbh,hans:generalization} succeed in characterizing quantum theory in terms of three information-theoretic constraints. We shall call this the CBH characterization theorem.

The constraints are these:

\begin{singlespacing}
\begin{enumerate}
\item No superluminal information transmission between two systems by measurement on one of them;
\item no broadcasting;
\item no unconditionally secure bit-commitment.
\end{enumerate} 
\end{singlespacing}
Let us briefly review these various terms.

First, the setting is to assume a $C^*$-algebraic characterization of physical theories (for a friendly introduction to this formalism, see for example \citet{gudder:1977}). A $C^*$-algebra is an involutive Banach algebra $\mathcal{B}$ over the complex numbers satisfying $\|A^{*}A\|=\|A\|^{2}$ for every $A\in \mathcal{B}$.

Some definitions: A \textit{complex algebra} is a complex vetor space with an identity and an associative, distributive product, $AB$. An \textit{involution} on a complex algebra $\mathcal{B}$ is a map $^{*}: \mathcal{B}\mapsto\mathcal{B}$, satisfying:
\[(A^{*})^{*}=A,\;\; (A+B)^{*}=A^{*}+B^{*},\;\; (\lambda A)^{*}=\lambda^{*}A^{*},\;\; (AB)^{*}=B^{*}A^{*},\;\; \forall A,B\in\mathcal{B}.\]  
A \textit{Banach algebra} is an algebra equipped with a norm such that $\|AB\|\leq \|A\|\|B\|$, complete in the norm topology. 

An element of $\mathcal{B}$ is \textit{self-adjoint} if $A^{*}=A$. A familiar example of a $C^{*}$-algebra is given by the set $\mathcal{B}(\mathcal{H})$ of bounded linear operators on a Hilbert space $\mathcal{H}$, where the involution operation $^{*}$ is the familiar adjoint $^{\dagger}$. The self-adjoint elements of a $C^{*}$-algebra are usually interpreted as observables. 

In a $C^{*}$-algebra, a \textit{state}, $\omega$, is a linear functional on the $C^{*}$-algebra that is i) \textit{positive}, $\omega(AA^{*})\geq 0$, and ii) \textit{normalized}, $\omega(\mathbf{1})=1$. The state is to be understood as ascribing expectation values to the elements of the algebra corresponding to observable quantities.

In this framework, the schematic picture of a physical theory involves `black box' preparation and measuring devices. A fixed preparation procedure in the lab will give rise to certain observed average values for measurements using a range of devices; systems prepared in this way will correspondingly be assigned a particular state, $\omega$. The measuring devices themselves are associated with elements of the algebra corresponding to observable quantities: we can imagine black boxes in the lab with the letters `$A$', `$B$', `$C$' and so on, inscribed on their surfaces, where $A$, $B$...are self-adjoint elements of a $C^{*}$-algebra.  

Finally, \citet{cbh} assume the most general form of dynamical evolution, \textit{viz.}, non-trace increasing completely positive maps (cf. Section~\ref{entanglement witnesses and the PPT condition}). 

By `a quantum theory', Clifton, Bub and Halvorson mean a theory formulated in $C^{*}$-algebraic terms for which the algebras of observables pertaining to distinct systems commute, for which the algebra of observables on an \textit{individual} system is non-commutative, and which allows space-like separated systems to be in entangled states. Roughly speaking, these  characteristics are associated respectively with the first, second and third information-theoretic constraints. Now, while there is clearly much, much more to quantum theory than these rather abstract algebraic features, it is nonetheless plausible to argue that together they do capture the distinctive structural features of the theory.

It is of course an important pre-supposition of the general argument that the $C^{*}$-algebraic approach be a sufficiently general one, and Clifton, Bub and Halvorson argue accordingly, e.g.:
\begin{quoting}
...it might seem that $C^{*}$-algebras offer no more than an abstract way of talking about quantum mechanics. In fact, the $C^{*}$-algebraic formalism provides a mathematically abstract characterization of a broad class of physical theories that includes all classical mechanical particle and field theories, as well as quantum mechanical theories. \citep[p.245]{bub:why}
\end{quoting}
Thus, as well as reflecting that the set of bounded operators on a Hilbert space is a $C^{*}$-algebra, and that via the Gelfand, Naimark and Segal (GNS) construction and the Gelfand-Naimark theorem, we know that every abstract $C^{*}$-algebra has a concrete faithful (i.e., isomorphic) representation as a $^{*}$-subalgebra of the bounded operators on some appropriate Hilbert space $\mathcal{H}$ \citep[cf.][]{cbh}, it is pertinent to point out that classical phase space theories may be formulated in $C^{*}$-algebraic terms, and moreover to note that it may be shown that every \textit{commutative} $C^{*}$-algebra may be given a phase space representation \citep[cf.][]{cbh,bub:why}. However, as we shall shortly see, some questions can nevertheless be raised about whether the starting assumption of a $C^{*}$-algebraic framework may perhaps be overly strong.   

Turning now to the constraints featuring in the characterization theorem. The very first, and a non-information theoretic one, is a constraint not yet mentioned, intended to capture the idea that if we have two sub-algebras $\mathcal{A}$ and $\mathcal{B}$ of a $C^{*}$-algebra $\mathcal{C}$, whose self-adjoint elements are to represent, respectively, the observables of two distinct systems A and B, then we need to ensure that $\mathcal{A}$ and $\mathcal{B}$ are distinct objects. \citet{cbh} adopt the notion of $C^{*}$-independence to this end, the criterion being that the preparation of any state of $\mathcal{A}$ has to be compatible with preparation of any state of $\mathcal{B}$. That is, for any state $\rho_{1}$ of $\mathcal{A}$ and for any state $\rho_{2}$ of $\mathcal{B}$, there is some joint state $\rho$ of the joint algebra $\mathcal{A}\vee\mathcal{B}$ such that $\rho|_{\mathcal{A}}=\rho_{1}$ and $\rho|_{\mathcal{B}}=\rho_{2}$. (The significance  of requiring a notion of independence of this sort is elaborated in \citet{halvorsonbub:smolin}.) 

The first of the information-theoretic constraints, no superluminal signalling via measurement, is fairly self-explanatory, corresponding to the no-signalling via entanglement feature in ordinary quantum mechanics. The requirement is that the state of system B, say, should be unaffected by any (non-selective) operation performed on the other system. \citet{cbh} show that this will hold \textit{iff} the algebras $\mathcal{A}$ and $\mathcal{B}$ commute (\textit{kinematic independence}).  

The property of \textit{no broadcasting}, the second of the three constraints, is a generalisation of the idea of no cloning appropriate to mixed states \citep{nobroadcasting}. The requirement on a cloning device was that it take as an input a system in any arbitrary state $\ket{\alpha}{}$ and return two systems, each in the state $\ket{\alpha}{}$. Now, one might consider instead a process which takes as an input a system in a state $\rho$ and returns as an output a pair of systems A and B with a joint state $\tilde{\rho}^{\text{AB}}$, which may not be equal to $\rho\otimes\rho$, but for which the reduced states of A and B \textit{are} equal to $\rho$, $\mathrm{Tr}_{\mathrm{B}}\tilde{\rho}^{\text{AB}}= \mathrm{Tr}_{\mathrm{A}}\tilde{\rho}^{\text{AB}}= \rho$. Such a process is termed \textit{broadcasting}. (Clearly, it represents a more general process only when the input state is mixed; for pure states it reduces to cloning.) \citet{nobroadcasting} showed that in quantum mechanics, broadcasting is possible for a set of states $\rho_{i}$ \textit{iff} they are commuting.

\citet{cbh} first generalise the notion of broadcasting to the setting of $C^{*}$-algebraic states, and then prove that if $\mathcal{A}$ and $\mathcal{B}$ are abelian, then there is an operation on $\mathcal{A}\vee\mathcal{B}$ that broadcasts all states of $\mathcal{A}$, while, conversely, if for each pair $\{\rho_{0},\rho_{1}\}$ of states of $\mathcal{A}$, there is an operation on $\mathcal{A}\vee\mathcal{B}$ that may broadcast this pair, then $\mathcal{A}$ is abelian.   

So, thus far it has been proved that for a $C^{*}$-algebraic theory, if it satisfies no-signalling and no-broadcasting, it must have algebras of observables that are non-commuting for individual systems, while observables for distinct systems commute; and conversely.

The third information-theoretic constraint---no bit-commitment---takes a little more explaining. A bit-commitment protocol is an information-theoretic protocol in which one party, Alice, provides another party, Bob, with an encoded bit value (0 or 1) in such a way that Bob may not determine the value of the bit unless Alice provides him with further information at a later stage (the `revelation' stage), yet in which the information that Alice initially gives to Bob is nonetheless sufficient for him to be sure that the bit value he obtains following revelation is indeed the one that Alice commited to initially. An illustrative analogy would be a case in which Alice chooses a bit value and writes it on a piece of paper. She then locks the piece of paper in a safe and delivers the safe to Bob, but keeps the key to the safe herself. Bob may not immediately determine the value of the bit as the paper is locked in the safe, but he does know that when Alice later gives him the key, the bit value he will learn after opening the safe and reading the paper is indeed the one that Alice wrote down earlier. An insecure bit-commitment protocol is one in which either party can cheat: Bob, by determining something about the encoded bit value prior to revelation, or Alice, by remaining free to reveal either bit value at will at the revelation stage.   

Bit-commitment is not unconditionally secure classically because the encrypted information that Alice initially provides to Bob will always display some bias towards the encoded bit value that will allow Bob to cheat. It was shown by \citet{lochau} and \citet{mayers} that bit-commitment is not secure in the quantum mechanical case either, but importantly, for a very different reason.

In ordinary quantum mechanics we are familiar with the idea of the ambiguity of density operators: quite different preparation procedures may give rise to the same density operator, and one will not be able to determine which preparation procedure was used by performing measurements on the systems prepared. This seems to suggest a way in which quantum bit-commitment might be possible. If Alice were to associate her commitment with two different preparations of a given density matrix, then Bob would not be able to determine anything about the bit value thus encoded; if Alice later tells him the preparation procedure she used, then we might be able to arrange things so that Bob can check that she is true to her word in having previously commited to a specific bit value.

An example might go like this. Consider a spin-1/2 system: a 50/50 mixture of spin-up and spin-down in the $z$-direction is indistinguishable from a 50/50 mixture of spin-up and spin-down in the $x$-direction---both give rise to the maximally mixed density operator $\frac{1}{2}\mathbf{1}$. Alice might associate the first type of preparation with a 0 commitment and the second with a 1 commitment. Bob, when presented with a system thus prepared will not be able to determine which procedure was used. Alice also needs to keep a record of which preparation procedure she employed, though, to form part of the evidence with which she will convince Bob of her probity at the revelation stage. Thus, for a 0 commitment, Alice could prepare a classically correlated state of the form:   
\[\text{0 commitment:}\;\;\;\;\; \rho^{12}_{0}=\frac{1}{2}\bigl(\ket{\up_{z}}{}\!\!\bra{}{\up_{z}}\otimes\ket{\up_{z}}{}\!\!\bra{}{\up_{z}} + \ket{\down_{z}}{}\!\!\bra{}{\down_{z}}\otimes\ket{\down_{z}}{}\!\!\bra{}{\down_{z}}\bigr),\]
whilst for a 1 commitment, she could prepare a state
\[\text{1 commitment:}\;\;\;\;\; \rho^{12}_{1}=\frac{1}{2}\bigl(\ket{\up_{x}}{}\!\!\bra{}{\up_{x}}\otimes\ket{\up_{x}}{}\!\!\bra{}{\up_{x}} + \ket{\down_{x}}{}\!\!\bra{}{\down_{x}}\otimes\ket{\down_{x}}{}\!\!\bra{}{\down_{x}}\bigr).\]
System 2 is then sent to Bob. 

At the revelation stage, Alice declares which bit value she commited to, and hence which preparation procedure she used. The protocol then proceeds in the following way: If she committed to 0, Alice and Bob both perform $\sigma_{z}$ measurements and Alice declares the result she obtains, which should be perfectly correlated with Bob's result, if she really did prepare state $\rho^{12}_{0}$. Similarly, if she committed to 1, Alice and Bob both perform $\sigma_{x}$ measurements and Alice declares her result, which again should be perfectly correlated with Bob's result, if in truth she did prepare state $\rho^{12}_{1}$. If the results reported by Alice and obtained by Bob don't correlate then Bob knows that Alice is trying to mislead him.

The trouble with this otherwise attractive protocol is that Alice is able to cheat freely by making use of what is known as an \textit{EPR cheating strategy}. Thus, rather than preparing one of the states $\rho^{12}_{0}$ or $\rho^{12}_{1}$ at the commitment stage, Alice can instead prepare an entangled state, such as the Bell state $\ket{\phi^{+}}{12}$. The reduced density operator for Bob's system will still be $\frac{1}{2}\mathbf{1}$, but Alice can now simply wait until the revelation stage to perform a suitable measurement on her half of the entangled pair and prepare Bob's system at a distance in whichever of the two different mixtures she chooses.    

It turns out that this sort of EPR cheating strategy will always be available for any quantum bit-commitment protocol (\citet{lochau, mayers}; see \citet{bub:commitment} for a detailed discussion): the possibility of preparing entangled states shared between Alice and Bob rules out unconditionally secure bit-commitment in quantum mechanics. The result in the general case relies upon the theorem of \citet{hjw}, prefigured in \citet{schrodinger:1936}, which tells us that for a bipartite quantum system, any mixture of states on one system may be prepared by performing a suitable measurement (which may involve an ancilla) on the other system, when the pair are in an appropriate entangled state (\textit{viz.}, one giving the correct reduced state for the first system). Following \citet{schrodinger1,schrodinger:1936} this phenomenon associated with entanglement is often called \textit{remote steering}.     

The intuitive r\^ole for the no bit-commitment axiom in an attempted information-theoretic characterization of quantum mechanics is then as follows. In quantum mechanics, the ambiguity of density operators seems to hold out the possibility of secure bit-commitment, but this possibility is vitiated by the fact that entanglement may exist between two widely separated parties. Now, we could consider a class of possible theories which were locally like quantum mechanics in that they allowed ambiguous mixtures to be prepared, yet in which entanglement between separated systems was ruled out, perhaps decaying over distance---such a theory was in fact entertained by \citet{schrodinger:1936} as a way of resolving the EPR dilemma. Call such a theory a \textit{Schr\"{o}dinger-type} theory. In a Schr\"{o}dinger-type theory, secure bit-commitment would be possible as the EPR cheating strategy, which relies on entangled states, cannot be employed. 
In order to cheat, we would need entanglement. 

But now suppose that in our attempted axiomatic characterisation we arrive at a class of theories which we know all to allow ambiguous mixtures. If we were then to add to our list of axioms the further requirement that bit-commitment should be impossible, then this would seem tantamount to picking out those theories that \textit{do} contain non-local entanglement, as, drawing on the analogy with the familiar quantum mechanical case, we might expect that entanglement is required to cheat. By insisting on no bit-commitment in our axioms, we rule out the Schr\"{o}dinger-type theories from our consideration.    

That is the intuitive idea. \citet{cbh} argue rigorously as follows. First they show that a $C^{*}$-algebra $\mathcal{A}$ is nonabelian \textit{iff} it allows ambiguous mixtures, i.e., distinct mixtures of pure states giving rise to the same mixed state. As in the spin 1/2 example given above, such mixtures may be used as the basis for Alice's bit commitment. They then prove that if Alice and Bob only have access to classically correlated states (convex combinations of product states), then the bit-commitment protocol based on these distinct mixtures will be secure: there is no classically correlated state that will allow Alice to change her commitment from 0 to 1 at the revelation stage. The contrapositive statement of this result is that if, for a theory in which the algebras of observables for individual systems are nonabelian, unconditionally secure bit-commitment is \textit{not} possible then \textit{entanglement between spatially separated systems must be allowed}. The converse, that for any quantum theory in the sense of \citet{cbh}, unconditionally secure bit-commitment is not possible, was proven by \citet{hans:generalization}.

The achievement of the CBH characterization theorem then, is, first of all, a formulation of the three information-theoretic constraints in the general setting of $C^{*}$-algebraic theories, followed by the main result of a characterization of quantum theory in terms of these three constraints: Any theory formulated in $C^{*}$-algebraic terms that satisfies the three information-theoretic constraints will take the form of a quantum theory; with a non-commuting algebra of observables for individual systems, kinematic independence for the algebras of space-like separated systems and the possibility of entanglement between space-like separated systems; while conversely, any $C^{*}$-algebraic theory with these distinctively quantum properties will satisfy the three information-theoretic constraints.   

How much light does this result shed on the nature or origin of quantum mechanics? Clifton, Bub and Halvorson suggest that
\begin{quoting}
The fact that one can characterize quantum theory...in terms of just a few simple information-theoretic principles...lends credence to the idea that an information-theoretic point of view is the right perspective to adopt in relation to quantum theory. \citep[p.4]{cbh}
\end{quoting}
Certainly, the CBH characterization theorem indicates that concentrating on some information-theoretic principles has proven fruitful in providing a novel axiomatisation of the theory, but is something more than this intended by the statement that `an information-theoretic point of view is the right perspective to adopt'? In particular, does the CBH characterization shed light on how we should understand the quantum formalism more broadly? Above all, does it have implications for the traditional interpretive questions in quantum mechanics; for the knotty problems of the meaning of the formalism? \citet{cbh} seem to suggest so:
\begin{quoting} 
We...suggest substituting for the conceptually problematic mechanical perspective on quantum theory an information-theoretic perspective. That is, we are suggesting that quantum theory be viewed, not as first and foremost a mechanical theory of waves and particles...but as a theory about the possibilities and impossibilities of information transfer. \citep[p.4]{cbh}
\end{quoting}
The thought is pursued further by \citet{bub:why}:
\begin{quoting}
Assuming the information-theoretic constraints are in fact satisfied in our world, no mechanical theory of quantum phenomena that includes an account of measurement interactions can be acceptable, \textit{and the appropriate aim of physics at the fundamental level becomes the representation and manipulation of information}. \citet[p.242]{bub:why}, {\small my emphasis}.
\end{quoting} 
We shall return presently to the question of the interpretational implications of the CBH characterization. First, let us consider some points relating to the $C^{*}$-algebraic starting point of the theorem.

\subsection{Some queries regarding the $C^{*}$-algebraic starting point}

It is of course evident that any axiomatic characterization of a physical theory has to start from somewhere, and as mentioned above, Clifton, Bub and Halvorson suggest that adopting a $C^{*}$-algebraic framework is an appropriately  neutral starting point. However, some questions can be raised about the strength of this starting assumption.

For some, the very fact that $C^{*}$-algebras make use of a \textit{complex} vector space, as opposed, say, to a real or quaternionic one, may already be to assume too much\footnote{Cf. \citet[p.5]{fuchs:2001}, for example; this complaint is noted in \citet{bub:why,hans:2004}.}. A second sort of worry is raised by the existence of various `toy-theories' that satisfy the three information-theoretic constraints of the CBH characterization theorem and yet are palpably \textit{not} quantum mechanics \citep{spekkens,smolin}. These toy theories are not counter-examples in the logical sense to the CBH theorem, as they fail to satisfy the requirements of the theorem: \citet{halvorsonbub:smolin} argue that Smolin's toy theory exhibits physical pathologies as it violates an analogue of the $C^{*}$-independence condition, and \citet{hans:2004} proves that Spekkens' toy-theory is not a $C^{*}$-algebraic theory. But if, from the point of view of the CBH characterization, what distinguishes Spekkens' theory, which satisfies the three information-theoretic constraints, from quantum mechanics, is the fact that it is not a $C^{*}$-algebraic theory, then this throws into stark relief the question of what the important physical, or information-theoretic, content of the initial $C^{*}$-algebraic assumption is.

We shall consider two further questions however. The first is discussed in some depth in \citet{hans:2004}, but it bears re-emphasising. It concerns the r\^ole that can be attributed to the no-bit commitment axiom when one starts in the $C^{*}$-algebraic setting.

\subsubsection{The r\^ole of no bit-commitment}

As we have noted, the intuitive r\^ole for the no bit-commitment axiom in the characterization theorem is to ensure that one arrives at theories which allow entanglement between separated systems. However, it is known \citep{landau:1987,bacciagaluppi:1994} that if the $C^{*}$-algebras $\mathcal{A}$ and $\mathcal{B}$ associated with two distinct (spatially separated) systems are kinematically independent and non-commutative, then it already follows automatically that there are entangled states for the joint system, in the $C^{*}$-algebraic framework. That is, if we assume no-signalling and no-broadcasting, then entanglement follows automatically, and a further axiom is not required. But this seems to indicate that the formal structure of $C^{*}$-algebras is not as neutral as one might suppose and is really doing a good deal of work in arriving at the distinctive quantum features we are seeking to derive.

This fact is already noted in \citet{cbh}. There the suggestion is made that the third axiom is required nonetheless, to ensure that the entangled states for spatially separated systems that arise are actually part of the physical state space, as opposed to being mere mathematical artefacts of the formalism. But this argument seems unconvincing. Whilst we are familiar with the idea that it may sometimes be necessary to place restrictions on the allowed states within a given state space (superselection rules and the like), the case we are now being asked to entertain is of a very different kind. It is not that we have a state space that we are restricting by adding a further clause---ruling certain states \textit{out}---rather, we have a particular state space postulated, and are being asked to consider having to rule certain states \textit{in} as physical. But ruling states \textit{in} rather than \textit{out} by axiom seems a funny game. Indeed, once we start thinking that some states may need to be ruled \textit{in} by axiom then where would it all end? Perhaps we would ultimately need a separate axiom to rule in \textit{every} state, and that can't be right. Thus the r\^ole that is supposed to be being played by the third axiom remains obscure.

One might try to re-phrase the argument so as not to appeal to the objectionable idea of the axiom being required to `rule states in'\footnote{Bub, personal communication.}. One might instead emphasize that the r\^{o}le of the no bit-commitment axiom is to rule out a certain class of theories---namely, Schr\"{o}dinger-type theories---that would still be on the table otherwise. But we should be clear in what sense the Schr\"{o}dinger-type theories are an option once one has postulated the first two information-theoretic axioms. We know that all ($C^{*}$-algebraic) theories consistent with the first two axioms allow entangled states between space-like separated systems, thus a Schr\"{o}dinger-type theory, which lacks such states, could only arise as the result of imposing \textit{further} restrictions on allowed theories that cut the entangled states out\footnote{N.B. a further option may be noted. It could be that the \textit{dynamics} is such as to lead to decay of entanglement on spatial separation---but to consider this possibility is, strictly speaking, to go beyond the remit of the CBH theorem which is intended to concern itself with the quantum mechanical \textit{kinematics}.}. Thus a Schr\"{o}dinger-type theory is only an option in the sense that we could arrive at such a theory by imposing further requirements to eliminate the entangled states that would otherwise occur naturally in the theory's state space. (Of course, such a theory would not be quantum mechanics, and in the light of the experimental violation of Bell inequalities, we know such a theory would not be empirically adequate, but that is by-the-by.)

Having postulated the first two axioms, the pertinent question to ask is whether the desired class of theories has then been delimited. The answer, given the $C^{*}$-algebraic setting, is indeed `yes'.
The fact that there may be other types of (perhaps rather gerrymandered) theory that could be reached by imposing further requirements of some sort would not seem to undermine this claim. We don't need to appeal to the no bit-commitment axiom to leave us only with quantum-type theories: all the theories before us (following the first two axioms) are of the desired type.

The no bit-commitment condition does not seem, then, to play a genuine r\^{o}le in characterizing quantum theory in a $C^{*}$-algebraic setting, but to figure more as a corollary: quantum theory may be characterized as a $C^{*}$-algebraic theory that abjures both superluminal signalling by measurement and broadcasting; having thus reached our desired class of theory, it transpires that this desired class will also be one for which secure bit-commitment is not possible. 
Note, though, that a scenario \textit{could} be imagined in which the no bit-commitment condition would play more of an active r\^{o}le. If, for some reason, we were unsure about whether a Schr\"{o}dinger-type theory or a quantum theory were the correct physical theory, then being informed by an oracle whether or not unconditionally secure bit-commitment was possible would be decisive: we would be saved the effort of having to go out into the world and perform Aspect experiments. But as this is not our position, the no bit-commitment axiom does not play an active r\^ole in picking out quantum theory.  

The position we have reached seems to be as follows. If one is attempting to provide a characterization of quantum mechanics in information-theoretic terms, it seems reasonable to desire an information-theoretic explanation of the existence of entanglement \citep{cbh,bub:why}. Starting from the $C^{*}$-algebraic setting of the CBH characterization theorem, however, entanglement just seems to spring automatically out of the mathematical machinery, when one would hope instead to be providing an information-theoretic explanation. We have seen, moreover, that the no bit-commitment condition is precluded from providing such an explanation in the context of the CBH theorem. How, then, might one proceed?

One option, as \citet[p.6]{bub:why} notes is to conjecture that in a weaker algebraic setting (e.g. Segal algebras) the existence of entangled states would \textit{not} follow from the first two information-theoretic axioms, but would require the imposition of the no bit-commitment axiom in addition. On the other hand, however, it is also conceivable that the intuitive argument outlined above linking no bit-commitment to the existence of entanglement might simply be misleading. Perhaps, in the end, it may turn out not to be possible to cash out the intuitive argument formally.

Another option, if one is after a proper information-theoretic explanation of the appearance of entanglement, would be to provide an information-theoretic reason for the initial choice of $C^{*}$-algebras as the mathematical framework. Then the fact that entanglement emerges naturally in the framework would not be worrying. However, in this case, it is not immediately obvious that one should expect such a reason to be based on the possibility of bit-commitment.     

\subsubsection{Additivity of expectation values}\label{additivity}

There is another reason why it may be suspected that adopting a $C^{*}$-algebraic approach is perhaps an overly restrictive starting point; why the framework may not be quite so neutral as it first appears. This concern centres on the nature of states in $C^{*}$-algebraic theories.

Ever since Bell's influential criticism of von Neumann's no-hidden variables theorem (\citet{bell:1966},\citet[pp.305-324]{vN}) it has been widely appreciated that it is an extremely strong assumption to adopt a requirement of additivity of expectation values for observable quantities. \textit{Vide} Bell:
\begin{quoting}
...the additivity of expectation values...is a quite peculiar property of quantum mechanical states, not to be expected \textit{a priori}. \citep[\S 3]{bell:1966}
\end{quoting} 
In particular, he goes on to note, when one is considering hidden variable theories:
\begin{quoting}
There is no reason to demand [expectation value additivity] individually of the hypothetical dispersion free states, whose function it is to reproduce the \textit{measureable} peculiarities of quantum mechanics \textit{when averaged over}. \citep[\S 3]{bell:1966}
\end{quoting}

These familiar observations are relevant to our concerns because the $C^{*}$-algebraic notion of state makes \textit{precisely this assumption}: states are \textit{linear} functionals of observables. In what follows I shall seek to elaborate this concern by adapting the methodology of Valentini.   

It is well known that in many ways, the Bohm theory is \textit{characteristic} of what a hidden variable theory for quantum mechanics must look like. We know, for example, that any acceptable hidden variable theory would have to be nonlocal and contextual; indeed it was the example of the Bohm theory that led Bell to pose the question of whether \textit{any} hidden variable theory replicating the predictions of quantum mechanics would have to be nonlocal. Now the Bohm theory reproduces the empirical predictions of quantum mechanics \textit{iff} the probability distribution $P$ for particle positions is given by $|\Psi|^{2}$. That $P$ equals $|\Psi|^{2}$ is an additional assumption in the standard Bohm theory; the (primary) r\^ole of the wavefunction as a guiding field is logically independent of its r\^ole in determining the distribution for particle position. \citet{bohm:1952} therefore explicitly countenanced the possibility that situations could arise in which $P$ would differ from $|\Psi|^{2}$ and thus empirical predictions would be expected that differ from those of quantum theory; in particular, violation of the position-momentum uncertainty principle becomes possible\footnote{`if the theory is generalized...The probability density of particles will cease to equal $|\Psi|^{2}$. Thus experiments would become conceivable that distinguish between $|\Psi|^{2}$ and this probability; and in this way we could obtain an experimental proof that the normal interpretation, which gives $|\Psi|^{2}$ \textit{only} a probability interpretation, must be inadequate.' \citep[I \S 9]{bohm:1952}}. However, he also went on to suggest that an argument could be given that the distribution $P$ can be expected to tend to $|\Psi|^{2}$ as a kind of equilibrium distribution.       

This thought was developed in detail by \citet{valentini:1991I} who showed that the relation $P=|\Psi|^{2}$ can indeed be derived as the `quantum equilibrium' distribution towards which systems will tend, as the result of a `subquantum $H$-theorem'. He also demonstrated that \textit{signal-locality} (the impossibility of superluminal signalling via measurement) and the uncertainty principle hold in general only in the equilibrium state, i.e., only if $P=|\Psi|^{2}$ \citep{valentini:1991II}. Thus the features of signal-locality and uncertainty can be understood to arise as effective features of an underlying nonlocal and deterministic theory, a pleasing result if one is exercised by the apparently conspiratorial fact that quantum mechanics (on many interpretations) gives rise to nonlocality, but only of a carefully restricted kind (`passion-at-a-distance'?) that does not permit signalling and hence avoids explicit conflict with relativity.   

More recently, \citet{valentini:signallocality} has shown that the r\^{o}le of the Bohm theory as a stereotype hidden variables theory extends further: it can be shown that for \textit{any} deterministic hidden variable theory, signal-locality will hold in general only in equilibrium. These facts are pertinent to our discussion of the axiomatic derivation of quantum mechanics from information-theoretic principles, as many of the principles appealed to will be, from the perspective of a deterministic hidden variable theory, merely contingent and accidental features of the equilibrium state. This factor leads \citet{valentini:sqinfodethv,valentini:sqinfo} to discuss the possibility of `sub-quantum' information processing that would be possible using non-equilibrium matter (perhaps matter left over from early stages of the life of the universe \citep{valentini:early}). 

In particular, out of equilibrium, instantaneous signalling would be possible, thus conflicting with the first of the three information-theoretic conditions of the CBH theorem; and it would also become possible to distinguish non-orthogonal states \citep[\S 5]{valentini:sqinfo}, leading to a violation of no-cloning and hence conflict with the no-broadcasting constraint of CBH.

Now we may adopt Valentini's framework of deterministic hidden variables theories which admit of an equilibrium distribution that ensures empirical agreement with standard quantum mechanics, along with non-equilibrium distributions that in general lead to violations of the quantum predictions, in order to elucidate the sense in which the assumption of linearity associated with the $C^{*}$-algebraic notion of state may be seen as problematic. In brief, the assumption of linearity, hence additivity of expectation values, rules out by \textit{fiat} the possibility of non-equilibrium deterministic hidden variables theories.
That is, one can show that additivity of expectation values can be expected to hold \textit{only in equilibrium} for such hidden variable theories.
Thus, by taking $C^{*}$-algebras as our theoretical starting point, we are immediately ruling out the possibility of deterministic hidden variables theories in the general case. But this is a big assumption. 

The relevant result is a straightforward generalisation of Bell's argument \textit{contra} von Neumann\footnote{See also \citet{valentini:additivity} for a closely related discussion.}. We will consider schematic hidden variables theories of the following sort. Assume (following \citet{bell:1966,bell:1982}) a function $f$ which determines the value of the outcome of an experiment measuring the quantum mechanical observable $A$, for an initial hidden variable $\lambda$ and quantum state $\ket{\psi}{}$. So $f$ is a function $f(\lambda, \ket{\psi}{}, A)$ whose range is the set of eigenvalues of $A$\footnote{Clearly, the mapping $f$ will in general also depend on the way in which the observable in question is measured (in order to avoid the sorts of problem made famous by Kochen-Specker). For example, in the Bohm theory, the mapping from the initial value of the hidden variable to determinate outcome depends on the measurement Hamiltonian. (Compare also \citet[p.6]{valentini:additivity}.)}. The expectation value of the observable $A$ will then be given in the usual way by averaging over the space $\Lambda$ of hidden variables:
\begin{equation}\label{expectation value}
\langle A\rangle=\int d\lambda P(\lambda)f(\lambda,\ket{\psi}{}, A),
\end{equation}
where $P(\lambda)$ is the probability distribution  for the hidden variables $\lambda$. (This distribution may also depend on the quantum state $\ket{\psi}{}$.) \textit{Ex hypothesi} there exists an equilibrium distribution $P_{\mathrm{eq}}(\lambda)$ for which eqn.~(\ref{expectation value}) will return the quantum expectation values.      

Now we know that the function $f$ will not be linear in the observable argument $A$, as the outcome of the measurement has to be one of the eigenvalues of the operator in question, and the eigenvalues of linearly related operators are not themselves linearly related \citep[cf.][]{bell:1966,bell:1982}. The requirement of additivity of expectation values is that
\[ \langle A + B\rangle= \langle A \rangle + \langle B \rangle;\]
in our deterministic hidden variable context this will become:
\begin{equation}\label{det hv additivity}
\int d\lambda P(\lambda) f(\lambda,\ket{\psi}{}, A+B) = \int d\lambda P(\lambda) \bigl[f(\lambda,\ket{\psi}{},A) + f(\lambda,\ket{\psi}{},B)\bigr].
\end{equation}
Now we know this equation holds for the equilibrium distribution $P_{\mathrm{eq}}(\lambda)$: it has to for empirical adequacy; but it can hold for arbitrary $P(\lambda)$ only if 
\[f(\lambda,\ket{\psi}{},A+B)=f(\lambda,\ket{\psi}{},A) + f(\lambda,\ket{\psi}{},B),\]
that is, only if $f$ is linear in the observable argument. But we know it isn't, hence expectation values won't be additive for general distributions for the hidden variables.

Thus we see that the assumptions involved in the $C^{*}$-algebraic notion of state are arguably overly strong when seeking to provide an axiomatic characterization or derivation of quantum mechanics. A large and potentially interesting class of theories is being ruled out by assumption. The requirement of expectation value additivity will not hold in general for a non-equilibrium deterministic hidden variable theory. Even if one is not particularly enamoured of hidden variables, this nonetheless serves as a vivid illustration of the fact that the assumption of states as linear functionals is a non-trivial one. 

Having presented this argument, however, it is important to note that there is a danger of a certain degree of failure of communication between a proponent of the argument and advocates of $C^{*}$-algebras as a comprehensive framework for describing physical theories. For, the latter will argue, there is surely no problem; the definite particle trajectories of the Bohm theory, for example, can happily be incorporated into the $C^{*}$-algebraic framework: the algebra of observables for the Bohm theory, in fact, will be the commutative algebra generated by the position observable \citep[cf.][pp.257-8]{bub:why}.
 
The source of the trouble is a possible equivocation over what is meant by `observables' by the two parties. In the argument that I have presented, `an observable' refers to a quantum mechanical observable; in concrete terms, to those quantities measured in the standard ways by quantum physicists in the lab. By contrast, when it is said that Bohm trajectories may be described in the terms of a $C^{*}$-algebraic theory, it is not \textit{these} observables which are the observables of the theory, hence my argument does not get a grip; but equally, the theory in question does not then (in general) assign values to the outcomes of the experiments we might expect to be interested in---those being performed by quantum physicists. 

If one is interested in a theory which assigns values to the outcomes of measurements that are performed by quantum physicists, i.e., to measurements of observables with the quantum structure (and such theories, of course, have a prominent history in discussion of the foundations of quantum mechanics), then the argument given above will apply; in the general case, expectation value additivity will not hold. Even if one is unmoved though and remains persuaded of the generality of the $C^{*}$-algebraic framework for all cases of interest, the argument described here remains important. It provides another example, to add to those already provided by Valentini, of where the assumptions involved in the CBH characterization theorem depend, from the point of view of a deterministic hidden variable theory, on a special feature of quantum equilibrium; that is, on contingent and accidental matters of fact that will not obtain in general.   

\subsection{Questions of Interpretation}

Perhaps the most intriguing question from the philosophical point of view is whether, or to what extent, the CBH characterization theorem has implications for the familiar interpretational questions of quantum mechanics. As we have noted, \citet{cbh} do seem to suggest that some implications of this nature are forthcoming. On reflection, however, this suggestion may appear somewhat surprising: the aim of their enterprise, after all, was to provide an axiomatic derivation of the mathematical structure of quantum theory; yet we are all too aware that this structure may be subject to interpretation in very many different ways (we saw, for example, an incomplete selection of views in Section~\ref{interpretations}).
One would think that to provide an axiomatic characterization of a particular mathematical structure is to do just that and no more. Surely, when faced with the same old structure before us once again, the standard range of interpretations will be as applicable as ever?

\citet{cbh} suggest, though, that their theorem intimates that quantum mechanics may be seen as a \textit{principle theory} and it is in this sense that an interpretation is provided. \citet{bub:why} adopts a rather different tack. I shall maintain against these arguments that the rather negative line of assessment just mooted regarding the interpretational implications of the CBH theorem is nevertheless on track.    

\subsubsection{Quantum mechanics as a principle theory?}

The distinction between principle and constructive theories is familiar from Einstein's discussions of his 1905 methodology in arriving at the correct form of relativistic kinematics\footnote{His most detailed presentation of the distinction is to be found in \citet{einstein:times}. See \citet{brownpooley:2001,brownpooley:2004} for recent discussions of the principle/constructive distinction in relativity; in particular for their emphasis that---as recognised by Einstein---principle theories lose out to constructive theories in terms of explanatory power. As they note \citep{brownpooley:2001}, while the distinction between principle and constructive theories is not absolute, it is nonetheless enlightening.}. The paradigm example of a principle theory is thermodynamics, which is to be contrasted with a \textit{constructive} theory such as the kinetic theory of gases. While constructive theories seek to `build up a picture of the more complex phenomena out of the materials of a relatively simple formal scheme from which they start out' \citep{einstein:times}, principle theories proceed from the basis of some well grounded phenomenological principles that are found to govern a class of physical processes of interest (e.g., the non-existence of perpetual motion machines of the first and second kind, in the case of thermodynamics), in order to derive constraints that all instances of such processes have to satisfy.  

As recounted in his \textit{Autobiographical Notes} \citep[p.49ff.]{einsteinauto} Einstein turned to the methodological example of thermodynamics as a \textit{faute de mieux}, given the confused state of knowledge in electrodynamics and mechanics at the turn of the 20th Century:
\begin{quoting}
Gradually I despaired of the possibility of discovering the true laws [of electrodynamics and mechanics] by means of constructive efforts based on the known facts. The longer and more desperately I tried, the more I came to the conviction that only the discovery of a universal formal principle could lead us to assured results. The example I saw before me was thermodynamics. \citep[p.49]{einsteinauto}
\end{quoting}
The Principle of Relativity and the Light Postulate became, of course, the principles that Einstein fixed upon; and these allowed him to derive the correct form of the co-ordinate transformations between inertial frames\footnote{Note, however, that it would be a mistake to construe special relativity \textit{purely} as a principle theory. Einstein was later to refer to the `sin' of treating rods and clocks as unanalysed bodies, as opposed to `moving atomic configurations' \citep[pp.55-7]{einsteinauto}; see also \citet[p.14]{pauli:relativity} in this regard. This point is elaborated in detail in \citet{brown:1993,brownpooley:2001,brownpooley:2004}.}.  

Now \citet{cbh} suggest that their theorem shows that quantum mechanics may be understood as a principle theory---where the relevant principles are information-theoretic---and that in this sense an interpretation of quantum mechanics is provided. One has arrived at a description of the conditions (\textit{viz.}, the obtaining of the three information-theoretic constraints) under which quantum theory will be true. To illuminate this sense of interpretation, they present an illustrative fable in which one imagines that relativity had originally been formulated geometrically by Minkowski as an algorithm for relativistic kinematics, and then Einstein came along and provided an interpretation of this algorithm by presenting his principle theory derivation of the Lorentz transformations. Similarly, the analogy goes, we have quantum mechanics as an algorithm for predicting the results of various experiments; and this algorithm now finds an interpretation in terms of the three information-theoretic constraints. We now understand how the world is organised so that quantum theory has to be true (or so the claim).

However, it may be doubted whether this approach provides us with a particularly interesting sense of `interpretation'. To pursue the analogy with relativity: Einstein showed us \textit{why} the co-ordinate transformations between inertial frames had to be the Lorentz transformations---if they were not then one or more of the principles (or the symmetry assumptions) in his derivation would have to be false. 
But this explanation, or interpretation, remains silent on a very important point. The fact that the Lorentz transformations are the correct transformations between inertial frames encodes a great deal of detail about the \textit{dynamical behaviour} of (ideal) rods and clocks---these are, after all, complex material bodies. Arguably, the fact that the speed of light, say, is measured to be the same in all inertial frames is ultimately to be explained in terms of the dynamical behaviour of rods and clocks---a constructive style of explanation \citep[cf.][]{brown:1993,brownpooley:2001,brownpooley:2004}. In any event, it is clear that if appeal to the principles of relativity is providing an \textit{interpretation} of the formulae of relativistic kinematics, it is an interpretation that glosses over a lot: there is a good deal more to be said about the conditions under which the Lorentz transformations constitute the correct transformations between inertial frames.  


Analogously, in the case of quantum mechanics, given the three information-theoretic constraints, the CBH theorem provides us with an explanation of why the states and observables in our theory have to take their characteristic quantum structure: if they did not, at least one of the assumptions would be false. But nothing is said about how the world should be understood if states and observables take on this form. 

By assumption, the world is such that the information-theoretic constraints are true, but this is too general and it says too little: it is consistent with a wide range of ways of understanding the quantum formalism.

To elaborate: If one were to adopt the proposal under discussion, that quantum mechanics should be seen as a principle theory, then the objects of the theory whose behaviour the principles constrain are preparation devices and measuring apparatuses, considered as unanalysed black boxes. (Recall the association of states with preparation devices and observables with measuring apparatuses in the $C^{*}$-algebraic setting, discussed earlier). From the information-theoretic principles, the general sorts of relations that should obtain between various preparations and measurements (and sequences of measurements) are derived. These principles are thought to provide an explanation (in some form) of why preparation devices and measuring apparatuses display the relations---in terms of observed relative frequencies of various experimental outcomes---that they do. 
Note that in saying this we are supposing what might be called a basic level of interpretation of our theory: we have related elements of the formalism (states, observables) with physical quantities (the statistical frequencies with which various outcomes of experiments may be expected). The main difficulty for the principle theory approach, construed as providing a putative interpretation of quantum theory, is that it doesn't involve anything going beyond this most elementary level of interpretation. 

However, typically when one is in concerned with the interpretation of a theory, and in particular, with the interpretation of \textit{quantum} theory, one is interested in the further question of how these reports posed in terms of experimental results are to be understood. Are they merely reports of brute regularities, for example---an instrumentalist view---or is something more realistic appropriate? Do measurements reveal pre-existing values, or contextually determined outcomes, or are they to be understood in some other way? And so on. This is the traditional battle-ground of interpretive questions in quantum theory; and \textit{something} needs to be said at this level, even if it is the bare claim that there is no more to be said (instrumentalism)\footnote{This recalls the earlier discussion of Bohr's semantic ascent (Section~\ref{ascent}): ascent notwithstanding, something still had to be said about about how claims made using the terms of the theory are to be understood.}. But the principle theory approach, as it only engages with the statistical relations between preparation devices and measuring apparatuses, says nothing.  

Of course, various different approaches might be taken to specifying what is involved in the interpretation of a theory. The route I have adopted here is close to that of \citet{redheadbook}. \citet[Chpt. 2]{redheadbook} distinguishes two senses of interpretation of a theory. To provide an interpretation in the first sense is to supply rules which correlate elements of the mathematics of a theory with physical quantities. In this bracket, for example, is what he terms the \textit{minimal instrumentalist interpretation} of quantum mechanics: the familiar rules that tell us what the possible results of measurements are in quantum mechanics and how the statistical frequencies may be calculated with which these measurement results will turn up when a measurement is repeated very many times on systems prepared in the same way.

An interpretation in the \textit{second} sense, he says, is:
\begin{quoting}
...some account of the nature of the external world and/or our epistemological relation to it that serves to \textit{explain} how it is that the statistical regularities predicted by the formalism with the minimal instrumentalist interpretation come out the way they do. \citep[p.44]{redheadbook}
\end{quoting} 
He goes on to note that we might simply accept the statistical regularities as brute facts, which is to take the instrumentalist view (theories in physics just \textit{are} instruments for expressing regularities between observations); but this is certainly to take a stance on interpretation in sense two\footnote{`Indeed we shall often refer to the formalism of QM plus the minimal instrumentalist interpretation \textit{in the first sense} as the minimal instrumentalist interpretation \textit{in the second sense}.' \citep[p.44]{redheadbook} Redhead's minimal instrumentalist interpretation (sense 2) is what I earlier termed a statistical interpretation.}. 

Now, the sense of `interpretation' associated with the principle theory approach is this: an argument is given for why we have one theory (which is already interpreted in Redhead's sense 1) rather than another; why the states and observables take one form, rather than another. But to repeat, this doesn't tell us anything about how the theory thus chosen should be interpreted in sense 2. It is only a minimal instrumentalist interpretation (in sense 1) linking the formalism to empirical predictions that is ever involved. In the thin sense in which an interpretation might be forthcoming from the principle theory approach, it is not a sense of interpretation that engages with the traditional problems of the meaning of the quantum formalism: with the question of how this familiar formalism is to be understood. Since the result of the CBH theorem is to recover the standard structure of quantum theory, the usual ranges of interpretive options will be open to us; and indeed one of these options must be taken, even if one adopts the principle theory viewpoint as advocated by \citet{cbh}. 
Thus, far from the CBH theorem motivating a principle theory viewpoint (`an information-theoretic perspective') that ameliorates the conceptual puzzles of quantum mechanics, we see that it simply fails to engage with these questions.

\subsubsection{Bub's 2004 argument: a problem of underdetermination}

More recently, \citet{bub:why} has adopted a rather different line of attack. He argues that in light of the CBH theorem we are not in fact free to adopt the full range of (sense 2) interpretations of the quantum formalism. Assuming that the information-theoretic constraints are satisfied in our world, he insists, no mechanical theory of quantum phenomena that includes an account of measurement interactions can be acceptable. Such accounts will face, in his view, a problem of in-principle underdetermination which renders them unacceptable:
\begin{quoting}
...a mechanical theory that purports to solve the measurement problem is not acceptable if it can be shown that, \textit{in principle}, the theory can have no excess empirical content over a quantum theory. \citep[p.261]{bub:why}
\end{quoting}

We need to examine how this problem of underdetermination is thought to arise, but first it will be useful to have a rough statement of how the different styles of interpretation one might be interested in are to be divided up.
For the purposes of this discussion, then, let us distinguish between those interpretations (in sense 2) that involve adding extra structure to the bare formalism to ensure a definite measurement outcome (this group would include the Bohm theory, hidden variables theories and the sorts of modal interpretation picked out by the Bub-Clifton uniqueness theorem \citep[Chpt. 4]{bub:1997}); those interpretations that appeal to a non-unitary dynamics (i.e., dynamical collapse theories \textit{ \`{a} la} GRW); and those that stick as closely as possible to the bare quantum formalism (e.g., instrumentalist views and modern versions of the Everett interpretation\footnote{Bub in fact appears to lump the Everett interpretation in with `extra structure' interpretations. While this may be appropriate for some attempts to cash out Everett's ideas, it is not for the more satisfactory (for this very reason!) modern versions of Everett, as formulated by Saunders, Wallace and company (see refs. in Section~\ref{interpretations}). This point is important for the conclusions that can be drawn from Bub's argument, as we will see below.}). 

It is the first group, Bub suggests, that will suffer from in-principle underdetermination, in light of the CBH theorem; while GRW approaches may conflict with the exact obtaining of the no bit-commitment axiom and are to be ruled out on that ground (spontaneous collapse might interfere with some efforts to cheat in bit-commitment \citep[p.256]{bub:why}). Let us now see how the underdetermination argument is supposed to run.

It is essential to recognise that the argument has two components. The first is the claim that follows from the CBH theorem, that if the information-theoretic constraints are satisfied in this world, then the empirical results we obtain will be those modelled by a quantum theory in the sense of \citet{cbh} (i.e., a theory with a non-commuting algebra of observables for individual systems, kinematic independence for distinct systems and entangled states across space-like separated systems). The second part of the argument is the assertion that the information theoretic constraints do hold in our world, both \textit{exactly} (with no exceptions) and as a \textit{matter of law}.

Now consider an `extra structure' interpretation, such as the Bohm theory. Bub views this as an extension of a quantum theory that seeks to describe the mechanics underlying the statistics of a $C^{*}$-algebraic quantum theory. However, if the information-theoretic constraints are to hold, then the empirical predictions of the Bohm theory, or any other such extension (`extra structure interpretation') must be just the same as the quantum theory. But now, if the information-theoretic constraints are both law-like and hold exactly, then in \textit{any} physically possible world, the empirical predictions of such an extension will be just the same as those of the bare quantum theory. In other words, it is physically impossible that there could be any evidence that would favour one such extension over another: there is in-principle underdetermination. Accordingly, the claim is, we should reject all such extensions\footnote{Bub emphasizes that the epistemological principle at work here is not the---implausible---claim that it is never rational to adopt one theory over an empirically equivalent rival, but the far weaker claim that if there could never, in any physically possible world, be evidence favouring one theory over another, then it would not be rational to believe either.}. It is for this reason that extra structure interpretations are not acceptable, for Bub.    

This argument fails, however. It has no dialectical power against extra-structure interpretations as it involves a \textit{petitio principii}. The crucial assumption, that the information-theoretic constraints are both law-like and hold exactly, is denied in the extra structure interpretations (at least in the case of the Bohm theory and hidden variables theories). We have seen how, in the case of the Bohm theory and deterministic hidden variables theories, the information-theoretic constraints and even the assumption of expectation value additivity hold, if they hold at all, merely as contingent and accidental (non law-like) matters of fact. From the point of view of these theories, the constraints certainly don't hold in all physically possible worlds, and they might not even hold under all conditions in \textit{this} world. Similarly, the argument against the GRW-type theories is also a \textit{petitio}; while the information-theoretic constraints are law-like in this case, they don't always hold exactly: there may sometimes be a violation of no bit-commitment. But one does not provide an argument against a position by simply insisting on an assumption that is inconsistent with it.

In all this, it is important to recognise that we only have reason to believe that the information-theoretic conditions obtain in the quantum context as they are consequences of the standard quantum formalism. The empirical evidence we have for them derives second-hand from the empirical evidence for quantum theory. The evidence for quantum theory doesn't settle the question of how the formalism is to be interpreted (if it did one wouldn't need to try to detour via the CBH theorem!), so the empirical evidence we have is consistent with various different views on what the status of the information-theoretic conditions should be. From the point of view of an `extra structure' interpretation such as the Bohm theory, they will, as we have said, be seen as contingent and accidental features that obtain in some conditions; from points of view that stick closely to the quantum formalism (instrumentalism, Everett), they will be understood as lawlike and exact. But if the status of the information-theoretic constraints is explicitly an interpretation dependent question, we may not appeal to an argument that essentially involves a controversial assumption about their status, in order to rule out certain forms of interpretation.         

Towards the end of his 2004 paper, Bub remarks that if one has succeeded in ruling out dynamical collapse theories and those interpretations that involve extra structure then 
\begin{quoting}
\textit{It follows that our measuring instruments ultimately remain black boxes at some level} that we represent in the theory simply as probabilistic sources of ranges of labelled events[...]i.e., effectively as sources of signals...\citep[p.261]{bub:why} {\small original emphasis}
\end{quoting}
Furthermore, he suggests:
\begin{quoting}
...this amounts to treating a quantum theory as \textit{a theory about the representation and manipulation of information}...[A] consequence of rejecting Bohm-type hidden variable theories or other `no collapse' theories is that we recognise information as a new sort of physical entity.... \citep[p.262]{bub:why}
\end{quoting} 

Regarding the first point, it is pertinent to note that if one accepts my broad three-way carving up of the different interpretational options, then even if one has somehow managed to rule out the first two sets of possibilities (extra structure and dynamical collapse), then this still leaves us with at least two options in the third category, that is, with some form of instrumentalism, or an Everettian approach. Now while instrumentalism may well be appropriately described in the terms Bub uses---measuring apparatuses that must remain as unanalysed black boxes---this characterization is by no means apt for the Everett interpretation. Here measurement is perfectly well analysable, as one particular sort of dynamical interaction amongst many, set within a realist view of the universal quantum state.    

On the second point, even if one has placed Everettian views to one side, 
it remains obscure in what sense quantum theory would have become a theory about the representation and manipulation of information,
if this is supposed to be more than a new way of describing an old instrumentalist view.
There is a simple difficulty, for instance, with trying to cash this idea out by suggesting that a measuring apparatus can be seen as a source of signals. If one has a signal, then it is intelligible to ask what the signals signify, or indicate (whether naturally or as a matter of convention). But what is a particular measurement outcome a signal of? It would seem that the only thing that \textit{could} be signified would be something about pre-existing hidden variables; and this, presumably, is not what is desired at all\footnote{It is a quite different matter, of course, to consider a measuring apparatus as an information source in the sense of information theory, for then one is considering compressing and transmitting the output of the source, while the physical constitution of the source itself is wholly irrelevant (but for this very reason, one will not find any implications for quantum ontology here). From the point of view of information theory, the outputs of an information source signify nothing and have no meaning, conventional or otherwise, but no more, then, are they, strictly speaking, signals. They are elements which have no semantic, nor even syntactic significance. This is just to repeat the familiar line that `information' in the technical sense is not a semantic notion. If something \textit{is} a source of signals then one might well be interested in applying communication theory to it and modelling it as an information source in the sense of that theory. But you don't \textit{make} something a source of signals by considering it as an information source.}. As for the inference to information as a new sort of physical entity, it was, of course, a large part of the trajectory of argument in Part 1 to make such a conception appear implausible, even downright mistaken. In combative mood, one might insist that to give an otherwise instrumentalist view of quantum mechanics a subject matter does not seem a sufficient reason to conclude that information, or quantum information, is an entity.

\end{doublespacing}

\chapter*{Envoi}
\markright{ENVOI}
\begin{doublespacing}

It is clear that there is a good deal more to be said in the tale of the significance of quantum information theory for the meaning of the quantum formalism; a tale which, to a considerable degree, is still being written.

In particular, considerations of time and space have precluded any discussion of what is perhaps the most radical proposal advanced so far: the quantum Bayesianism of Caves, Fuchs and Schack \citep{fuchs:paulian,cfs:definetti,fuchs:only,fuchs:compatibility,fs:unknown04}. Let me, then, by way of a send-off, provide the briefest sketch of this position, locating it with respect to the discussion of the previous two chapters. (I concentrate on the position as advocated by Fuchs.)   

The quantum Bayesian approach is characterized by its non-realist view of the quantum state: the quantum state ascribed to an individual system is understood to represent a compact summary of an agent's degrees of belief about what the results of measurement interventions on a system will be. The probability ascriptions arising from a particular state are understood in a purely subjective, Bayesian manner. Then, just as with a subjective Bayesian view of probability there is no right or wrong about what the probability of an event is, with the quantum Bayesian view of the state, there is no right or wrong about what the quantum state assigned to a system is\footnote{The fact that scientists in the lab tend to agree about what states should be assigned to systems is then explained by providing a subjective `surrogate' for objectivity, along the lines that de Finetti provided for subjective probability: an explanation why different agents' degrees of beliefs may be expected to come into alignment given enough data, in suitable circumstances \citep{cfs:definetti}.}.  
The approach thus figures as the terminus of the tradition which has sought to tie the quantum state to cognitive states, but now the cognitive state invoked is belief, not knowledge, and, crucially, the problems raised by factivity are thus avoided.

Importantly, however, this non-realist view of the quantum state is not the \textit{end point} of the proposal, but merely its \textit{starting point}. (The aim, then is for more than a new formulation of instrumentalism.) 
The hope expressed is that when the correct view is taken of certain elements of the quantum formalism (\textit{viz.} quantum states and operations) it will be possible to `see through' the quantum formalism to the real ontological lessons it is trying to teach us\footnote{`[O]ne...might say of quantum theory, that in those cases where it is not just Bayesian probability theory full stop, it is a theory of stimulation and response \citep{fuchs:what,fuchs:paulian}. The agent, through the process of quantum measurement stimulates the world external to himself. The world, in return, stimulates a response in the agent that is quantified by a change in his beliefs---i.e., by a change from a prior to a posterior quantum state. Somewhere in the structure of those belief changes lies quantum theory's most direct statement about what we believe of the world as it is without agents.' \citep{fs:unknown04}}. Given the point of departure of a Bayesian view of the state, and using techniques from quantum information, the aim is to winnow the objective elements of quantum theory (reflecting physical facts about the world) from the subjective (to do with our reasoning). Ultimately, the hope is to show that the mathematical structure of quantum mechanics is largely forced on us, by demonstrating that it represents the only, or, perhaps, simply the most natural, framework in which intersubjective agreement and empirical success can be achieved given the basic fact (much emphasized in the Copenhagen tradition) that in the quantum domain, it seems that the ideal of a detached observer may not be obtained.

One of the main attractions of this approach, therefore, is that it aims to fill-in an important lacuna associated with many views in the Copenhagen tradition: It is all very well, after all, adopting some non-realist view of the quantum formalism, but, one may ask, why is it that our best theory of the very small takes such a form that it needs to be interpreted in this manner? Why are we forced to a theory that does not have a straightforward realist interpretation? Why is this the best we can do? The programme of Caves, Fuchs and Schack sets out its stall to make progress with these questions, hoping to arrive at some simple physical statements which capture what it is about that world that forces us to a theory with the structure of quantum mechanics. 
Although the aim is to seek a transparent conceptual basis for quantum mechanics, there is no claim that the theory should be understood as a principle theory. In further contrast to the CBH approach, rather than seeking to provide an axiomatisation of the quantum formalism which might be interpreted in various ways, the idea instead is to take one particular interpretive stance and see whether this leads us to a perspicuous axiomatisation. 

Now Fuchs' direct arguments for the non-objective view of the quantum state are not, we may note, logically compelling \citep[e.g.][\S 3]{fuchs:only}; they are plausibility arguments based on the oddity of nonlocality in the EPR scenario; and those of a more realist bent might simply accept the nonlocality associated with collapse or hidden variables, or move to a realist view such as Everett that avoids the problem. But this is no real objection to the approach. The quantum Bayesian view is presented as a research programme: when this view of the quantum state and the quantum formalism is adopted, where does it take us? The proof of the pudding, ultimately, will be in the eating. Meanwhile, the approach is to be applauded for providing a consistent way, perhaps the \textit{only} consistent way, of fruitfully developing the old line of thought that links the quantum state to information. But, finally, it might turn out to be that in the end, taking the Bayesian route does cause us to give up too much of what one needs as objective in quantum theory. These matters deserve further discussion.    

\end{doublespacing}

\backmatter

\end{document}